\documentclass[a4paper,11pt]{article}
\pdfoutput=1 

\usepackage{jheppub} 

\usepackage{amscd}
\usepackage{amsfonts}
\usepackage{amsmath}
\usepackage{amssymb}
\usepackage{amsthm}
\usepackage{accents}
\usepackage{youngtab}
\usepackage{braket}
\usepackage{titlesec}
\usepackage{cancel}
\usepackage[svgnames]{xcolor}
\usepackage{ytableau}
\usepackage{here}
\usepackage{tikz}
 \usetikzlibrary{patterns,arrows,calc,tikzmark}
\usepackage{feynmf}
\usepackage{mathtools}
\usepackage{pb-diagram}
\usepackage{simplewick}
\usepackage[labelformat=parens,subrefformat=parens]{subcaption}
\usepackage{comment}
\usepackage{ascmac} 
\def\Abox{\tikz[scale=0.007cm] \draw (0,0) rectangle (1,1);}
\DeclareMathOperator*{\Res}{Res}

\newcommand{\ba}{\begin{eqnarray}}
\newcommand{\ea}{\end{eqnarray}}

\titleclass{\subsubsubsection}{straight}[\subsection]

\newcounter{subsubsubsection}[subsubsection]
\renewcommand\thesubsubsubsection{\thesubsubsection.\arabic{subsubsubsection}}

\titleformat{\subsubsubsection}
  {\normalfont\normalsize\bfseries}{\thesubsubsubsection}{1em}{}
\titlespacing*{\subsubsubsection}
{0pt}{3.25ex plus 1ex minus .2ex}{1.5ex plus .2ex}

\makeatletter
\renewcommand\paragraph{\@startsection{paragraph}{5}{\z@}%
  {3.25ex \@plus1ex \@minus.2ex}%
  {-1em}%
  {\normalfont\normalsize\bfseries}}
\renewcommand\subparagraph{\@startsection{subparagraph}{6}{\parindent}%
  {3.25ex \@plus1ex \@minus .2ex}%
  {-1em}%
  {\normalfont\normalsize\bfseries}}
\def\toclevel@subsubsubsection{4}
\def\toclevel@paragraph{5}
\def\toclevel@paragraph{6}
\def\l@subsubsubsection{\@dottedtocline{4}{7em}{4em}}
\def\l@paragraph{\@dottedtocline{5}{10em}{5em}}
\def\l@subparagraph{\@dottedtocline{6}{14em}{6em}}
\makeatother

\makeatletter
\@addtoreset{equation}{subsection}
\makeatother

\graphicspath{{./figures/}}

\title{Shifted Quiver Quantum Toroidal Algebra and Subcrystal Representations}

\author[a]{Go Noshita,}
\author[a]{Akimi Watanabe}

\affiliation[a]{Department of Physics, The University of Tokyo,\\ 7-3-1 Hongo, Bunkyo-ku, Tokyo 113-0033, Japan}

\emailAdd{noshita@hep-th.phys.s.u-tokyo.ac.jp}
\emailAdd{awatanabe@hep-th.phys.s.u-tokyo.ac.jp}

\abstract{Recently, new classes of infinite-dimensional algebras, quiver Yangian (QY) and shifted QY, were introduced, and they act on BPS states for non-compact toric Calabi-Yau threefolds. In particular,  shifted QY acts on general subcrystals of the original BPS crystal.  A trigonometric deformation called quiver quantum toroidal algebra (QQTA) was also proposed and shown to act on the same BPS crystal. Unlike QY, QQTA has a formal Hopf superalgebra structure which is useful in deriving representations. 

In this paper, we define the shifted QQTA and study a class of their representations. We define  1d and 2d subcrystals of the original 3d crystal by removing a few arrows from the original quiver diagram and show how the shifted QQTA acts on them. We construct the 2d crystal representations from the 1d crystal representations by utilizing a generalized coproduct acting on different shifted QQTAs. We provide a detailed derivation of subcrystal representations of $\mathbb{C}^{3}$, $\mathbb{C}^{3}/\mathbb{Z}_{n}(n\geq 2)$, conifold, suspended pinch point, and $\mathbb{C}^{3}/(\mathbb{Z}_{2}\times\mathbb{Z}_{2})$.}

\begin{document} 
\maketitle

\section{Introduction and summary}   
Infinite-dimensional alebra has been one of the most powerful tools to study supersymmetric gauge theories \cite{Alday2010,Nakajima_Heisenberg,maulik2018quantum,schiffmann2012cherednik,Tsymbaliuk:2014}. In particular, quantum toroidal algebras \cite{feigin2011quantum,Feigin2011,Feigin_2012,Ding:1996mq,Miki2007,Feigin:2015raa,feigin2013representations,Feigin_2019glmn,bezerra2019quantum,bezerra2021representations} and their truncations \cite{bershtein2018plane, Shiraishi:1995rp,Feigin:1995sf,Awata:1995zk,Awata:1996dx, FHSSY:2010, Miki2007, Kojima2019, Kojima2021, Harada:2021xnm} have played significant roles in the context of 5d AGT correspondence \cite{Awata_2010,awata2010five,Yanagida_2010,awata2011notes,Awata_2016,Awata_2017,Awata:2016mxc,Awata_2018} and topological vertex \cite{Iqbal_2009,Aganagic:2003db,Bourgine:2017jsi,Bourgine_2020,Bourgine_2019,Bourgine_2016,Bourgine_2017,Bourgine:2021nyw,Bourgine:2015szm,Bourgine:2021yba,Bourgine:2021gnb,Awata_2016,Awata_2017RTT,Awata_2017,Awata_2018,Awata:2011ce,zenkevich2019mathfrakgln,zenkevich2020mixed,zenkevich2021higgsed,Cheewaphutthisakun:2021cud,Mironov:2016yue,Ghoneim:2020sqi} (see also \cite{Gaiotto:2017euk,Prochazka:2015deb,Prochazka:2017qum,Prochazka:2018tlo} for the degenerate case). 

Recently, infinite-dimensional algebras associated with toric Calabi-Yau threefolds called quiver Yangian \cite{Li:2020rij,Galakhov:2020vyb} was introduced (see also \cite{Rapcak_2019,rapcak2020cohomological} for similar directions). As a generalization, a shifted version of quiver Yangian was defined \cite{galakhov2021shifted}. Quiver Yangian acts on three-dimensional BPS crystals \cite{Ooguri_2009}, while shifted quiver Yangian acts on the subcrystal of the mother BPS crystal. A trigonometric deformation of quiver Yangian, which we call quiver quantum toroidal algebra (QQTA), was introduced in our previous paper \cite{Noshita:2021ldl}. QQTA acts on the same three-dimensional BPS crystal as the quiver Yangian. Thus, it is natural to consider its subcrystals and the action of the algebra. 

The goal of this paper is to study a shifted version of QQTA in (\ref{eq:defofshiftedQuiverAlgebra}) and two classes of their subcrystal representations. We will call them ``one-dimensional" and ``two-dimensional" crystal representations.\footnote{The names ``one-dimensional" and ``two-dimensional" do not mean they are finite-dimensional representations. They come from the crystal picture of the bases of the module. Following the case of quantum toroidal $\mathfrak{gl}_1$, these can be called vector and Fock representations, respectively.} In deriving these representations, the concept of brane tiling \cite{Franco:2005rj} will play a significant role. Let us give motivations and summarize what is done in this paper. The three-dimensional BPS crystal representation is a generalization of the MacMahon representation \cite{Feigin_2012} of quantum toroidal $\mathfrak{gl}_{1}$ \cite{Ding:1996mq, Miki2007, feigin2011quantum}. Quantum toroidal $\mathfrak{gl}_{1}$ has other representations such as vector representations and Fock representations \cite{feigin2011quantum}, and their crystal pictures are one-dimensional and two-dimensional. An interesting feature of these representations is that Fock representations are obtained by taking tensor products of vector representations, while MacMahon representations are obtained by taking tensor products of Fock representations. From this point of view, other quantum toroidal algebras are expected to have one-dimensional and two-dimensional crystal representations. Another expectation is that higher-dimensional crystal representations should be obtained from tensor products of lower-dimensional crystal representations.

\begin{figure}
    \centering
    \includegraphics[width=12cm]{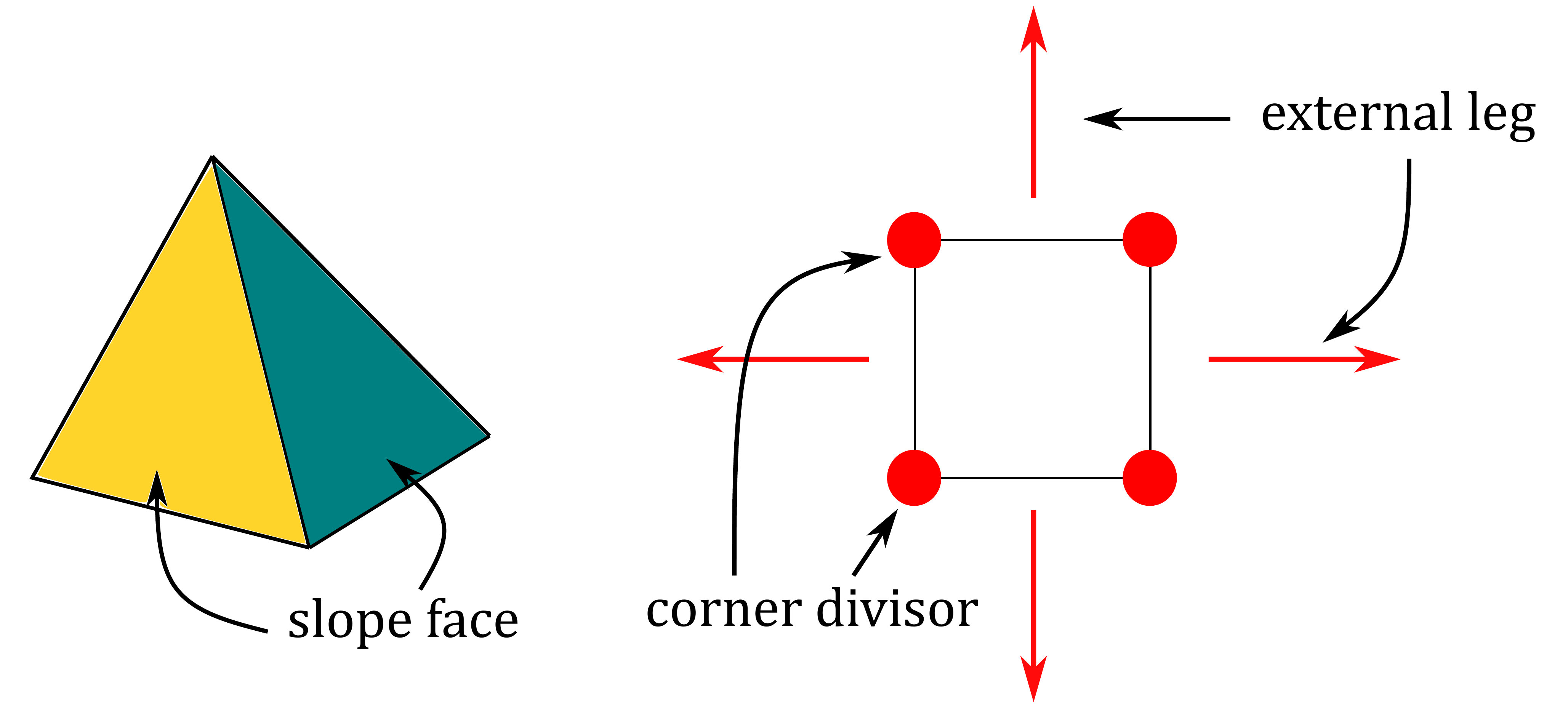}
    \caption{Three-dimensional crystal (pyramid partition) and toric diagram of the conifold geometry. Each of the slope face corresponds to the corner divisor of the toric diagram. External legs of the toric diagram are lines dual to the edges of the toric diagram.}
    \label{fig:Introduction_slopeface}
\end{figure}
The two-dimensional crystal \cite{Nishinaka_2011, Nishinaka_2012, Nishinaka_2014} we focus on is a reduction of the three-dimensional crystal of \cite{Ooguri_2009}. The three-dimensional crystal \cite{Ooguri_2009} physically corresponds to D2-D0 states bound to a D6-brane on a toric Calabi-Yau threefold, while the two-dimensional crystal corresponds to D2-D0 states bound to a D4-brane wrapping a divisor of the threefold. This two-dimensional crystal model reproduces the BPS index of the D4-D2-D0 states, and it is a ``slope face" of the original three-dimensional crystal (see Figure \ref{fig:Introduction_slopeface}). The two-dimensional crystals are associated with corner divisors of the toric diagram, and the shapes are determined by the external legs of the dual web diagram surrounding them. The quiver diagram for this crystal is obtained by removing a few arrows from the original quiver diagram. The removed arrows correspond to the unique perfect matching \cite{Hanany_2007} of the corner divisor. Although how to obtain the crystal shape is already known, it is unknown what kind of algebra acts on it, so we provide an answer to it. 

\begin{figure}
    \centering
    \includegraphics[width=10cm]{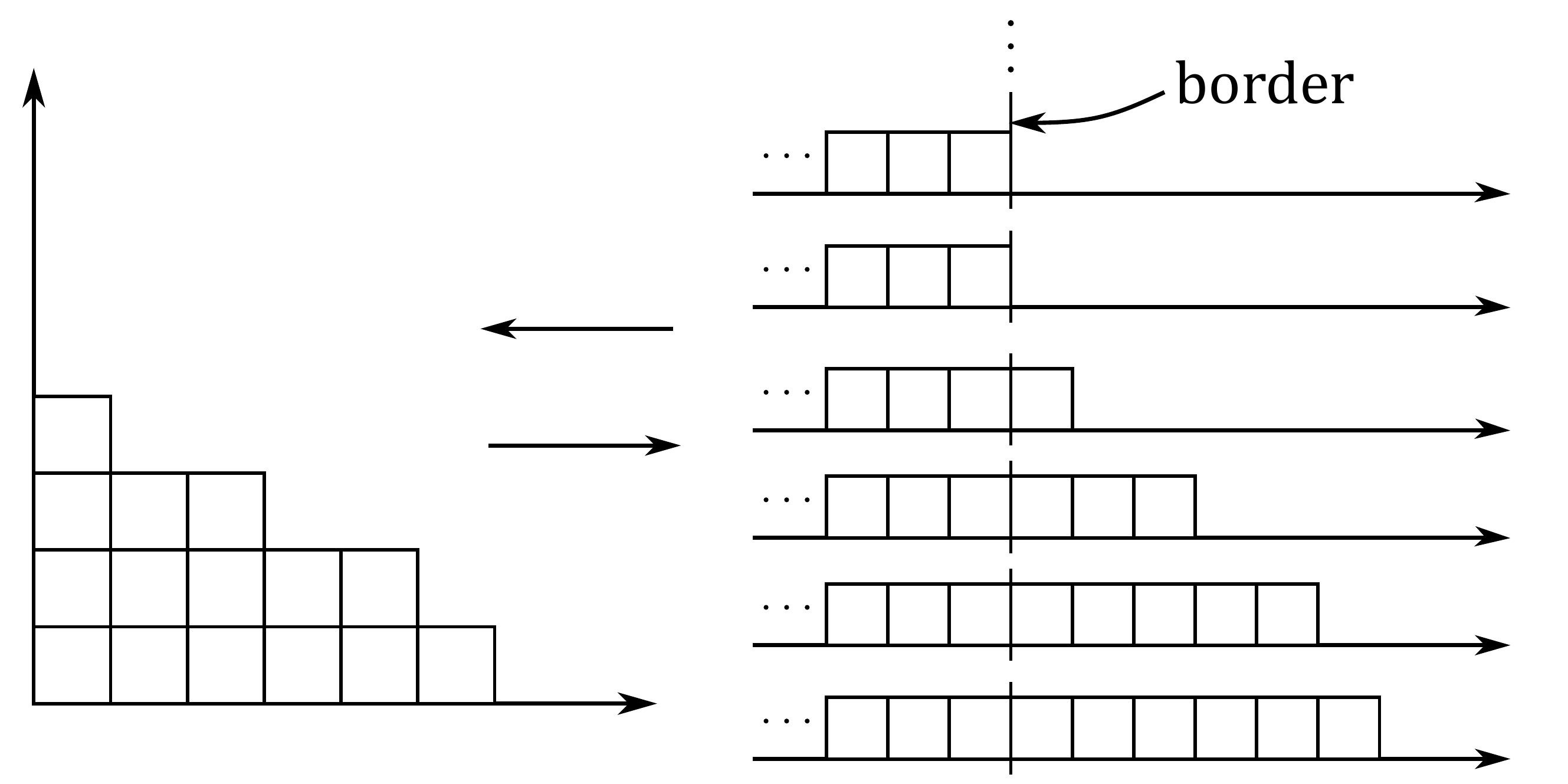}
    \caption{Two-dimensional crystal and one-dimensional crystal for $\mathbb{C}^{3}$-geometry. Each layer of the two-dimensional crystal can be understood as a one-dimensional crystal by extending it left to the border. By taking tensor products of these one-dimensional crystal representations, we obtain the two-dimensional crystal representation. }
    \label{fig:Introduction2d_1d}
\end{figure}

We further generalize the discussions on two-dimensional crystals and construct a one-dimensional crystal. Its shape is a reduced and extended version of the two-dimensional crystal.  We can associate it with an external leg of the toric dual web diagram. The crystal picture is obtained by projecting the two-dimensional crystal to one of the axes and extending it left to the origin periodically. The basis of this representation can be understood as a semi-infinite row of atoms (see Figure \ref{fig:Introduction2d_1d} for the $\mathbb{C}^{3}$ case). In deriving the shape of the one-dimensional crystal from the toric diagram, we will see they have connections with perfect matchings. Although the physical meaning of this crystal is not clear, we expect they have relations to vortex moduli space. We leave it for future work.

Since the one-dimensional and two-dimensional crystals above are subcrystals of the original three-dimensional crystal, one might think they should be representations of the quiver quantum toroidal algebra. However, we will see this is not true and that we have to use ``shifted quiver quantum toroidal algebra" in (\ref{eq:defofshiftedQuiverAlgebra}).\footnote{See \cite{negut2021agt} and references therein for the definition of shifted quantum toroidal $\mathfrak{gl}_{n}$.}
We will see that this is because the vacuum charge function does not have the same number of poles and zeros anymore.  

The shifted quiver quantum toroidal algebra we define has a structure similar to the Hopf superalgebra structure of QQTA \cite{Noshita:2021ldl}. It has a generalized coproduct, counit, and antipode. However, this time the coproduct and antipode will be maps between \textit{different} shifted quantum toroidal algebras. In constructing subcrystal representations explicitly, we start from the one-dimensional crystal. Using the shape of this crystal, we can derive the action of the algebra. Then, utilizing the generalized coproduct and taking tensor products of this one-dimensional crystal representation, we derive the two-dimensional crystal representation (see Figure \ref{fig:Introduction2d_1d} for the $\mathbb{C}^{3}$-geometry case).  

This paper is organized as follows. In section \ref{sec:QQTA_review}, we review the definition of the quiver quantum toroidal algebra defined in \cite{Noshita:2021ldl}. In section \ref{sec:quantum_toroidalgl1}, we review the properties and representations of the quantum toroidal $\mathfrak{gl}_{1}$. The way the representations are derived will be essential in the later sections. We will introduce special subcrystals (one-dimensional and two-dimensional) of the original three-dimensional crystal in section \ref{sec:subcrystal_rep}. We will also introduce shifted QQTA and study its generalized coproduct, counit, antipode structure. We also derive the crystal representations of the conifold. In section \ref{sec:examples}, we will utilize the generalized coproduct structure and derive two-dimensional crystal representations from one-dimensional crystal representations. Subcrystal representations of $\mathbb{C}^{3}$, $\mathbb{C}^{3}/\mathbb{Z}_{n}$ $(n\geq 2)$, suspended pinch point, and $\mathbb{C}^{3}/(\mathbb{Z}_{2}\times\mathbb{Z}_{2})$ are derived in detail. In section \ref{sec:summary}, we give some discussions for future work. Supplementary materials are in the appendices.

\paragraph{Note added.} 
While preparing this paper, new preprints \cite{galakhov2021shifted,Galakhov:2021vbo} appeared in the arXiv.
In section 2 of \cite{Galakhov:2021vbo}, they defined the trigonometric deformation of quiver Yangian, which is the same as what we call shifted QQTA in our paper. While they also discuss the representations of crystals, we investigate in detail the relation between 1d and 2d crystal representations.

\section{Review: Quiver quantum toroidal algebra}\label{sec:QQTA_review}
Let us review the quiver quantum toroidal algebra defined in \cite{Noshita:2021ldl}.  The algebra is defined from the quiver diagram, which is derived from the toric Calabi-Yau geometry. 

We denote a quiver diagram as $Q=(Q_{0},Q_{1},Q_{2})$, where $Q_{0}$ is a set of vertices, $Q_{1}$ is a set of arrows between vertices, and $Q_{2}$ is the set of closed loops surrounded by $Q_{1}$ (see section 2 and Appendix B in \cite{Noshita:2021ldl}). We use
\begin{align}
    Q_{0}=\{i\}_{i\in Q_{0}},\quad Q_{1}=\{I\}_{I\in Q_{1}}, \quad Q_{2}=\{L\}_{L\in Q_{2}},
\end{align}
namely $i,j,..$ are used to label vertices and $I,J,..$ are used to label arrows of the quiver diagram. The number of vertices and arrows are denoted by $|Q_{0}|$ and $|Q_{1}|$ respectively.

There are two types of quivers, which can be derived from the geometry: ``symmetric" quiver and ``asymmetric" quiver. The symmetric quiver is a quiver with the condition,
\begin{align}
    |i\rightarrow j|=|j\rightarrow i|,\quad \forall i,j\in Q_{0},
\end{align}
while the asymmetric quiver is a quiver that does not satisfy the above condition. $|i\rightarrow j|$ is the number of arrows from vertex $i$ to vertex $j$. The symmetric quiver corresponds to toric Calabi-Yau threefolds without compact 4-cycles. As \cite{Noshita:2021ldl}, we will focus on the symmetric quiver in this whole paper. Similar discussions are expected to be true for the asymmetric quiver.

A parameter $q_{I}$ is associated with each arrow of the quiver diagram. We impose the following conditions:
\begin{screen}
\begin{align}
&\text{loop constraint}:\quad \prod_{I\in L}q_{I}=1,\quad L\in Q_{2}\label{eq:loop_constraint}\\
&\text{vertex constraint}:\quad \prod_{I\in i}q_{I}^{\text{sign}_{i}I}=1,\quad i\in Q_{0}\label{eq:vertex_constraint}
\end{align}
\end{screen}
where
\begin{displaymath}
    \text{sign}_{i}(I)\equiv
    \begin{dcases}
    +1\quad(s(I)=i,\quad t(I)\neq i),\\
    -1\quad(s(I)\neq i,\quad t(I)= i),\\
    0\quad(\text{otherwise})
    \end{dcases}
\end{displaymath}
and $\prod_{I\in i}$ means the product of all arrows going in and out of vertex $i$.
Here and below, $s(I)$ and $t(I)$ denote the starting and ending vertex of the arrow $I$, respectively.
After imposing loop and vertex constraints above, we get only two independent parameters. 

The quiver quantum toroidal algebra  $\ddot{\mathcal{U}}_{Q}$ is generated by $E_{i,k},F_{i,k},K^{\pm}_{i,\pm r}$, and invertible elements $K^{\pm}_{i,0},\;C$, where $i\in Q_{0}$, $k\in\mathbb{Z}$ and $r\in\mathbb{Z}^{\times}$.\\
The defining relations are given in terms of generation series 
\begin{align}
    E_{i}(z)=\sum_{k\in\mathbb{Z}}E_{i,k}z^{-k},\quad F_{i}(z)=\sum_{k\in\mathbb{Z}}F_{i,k}z^{-k},\quad K_{i}^{\pm}(z)=\sum_{r\geq 0}K^{\pm}_{i,\pm r}z^{\mp r}.\label{eq:QQTAgenerator}
\end{align}
Generators have $\mathbb{Z}_{2}$ grading, which are denoted by $|E_{i,k}|=|F_{i,k}|=|i|$. The grading rule is 
\begin{align}
    |i|=\begin{cases}
    0\quad (\exists I\in Q_{1}\quad \text{such that}\quad s(I)=t(I)=i),\\ 
    1\quad (\text{otherwise}),
    \end{cases}\label{eq:statistics}
\end{align}
where the operators are bosonic when $|i|=0$ and fermionic when $|i|=1$. In other words, the operator is bosonic when there is a loop, otherwise it is fermionic. We note the generators $K_{i}^{\pm}(z)$ are set to be bosonic.

The defining relations are given as follows:
\begin{screen}
\begin{align}
\begin{split}
    K^{+}_{i,0}K_{i,0}^{-}&=K_{i,0}^{-}K_{i,0}^{+}=1,\\
    C^{-1}C&=CC^{-1}=1,\\
    K_{i}^{\pm}(z)K_{j}^{\pm}(w)&=K_{j}^{\pm}(w)K_{i}^{\pm}(z),\\
    K_{i}^{-}(z)K_{j}^{+}(w)&=\frac{\varphi^{j\Rightarrow i}(z,Cw)}{\varphi^{j\Rightarrow i}(Cz,w)}K_{j}^{+}(w)K_{i}^{-}(z),\\
    K_{i}^{\pm}(C^{\frac{1\mp1}{2}}z)E_{j}(w)&=\varphi^{j\Rightarrow i}(z,w)E_{j}(w)K_{i}^{\pm}(C^{\frac{1\mp1}{2}}z),\\
    K_{i}^{\pm}(C^{\frac{1\pm1}{2}}z)F_{j}(w)&=\varphi^{j\Rightarrow i}(z,w)^{-1}F_{j}(w)K_{i}^{\pm}(C^{\frac{1\pm1}{2}}z),\\
    [E_{i}(z),F_{j}(w)]=\delta_{i,j}&\left(\delta\left(\frac{Cw}{z}\right)K_{i}^{+}(z)-\delta\left(\frac{Cz}{w}\right)K_{i}^{-}(w)\right),\\
    E_{i}(z)E_{j}(w)&=(-1)^{|i||j|}\varphi^{j\Rightarrow i}(z,w)E_{j}(w)E_{i}(z),\\
    F_{i}(z)F_{j}(w)&=(-1)^{|i||j|}\varphi^{j\Rightarrow i}(z,w)^{-1}F_{j}(w)F_{i}(z),
\end{split}\label{eq:defofQuiverAlgebra}
\end{align}
\end{screen}
The commutator above must be understood in the usual superalgebra sense.

The function $\varphi^{i\Rightarrow j}(z,w)$ is defined to be 
\begin{align}
    \varphi^{i\Rightarrow j}(z,w)=\frac{\prod_{I\in\{j\rightarrow i\}}(q_{I}^{1/2}z-q_{I}^{-1/2}w)}{\prod_{I\in\{i\rightarrow j\}}(q_{I}^{-1/2}z-q_{I}^{1/2}w)}=\frac{\prod_{I\in\{j\rightarrow i\}}\phi(q_{I};z,w)}{\prod_{I\in\{i\rightarrow j\}}\phi(q_{I}^{-1};z,w)},\label{eq:defstruc}
\end{align}
where $\{i\rightarrow j\}$ are the arrows from vertex $i$ to $j$ and 
\begin{align}
    \phi(p;z,w)=p^{1/2}z-p^{-1/2}w.
\end{align} See Appendix \ref{sec:appendix_notation} for the convention we use. We call (\ref{eq:defstruc}) ``bond factors" following the original terminology in \cite{Li:2020rij}. When there are no arrows between the two vertices the bond factor is trivial:
\begin{equation}
    \varphi^{i\Rightarrow j}(z,w)=1.
\end{equation}

\paragraph{Hopf superalgebra structure}
The algebra (\ref{eq:defofQuiverAlgebra}) has a Hopf superalgebra structure. Recall that a Hopf algebra is equipped with a unit, a counit, a product, a coproduct, and an antipode following appropriate properties (see section 4.4 in \cite{Noshita:2021ldl}).

The coproduct formula is
\begin{align}
\begin{split}
&\Delta: \ddot{\mathcal{U}}_{Q}\rightarrow\ddot{\mathcal{U}}_{Q}\otimes \ddot{\mathcal{U}}_{Q}, \\
    &\Delta E_{i}(z)=E_{i}(z)\otimes 1+K_{i}^{-}(C_{1}z)\otimes E_{i}(C_{1}z),\\
    &\Delta F_{i}(z)=F_{i}(C_{2}z)\otimes K_{i}^{+}(C_{2}z)+1\otimes F_{i}(z),\\
    &\Delta K_{i}^{+}(z)=K_{i}^{+}(z)\otimes K_{i}^{+}(C_{1}^{-1}z),\\
    &\Delta K_{i}^{-}(z)=K_{i}^{-}(C_{2}^{-1}z)\otimes K_{i}^{-}(z),\\
    &\Delta C= C\otimes C.
    \end{split}\label{eq:coproduct}
    \end{align}
The counit formula is 
\begin{align}
\begin{split}
&\epsilon:\ddot{\mathcal{U}}_{Q}\rightarrow \mathbb{C},\\
&\epsilon(E_{i}(z))=\epsilon(F_{i}(z))=0,\\
&\epsilon(K_{i}^{\pm}(z))=\epsilon(C)=1.
\end{split}\label{eq:counit}
\end{align}
The antipode formula is 
\begin{align}
\begin{split}
&S:\ddot{\mathcal{U}}_{Q}\rightarrow \ddot{\mathcal{U}}_{Q},\\
&S(E_{i}(z))=-(K_{i}^{-}(z))^{-1}E_{i}(C^{-1}z),\\
&S(F_{i}(z))=-F_{i}(C^{-1}z)(K_{i}^{+}(z))^{-1},\\
&S(K_{i}^{\pm}(z))=(K_{i}^{\pm}(Cz))^{-1},\\
&S(C)=C^{-1}.
\end{split}\label{eq:antipode}
\end{align}

\paragraph{Three-dimensional crystal representation}
After setting $C=1$, we obtain one natural representation on the three-dimensional BPS crystal of \cite{Ooguri_2009}. The representation is 
\begin{align}
\begin{split}
&K_{i}^{\pm}(z)\ket{\Lambda}=\left[\Psi_{\Lambda}^{(i)}(z,u)\right]_{\pm}\ket{\Lambda},\\
&E_{i}(z)\ket{\Lambda}=\sum_{\fbox{$i$}\in\text{Add}(\Lambda)}\pm\sqrt{p^{(i)}\underset{x=uq(\fbox{$i$})}{\Res}\Psi_{\Lambda}^{(i)}(x,u)}\delta\left(\frac{z}{uq(\fbox{$i$})}\right)\ket{\Lambda+\fbox{$i$}},\\
&F_{i}(z)\ket{\Lambda}=\sum_{\fbox{$i$}\in\text{Rem}(\Lambda)}\pm\sqrt{q^{(i)}\underset{y=uq(\fbox{$i$})}{\Res}\Psi_{\Lambda}^{(i)}(y,u)}\delta\left(\frac{z}{uq(\fbox{$i$})}\right)\ket{\Lambda-\fbox{$i$}},
\end{split}\label{eq:summaryofalgebraansatz}
\end{align}
where $q^{(i)}=1$, $p^{(i)}=\varphi^{i\Rightarrow i}(1,1)$, and $\left[f(z)\right]_{\pm}$ means formal expansion of $f(z)$ in $z^{\mp k},k\geq0$.
The $\pm$ signs in the action of $E_i(z)$ and $F_i(z)$ are determined by the crystal shapes (see \cite{Li:2020rij}). We will see some examples in Section \ref{sec:examples}.
Note that the residue here is different from the original definition (see (\ref{eq:residue}) and \cite{Noshita:2021ldl}). We also note
\begin{align}
    q(\fbox{$i$})\equiv\prod_{I\in\text{path}[\mathfrak{o}\rightarrow \fbox{$i$}]}q_{I},
\end{align}
namely the coordinate function for $\hbox{$i$}$ is the product of all charges along the path from the origin $\mathfrak{o}$. $\Lambda$ is the crystal configuration satisfying the melting rule.
The melting rule specifies the possible places where we can add or remove atoms from the crystal, and the loop constraint (\ref{eq:loop_constraint}) determines it. For more information on the melting rule, see \cite{Ooguri_2009} and \cite{Li:2020rij}.
Sets $\text{Add}(\Lambda)$ and $\text{Rem}(\Lambda)$ denote the sets of atoms that can be added to the crystal configuration $\Lambda$, respectively. The charge function $\Psi_{\Lambda}^{(i)}(z,u)$ and the vacuum charge function $\psi_{\emptyset}^{(i)}(z,u)$ are defined as 
\begin{align}
 \begin{split}
&\Psi_{\Lambda}^{(i)}(z,u)=\psi_{\emptyset}^{(i)}(z,u)\prod_{j\in Q_{0}}\prod_{\fbox{$j$}\in \Lambda}\varphi^{j\Rightarrow i}(z,uq(\fbox{$j$})),\\
&\psi_{\emptyset}^{(i)}(z,u)=(\psi_{\emptyset}(z,u))^{\delta_{i,a}},\\
& \psi_{\emptyset}(z,u)=\frac{K^{-1/2}z-K^{1/2}u}{z-u},\quad K\in \mathbb{C},
\end{split}\label{eq:charge_function}
\end{align}
where $a$ is the color of the atom in the origin.

\section{Review: Quantum toroidal \texorpdfstring{$\mathfrak{gl}_1$}{gl1}}\label{sec:quantum_toroidalgl1}
Let us review the simplest but important example of quiver quantum toroidal algebra, quantum toroidal $\mathfrak{gl}_{1}$ \cite{feigin2011quantum,Feigin2011,Feigin_2012,Ding:1996mq,Miki2007,Feigin:2015raa}. It is the quantum toroidal algebra associated with $\mathbb{C}^{3}$. It has two central elements $C,K^{-}$, and specific values of them give various representations of the algebra. We will focus mainly on representations when the central charge is $C=1$. These are called vertical representations in the literature.  On the other hand, representations with $C\neq 1$ are called horizontal representations. We will not discuss about these representations, so see \cite{bershtein2018plane, Shiraishi:1995rp,Feigin:1995sf,Awata:1995zk,Awata:1996dx, FHSSY:2010, Miki2007, Kojima2019, Kojima2021, Harada:2021xnm} for details. See also \cite{Awata2019} for a good review.
\subsection{Definition}
\begin{figure}[H]
    \centering
    \includegraphics[width=10cm]{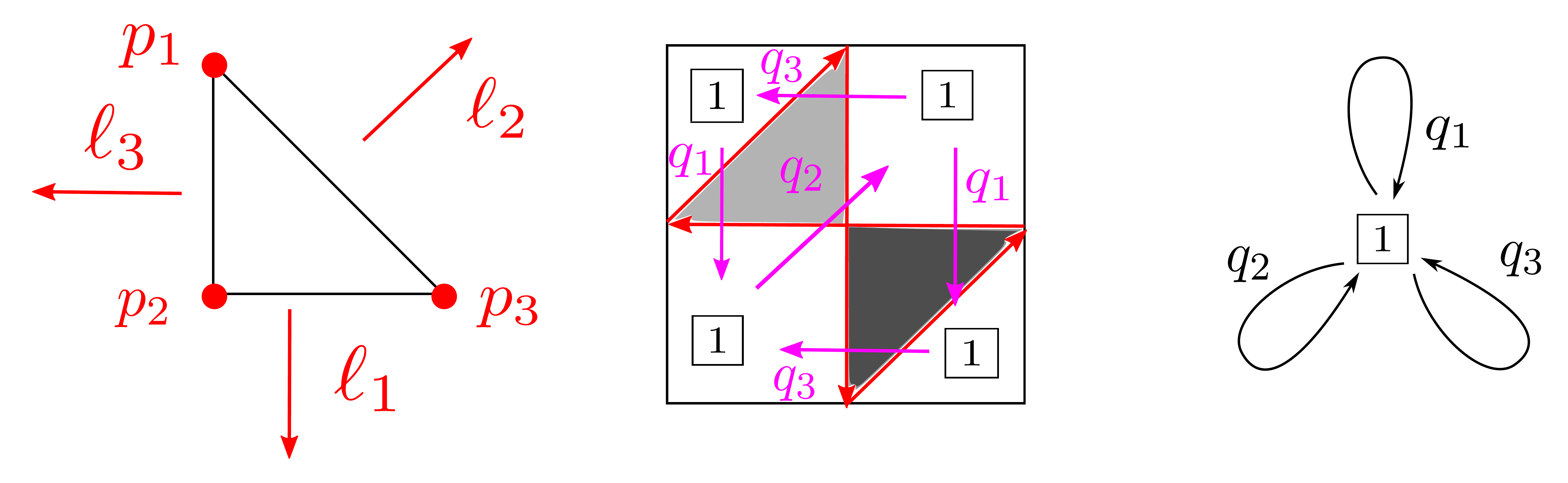}
    \caption{Toric diagram, periodic quiver, and quiver diagram of $\mathbb{C}^{3}$. The three lattice points of the toric diagram are denoted $p_{1}=(0,1)$, $p_{2}=(0,0)$, and $p_{3}=(1,0)$. All of them are corner divisors. The three external legs are denoted $\ell_{1}$, $\ell_{2}$, and $\ell_{3}$.}
    \label{fig:gl1_toricdiagram}
\end{figure}
The quiver diagram and periodic quiver diagram are in Figure \ref{fig:gl1_toricdiagram}.
First of all, we only have one vertex and three loops and thus the operators are all bosonic.
The loop constraint (\ref{eq:loop_constraint}) and vertex constraint (\ref{eq:vertex_constraint}) gives $q_{1}q_{2}q_{3}=1.$ 
The bond factor is
\begin{align}
    \varphi^{1\Rightarrow 1}(z,w)=\frac{\prod_{i=1}^{3}(q_{i}^{1/2}z-q_{i}^{-1/2}w)}{\prod_{i=1}^{3}(q_{i}^{-1/2}z-q_{i}^{1/2}w)}=\frac{\prod_{i=1}^{3}\phi(q_{i};z,w)}{\prod_{i=1}^{3}\phi(q_{i}^{-1};z,w)}.
\end{align}
The generators and defining relations of the algebra are as follows:
\begin{align}
\begin{split}
 E(z)=\sum_{k\in\mathbb{Z}}E_{k}z^{-k},\quad F(z)=\sum_{k\in\mathbb{Z}}&F_{k}z^{-k},\quad K^{\pm}(z)=K^{\pm}\exp\left(\pm\sum_{r=1}^{\infty}H_{\pm r}z^{\mp r}\right),\\
    K^{+}K^{-}&=K^{-}K^{+}=1,\\
    K^{\pm}(z)K^{\pm}(w)&=K^{\pm}(w)K^{\pm}(z),\\
    K^{-}(z)K^{+}(w)&=K^{+}(w)K^{-}(z),\\
    K^{\pm}(z)E(w)&=\varphi^{1\Rightarrow 1}(z,w)E(w)K^{\pm}(z),\\
    K^{\pm}(z)F(w)&=\varphi^{1\Rightarrow 1}(z,w)^{-1}F(w)K^{\pm}(z),\\
    [E(z),F(w)]=&\left(\delta\left(\frac{w}{z}\right)K^{+}(z)-\delta\left(\frac{z}{w}\right)K^{-}(w)\right),\\
    E(z)E(w)&=\varphi^{1\Rightarrow 1}(z,w)E(w)E(z),\\
    F(z)F(w)&=\varphi^{1\Rightarrow 1}(z,w)^{-1}F(w)F(z),
    \end{split}\label{eq:gl1definingrelation}
\end{align}
where we set $C=1$ in (\ref{eq:defofQuiverAlgebra}). Obviously, this algebra is symmetric in the permutation of $q_{1},q_{2},q_{3}$, which is called ``triality" in the literature. As one can see from (\ref{eq:gl1definingrelation}), $K(z)$ commutes with each other, and thus we can use simultaneous eigenstates of $K(z)$ as bases to construct representations. We note that $K^{-}=(K^{+})^{-1}$ is another central element and the value of it gives various representations.

We also note that the coproduct structure is written as 
\begin{align}
\begin{split}
    &\Delta E(z)=E(z)\otimes 1+K^{-}(z)\otimes E(z),\\
    &\Delta F(z)=F(z)\otimes K^{+}(z)+1\otimes F(z),\\
    &\Delta K^{\pm}(z)=K^{\pm}(z)\otimes K^{\pm}(z),\\
    &\Delta C= C\otimes C.
\end{split}\label{eq:gl1_coproduct}
\end{align}

\subsection{Representations}
Representations of quantum toroidal $\mathfrak{gl}_{1}$ are illustrated with 1d, 2d, or 3d Young diagrams. First, we construct the most basic representation called vector representation from 1d Young diagrams. Then, we use the coproduct structure of the algebra to construct Fock representation, which is characterized by 2d Young diagrams. 
\paragraph{Representation by 1d Young diagrams}
We start with the vector representation, whose basis corresponds to a 1d Young diagram. 

The central elements of this representation are $C=1, K^-=1$. We denote the basis as $[u]_j\; (j\in \mathbb{Z})$.
Corresponding to the triality of the quantum toroidal $\mathfrak{gl}_1$, we have three vector representations.
Here we construct the representation with $q_1$, and the other two can be obtained by replacing $q_i$ cyclically.

The actions of $K^\pm(z), E(z), F(z)$ on the basis are defined as follows, which satisfies the relations of the algebra \cite{feigin2011quantum}.\footnote{As explained below (\ref{eq:summaryofalgebraansatz}), $\left[f(z)\right]_\pm$ means formal expansion of $f(z)$ in $z^{\mp k}$, $k\geq 0$.}
\begin{align}
\begin{split}
    K^{\pm}(z)[u]_{j}&=\left[\Psi_{[u]_{j}}(z)\right]_{\pm}[u]_{j},\\
    E(z) [u]_j &= \mathcal{E} \delta(q_1^{j+1}u/z) [u]_{j+1},\\
    F(z) [u]_{j+1} &= \mathcal{F} \delta(q_1^{j+1}u/z) [u]_j,
    \end{split}\label{eq:gl1_vectorrep}
\end{align}
where $\Psi_{[u]_{j}}(z)$ is\footnote{$\Psi_{[u]_{j}}(z)$ has poles at $u q_1^j$ and $u q_1^{j+1}$ and zeros at $u q_1^j q_2^{-1}$ and $u q_1^j q_3^{-1}$. These zeros cancel possible new poles. Actually, $\varphi^{1\Rightarrow 1}(z, u q_1^{j+1})$ has poles at $u q_1^{j+2}$, $u q_1^{j+1} q_2 = u q_1^j q_3^{-1}$, and $u q_1^{j+1} q_3 = u q_1^j q_2^{-1}$ by the constraints $q_1q_2q_3=1$, and two of them are cancelled.}
\begin{equation}
    \Psi_{[u]_{j}}(z)=\frac{\phi(q_{2}q_{1}^{-j};z,u)\phi(q_{3}q_{1}^{-j};z,u)}{\phi(q_{1}^{-j};z,u)\phi(q_{1}^{-1}q_{1}^{-j};z,u)}.
\end{equation}
Actually, this can be written in infinite products of the bond factor:
\begin{equation}
    \Psi_{[u]_{j}}(z)=\prod_{l\leq j}\varphi^{1\Rightarrow 1}(z,uq_{1}^{l}).\label{eq:infiniteproductgl11d}
\end{equation}
$\mathcal{E}$ and $\mathcal{F}$ satisfy the following relation:
\begin{equation}
    \mathcal{E} \mathcal{F}=\frac{\phi(q_{2};1,1)\phi(q_{3};1,1)}{\phi(q_{1}^{-1};1,1)}.
\end{equation}
The exact value of $\mathcal{E}$ and $\mathcal{F}$ is not important, so we will not be careful of it from now on. 

\begin{figure}[h]
    \centering
    \includegraphics[width=9cm]{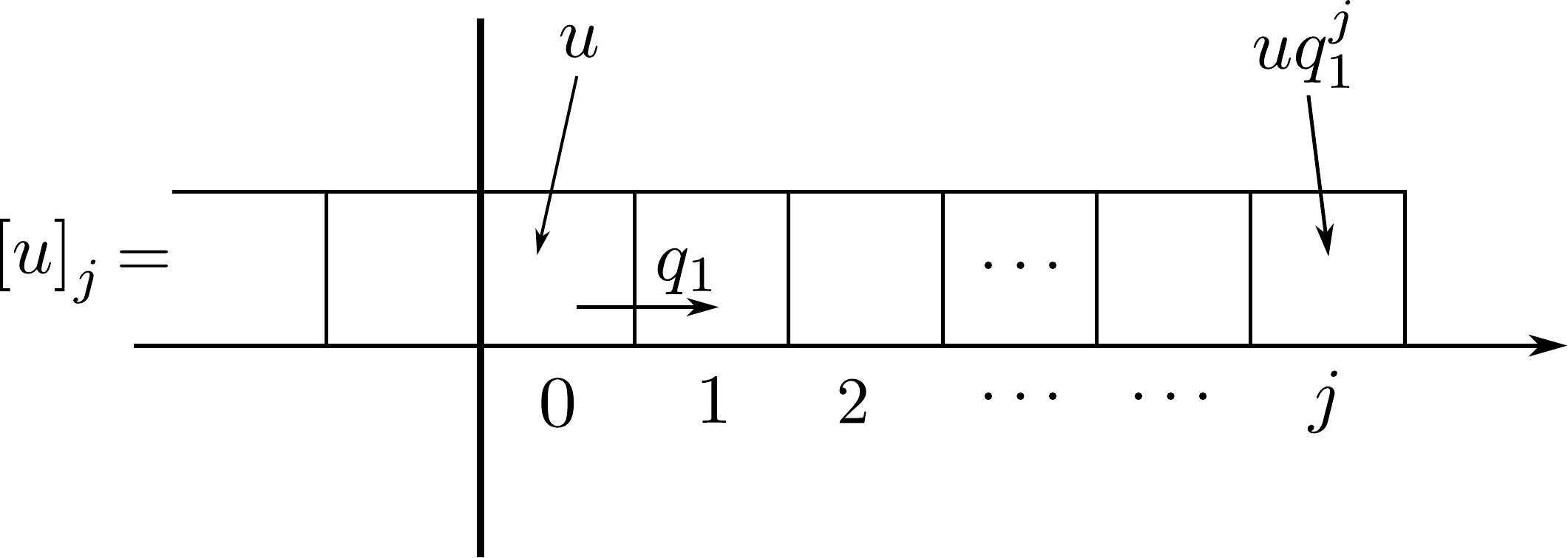}
    \caption{Interpretation of vector representation basis as a 1d-sequence of boxes. We call it the 1d Young diagram. $[u]_{j}$ has $j+1$ boxes right to the border. We note the boxes are labeled with numbers starting from $0,1,$ at the right of the border and extended left to the border.}
    \label{fig:vec_rep}
\end{figure}

We can interpret $[u]_j$ as a one-dimensional sequence of boxes from $-\infty$ to $j$, as shown in Figure \ref{fig:vec_rep}.
Since we are considering the simultaneous eigenstates of $K^\pm(z)$, the actions of $K^+(z)$ and $K^-(z)$ do not change the number of boxes.
On the other hand, $E(z)$ plays the role of the creation operator, which increases the number of boxes by one, and $F(z)$ plays the role of the annihilation operator, which decreases the number of boxes by one.
\paragraph{Representation by 2d Young diagrams}
In the vector representation, we arrange boxes in one direction, and the number of boxes right to the border labels each state.
We can construct a representation called the Fock representation by combining multiple bases of this vector representation $[u]_j$ using tensor products. \\
Similar to the vector representation, we have three Fock representations due to triality. Their central charges are $C=1$, $K^{-}=q_{i}^{1/2}$ $(i=1,2,3)$. The bases of these representations are illustrated as two-dimensional Young diagrams. These representations are obtained by taking tensor products of the vector representations using the coproduct (\ref{eq:gl1_coproduct}). 

\begin{figure}[H]
    \centering
    \includegraphics[width=8cm]{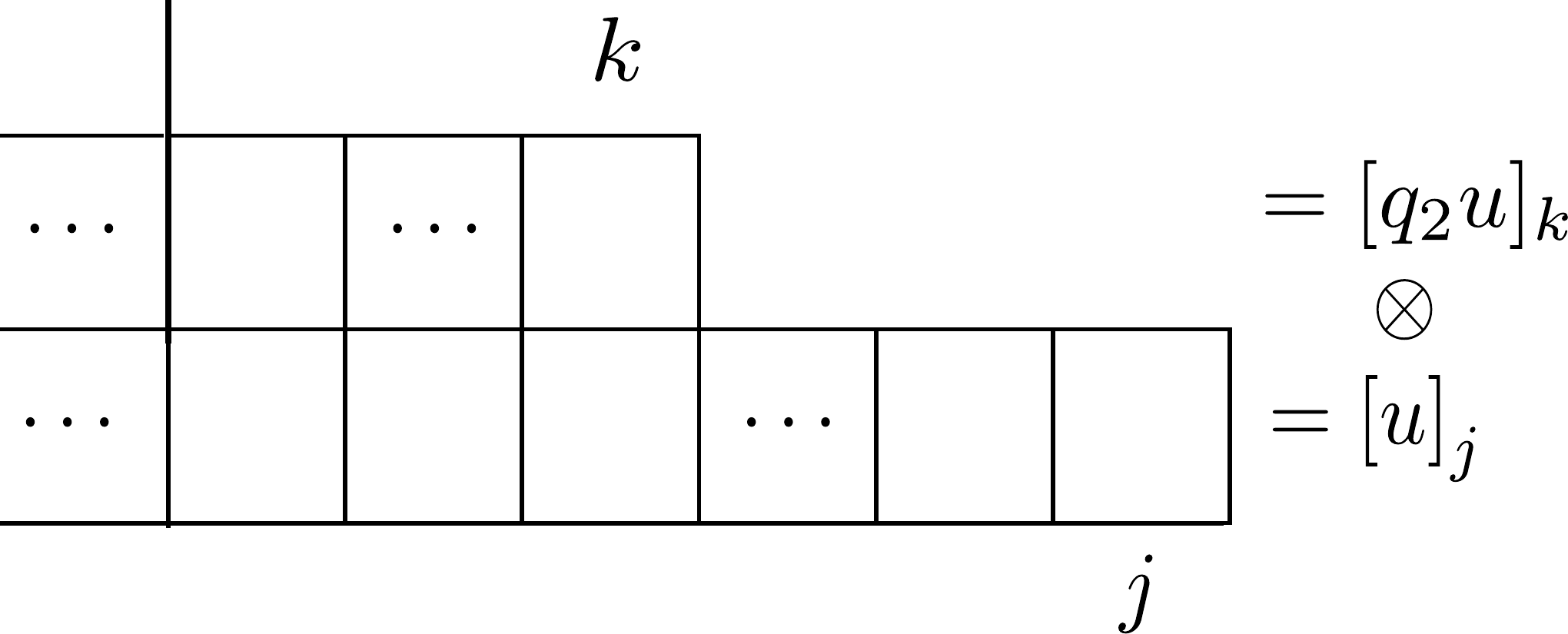}
    \caption{The basis using the tensor product of two vector representation. Each row is a 1d Young diagram, and the combination represents the tensor product of the two vector representations.}
    \label{fig:tensor_2vec}
\end{figure}

Let us construct the Fock representation with central charges $C=1$, $K^{-}=q_{3}^{1/2}$ in this section. First, we consider the tensor product of the two vector representations.
The bases are denoted as
\begin{equation}\label{eq:basis_2vec}
    [u]_j \otimes [q_2 u]_k,\quad j,k\in\mathbb{Z}.
\end{equation}
We used $q_2$ to translate, and it is different from $q_1$ characterizing the vector representation.
This can be interpreted as a state with $j+1$ boxes in the first row from the bottom and $k+1$ boxes in the second row as shown in Figure \ref{fig:tensor_2vec}.

The actions of $E(z), F(z), K^\pm(z)$ on this basis are defined by the coproduct as equation (\ref{eq:gl1_coproduct}) with $C=1$.
 
The first term of $\Delta E(z)$ always adds a box to the first row of Figure \ref{fig:tensor_2vec}, while the second term adds a box to the second row depending on the nontrivial coefficient coming from $K^{-}(z)$. By using (\ref{eq:gl1_vectorrep}) and (\ref{eq:gl1_coproduct}), we obtain
\begin{align}
\begin{split}
    K^{-}(z)\otimes E(z)[u]_{j}\otimes [q_{2}u]_{j}=0.\label{eq:gl1_E_Young}
    \end{split}
\end{align}
Similarly, the second term of $\Delta F(z)$ always removes a box from the second row of Figure \ref{fig:tensor_2vec}, while the first term removes a box from the first row depending on the nontrivial coefficient coming from $K^{+}(z)$. The nontrivial coefficient becomes zero when the length of the two rows are the same:
\begin{align}
    F(z)\otimes K^{+}(z) [u]_{j}\otimes [q_{2}u]_{j}=0.\label{eq:gl1_F_Young}
\end{align}
We obtain the Young diagram condition from (\ref{eq:gl1_E_Young}) and (\ref{eq:gl1_F_Young}), namely $[u]_{j}\otimes[q_{2}u]_{k}(j\geq k)$ forms a submodule. 

The same can be done for tensor products of $N$ vector representations.
In this case, the basis can be written using the Young diagram $\lambda=(\lambda_1, \lambda_2,\cdots, \lambda_N)\hspace{1mm}(\lambda_{1}\geq\lambda_{2}\geq\lambda_{3}\ldots)$ as
\begin{align}
    \bigotimes_{j=1}^N [q_2^{j-1}u]_{\lambda_j-1}.
\end{align}
Actions of $E(z), F(z), K^\pm(z)$ on this basis are expressed using the coproduct $N-1$ times as
\begin{align}
\begin{split}
    \Delta^{(N-1)}(K^\pm(z)) &= {K^\pm(z)}^{\otimes N},\\
    \Delta^{(N-1)}(E(z)) &= \sum_{k=1}^N {K^-(z)}^{\otimes k-1}\otimes E(z) \otimes 1^{\otimes N-k},\\
    \Delta^{(N-1)}(F(z)) &= \sum_{k=1}^N 1^{\otimes k-1}\otimes F(z) \otimes {K^+(z)}^{\otimes N-k}.
    \end{split}\label{eq:Ncoproduct}
\end{align}
Again, the $k$-th term of $E(z)$ adds one box to the $k$-th row of the basis, and the $k$-th term of $F(z)$ removes one box from the $k$-th row of the basis.
States satisfying the Young diagram condition only remain due to (\ref{eq:gl1_E_Young}) and (\ref{eq:gl1_F_Young}). Other states do not appear because the coefficients will be zero.

\begin{figure}[h]
    \centering
    \includegraphics[width=6cm]{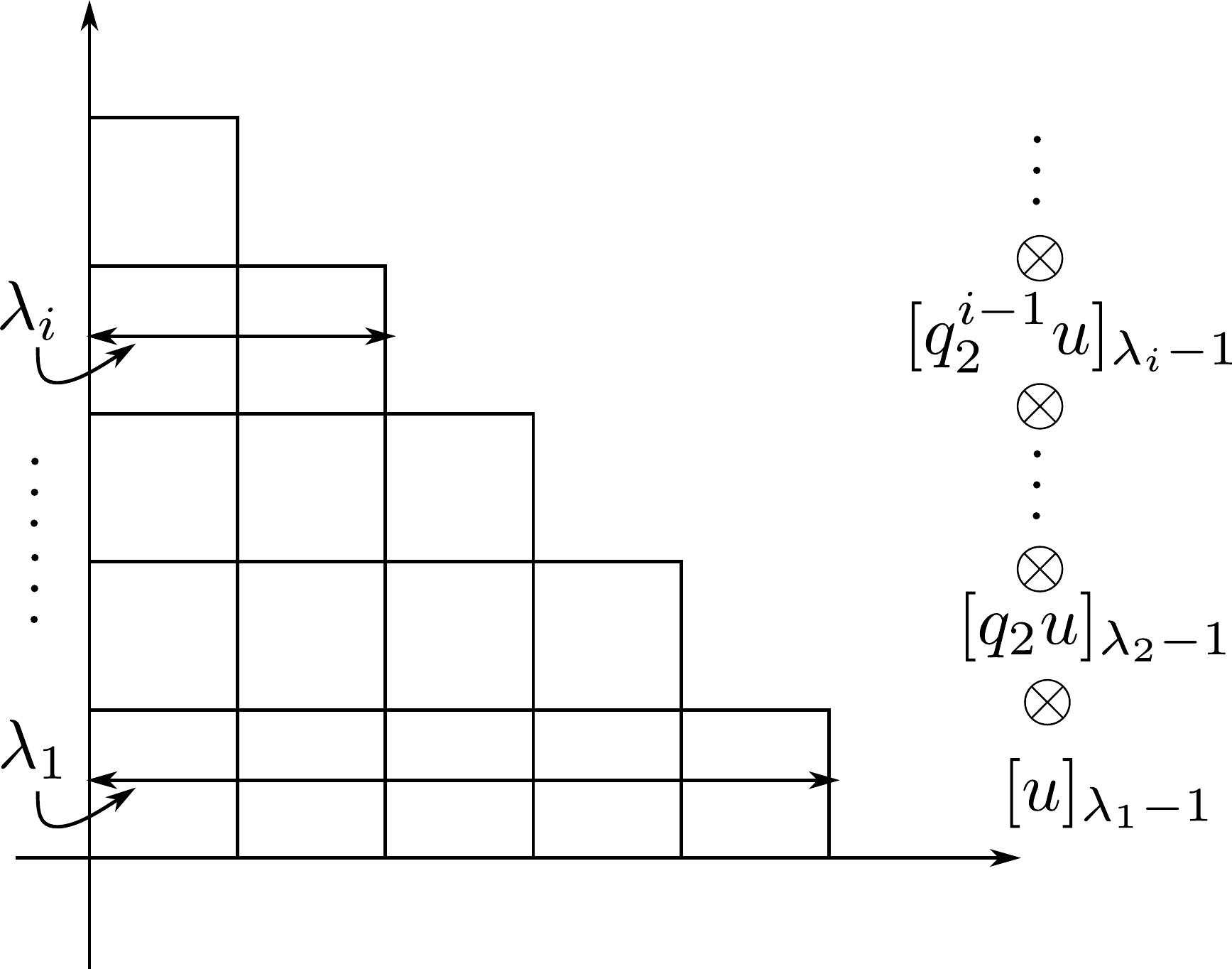}
    \caption{Young diagram and the basis of Fock representation. This is the generalization of Figure \ref{fig:tensor_2vec} to the infinite number of rows.}
    \label{fig:Fock_rep_inf}
\end{figure}

To take the limit $N\to \infty$, we need to reguralize the actions.
The basis is defined as
\begin{equation}
    \ket{\lambda}=\bigotimes_{j=1}^\infty [q_2^{j-1} u]_{\lambda_j-1}\label{eq:lambda_inf},
\end{equation}
where $\lambda_n=0\;(n>\ell(\lambda))$ and $\lambda=(\lambda_1, \lambda_2, \cdots, \lambda_{\ell(\lambda)})$. $\ell(\lambda)$ is the length of the Young diagram.
This state is illustrated as in Figure \ref{fig:Fock_rep_inf}.
The first row from the bottom has $\lambda_1$ boxes, the second row has $\lambda_2$ boxes and so on.

The actions on the basis are as follows:
\begin{align}
\begin{split}
     K(z)\ket{\lambda}&=\frac{\phi(q_{2}^{1-\ell(\lambda)}q_{1};z,u)}{\phi(q_{2}^{-\ell(\lambda)};z,u)}\prod_{i=1}^{\ell(\lambda)}\Psi_{[uq_{2}^{i-1}]_{\lambda_{i}-1}}(z)\ket{\lambda},\\
    E(z)\ket{\lambda}&=\mathcal{E}\sum_{i=1}^{\ell(\lambda)+1}\prod_{l=1}^{i-1}\left[\Psi_{[uq_{2}^{i-1}]_{\lambda_{i}-1}}(uq_{2}^{i-1}q_{1}^{\lambda_{i}})\right]_{-}\delta\left(\frac{z}{uq_{2}^{i-1}q_{1}^{\lambda_{i}}}\right)\ket{\lambda+\Abox_{i}},\\
    F(z)\ket{\lambda}&=\mathcal{F}\sum_{i=1}^{\ell(\lambda)}\frac{\phi(q_{1}^{2-\lambda_{i}}q_{2}^{2-i-\ell(\lambda)};1,1)}{\phi(q_{1}^{1-\lambda_{i}}q_{2}^{1-i-\ell(\lambda)};1,1)}\\
    &\hspace{1cm}\times\prod_{j=i+1}^{\ell(\lambda)}\left[\Psi_{[uq_{2}^{j-1}]_{\lambda_{j-1}}}(uq_{2}^{i-1}q_{1}^{\lambda_{i}-1})\right]_{+}\delta\left(\frac{z}{uq_{2}^{i-1}q_{1}^{\lambda_{i}-1}}\right)\ket{\lambda-\Abox_{i}},
    \end{split}
\end{align}
where $\Abox_{i}$ is a box at the $i$-th entry. From now on, we omit the $[\quad]_{\pm}$ when it is obvious.

The actions of $K(z)$ on the vacuum configuration $\ket{\emptyset}\equiv\otimes_{j=1}^{\infty}[q_{2}^{j-1}u]_{\lambda_{j}-1}$ can be read of:
\begin{align}
\begin{split}
    K(z)\ket{\emptyset}=\frac{\phi(q_{3}^{-1};z,u)}{\phi(1;z,u)}\ket{\emptyset}.
\end{split}\label{eq:gl1_vacuum_charge}
\end{align}
This gives us the nontrivial central charge $K^{-}=q_{3}^{1/2}$.

The regularization process will be important later, so let us look at the derivation of the action of $K(z)$.
Using (\ref{eq:Ncoproduct}), the action of $K(z)$ on $N$ tensor products is written as 
\begin{align}
     K(z)\ket{\lambda}=\prod_{i=1}^{\ell(\lambda)}\Psi_{[uq_{2}^{i-1}]_{\lambda_{i}-1}}(z)\prod_{j=\ell(\lambda)+1}^{\infty}\Psi_{[uq_{2}^{j-1}]_{-1}}(z)\ket{\lambda}.
 \end{align}
In this case, infinite product involves but we can formally regularize this by specifying the order of infinite products:
\begin{align}
\begin{split}
    \prod_{j=\ell(\lambda)+1}^{\infty}\Psi_{[uq_{2}^{j-1}]_{-1}}(z)&=\prod_{j=\ell(\lambda)+1}^{\infty}\frac{\phi(q_{2}^{2-j}q_{1};z,u)\phi(q_{2}^{-j};z,u)}{\phi(q_{1}q_{2}^{1-j};z,u)\phi(q_{2}^{1-j};z,u)}\\
    &=\frac{\phi(q_{2}^{1-\ell(\lambda)}q_{1};z,u)}{\phi(q_{2}^{-\ell(\lambda)};z,u)}\prod_{j=\ell(\lambda)+1}^{\infty}\frac{\bcancel{\phi(q_{2}^{1-j}q_{1};z,u)}}{\bcancel{\phi(q_{1}q_{2}^{1-j};z,u)}}\frac{\cancel{\phi(q_{2}^{-j};z,u)}}{\cancel{\phi(q_{2}^{-j};z,u)}}\\
    &=\frac{\phi(q_{2}^{1-\ell(\lambda)}q_{1};z,u)}{\phi(q_{2}^{-\ell(\lambda)};z,u)}.
    \end{split}\label{eq:infiniteproductgl1}
\end{align}

\section{Subcrystal representations}\label{sec:subcrystal_rep}
In this section, we focus on two special crystal representations: one-dimensional crystal and two-dimensional crystal. The two-dimensional crystals are associated with the corner divisors of the toric diagram, while the one-dimensional crystals are associated with the external lines of the toric diagram. The two-dimensional crystals we focus on were originally defined in \cite{Nishinaka_2011,Nishinaka_2012,Nishinaka_2014}. The one-dimensional crystals we introduce in this section are a reduced version of them.
Similar to the previous section, a quantum toroidal algebra acts on this crystal. This quantum toroidal algebra is not the same as the one defined in (\ref{eq:defofQuiverAlgebra}) after setting $C=1$ but are shifted quantum toroidal algebras.\footnote{We call the algebra (\ref{eq:defofQuiverAlgebra}) in the previous section ``unshifted" quantum toroidal algebra.}

We first introduce the one-dimensional and two-dimensional crystals in section \ref{sec:subcrystal_subquiver}. Starting from the original periodic quiver diagram of the three-dimensional crystal and removing some arrows from the diagram, we obtain various subquivers. These subquivers determine the shape of one-dimensional and two-dimensional crystals. Readers interested in only the algebraic structure can accept the shape of crystals and read from the next subsection (see \cite{Nishinaka_2011,Nishinaka_2012,Nishinaka_2014} for a detailed derivation of the two-dimensional crystals). In section \ref{sec:algebra_subcrystal}, we introduce a generalized version of the QQTA defined in section \ref{sec:QQTA_review} and give a discussion that they act on the subcrystals defined in section \ref{sec:subcrystal_subquiver}. We also derive the vacuum charge function of the two-dimensional crystals and translate it to the brane tiling language. The algebra introduced includes shift parameters defined by the difference between the number of poles and zeros of the vacuum charge function. In section \ref{sec:shift_coproduct}, we introduce a generalized coproduct and antipode structure. These maps act on \textit{different} shifted quantum toroidal algebras. In particular, the generalized coproduct structure will be important in deriving the two-dimensional crystal representations from the one-dimensional crystal representations. In section \ref{sec:conifolddetail}, we give a detailed derivation for the conifold case. Other examples are in section \ref{sec:examples}. 
\subsection{Crystals and quivers}\label{sec:subcrystal_subquiver}
We briefly review the constructions of two-dimensional crystal models associated with D4-D2-D0 bound states in the toric Calabi-Yau three-fold introduced in \cite{Nishinaka_2014} (see it for details). We then further reduce the crystal to a one-dimensional crystal.

\begin{figure}[H]
    \centering
    \includegraphics[width=12cm]{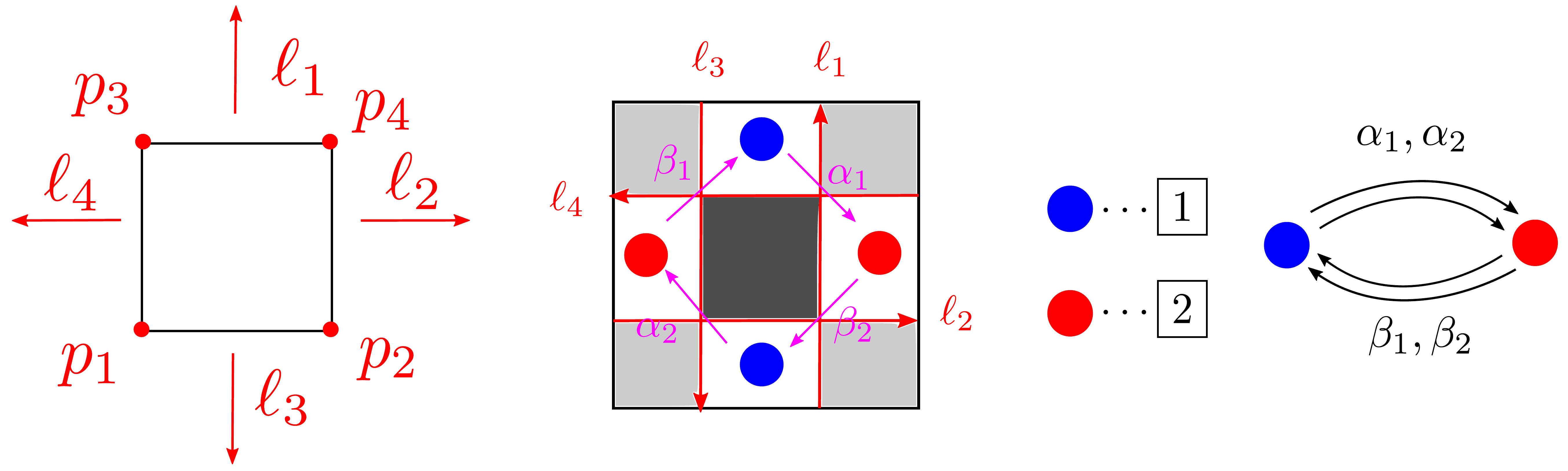}
    \caption{The toric diagram, periodic quiver (brane tiling) and quiver diagram for the conifold geometry. The lattice points of the toric diagram are denoted as $p_{1}=(0,0),p_{2}=(1,0),p_{3}=(0,1)$, and $p_{4}=(1,1)$. External legs are denoted as $\ell_{1},\ell_{2},\ell_{3}$, and $\ell_{4}$. Bifundamental chiral fields are denoted $\alpha_{1},\alpha_{2},\beta_{1},\beta_{2}$.} 
    \label{fig:conifold-brane-tiling}
\end{figure}
D2-D0 states are described by the quiver $Q=(Q_{0},Q_{1},Q_{2})$ dual to the Calabi-Yau geometry, where $Q_{0},Q_{1},Q_{2}$ are the set of vertices, arrows, and faces\footnote{Faces are identified with the loops in section \ref{sec:QQTA_review}. We do not distinguish them.}, respectively. Vertices correspond to gauge groups, arrows correspond to the bifundamental chiral multiplets, and faces correspond to the superpotential of the quiver gauge theory. The information of both the gauge theory and the dual geometry are encoded in a dual graph called brane tiling \cite{Franco:2005sm} (see Appendix \ref{sec:appendix-branetiling}). As an example, let us consider the conifold (see Figure \ref{fig:conifold-brane-tiling}). There are two faces $Q_{2}=\{\alpha_{1}\beta_{2}\alpha_{2}\beta_{1},\alpha_{1}\beta_{1}\alpha_{2}\beta_{2}\}$, where we are identifying the faces with the sequence of arrows. The superpotential\footnote{See Appendix \ref{sec:appendix-branetiling} and (\ref{eq:superpotentialnotation}) for how to derive the superpotential from the brane tiling.} is read as
\begin{align}
    W_{0}=\text{Tr}\,(\alpha_{1}\beta_{1}\alpha_{2}\beta_{2}-\alpha_{1}\beta_{2}\alpha_{2}\beta_{1}).
\end{align}
The F-term relations $\partial W/\partial X_{A},\,(X_{A}\in Q_{1})$ and the so-called D-term relations\footnote{We will not discuss about the D-term relations so see \cite[Sec.2.3]{Nishinaka_2014}} give the vacuum moduli space.
\begin{figure}[t]
    \centering
    \includegraphics[width=5cm]{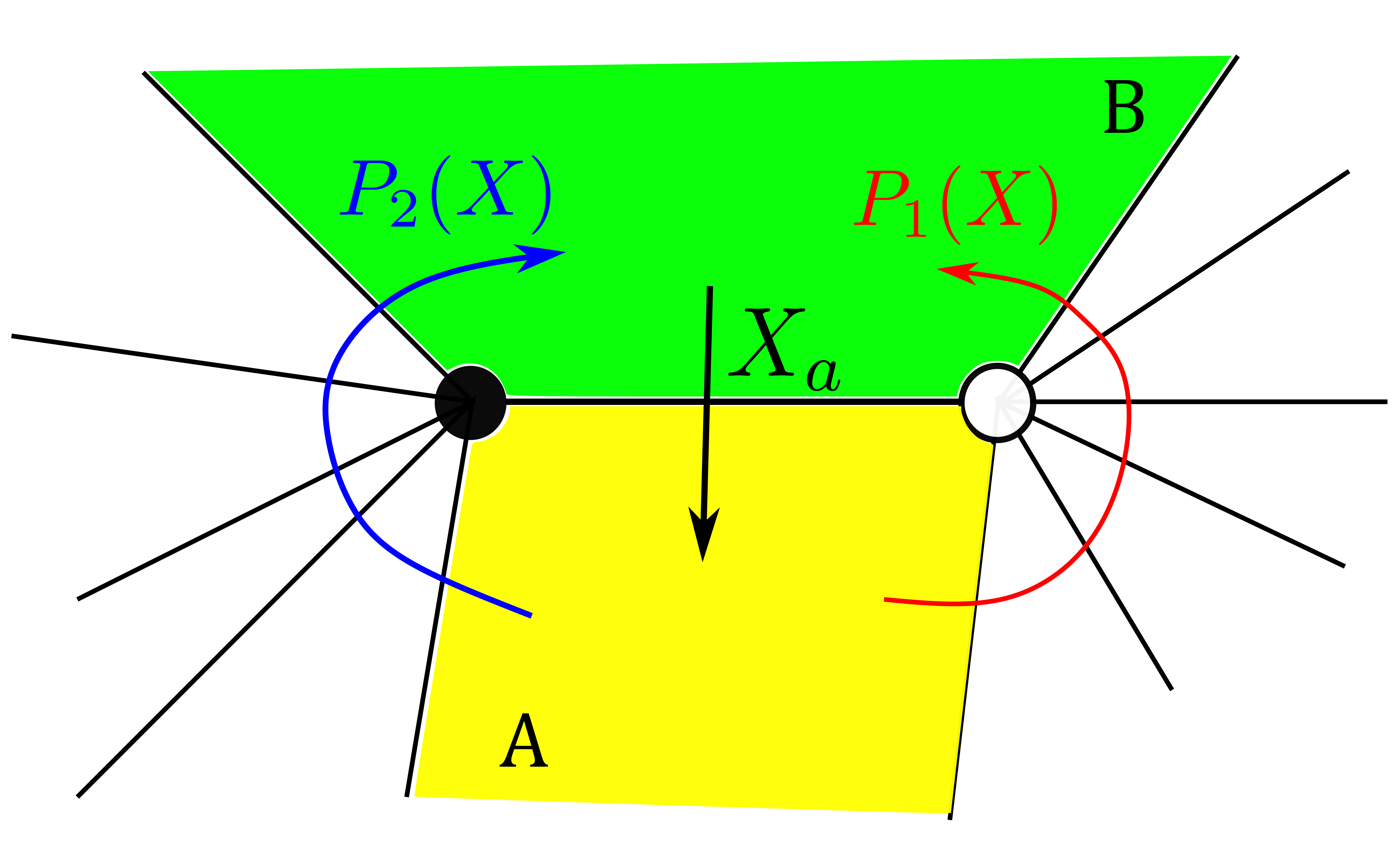}
    \caption{Graphical interpretation of F-term condition of (\ref{eq:Ftermsuperpotential}) in the bipartite graph. The edge connecting the black and white node is dual to the arrow $X_{a}$. The bifundamental $X_{a}$ connects two regions A (yellow) and B (green). The red path asoociated with a monomial $P_{1}(X)$ and the blue path associated with a monomial $P_{2}(X)$ connect the two regions.}
    \label{fig:Ftermfigure}
\end{figure}
\begin{figure}[t]
    \centering
    \includegraphics[width=5cm]{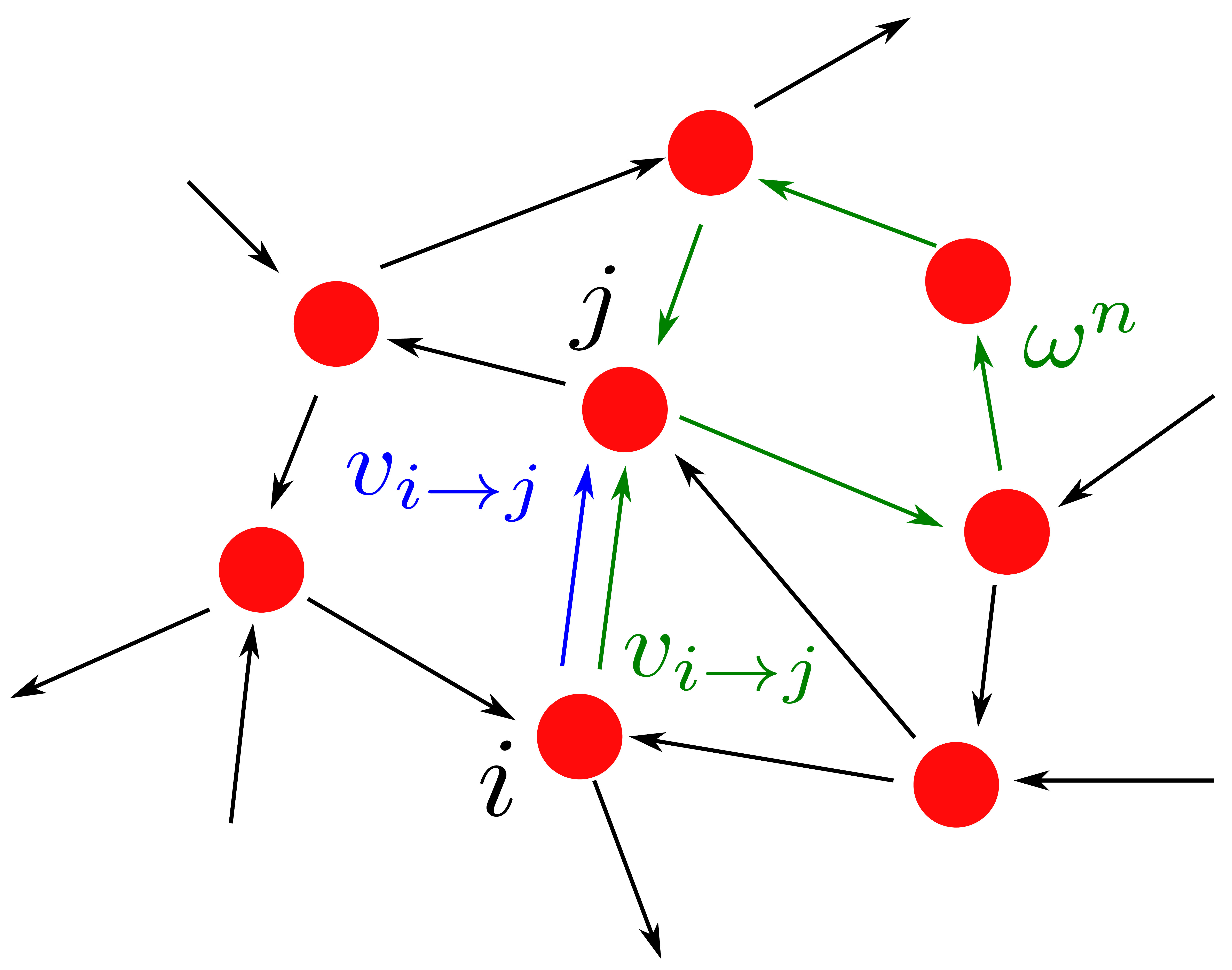}
    \caption{Two paths between quiver nodes $i,j\in Q_{0}$. The blue path is the shortest path $v_{i\rightarrow j}$ and the green path is the path decomposed into $v_{i\rightarrow j}$ and the loop part $\omega^{n}$.}
    \label{fig:Ftermloop}
\end{figure}

When the dual geometry is a toric Calabi-Yau, the superpotential obeys the condition that every chiral field appears in the superpotential only twice \cite{Franco:2005rj}. Focusing on terms including a bifundamental field $X_{a}\in Q_{0}$, we have
\begin{align}
    W_{0}=X_{a}P_{1}(X)-X_{a}P_{2}(X)+\cdots,\quad X_{a}\in Q_{1},\label{eq:Ftermsuperpotential}
\end{align}
where $P_{1}(X)$ and $P_{2}(X)$ are products of bifundamental fields not including $X_{a}$ (see Figure \ref{fig:Ftermfigure}). The F-term relation of $X_{a}$ gives the condition $P_{1}(X)=P_{2}(X)$. Generally, the F-term relations mean that as long as the starting point and ending point are determined, the path between them are all equivalent. In particular, for any point $i,j\in Q_{0}$, there is a shortest path $v_{i\rightarrow j}$ such that any path from $i$ to $j$ is F-term equivalent to $v_{i\rightarrow j}\omega^{n}$, where $n$ is an integer and $\omega$ is a loop of one face (see Figure \ref{fig:Ftermloop}).

\paragraph{Three-dimensional crystal \cite{Ooguri_2009}}
\begin{figure}[t]
    \centering
    \includegraphics[width=8cm]{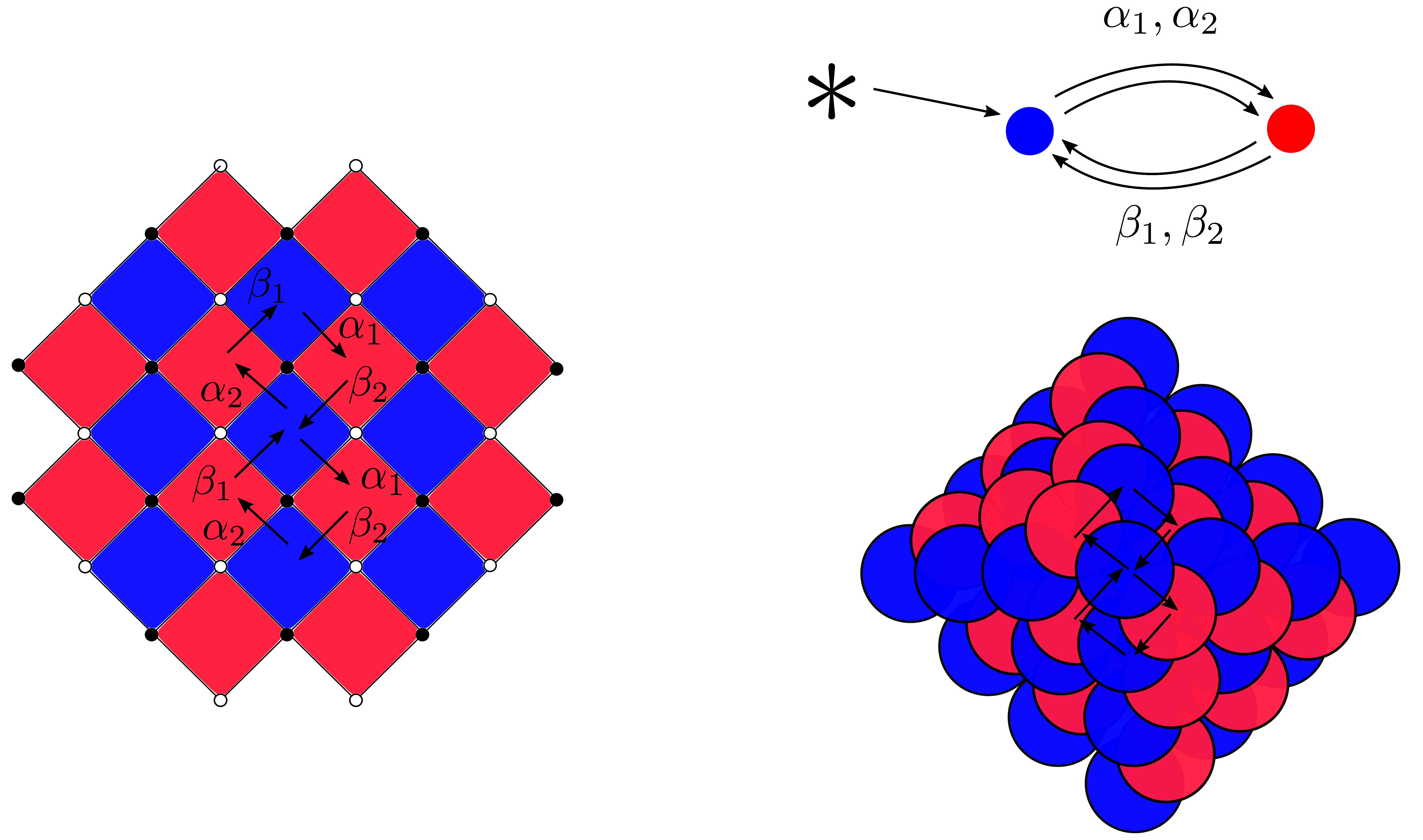}
    \caption{Enhanced quiver diagram, universal covering of it and the three-dimensional crystal of the conifold geometry.}
    \label{fig:conifold3dcrystal}
\end{figure}
To consider D6-D2-D0 states, we need to wrap the Calabi-Yau with a D6-brane. This process enhances the quiver by adding a flavor node $*$ and an edge $I:*\rightarrow i_{0},\,(i_{0}\in Q_{0})$ to the quiver:
\begin{align}
    Q_{0}\rightarrow Q_{0}\cup\{*\},\quad Q_{1}\rightarrow Q_{1}\cup \{I:*\rightarrow i_{0}\},
\end{align}
where $i_{0}$ is called a reference node. The superpotential and the F-term conditions do not change. Starting from this quiver, we can construct a three-dimensional crystal \cite{Ooguri_2009}. Each of the atoms in the crystal is identified with a path in the universal covering of the quiver. The process is the following (see Figure \ref{fig:conifold3dcrystal} for the conifold case):
\begin{enumerate}\label{3dcrystalshape}
\renewcommand{\theenumi}{A\arabic{enumi}}
    \item Using the fact that the quiver $Q$ is drawn on a $\mathbb{T}^{2}$, we can extend it and consider a universal covering on $\mathbb{R}^{2}$, which we denote it $\Tilde{Q}$. We define a projection map\footnote{The vertices in $\Tilde{Q}$ are all distinguished. The projection to $Q$ gives a color to it. Namely, the colors of the atoms in the crystals are determined from the vertices of $Q$.} $\pi:\Tilde{Q}\rightarrow Q$. This defines the colors of the atoms. Namely, the three-dimensional crystal is a crystal with \emph{colored} atoms.
    \item Place an atom with color $\pi(i_{0})$ on the reference node $i_{0}\in\Tilde{Q}$. The reference node gives the origin of the three-dimensional crystal.
    \item If there is an edge $i_{0}\rightarrow j$ for $j\in \Tilde{Q}$, place an atom with the color $\pi(j)$ next to the origin in the reference node $i_{0}$. These atoms are placed below the origin atom.
    \item Follow the same step above and continue placing atoms if they are connected by arrows to the atoms already placed. In the case when one comes back to the node on which an atom is already placed, we use the following rules:
    \begin{itemize}
        \item Every path from the origin $i_{0}\rightarrow j$ is written as $v_{i_{0}\rightarrow j}\,\omega^{n}$, where $v_{i_{0}\rightarrow j}$ is the shortest path from $i_{0}$ to $j$ and $\omega$ is the loop (see Figure \ref{fig:Ftermloop}). For this kind of path, place an atom at the $n$-th place under the first atom on the node $j$. This integer $n$ is called the depth of the atom.
        \item If there is already an atom in the $n$-th place, do not place any atom anymore.
    \end{itemize}
\end{enumerate} 

 Actually, this crystal corresponds to one D6-brane with no D0, D2 charges. Crystals associated with nonzero D0, D2 charges are subcrystals called molten crystals. It is obtained from the melting rule in \cite{Ooguri_2009}. As was shown in \cite{Li_2020,Galakhov:2020vyb, Noshita:2021ldl}, quiver Yangian and quiver quantum toroidal algebra act on this molten crystal.

\paragraph*{Two-dimensional crystal \cite{Nishinaka_2014}}
\begin{figure}[t]
    \centering
    \includegraphics[width=14cm]{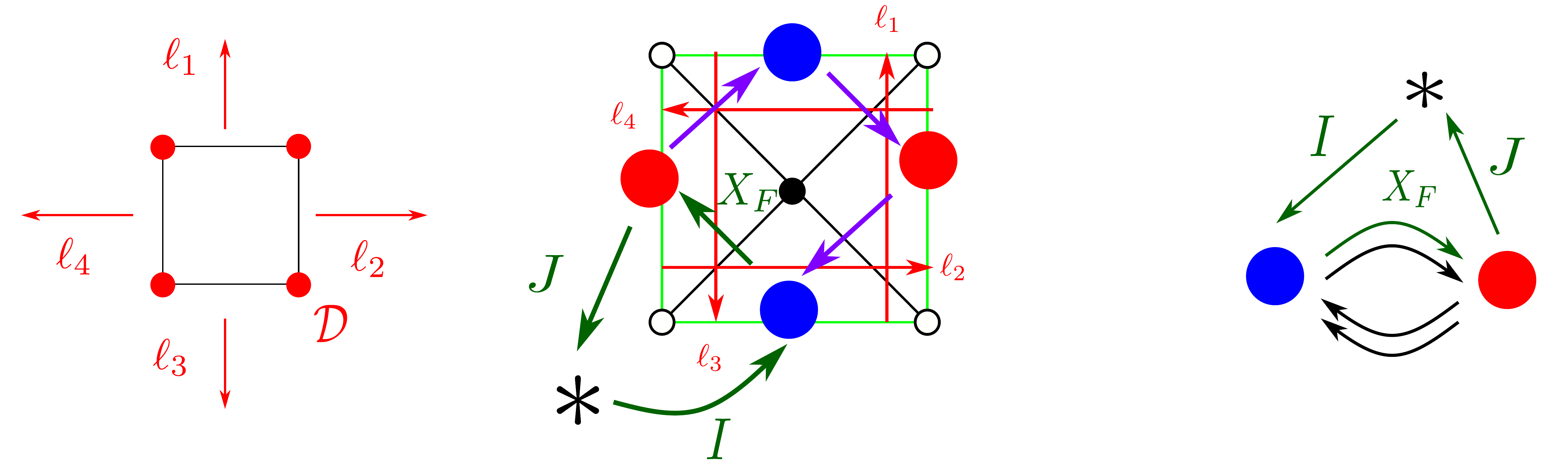}
    \caption{Quiver of the conifold with a D4-brane wrapping the toric divisor $\mathcal{D}=p_{2}$. Left: The divisor is surrounded by external legs $\ell_{2},\ell_{3}$. Middle: The flavor node $*$ is drawn outside the brane tiling for convenience. Two arrows $I,J$ are added to the quiver diagram. The bifundamental field $X_{F}$ for this case is $X_{F}=\alpha_{2}$. Right: The enhanced quiver, where the green arrows introduce a new term $W_{\text{flavor}}=\text{Tr}(JX_{F}I)$ to the superpotential. }
    \label{fig:conifoldD4quiver}
\end{figure}
\begin{figure}[t]
    \centering
    \includegraphics[width=10cm]{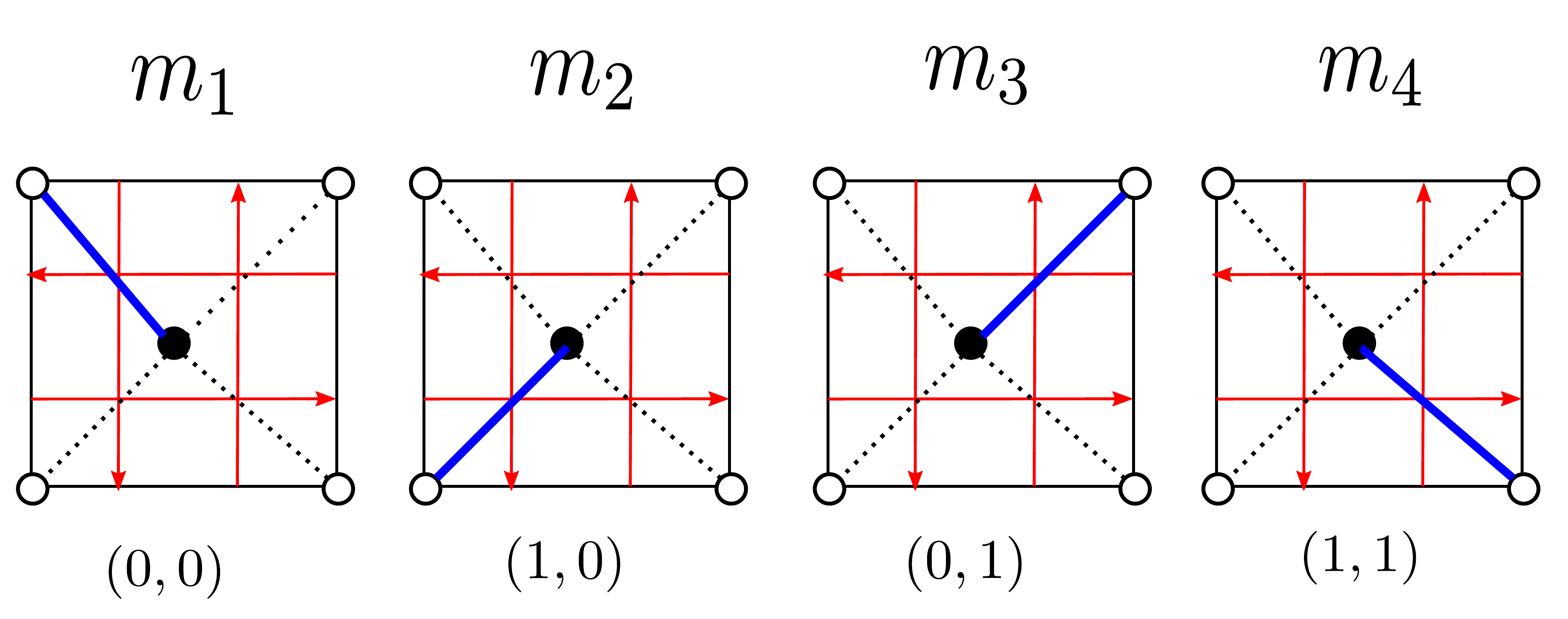}
    \caption{Perfect matchings of conifold geometry. Each lattice point of the toric diagram is a corner divisor and thus has a unique perfect matching. We denote the perfect matching as $m_{i}$ for the corner divisor $p_{i}$, where $i=1,2,3,4$. We note $m_{1}=\{\beta_{1}\}, m_{2}=\{\alpha_{2}\}, m_{3}=\{\alpha_{1}\}$, and $m_{4}=\{\beta_{2}\}$.}
    \label{fig:gl_(1,1)perfectmatching}
\end{figure}
Let us now consider the D0-D2 states bound to a D4-brane wrapping a corner toric divisor $\mathcal{D}$ of the Calabi-Yau. The flavor D4-brane adds a flavor node $*$ to the quiver, similar to the D6-brane. However, in this case, the D4-brane will also add arrows $I:*\rightarrow i$ and $J:j\rightarrow *$, where $i,j\in Q_{0}$ are regions adjacent to the place of the D4-brane. These arrows give an additional term to the superpotential as 
\begin{align}
    W_{\text{flavor}}=\text{Tr} (JX_{F}I),
\end{align}
where $X_{F}$ is the bifundamental field connecting the two vertices $i,j$. The quiver will be modified as 
\begin{align}
Q_{0}\rightarrow Q_{0}\cup \{*\},\quad Q_{1}\rightarrow Q_{1}\cup\{I:*\rightarrow i, J:j\rightarrow *\},\quad Q_{2}\rightarrow Q_{2}\cup\{JX_{F}I\}.    
\end{align} See Figure \ref{fig:conifoldD4quiver} for an example of the conifold. 

The total superpotential is $W=W_{0}+W_{\text{flavor}}$ now, and the F-term relations are modified as
\begin{align}
\begin{split}
    &\frac{\partial W}{\partial I}=\frac{\partial W}{\partial J}=0 \Leftrightarrow JX_{F}=X_{F}I=0,\\
    &\frac{\partial W}{\partial X_{F}}\Leftrightarrow \frac{\partial W_{0}}{\partial X_{F}}+IJ=0,\\
    &\frac{\partial W_{0}}{\partial X_{a}}=0,\quad X_{a}\neq X_{F}.
\end{split}
\end{align}
It was shown in \cite{Nishinaka_2014} that special constraints are imposed on the field configurations at the supersymmetric vacua. A subset of the arrows called ``perfect matching" plays a role in these constraints. Perfect matching is a subset of the edges connecting black and white points of the graph, such that any black or white point is contained only once (see Figure \ref{fig:gl_(1,1)perfectmatching} for an example of the conifold). One of the interesting properties of the perfect matching is that there is a surjective map from the perfect matchings to the lattice points of the toric diagram (we will not discuss this, so see \cite{kenyon2003introduction,Hanany_2007}). Generally, the corresponding perfect matching is not unique, but for corner divisors, they are unique \cite{broomhead2010dimer,Gulotta_2008}. 

The F-terms constraints imposed on the moduli space are
\begin{align}
    J=0,\quad X_{a}=0\hspace{2mm} (X_{a}\in m_{\mathcal{D}}),\label{eq:FtermD4}
\end{align}
where $m_{\mathcal{D}}$ is the unique perfect matching associated with the corner divisor $\mathcal{D}$ \cite[Sec.3]{Nishinaka_2014}. These constraints imply that the moduli space of the D4-brane system $\mathcal{M}_{\text{D4}}$ is a subspace of the moduli space $\mathcal{M}_{\text{D6}}$ of the D6-brane system:
\begin{align}
    \mathcal{M}_{\text{D4}}\subset\mathcal{M}_{\text{D6}}. \label{eq:D4inD6}
\end{align}
\begin{figure}[t]
    \centering
    \includegraphics[width=14cm]{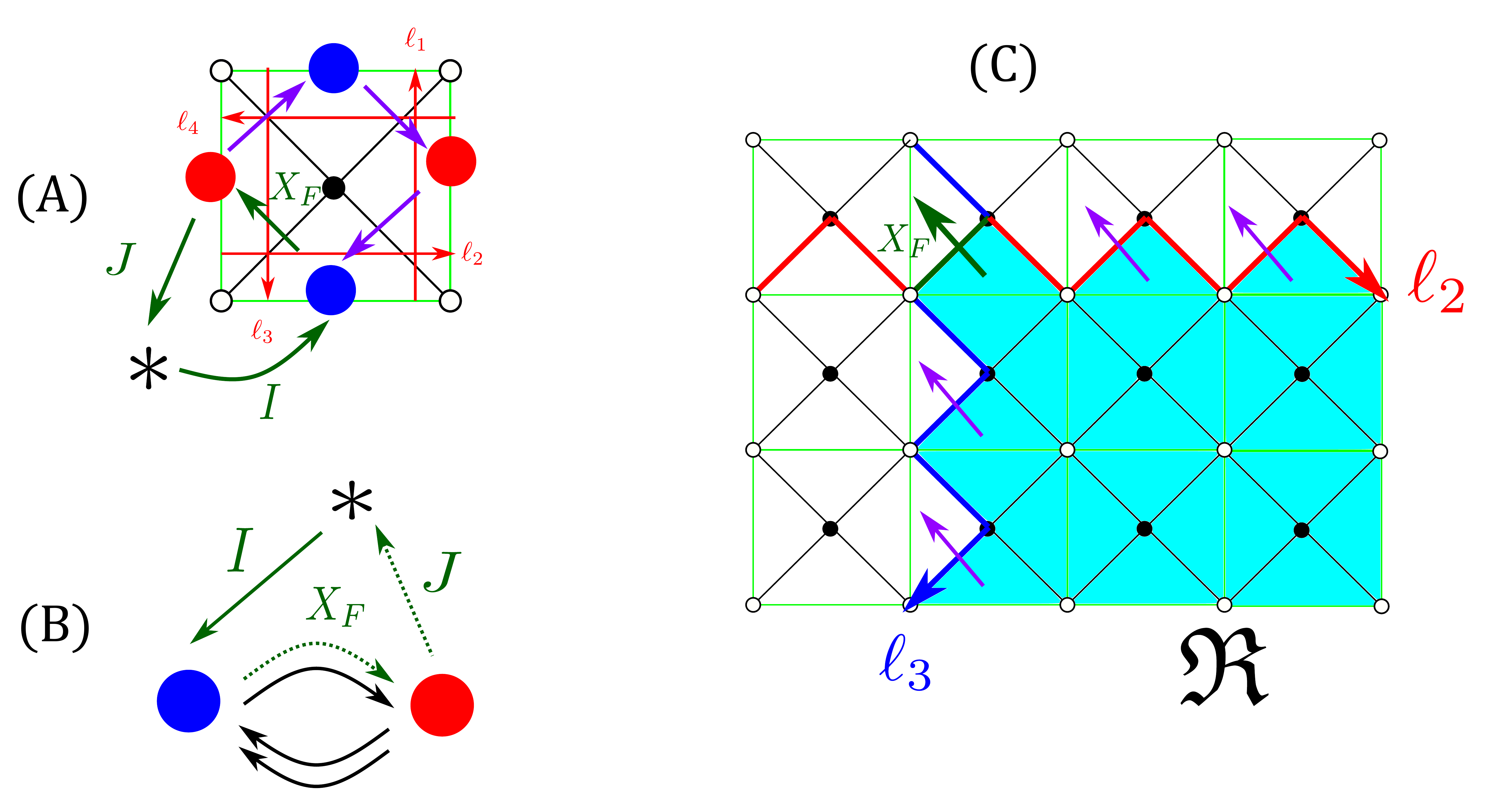}
    \caption{Two-dimensional crystal of the conifold with a D4-brane wrapping the divisor $\mathcal{D}=(1,0)$. (A) Brane tiling of the setup. (B) Quiver of the setup. The dashed arrows $J,X_{F}$ are zero in the supersymmetric vacua, and thus, the quiver is understood as a subquiver of the D6-D2-D0 case. (C) Region of the two-dimensional crystal in the universal covering. The light blue region $\mathfrak{R}$ is surrounded by the external legs $\ell_{2},\ell_{3}$. The green and purple arrows are outgoing arrows.}
    \label{fig:conifold2dcrystalshape}
\end{figure}
 Similar to the three-dimensional case, we can construct a two-dimensional crystal describing D0-D2-D4 bound states. As one can expect from the equation (\ref{eq:D4inD6}), the crystal is a \textbf{subcrystal} of the three-dimensional crystal. Actually, the crystal is obtained from the original crystal by eliminating all the arrows associated with the perfect matching $m_{\mathcal{D}}$.  Moreover, the crystal lies in a two-dimensional plane at depth zero, and it is a surface of the three-dimensional crystal. We will call the quiver obtained by eliminating the arrows of $m_{\mathcal{D}}$ from the quiver of the D6-D2-D0 setup a \textbf{subquiver}. We will not write down anymore the arrow $J$ because it is always zero in the supersymmetric vacua. We will also not write down the flavor node $*$ anymore when it is obvious in the next sections (recall that the flavor node just determines the color of the origin atom). The arrows of $m_{\mathcal{D}}$ will be drawn in dashed lines. The \textbf{subcrystal} and \textbf{subquiver} is obtained as follows:
 \begin{enumerate}
 \renewcommand{\theenumi}{B\arabic{enumi}}
\item Similar to the three-dimensional crystal, we can consider the universal covering $\Tilde{Q}$ of the quiver. The projection $\pi:\Tilde{Q}\rightarrow Q$ defines the color of the atoms.
\item Choose one corner divisor and remove arrows from the quiver diagram associated with the perfect matching of that corner divisor. This condition comes from (\ref{eq:FtermD4}). After this process, we obtain the \textbf{subquiver}. For the conifold case with a D4-brane wrapping the divisor $\mathcal{D}=p_{2}=(1,0)$, the perfect matching is $m_{\mathcal{D}}=m_{2}=\{\alpha_{2}\}$ (see Figure \ref{fig:gl_(1,1)perfectmatching}). The subquiver is Figure \ref{fig:conifold2dcrystalshape} (B). The dashed arrows are the arrows whose chiral fields are zero in the supersymmetric vacua. From the next section, we will not draw $*,I,J$ anymore but only the dashed arrows in $m_{\mathcal{D}}$.
\item The two-dimensional crystal is the region in the periodic quiver diagram, surrounded by two zig-zag paths of the external legs surrounding the corner divisor. The zig-zag path here is an oriented path in the periodic quiver diagram that turns maximally right at a black vertex and maximally left at a white vertex. There is a one-to-one correspondence between zig-zag paths and external legs of the toric web diagram \cite{Hanany_2007}. For the conifold case, see Figure \ref{fig:conifold2dcrystalshape}. The two external legs $\ell_{2},\ell_{3}$ surrounding the divisor are drawn in red and blue in Figure \ref{fig:conifold2dcrystalshape} (C). They divide the universal covering $\Tilde{Q}$ into four regions, and the lower right region is denoted as $\mathfrak{R}$. This region determines the shape of the crystal.
\item Generally, the arrows of $m_{\mathcal{D}}$ are arrows going out of the region $\mathfrak{R}$. This means that the crystal obtained from the subquiver is restricted to the region $\mathfrak{R}$ \cite{Nishinaka_2014,broomhead2010dimer,Gulotta_2008}. For the conifold case, the green arrow $X_{F}$ and the purple arrows in Figure \ref{fig:conifold2dcrystalshape} (C) are the outgoing arrows, and they are eliminated. We also note that the two zig-zag paths surrounding the corner divisor share a unique edge, which is the green arrow in Figure \ref{fig:conifold2dcrystalshape} (C). We will see later that it defines the vacuum charge function of this two-dimensional crystal representation.      
\begin{figure}[t]
    \centering
    \includegraphics[width=12cm]{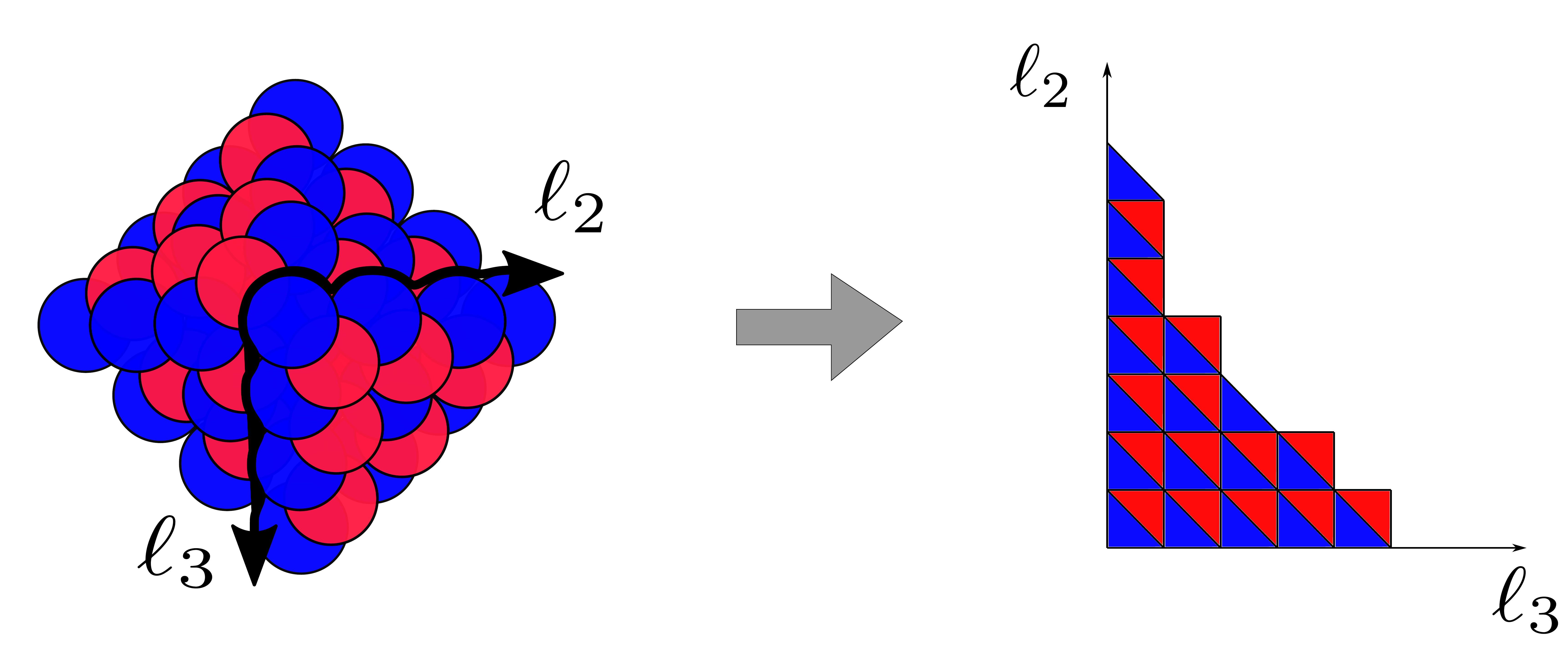}
    \caption{The lift up of the region $\mathfrak{R}$ in the three-dimensional crystal. The right figure is a convenient way to draw the crystal shape in the plane.}
    \label{fig:conifold2dcrystalliftup}
\end{figure}
\item The crystal is a lift-up of this region to the periodic quiver to the three-dimensional crystal. From the three-dimensional crystal viewpoint, it corresponds to the ``slope face" of the crystal (see Figure \ref{fig:conifold2dcrystalliftup} for the conifold case). 
\end{enumerate}

Similar to the three-dimensional crystal case, molten crystals of this crystal are the crystals with nonzero D0-D2 charges. Later, we will show that a slightly generalized version of the quiver quantum toroidal algebra with the \textit{same} bond factors acts on this crystal. This is because the moduli space $\mathcal{M}_{\text{D4}}$ can be embedded in $\mathcal{M}_{\text{D6}}$ as (\ref{eq:D4inD6}), and thus, we can use the same bond factors coming from the D6 setup.

\paragraph{One-dimensional crystal}
We would like to further reduce the crystal and construct a one-dimensional crystal. Unfortunately, we do not have any physical interpretations for this crystal, so we will just write down the procedure to obtain \textbf{subquivers} and \textbf{subcrystals}. Strictly speaking, the crystal constructed here is \textit{not} a subcrystal of the two and three-dimensional crystal. It is rather a reduced and extended crystal of the two-dimensional crystal\footnote{Since this crystal is not a subcrystal of the three-dimensional crystal in a strict sense, it is not obvious to expect that the bond factors are the same as the two, three-dimensional crystal case. However, we will see later that we can use the same bond factors to derive representations acting on this crystal.}. The procedure is as follows:
\begin{enumerate}
\renewcommand{\theenumi}{C\arabic{enumi}}
    \item One-dimensional crystals are associated with the external legs of the toric web diagram. 
    Choose one external leg surrounded by two divisors. Take the union set of the perfect matchings of these divisors and remove them from the quiver diagram. Then we obtain the subquiver. Let us consider the one-dimensional crystal associated with the external leg $\ell_{2}$ of the conifold. It is surrounded by two divisors $p_{2}$ and $p_{4}$ (see Figure \ref{fig:conifold-brane-tiling}). Removing $m_{2}\cup m_{4}=\{\alpha_{2},\beta_{2}\}$, we obtain the subquiver in Figure \ref{fig:conifold1dcrystal3dsubcrystal}.
    \begin{figure}[t]
    \centering
    \includegraphics[width=10cm]{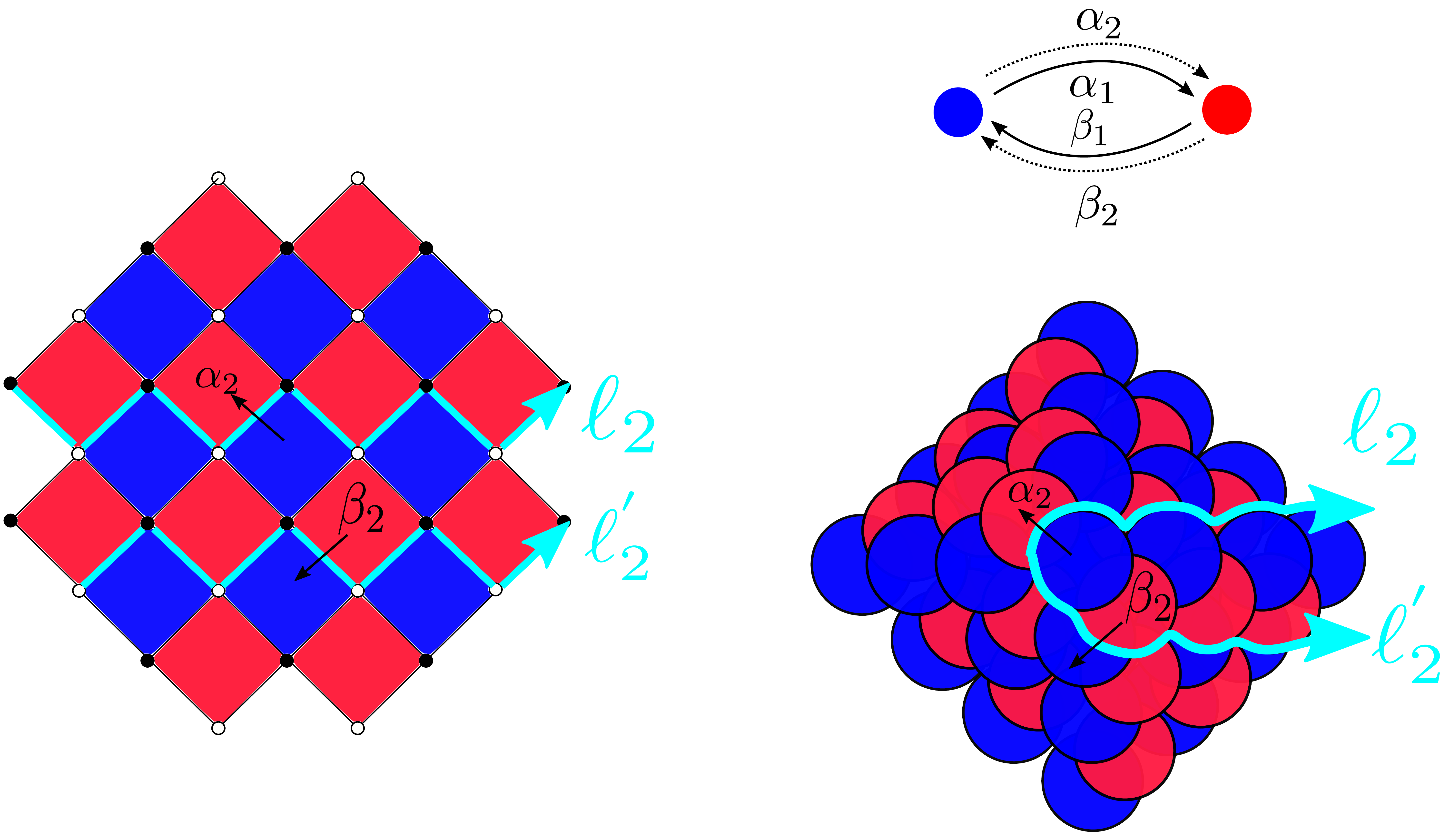}
    \caption{The region surrounded by the two external legs $\ell_{2}$ and $\ell_{2}'$. The $\ell_{2}'$ here is the same external leg as $\ell_{2}$ in the toric diagram but different in the universal covering. The subquiver is obtained by removing $m_{2}\cup m_{4}=\{\alpha_{2},\beta_{2}\}$ because the external leg $\ell_{2}$ is surrounded by divisors $p_{2},p_{4}$. The uplift to the three-dimensional crystal is the right down figure.}
    \label{fig:conifold1dcrystal3dsubcrystal}
\end{figure}
\begin{figure}[t]
    \centering
    \includegraphics[width=10cm]{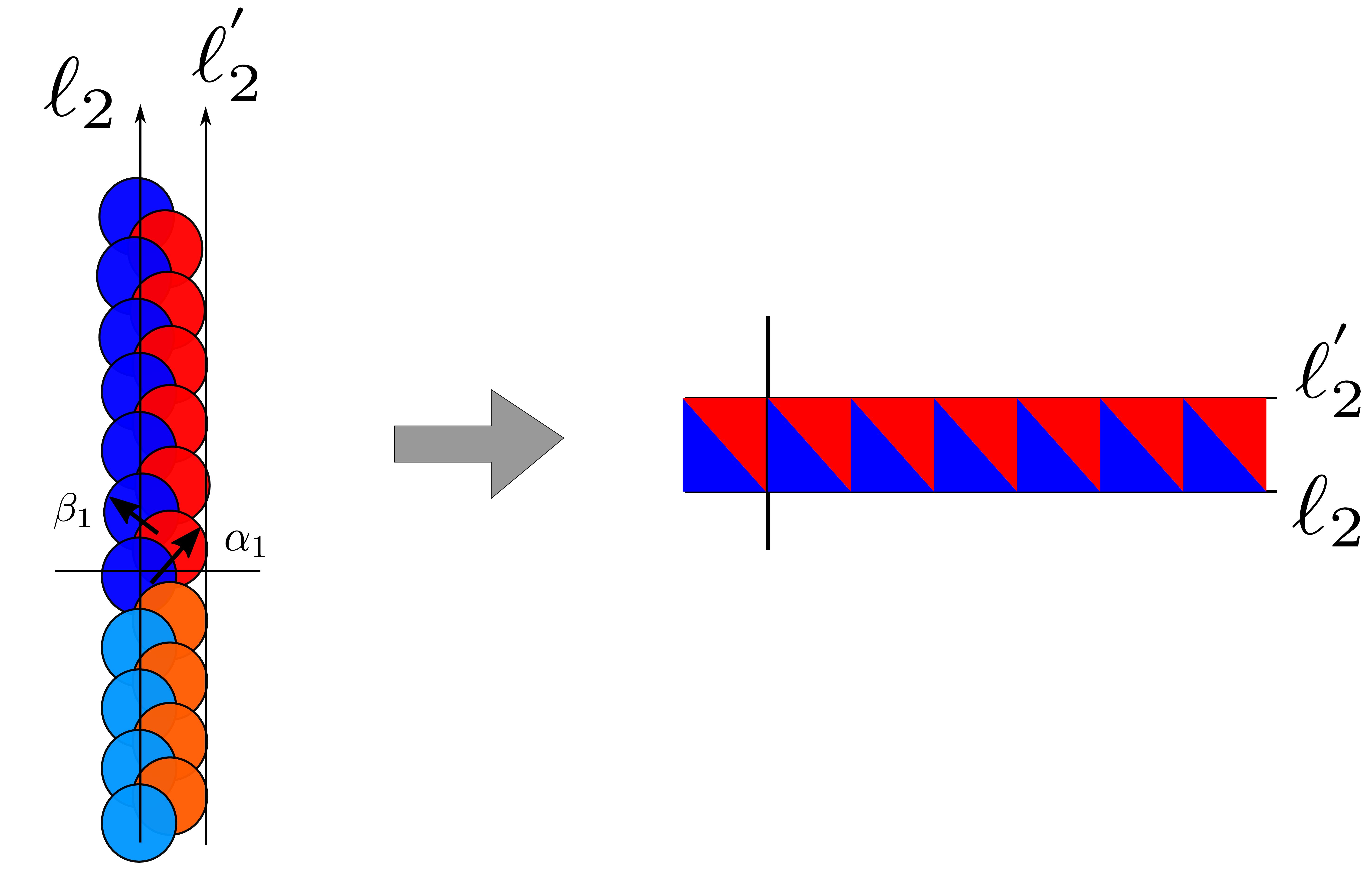}
    \caption{The one-dimensional crystal associated with $\ell_{2}$. The orange and light blue atoms are atoms originally not in the three-dimensional crystal but added to obtain the one-dimensional crystal. We can redraw this crystal as a sequence of boxes with blue and red triangles inside it.}
    \label{fig:conifold1dcrystalvector}
\end{figure}
    \item The one-dimensional crystal is the region surrounded by two parallel zig-zag paths. They may be zig-zag paths associated with the same external leg or zig-zag paths associated with different external legs. For the conifold case, the region surrounded by the two external legs $\ell_{2},\ell_{2}'$ in Figure \ref{fig:conifold1dcrystal3dsubcrystal} is the crystal. The crystal shape can be also written as Figure \ref{fig:conifold1dcrystalvector}.
\end{enumerate}

\subsection{Shifted algebra for subcrystal}\label{sec:algebra_subcrystal}
In this section, we consider what algebra acts on these subcrystals. It is natural to expect that the quiver quantum toroidal algebra in (\ref{eq:defofQuiverAlgebra}) with the condition $C=1$ acts on this crystal, but it seems this is not true. The correct algebra is the shifted quiver quantum toroidal algebra $\ddot{\mathcal{U}}_{Q}^{\mathbf{r}}$ whose definition is:
\begin{screen}
\begin{align}
\begin{split}
    E_{i}(z)=\sum_{k\in\mathbb{Z}}E_{i,k}z^{-k},\quad F_{i}(z)&=\sum_{k\in\mathbb{Z}}F_{i,k}z^{-k},\quad K_{i}^{\pm}(z)=\sum_{r\geq 0}K^{\pm}_{i,\pm r}z^{\mp r}.\\
\mathbf{r}=(r_{i})_{i\in Q_{0}}&,\quad r_{i}\in\mathbb{Z}\\
    K_{i}^{\pm}(z)K_{j}^{\pm}(w)&=K_{j}^{\pm}(w)K_{i}^{\pm}(z),\\
    K_{i}^{-}(z)K_{j}^{+}(w)&=K_{j}^{+}(w)K_{i}^{-}(z),\\
    K_{i}^{\pm}(z)E_{j}(w)&=\varphi^{j\Rightarrow i}(z,w)E_{j}(w)K_{i}^{\pm}(z),\\
    K_{i}^{\pm}(z)F_{j}(w)&=\varphi^{j\Rightarrow i}(z,w)^{-1}F_{j}(w)K_{i}^{\pm}(z),\\
    [E_{i}(z),F_{j}(w)]=\delta_{i,j}&\delta\left(\frac{w}{z}\right)\left(z^{r_{i}}K_{i}^{+}(z)-K_{i}^{-}(w)\right),\\
    E_{i}(z)E_{j}(w)&=(-1)^{|i||j|}\varphi^{j\Rightarrow i}(z,w)E_{j}(w)E_{i}(z),\\
    F_{i}(z)F_{j}(w)&=(-1)^{|i||j|}\varphi^{j\Rightarrow i}(z,w)^{-1}F_{j}(w)F_{i}(z).
\end{split}\label{eq:defofshiftedQuiverAlgebra}
\end{align}
\end{screen}
    We loosen the defining relations and do not impose the condition $K_{i,0}^{+}K_{i,0}^{-}=1$. As one can see, the mode expansion of the generator $K_{i}^{+}(z)$ only changes. We will see later in various examples that this comes from the fact that the charge functions $\Psi^{(s)}_{\Lambda}(z)$ will not have the same number of poles and zeros. We can absorb the $z^{r_{i}}$ part to $K_{i}^{+}(z)$ and redefine it as a new current.\footnote{Later, we will redefine this new current as $\Tilde{K}^{+}_{i}(z)=z^{r_{i}}K^{+}_{i}(z)$. The mode expansion will be $\Tilde{K}_{i}^{+}(z)=\sum_{r\geq 0}\frac{K^{+}_{i,r}}{z^{r-r_{i}}}$ and the modes will be shifted. This is the reason it is called ``shifted" quantum algebra.} After doing this, the defining relations do not change. As long as we derive representations in the current realization, we do not have to be so careful about this part. We also note that the quantum toroidal algebra in (\ref{eq:defofQuiverAlgebra}) is a shifted quantum toroidal algebra with the shift parameters $\mathbf{r}=\mathbf{0}$, and we will call it ``unshifted quantum toroidal algebra". We denote the unshifted quiver quantum toroidal algebra as $\ddot{\mathcal{U}}^{\mathbf{0}}_{Q}\equiv\ddot{\mathcal{U}}_{Q}$.

The definition for the shifted quantum toroidal $\mathfrak{gl}_{n}$ is in \cite{negut2021agt}. See also the references therein. Some examples for the shifted Yangian case were discussed previously in \cite{rapcak2020cohomological,kodera2016quantized}.

Explicit representations for the one-dimensional and two-dimensional crystal mentioned above will be derived in the next section. But before going to the next section, let us derive the charge function of the vacuum configuration of the two-dimensional crystal by considering the relation with the three-dimensional crystal representation (\ref{eq:summaryofalgebraansatz}). The vacuum configuration $\ket{\emptyset}$ is the crystal configuration with no atoms. The action of the generators are summarized in (\ref{eq:summaryofalgebraansatz}) and the basic strategy of this derivation was that the charge function contains poles where atoms are addable and removable. The action of $K_{s}(z)$,\footnote{The action of $K_{s}^{\pm}(z)$ must be understood as a formal expansion of $z^{\mp k}$ $(k\geq 0)$ but we will omit the $\pm$ when it is obvious. } on the vacuum $\ket{\emptyset}$ is 
\begin{align}
K_{s}(z)\ket{\emptyset}=
    \psi_{\emptyset}^{(s)}(z,u)\ket{\emptyset}=(\psi_{\emptyset}(z,u))^{\delta_{s,a}}\ket{\emptyset},
\end{align}
where \begin{align}
    \psi_{\emptyset}(z,u)=\frac{K^{-1/2}z-K^{1/2}u}{z-u}=\frac{\phi(K^{-1};z,u)}{\phi(1;z,u)}.
\end{align}
We note $\phi(p;z,u)=p^{1/2}z-p^{-1/2}u$ and that we set the atom at the origin to have color $a$. 
The denominator of this vacuum charge function says that we can add an atom with color $a$ to the origin. Since the charge function of general crystal configurations is obtained by multiplying bond factors $\varphi^{s\Rightarrow j}(z,uq(\fbox{$j$}))$ to the vacuum charge, nothing will happen if $K$ is generic. If $K$ cancels a pole of this bond factor, then the crystal will stop its growth, and we obtain a truncation of the algebra. Thus, the numerator of the vacuum charge function gives information on where the crystal can stop its growth. Since the action of $E_{s}(z)$ and $F_{s}(z)$ are determined by the pole structure of $\Psi^{(s)}_{\Lambda}(z)$, the crystal will terminate its growth only at an atom of color $a$ even if we set $K$ to special parameters.

To generalize this situation and make the crystal terminate its growth at any color of an atom, we need to add numerators to the vacuum charge function such as 
\begin{align}
    \psi^{(s)}_{\emptyset}(z,u)=\frac{\phi(q(\fbox{$b$})^{-1};z,u)^{\delta_{s,b}}}{\phi(1;z,u)^{\delta_{s,a}}}.
\end{align}
For example, in this case, we can put an atom with color $a$ to the origin, and the crystal will stop its growth at an atom at $q(\fbox{$b$})$ whose color is $b$. As one can see, the charge function does not have the same number of zeros and poles anymore. This is the reason why the algebra acting on subcrystals should be shifted quantum toroidal algebras. 
Although we considered only adding numerators, we can add denominators to the charge function, and this will give extra atoms from which the crystal can start its growth. We only consider the case when there is only one numerator and one denominator in the charge function in this paper.   

Let us use the above method and guess the vacuum charge function of the two-dimensional crystal representation we mentioned above. The subquiver of this crystal is obtained by removing arrows of the perfect matching associated with the corner divisor we are focusing on. Since this unique arrow is included in the perfect matching and connected to the origin atom, the crystal should stop its growth at the atom connected by this unique arrow. Therefore, we get the following claim: Let $q_{\mathfrak{m}}$ be the parameter associated with the unique arrow the zig-zag paths share. When the origin atom is color $a$, and the unique arrow connects this atom to another atom with color $b$, the vacuum charge function is 
\begin{screen}
\begin{align}
\begin{split}
    z^{r_{s}}K^{+}_{s}(z)\ket{\emptyset}=\left[\frac{\phi(q_{\mathfrak{m}}^{-1};z,u)^{\delta_{s,b}}}{\phi(1;z,u)^{\delta_{s,a}}}\right]_{+}\ket{\emptyset},\\
    K^{-}_{s}(z)\ket{\emptyset}=\left[\frac{\phi(q_{\mathfrak{m}}^{-1};z,u)^{\delta_{s,b}}}{\phi(1;z,u)^{\delta_{s,a}}}\right]_{-}\ket{\emptyset}.
\end{split} \label{eq:vacuum_charge_function}
\end{align}
\end{screen}
We will see this is indeed true by taking tensor products of one-dimensional crystal representations in section \ref{sec:examples}.

The shift $\mathbf{r}=(r_{i})_{i\in Q_{0}}$ of the algebra can be obtained as 
\begin{align}
    r_{i}=(\#\hspace{2mm}\text{of}\hspace{2mm}\text{zeros of}\hspace{2mm} \psi^{(i)}_{\emptyset}(z,u))-(\#\hspace{2mm}\text{of}\hspace{2mm}\text{poles of}\hspace{2mm}\psi^{(i)}_{\emptyset}(z,u)).\label{eq:shiftdefQQTA}
\end{align}
In the above case, the shift parameters are 
\begin{align}
    r_{a}=-1,\quad r_{b}=1,\quad r_{c}=0\;(c\neq a, b).\label{eq:shiftparameters}
\end{align}
when $a\neq b$ and 
\begin{align}
    r_{a}=0,\quad r_{c}=0\, (c\neq a)
\end{align}
when $a=b$.
This is to match the degrees of $z$ of both hand sides of (\ref{eq:vacuum_charge_function}) (see Appendix \ref{sec:shiftderiv}).
\subsection{Coproduct structure}\label{sec:shift_coproduct}
As discussed in \cite{Noshita:2021ldl}, the unshifted quiver quantum toroidal algebra $\ddot{\mathcal{U}}^{\mathbf{0}}_{Q}$ is a Hopf superalgebra and possesses a coproduct (\ref{eq:coproduct}), a counit (\ref{eq:counit}), and an antipode (\ref{eq:antipode}). In particular, the map $\Delta:\ddot{\mathcal{U}}^{\mathbf{0}}_{Q}\rightarrow \ddot{\mathcal{U}}^{\mathbf{0}}_{Q}\otimes\ddot{\mathcal{U}}^{\mathbf{0}}_{Q}$ defines a coproduct structure. In the quantum toroidal $\mathfrak{gl}_{1}$ case, this was essential in deriving representations (see section \ref{sec:quantum_toroidalgl1}). Similar to this situation, it is natural to expect that general subcrystals of the original three-dimensional crystal can be obtained by taking tensor products of representations of lower-dimensional crystals. Then, one would like to ask whether the shifted quiver quantum toroidal algebra has a similar property. We show that this is affirmative.

Let us look at the definition (\ref{eq:defofshiftedQuiverAlgebra}) carefully. As mentioned, by absorbing the $z^{r_{i}}$ into $K^{+}_{i}(z)$ as $\widetilde{K}^{+}_{i}(z)=z^{r_{i}}K^{+}_{i}(z)$ while keeping $K_{i}^{-}(z)$ unchanged as $\widetilde{K}^{-}_{i}(z)=K^{-}_{i}(z)$, the definition of $\ddot{\mathcal{U}}^{\mathbf{r}}_{Q}$ can be rewritten as: 
\begin{align}
\begin{split}
    \widetilde{K}_{i}^{\pm}(z)\widetilde{K}_{j}^{\pm}(w)&=\widetilde{K}_{j}^{\pm}(w)\widetilde{K}_{i}^{\pm}(z),\\
    \widetilde{K}_{i}^{-}(z)\widetilde{K}_{j}^{+}(w)&=\widetilde{K}_{j}^{+}(w)\widetilde{K}_{i}^{-}(z),\\
    \widetilde{K}_{i}^{\pm}(z)E_{j}(w)&=\varphi^{j\Rightarrow i}(z,w)E_{j}(w)\widetilde{K}_{i}^{\pm}(z),\\
    \widetilde{K}_{i}^{\pm}(z)F_{j}(w)&=\varphi^{j\Rightarrow i}(z,w)^{-1}F_{j}(w)\widetilde{K}_{i}^{\pm}(z),\\
    [E_{i}(z),F_{j}(w)]=\delta_{i,j}&\delta\left(\frac{w}{z}\right)\left(\widetilde{K}_{i}^{+}(z)-\widetilde{K}_{i}^{-}(w)\right),\\
    E_{i}(z)E_{j}(w)&=(-1)^{|i||j|}\varphi^{j\Rightarrow i}(z,w)E_{j}(w)E_{i}(z),\\
    F_{i}(z)F_{j}(w)&=(-1)^{|i||j|}\varphi^{j\Rightarrow i}(z,w)^{-1}F_{j}(w)F_{i}(z).
\end{split}\label{eq:shiftedQuiverAlgebra_for_rep}
\end{align}

We can generalize the normal coproduct of the unshifted quiver quantum toroidal algebra to a coproduct that gives tensor products of different shifted quantum toroidal algberas as the following.
\begin{align}
\begin{split}
&\Delta_{\mathbf{r},\mathbf{r'}}:\ddot{\mathcal{U}}^{\mathbf{s}}_{Q}\rightarrow \ddot{\mathcal{U}}^{\mathbf{r}}_{Q}\otimes \ddot{\mathcal{U}}^{\mathbf{r'}}_{Q},\quad \mathbf{s}=\mathbf{r}+\mathbf{r'},\\
    &\Delta_{\mathbf{r},\mathbf{r'}} E_{i}(z)=E_{i}(z)\otimes 1+\widetilde{K}_{i}^{-}(z)\otimes E_{i}(z),\\
    &\Delta_{\mathbf{r},\mathbf{r'}} F_{i}(z)=F_{i}(z)\otimes \widetilde{K}_{i}^{+}(z)+1\otimes F_{i}(z),\\
    &\Delta_{\mathbf{r},\mathbf{r'}} \widetilde{K}_{i}^{+}(z)=\widetilde{K}_{i}^{+}(z)\otimes \widetilde{K}_{i}^{+}(z),\\
    &\Delta_{\mathbf{r},\mathbf{r'}} \widetilde{K}_{i}^{-}(z)=\widetilde{K}_{i}^{-}(z)\otimes \widetilde{K}_{i}^{-}(z).\label{eq:shifted_coproduct_tilde}
\end{split}
\end{align}
We abuse the terminology and still call this map a coproduct. This map implies that by using tensor products of representations of shifted quantum toroidal algebras with shift parameters $\mathbf{r}$ and $\mathbf{r'}$, we can obtain representations of shifted quantum toroidal algebra with shift parameter $\mathbf{s}=\mathbf{r}+\mathbf{r'}$. 
We check the well-definedness of (\ref{eq:shifted_coproduct_tilde}) in Appendix \ref{sec:appendix_welldef}.

We note that during the calculation, we use the following property of tensor products of superalgebras:
\begin{align}
    (x\otimes y)(z\otimes w)=(-1)^{|y||z|}xz\otimes yw.
\end{align}
We also have the following property:
\begin{align}
    (1\otimes \Delta_{\mathbf{r_{2}},\mathbf{r_{3}}})\circ\Delta_{\mathbf{r_{1}},\mathbf{r_{2}}+\mathbf{r_{3}}}=(\Delta_{\mathbf{r_{1},\mathbf{r_{2}}}}\otimes 1)\circ \Delta_{\mathbf{r_{1}}+\mathbf{r_{2}},\mathbf{r_{3}}}.\label{eq:shifted_coassoc}
\end{align}
Using this, we can define 
\begin{align}
    \Delta^{(2)}_{\mathbf{r_{1}},\mathbf{r_{2}},\mathbf{r_{3}}}\equiv (1\otimes \Delta_{\mathbf{r_{2}},\mathbf{r_{3}}})\circ\Delta_{\mathbf{r_{1}},\mathbf{r_{2}}+\mathbf{r_{3}}}=(\Delta_{\mathbf{r_{1},\mathbf{r_{2}}}}\otimes 1)\circ \Delta_{\mathbf{r_{1}}+\mathbf{r_{2}},\mathbf{r_{3}}}.
\end{align}
Using (\ref{eq:shifted_coassoc}), we can multiply the generalized coproduct $N-1$ times and obtain:
\begin{align}
\begin{split}
&\Delta_{\mathbf{r_{1}},\mathbf{r_{2},...,\mathbf{r}_{N}}}^{(N-1)}: \ddot{\mathcal{U}}^{\mathbf{r}}_{Q}\rightarrow \ddot{\mathcal{U}}^{\mathbf{r_{1}}}_{Q}\otimes \ddot{\mathcal{U}}^{\mathbf{r_{2}}}_{Q}\otimes\cdots \otimes\ddot{\mathcal{U}}^{\mathbf{r_{N}}}_{Q},\quad \mathbf{r}=\sum_{i=1}^{N}\mathbf{r}_{i},\\
    &\Delta_{\mathbf{r_{1},\mathbf{r_{2}},..,\mathbf{r_{N}}}}^{(N-1)}(\widetilde{K}_{s}^{\pm}(z))=\underbrace{\widetilde{K}_{s}^{\pm}(z)\otimes\cdots \otimes \widetilde{K}_{s}^{\pm}(z)}_{N},\\
    &\Delta_{\mathbf{r_{1},\mathbf{r_{2}},..,\mathbf{r_{N}}}}^{(N-1)}(E_{s}(z))=\sum_{i=1}^{N}\underbrace{\widetilde{K}_{s}^{-}(z)\otimes \cdots \otimes \widetilde{K}_{s}^{-}(z)}_{i-1}\otimes E_{s}(z)\otimes \underbrace{1\otimes \cdots\otimes 1}_{N-i},\\
    &\Delta_{\mathbf{r_{1}},\mathbf{r}_{2},...,\mathbf{r_{N}}}^{(N-1)}(F_{s}(z))=\sum_{i=1}^{N}\underbrace{1\otimes \cdots \otimes 1}_{i-1}\otimes F_{s}(z)\otimes \underbrace{\widetilde{K}_{s}^{+}(z)\otimes \cdots\otimes \widetilde{K}_{s}^{+}(z)}_{N-i},
    \end{split}\label{eq:tensorproduct_coproduct}
\end{align}
where $\Delta_{\mathbf{r_{1}},\mathbf{r_{2}},...,\mathbf{r_{j+1}}}^{(j)}=(1\otimes\Delta_{\mathbf{r_{j}},\mathbf{r_{j+1}}})\circ\Delta_{\mathbf{r_{1},\mathbf{r_{2},...,\mathbf{r_{j}}+\mathbf{r_{j+1}}}}}^{(j-1)}$ for any $j$.

We note that the formulas for the original coproduct is obtained by setting
\begin{align}
\mathbf{r_{i}}=0,\quad (i=1,..,N),\label{eq:unshift_coproduct}
\end{align}
which gives $\mathbf{r}=0$.
We will see in the next section that even though we set (\ref{eq:unshift_coproduct}), after taking the limit $N\rightarrow \infty$, there is a possibility that the representation gains nontrivial shifts. This is similar to the situation of the vector representation of quantum toroidal $\mathfrak{gl}_{1}$. Namely, after taking infinite tensor products of the vector representations with trivial central charges, the resulting representation gains a nontrivial central charge and becomes a Fock representation.

We can also define the generalized antipode and counit structure similarly. However, in this case, the antipode maps a representation of a shifted quantum toroidal algebra to a representation of a \textit{different} shifted quantum toroidal algebra. The shift parameter changes from $\mathbf{r}$ to $\mathbf{-r}$.\\
The generalized antipode map is defined as 
\begin{align}
\begin{split}
&S:\ddot{\mathcal{U}}^{\mathbf{r}}_{Q}\rightarrow \ddot{\mathcal{U}}^{\mathbf{r'}}_{Q},\quad \mathbf{r'}=-\mathbf{r},\\
&S(E_{i}(z))=-(\widetilde{K}_{i}^{-}(z))^{-1}E_{i}(z),\\
&S(F_{i}(z))=-F_{i}(z)(\widetilde{K}_{i}^{+}(z))^{-1},\\
&S(\widetilde{K}_{i}^{\pm}(z))=(\widetilde{K}_{i}^{\pm}(z))^{-1}.
\end{split}\label{eq:shift_antipode}
\end{align}
The $\widetilde{K}_{i}^{+}(z)$ of the left hand side of the last equation of (\ref{eq:shift_antipode}) should be understood as $\widetilde{K}_{i}^{+}(z)=z^{r_{i}}K_{i}^{+}(z)$, while the right hand side is understood as $\widetilde{K}_{i}^{+}(z)=z^{r'_{i}}K_{i}^{+}(z)$. Comparing both hand sides, we obtain $\mathbf{r'}=-\mathbf{r}$. Using this map, we can obtain representations of different shifted quantum toroidal algebra. We note that the original antipode is obtained by specializing the shift parameters to $\mathbf{r'}=\mathbf{r}=0$.

The counit map does not change and is defined as:
\begin{align}
\begin{split}
&\epsilon:\ddot{\mathcal{U}}^{\mathbf{r}}_{Q}\rightarrow \mathbb{C},\\
&\epsilon(E_{i}(z))=\epsilon(F_{i}(z))=0,\\
&\epsilon(\widetilde{K}_{i}^{\pm}(z))=1.\\
\end{split}\label{eq:shift_counit}
\end{align}

\paragraph{Comments on superalgebra}
When we consider the above coproduct action on vectors, we have to be aware of the parity assignments. We have to assign parities to the vectors of the vector space. The action of the tensor product of the generators should obey the following:
\begin{align}
\begin{split}
    &(x\otimes y)\,(v\otimes w), \quad x,y\in\mathcal{E},\quad v\in V,\quad w\in W,\\
    =&(-1)^{|y||v|}xv\otimes yw,
\end{split}
\end{align}
where $\mathcal{E}$ is the algebra and $V,W$ are some modules.
Thus for example,
\begin{align}
\begin{split}
    &(\underbrace{1\otimes \cdots \otimes 1}_{i-1}\otimes E_{s}(z)\otimes \underbrace{1\otimes \cdots\otimes 1}_{N-i})\,(v_{1}\otimes \cdots\otimes v_{i-1}\otimes v_{i}\otimes v_{i+1}\otimes\cdots\otimes v_{N})\\
    =&(-1)^{|s|(\sum_{j=1}^{i-1}|v_{j}|)}v_{1}\otimes \cdots\otimes v_{i-1}\otimes E_{s}(z) v_{i}\otimes v_{i+1}\otimes\cdots\otimes v_{N},
\end{split}
\end{align}
where $|v_{i}|$ are the parities of the vectors $v_{i}$.
We also note that these parities are not determined uniquely. For example, when 
\begin{align}
    E_{i}(z)v_{1}=v'_{1} 
\end{align}
is satisfied, the parities should obey 
\begin{align}
    |v'_{1}|-|v_{1}|=|i|.\label{eq:parityrelative}
\end{align}
This comes from the comparison of the parities of both hand sides.
\subsection{Example of conifold and quantum toroidal \texorpdfstring{$\mathfrak{gl}_{1|1}$}{gl11}}\label{sec:conifolddetail}
Representations of quantum toroidal $\mathfrak{gl}_{m|n}\hspace{1mm}(m\neq n)$ have been studied in \cite{Feigin_2019glmn,bezerra2019quantum,bezerra2021representations}. Representations were not studied for the $m=n$ case. In this section, we study when $m=n=1$. We derive the one and two-dimensional crystal representations for the conifold in detail. Other examples are in the next section.
\subsubsection{Definition of algebra}


We derive the defining algebras of quantum toroidal $\mathfrak{gl_{1|1}}$ which is the quantum toroidal algebra associated with the conifold geometry. The toric diagram, periodic quiver diagram, and quiver diagram of it are written in Figure \ref{fig:conifold-brane-tiling}. The quiver diagram has two vertices, and we color them blue and red. There are two arrows from one of the vertices to the other one. Vertex 1 is blue, and vertex 2 is red. Since there are no loops for both vertices, they are fermionic:
\begin{align}
    |1|=|2|=1.
\end{align}By assigning 4 parameters $\alpha_{1},\alpha_{2},\beta_{1},\beta_{2}$ to the arrows between the vertices and using the loop constraint and vertex constraint,
we obtain two independent parameters 
\begin{align}
    \alpha_{1}=q_{1},\quad \alpha_{2}=q_{1}^{-1}, \quad \beta_{1}=q_{2},\quad \beta_{2}=q_{2}^{-1}.
\end{align}
Using these, the bond factors can be read of from (\ref{eq:defstruc}):  
\begin{align}
\begin{split}
    &\varphi^{1\Rightarrow 1}(z,w)=\varphi^{2\Rightarrow 2}(z,w)=1,\\
    &\varphi^{1\Rightarrow 2}(z,w)=\frac{(q_{2}^{1/2}z-q_{2}^{-1/2}w)(q_{2}^{-1/2}z-q_{2}^{1/2}w)}{(q_{1}^{-1/2}z-q_{1}^{1/2}w)(q_{1}^{1/2}z-q_{1}^{-1/2}w)}=\frac{\phi(q_{2};z,w)\phi(q_{2}^{-1};z,w)}{\phi(q_{1};z,w)\phi(q_{1}^{-1};z,w)},\\
    &\varphi^{2\Rightarrow 1}(z,w)=\frac{(q_{1}^{1/2}z-q_{1}^{-1/2}w)(q_{1}^{-1/2}z-q_{1}^{1/2}w)}{(q_{2}^{1/2}z-q_{2}^{-1/2}w)(q_{2}^{-1/2}z-q_{2}^{1/2}w)}=\frac{\phi(q_{1};z,w)\phi(q_{1}^{-1};z,w)}{\phi(q_{2};z,w)\phi(q_{2}^{-1};z,w)}.
    \end{split}
\end{align}

\subsubsection{One-dimensional crystal representation}
\begin{figure}[ht]
    \centering
    \includegraphics[width=11cm]{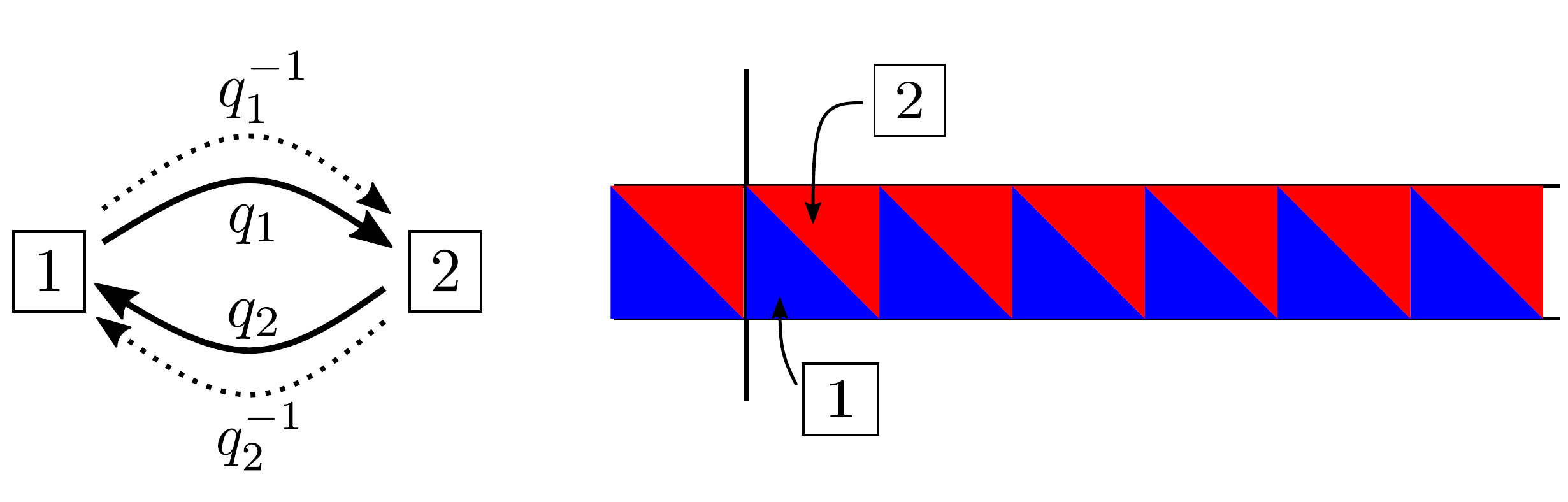}
    \caption{Subquiver and shape of one-dimensional crystal associated with external leg $\ell_{2}$. The external leg is surrounded by two divisors $p_{2}$ and $p_{4}$ so the union set of the perfect matching is $m_{2}\cup m_{4}=\{\alpha_{2},\beta_{2}\}$. We note $\alpha_{2}=q_{1}^{-1}$ and $\beta_{2}=q_{2}^{-1}$.  } 
    \label{fig:gl_(1,1)_subquiver_1dimcrystalshape}
\end{figure}
In this section we construct the one-dimensional crystal representation of Figure \ref{fig:conifold1dcrystalvector} and  \ref{fig:gl_(1,1)_subquiver_1dimcrystalshape}.

To construct the representation, we assign coordinates to the triangles. We assign coordinates $q_{1}^{j}q_{2}^{j}$ to the blue triangles, and $q_{1}^{j+1}q_{2}^{j}$ to the red triangles, where $j\in\mathbb{Z}$ is the number of blue triangles on the right of the border counted as $0,1,...$ and continuously extended to the left of the border. This is illustrated in Figure \ref{fig:gl_(1,1)coordinate}.

\begin{figure}[h]
      \begin{minipage}{1\hsize}
        \centering
        \includegraphics[width=8cm]{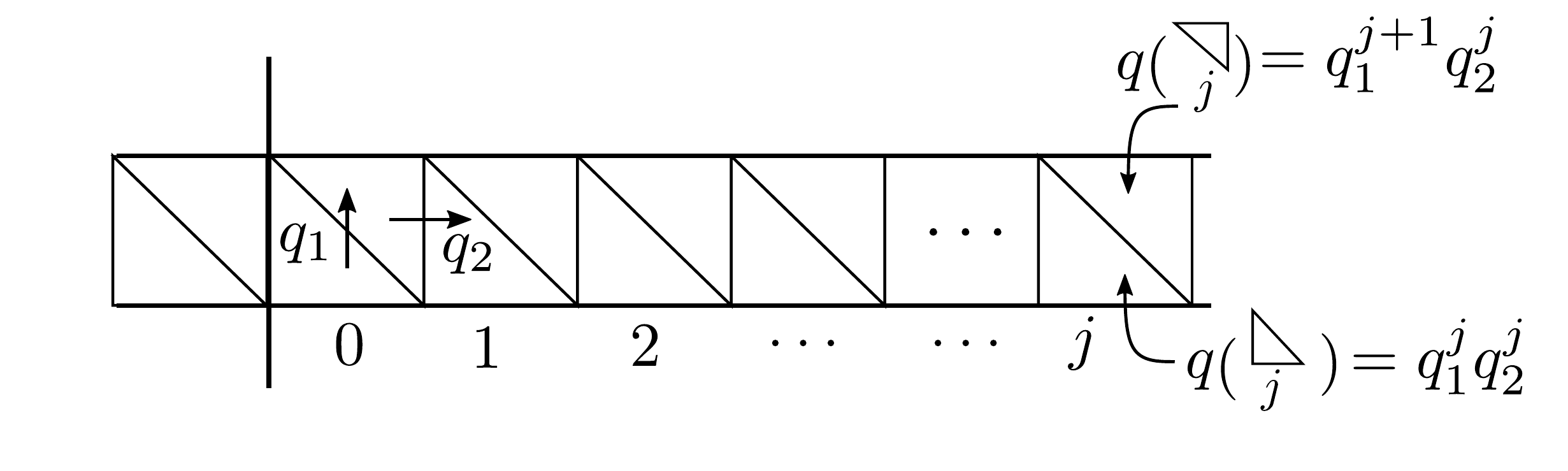}
        \subcaption{Coordinates of triangles of $\mathfrak{gl}_{1|1}$. }\label{fig:gl_(1,1)coordinate}
      \end{minipage}\\
      \begin{minipage}{1\hsize}
        \centering
       \includegraphics[width=9cm]{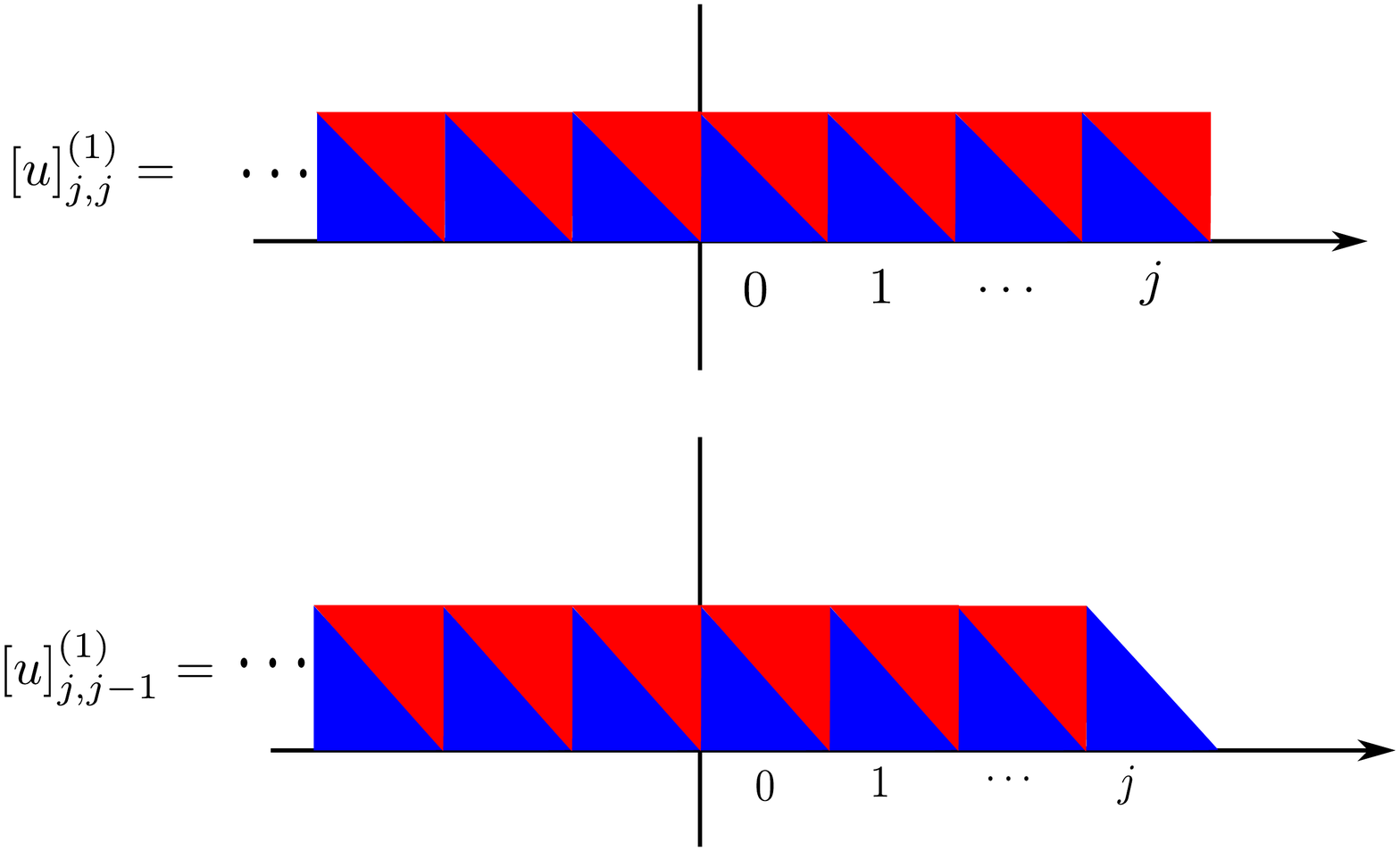}
       \subcaption{Two types of vectors in $V^{(\ell_{2})}(u)$.}\label{fig:gl_(1,1)vectorcolor}
      \end{minipage}
\caption{Coordinates and basis of the one-dimensional crystal representation. (a) The triangle with the oblique side up is the blue triangle and the triangle with the oblique side down is the red triangle. The blue triangle has coordinate $q_{1}^{j}q_{2}^{j}$, while the red triangle has coordinate $q_{1}^{j+1}q_{2}^{j}$. (b) $\left[u\right]^{(1)}_{j,j}$ has $j+1$ blue triangles and $j+1$ red triangles right to the border, while $\left[u\right]^{(1)}_{j,j-1}$ has $j+1$ blue and $j$ red triangles. }
\end{figure}

The bases of this representation are $[u]_{j,j}^{(1)}$ and $[u]_{j,j-1}^{(1)}$ where $j\in\mathbb{Z}$. The vectors can be illustrated as Figure \ref{fig:gl_(1,1)vectorcolor}.  We denote the vector space spanned by these bases $V^{(\ell_{2})}(u)$. Vector $[u]_{j,j}^{(1)}$ represents a row of triangles with $j+1$ blue triangles and $j+1$ red triangles, while vector $[u]_{j,j-1}^{(1)}$ represents a row of triangles with $j+1$ blue and $j$ red triangles. The upper index $(1)$ is the origin vertex number, and $u$ is the spectral parameter. The melting rule of \cite{Nishinaka_2011,Nishinaka_2012,Nishinaka_2014} claims that to add a blue triangle 
to the partition, we need to have a red triangle at the left of it, and to add a red triangle, we need to have a blue triangle at the left of it. From this observation, the above vectors are the only possible ones. The removing rules are also the same: to remove a blue triangle, we can not have a red triangle to its right, and to remove a red triangle, we can not have a blue triangle to its right.
Thus, we can say $E_{1}(z)$ $(E_{2}(z))$ adds a blue (red) triangle to the partition, $F_{1}(z)$ $(F_{2}(z))$ removes a blue(red) triangle from the partition and $K_{i}^{\pm}(z)$ acts diagonally. In the above convention, $E_{1}(z)$ $(F_{1}(z))$ increases (decreases) the first subscript of $[u]_{i,j}^{(1)}$ and $E_{2}(z)$ $(F_{2}(z))$ acts on the second subscript:
\begin{align}
\begin{split}
    E_{1}(z)\begin{cases}[u]^{(1)}_{k,k}\\ [u]^{(1)}_{k,k-1}\end{cases}&=\begin{cases}\mathcal{E}_{1}\left([u]^{(1)}_{k,k}\right)\delta\left(\frac{z}{u(q_{1}q_{2})^{k+1}}\right)[u]^{(1)}_{k+1,k},\\0,\end{cases}\\ E_{2}(z)\begin{cases}[u]^{(1)}_{k,k}\\
    [u]^{(1)}_{k,k-1}\end{cases}&=\begin{cases}0,\\\mathcal{E}_{2}\left([u]^{(1)}_{k,k-1}\right)\delta\left(\frac{z}{uq_{1}(q_{1}q_{2})^{k}}\right)[u]^{(1)}_{k,k},\end{cases}\\
    F_{1}(z)\begin{cases}[u]^{(1)}_{k,k}\\ [u]^{(1)}_{k,k-1}\end{cases}&=\begin{cases} 0,\\ \mathcal{F}_{1}\left([u]^{(1)}_{k,k-1}\right)\delta\left(\frac{z}{u(q_{1}q_{2})^{k}}\right)[u]^{(1)}_{k-1,k-1},\end{cases}\\
    F_{2}(z)\begin{cases} [u]^{(1)}_{k,k}\\ [u]^{(1)}_{k,k-1} \end{cases}&=\begin{cases} \mathcal{F}_{2}\left([u]^{(1)}_{k,k-1}\right) \delta\left(\frac{z}{uq_{1}(q_{1}q_{2})^{k}}\right)[u]^{(1)}_{k,k-1},\\ 0,\end{cases}
    \end{split}\label{eq:gl_(1,1)_1dimrep_EFdef}\\
\begin{split}
    K_{i}^{\pm}(z)\begin{cases}[u]^{(1)}_{k,k}\\ [u]^{(1)}_{k,k-1}\end{cases}&=\begin{cases}\left[\Psi_{[u]^{(1)}_{k,k}}^{(i)}(z)\right]_{\pm}[u]^{(1)}_{k,k},\\ \left[\Psi_{[u]^{(1)}_{k,k}}^{(i)}(z)\right]_{\pm}[u]^{(1)}_{k,k-1}.\end{cases}
    \end{split}\label{eq:gl_(1,1)_1dimrep_Kdef}
\end{align}

From the KE relations of (\ref{eq:shiftedQuiverAlgebra_for_rep}) 
we obtain the following recursion formulas:
\begin{align}
    \frac{\Psi_{[u]^{(1)}_{k+1,k}}^{(i)}(z)}{\Psi_{[u]^{(1)}_{k,k}}^{(i)}(z)}=\varphi^{1\Rightarrow i}(z,uq_{1}^{k+1}q_{2}^{k+1}),\quad \frac{\Psi_{[u]^{(1)}_{k,k}}^{(i)}(z)}{\Psi_{[u]^{(1)}_{k,k-1}}^{(i)}(z)}=\varphi^{2\Rightarrow i}(z,uq_{1}^{k+1}q_{2}^{k}).
\end{align}
By direct calculation, we obtain 
\begin{align}
\begin{split}
\Psi_{[u]^{(1)}_{k,k-1}}^{(1)}(z)=\frac{\phi(q_{1}^{-k-1}q_{2}^{-k+1};z,u)}{\phi(q_{1}^{-k}q_{2}^{-k};z,u)},\quad &
\Psi_{[u]^{(1)}_{k,k}}^{(1)}(z)=\frac{\phi(q_{1}^{-2-k}q_{2}^{-k};z,u)}{\phi(q_{1}^{-k-1}q_{2}^{-k-1};z,u)},\\
\Psi_{[u]^{(1)}_{k,k-1}}^{(2)}(z)=\frac{\phi(q_{1}^{-k}q_{2}^{-1-k};z,u)}{\phi(q_{1}^{-1-k}q_{2}^{-k};z,u)},\quad &
\Psi_{[u]^{(1)}_{k,k}}^{(2)}(z)=\frac{\phi(q_{1}^{-k}q_{2}^{-1-k};z,u)}{\phi(q_{1}^{-1-k}q_{2}^{-k};z,u)},
\end{split}\label{eq:gl_(1,1)chargefunction}
\end{align}
where we used an analogue of (\ref{eq:infiniteproductgl11d}).
We leave the coefficient factors undetermined because we do not use them. As one can see, since the charge functions have the same numbers of poles and zeros, this is a representation of the unshifted quiver quantum toroidal algebra. The shift parameters are 
\begin{align}
    r_{1}=0,\quad r_{2}=0.
\end{align}

\paragraph*{Parity assignments}
From the above equations, we obtain the following parity condition 
\begin{align}
    \begin{split}
        |[u]^{(1)}_{k+1,k}|=|[u]^{(1)}_{k,k}|+|1|,\\
        |[u]^{(1)}_{k,k}|=|[u]^{(1)}_{k,k-1}|+|2|.
    \end{split}
\end{align}
Since $|1|=|2|=1$, these give 
\begin{align}
    |[u]_{k+1,k+1}|=|[u]_{k,k}|,\quad |[u]_{k+1,k}|=|[u]_{k,k-1}|.
\end{align}

\subsubsection{Two-dimensional crystal representation}
\begin{figure}[ht]
    \centering
    \includegraphics[width=10cm]{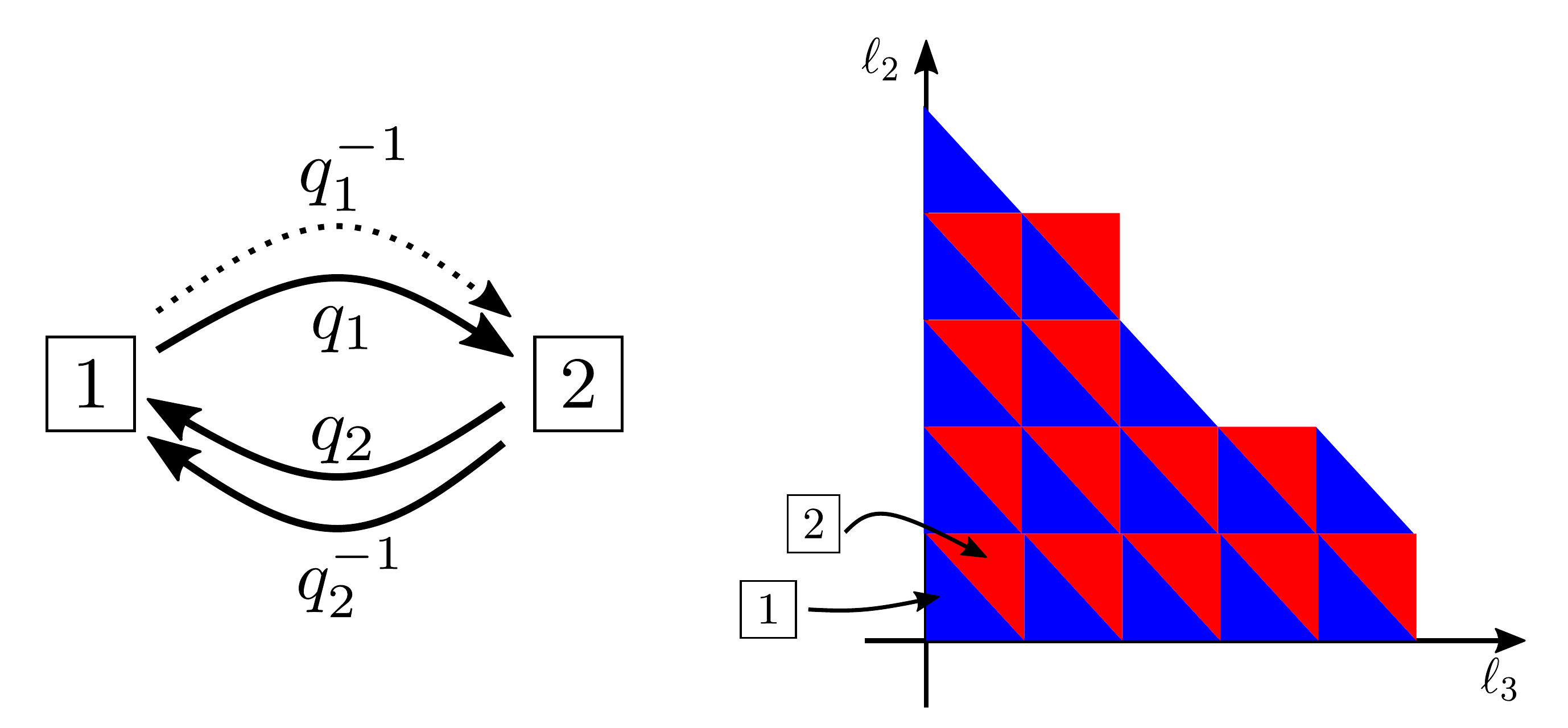}
    \caption{Subquiver and shape of the two-dimensional crystal associated with divisor $p_{2}=(1,0)$. It is obtained by removing arrows of the corresponding perfect matching $m_{2}=\{\alpha_{2}\}$ from the original quiver diagram. We note $\alpha_{2}=q_{1}^{-1}$. The labels $\ell_{2}$ and $\ell_{3}$ come from the fact that the divisor is surrounded by two external legs $\ell_{2}$ and $\ell_{3}$ of the toric diagram.}
    \label{fig:gl_(1,1)2dim_subquiver_crystalshape}
\end{figure}

In this section we derive the two-dimensional crystal representation in Figure \ref{fig:conifold2dcrystalliftup} and  \ref{fig:gl_(1,1)_subquiver_1dimcrystalshape} and by taking tensor products of $V^{(\ell_{2})}(u)$.

We set the parity assignments as 
\begin{align}
    \begin{split}
    |[u]_{k,k}|=0,\quad |[u]_{k+1,k}|=1.    
    \end{split}
\end{align}

We can similarly assign coordinates to the triangles in the two-dimensional plane as illustrated in Figure \ref{fig:gl_(1,1)_2dimcrystal_coordinates}. The triangle whose oblique side is looking up with horizontal parameter $i$ and vertical parameter $j$ has coordinate $q_{1}^{i+j}q_{2}^{i-j}$, while the triangle whose oblique side is looking down has coordinate $q_{1}^{i+j+1}q_{2}^{i-j}$.
\begin{figure}[H]
    \centering
    \includegraphics[width=10cm]{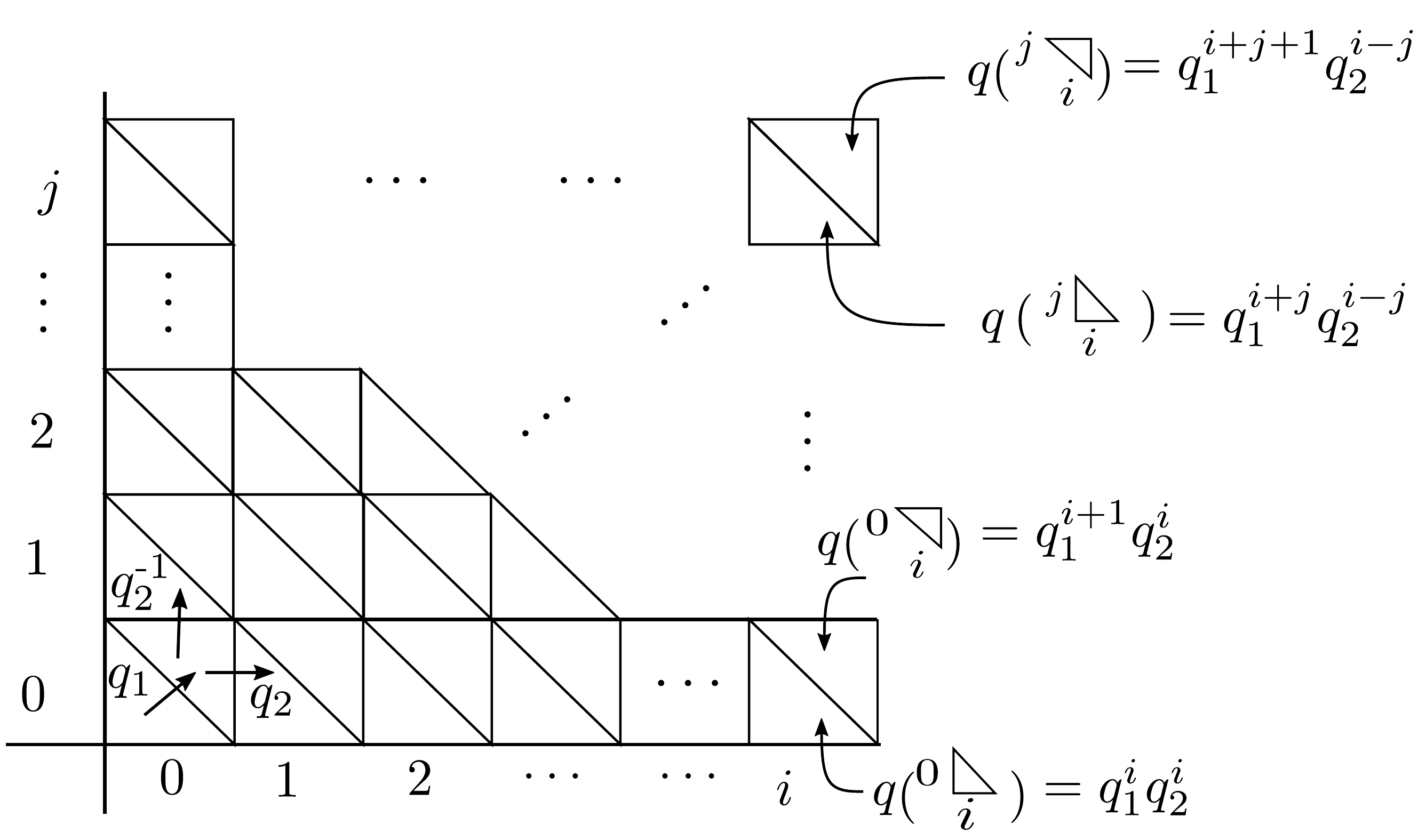}
    \caption{Coordinates of two-dimensional crystal. Each triangle is inside a box which has horizontal and vertical coordinates $(i,j)$. The blue triangle with horizontal parameter $i$ and vertical parameter $j$ has coordinate $q_{1}^{i+j}q_{2}^{i-j}$, while the red triangle has $q_{1}^{i+j+1}q_{2}^{i-j}$. }
    \label{fig:gl_(1,1)_2dimcrystal_coordinates}
\end{figure}

Let us consider actions of generators on tensor product $V^{(\ell_{2})}(u)\otimes V^{(\ell_{2})}(v)$. 
This vector space consists of four types of vectors:
\begin{align}
\begin{split}
    [u]_{i,i}^{(1)}\otimes[v]_{j,j}^{(1)},\quad &[u]_{i,i}^{(1)}\otimes[v]_{j,j-1}^{(1)},\\
    [u]_{i,i-1}^{(1)}\otimes[v]_{j,j}^{(1)},\quad 
    &[u]_{i,i-1}^{(1)}\otimes[v]_{j,j-1}^{(1)}.
    \end{split}
\end{align}
We consider the action of $E_{s}(z)$ on each of this vectors and see when the action is ill-defined and when the action makes a subspace. Let us consider the action of $E_{s}(z)$ on these vectors using 
\begin{align}
    \Delta(E_{s}(z))=E_{s}(z)\otimes 1+K_{s}^{-}(z)\otimes E_{s}(z).\label{eq:gl1|1coproduct}
\end{align}
Note that the nontrivial part comes from the second term of (\ref{eq:gl1|1coproduct}). The fist term actions are: $E_{1}(z)\otimes 1$ acts as zero when the first tensor component is $[u]_{i,i-1}^{(1)}$, $E_{2}(z)\otimes 1$ acts as zero when the first tensor component is $[u]_{i,i}^{(1)}$, otherwise the terms will not be zero due to (\ref{eq:gl_(1,1)_1dimrep_EFdef}).

We set $v=uq_{1}q_{2}^{-1}$. The reason of this choice can be seen from Figure \ref{fig:gl_(1,1)_2dimcrystal_coordinates}. 
The action of the second term of $\Delta(E_{s}(z))$ can be summarized as follows:
\begin{enumerate}
\item $[u]_{i,i}^{(1)}\otimes[v]_{j,j}^{(1)}$\\
   The action of $K_{1}^{-}(z)\otimes E_{1}(z)$ on $[u]_{i,i}^{(1)}\otimes[v]_{j+1,j}^{(1)}$ becomes zero only when $i=j$. On the other hand, $K_{2}^{-}(z)\otimes E_{2}(z)$ acts as zero always because of (\ref{eq:gl_(1,1)_1dimrep_EFdef}). Thus, vectors with the condition $i\geq j$ span a submodule. See Figure \ref{fig:submoduleoftwovector}.
\item $[u]_{i,i}^{(1)}\otimes[v]_{j,j-1}^{(1)}$\\
    $K_{1}^{-}(z)\otimes E_{1}(z)$ always acts as zero, while the action of $K_{2}^{-}(z)\otimes E_{2}(z)$ will not be zero and it will always extend the second tensor component. See Figure \ref{fig:submoduleoftwovector}.
 \begin{figure}[H]
    \centering
    \includegraphics[width=15cm]{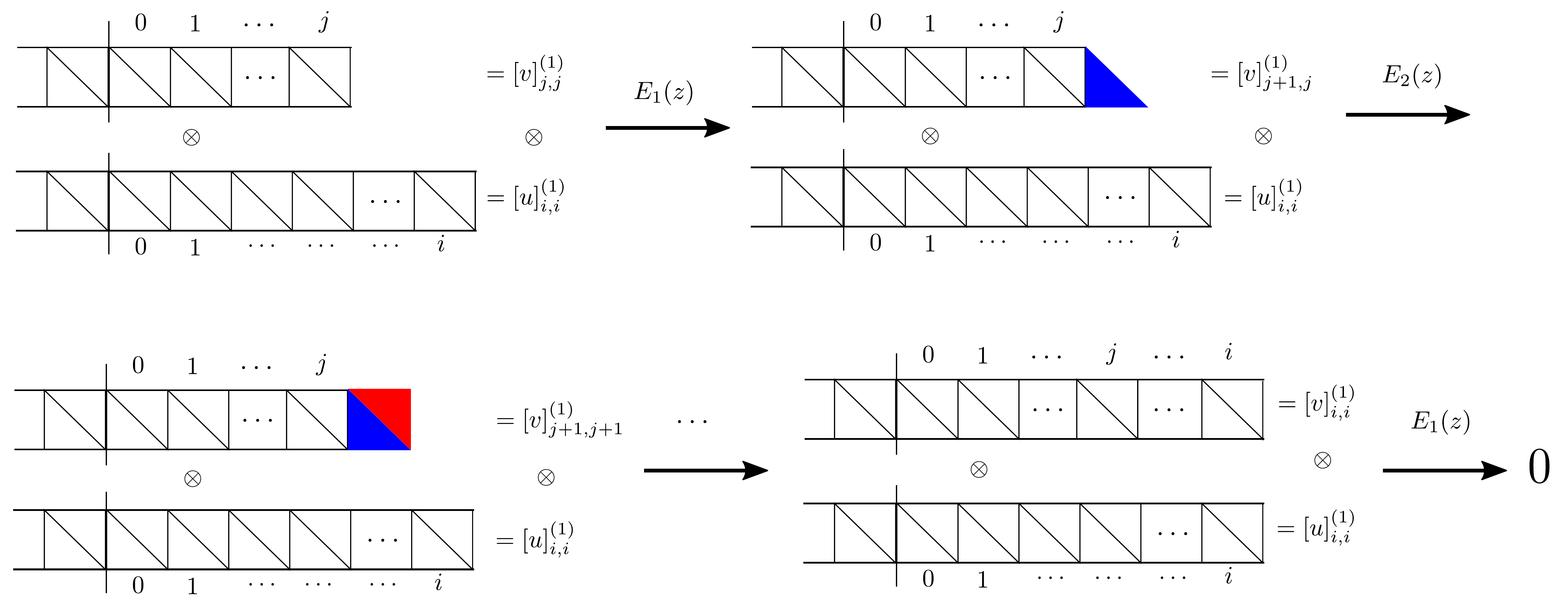}
    \caption{Sequence of actions of $E_{1}(z)$ and $E_{2}(z)$ on the second tensor components of  $[u]_{i,i}^{(1)}\otimes[v]_{j,j}^{(1)}$ and $[u]_{i,i}^{(1)}\otimes[v]_{j,j-1}^{(1)}$.} 
    \label{fig:submoduleoftwovector}
\end{figure}
\item $[u]_{i,i-1}^{(1)}\otimes[v]_{j,j}^{(1)}$\\
The action of $K_{1}^{-}(z)\otimes E_{1}(z)$ is well defined and becomes zero only when $i-j=1$. On the other hand, $K_{2}^{-}(z)\otimes E_{2}(z)$ always acts as zero on this vector. Thus, vectors with the condition $i>j$ span a sub-module. See Figure \ref{fig:submoduleoftwovector2}.
\item $[u]_{i,i-1}^{(1)}\otimes[v]_{j,j-1}^{(1)}$\\
$K_{1}^{-}(z)\otimes E_{1}(z)$ always acts as zero, while $K_{2}^{-}(z)\otimes E_{2}(z)$ always gives a nonzero contribution and extends the second tensor component. See Figure \ref{fig:submoduleoftwovector2}.
     \begin{figure}[H]
    \centering
    \includegraphics[width=15cm]{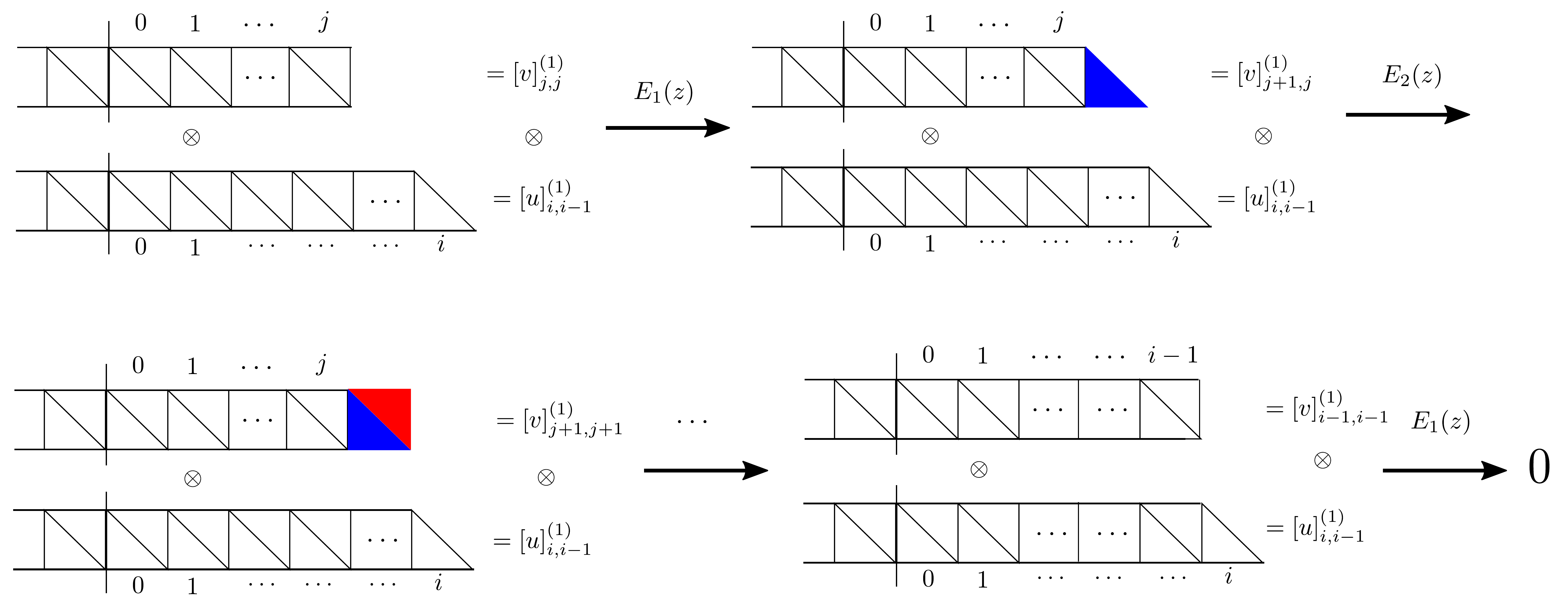}
    \caption{Sequence of actions of $E_{1}(z)$ and $E_{2}(z)$ on the second tensor components of  $[u]_{i,i-1}^{(1)}\otimes[v]_{j,j}^{(1)}$ and $[u]_{i,i-1}^{(1)}\otimes[v]_{j,j-1}^{(1)}$.} 
    \label{fig:submoduleoftwovector2}
    \end{figure}
\end{enumerate}
Figure \ref{fig:submoduleoftwovector} and Figure \ref{fig:submoduleoftwovector2} indeed reproduce the melting rule proposed in \cite{Nishinaka_2011,Nishinaka_2012,Nishinaka_2014}.

We consider next the action of the generators on arbitrary numbers of tensor products. The representation we are considering here is $V^{(\ell_{2})}(u)\otimes V^{(\ell_{2})}((q_{1}q_{2}^{-1})u)\otimes\cdots\otimes V^{(\ell_{2})}((q_{1}q_{2}^{-1})^{r-1}u)$. From now we omit the superscript (1).
For each row we assign two numbers, $\lambda_{i}\in\mathbb{Z}_{\geq0}$ and $\bar{\sigma}_{i}\in\mathbb{Z}_{2}=\{0,1\}$. $\lambda_{i}$ is the length as in Figure \ref{fig:gl_(1,1)_Youngdiagram}, and $\bar{\sigma}_{i}$ is the signature of this row which determines the shape of the row. We denote the sequence of these numbers as
\begin{align}
    (\lambda,\bar{\sigma})=((\lambda_{1},\bar{\sigma}_{1}),(\lambda_{2},\bar{\sigma}_{2}),....,(\lambda_{r},\bar{\sigma}_{r})).
\end{align}
The correspondence of the generalized partitions and vectors is illustrated in Figure \ref{fig:gl_(1,1)_Youngdiagram}. We denote $(\lambda,0)$ as a row of triangles with length $\lambda$. There are $\lambda$ blue and red triangles. We denote $(\lambda,1)$ as a row with length $\lambda$, defined as the length of the bottom side of the row. There are $\lambda$ blue and $\lambda-1$ red triangles.
\begin{figure}[t]
    \begin{tabular}{c}
      \begin{minipage}{0.95\hsize}
        \centering
        \includegraphics[width=9cm]{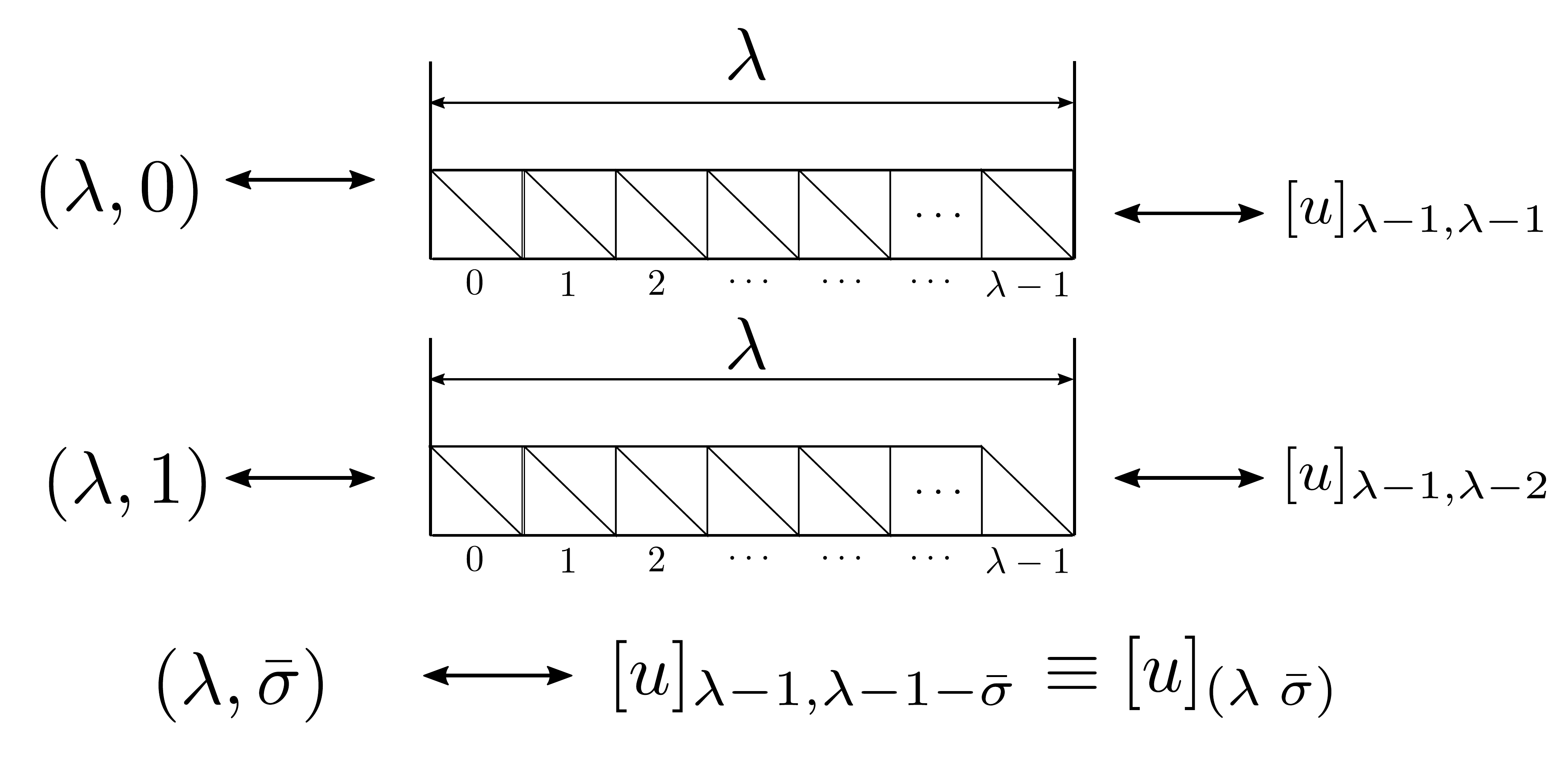}
        \subcaption{Generalization of Young diagram and correspondence with vectors.}\label{fig:gl_(1,1)_Youngdiagram}
      \end{minipage}\\
      \begin{minipage}{0.95\hsize}
        \centering
       \includegraphics[width=9cm]{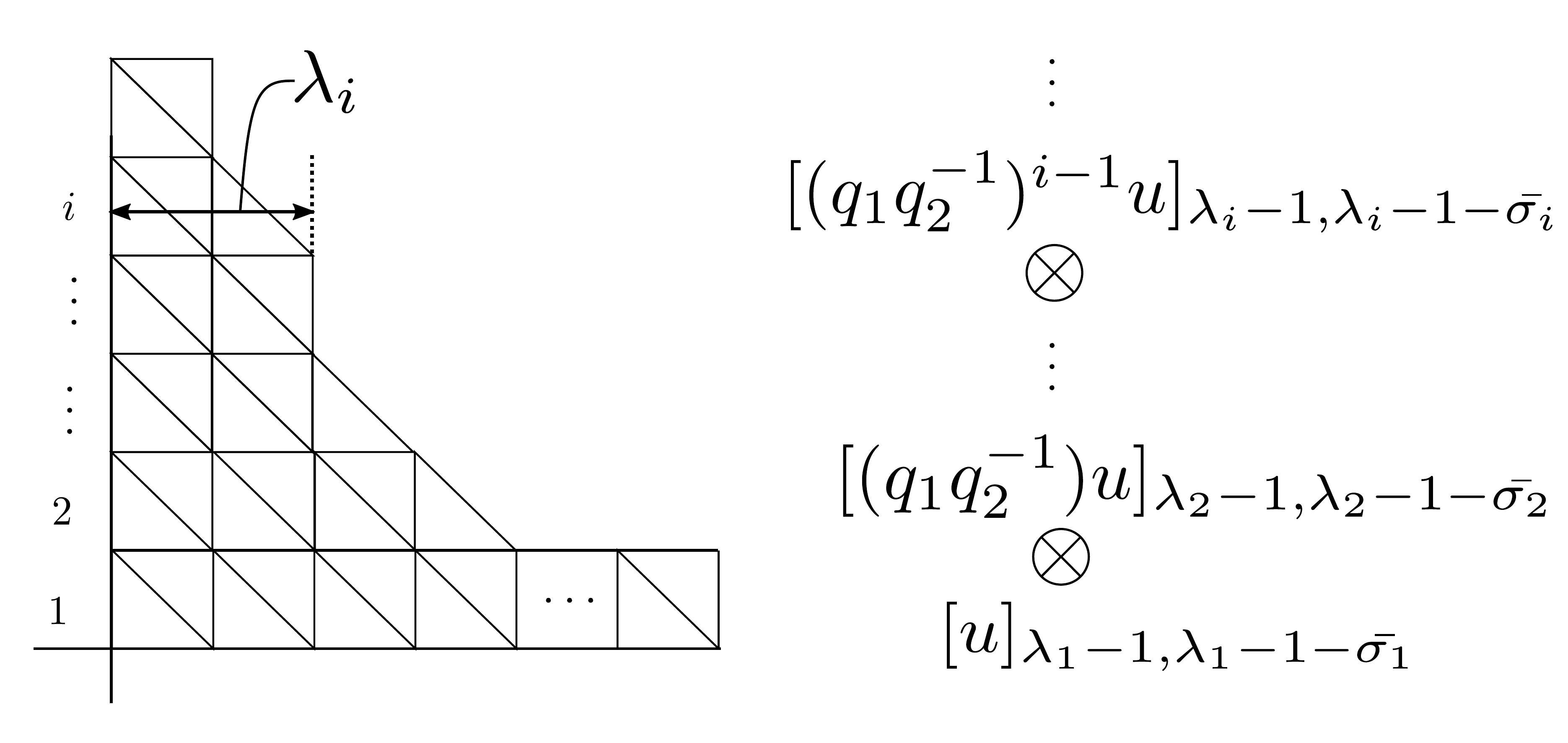}
       \subcaption{An example of the vectors in $\otimes V^{(\ell_{2})}(u)$.}\label{fig:gl_(1,1)_tensor_product_vector}
      \end{minipage}
    \end{tabular}
\caption{Generalization of Young diagram and tensor products. (a) $[u]_{(\lambda,0)}$ has $\lambda$ blue and $\lambda$ red triangles, while $[u]_{(\lambda,1)}$ has $\lambda$ blue and $\lambda-1$ red triangles in the right of the border. (b) Triangle partition model \cite{Nishinaka_2011} can be obtained by taking tensor products of $[(q_{1}q_{2}^{-1})^{i-1}u]_{(\lambda_{i},\bar{\sigma}_{i})}$. }
\end{figure}

The configuration in Figure \ref{fig:gl_(1,1)_tensor_product_vector} can be defined as
\begin{align}
    \ket{\lambda,\bar{\sigma}}&\equiv \prod_{i=1}^{r}\otimes[(q_{1}q_{2}^{-1})^{i-1}u]_{\lambda_{i}-1,\lambda_{i}-1-\bar{\sigma}_{i}}\nonumber=\prod_{i=1}^{r}\otimes[(q_{1}q_{2}^{-1})^{i-1}u]_{(\lambda_{i},\bar{\sigma}_{i})}\nonumber\\
    &\in V^{(1)}(u)\otimes V^{(1)}((q_{1}q_{2}^{-1})u)\otimes\cdots\otimes V^{(1)}((q_{1}q_{2}^{-1})^{r-1}u).
\end{align}
The melting rule is, for $i<j$
\begin{align}
\begin{split}
    \quad(\lambda_{i},0)\geq(\lambda_{j},0),\quad &(\lambda_{i},0)\geq(\lambda_{j},1),\\
(\lambda_{i},1)>(\lambda_{j},0),\quad &(\lambda_{i},1)>(\lambda_{j},1),
    \end{split}
\end{align}
where $(\lambda_{i},\bar{\sigma}_{i})>(\lambda_{j},\bar{\sigma}_{j})$ means $\lambda_{i}>\lambda_{j}$ and $(\lambda_{i},\bar{\sigma}_{i})\geq(\lambda_{j},\bar{\sigma}_{j})$ means $\lambda_{i}\geq\lambda_{j}$.

From now on, we write $E_{2}(z)\equiv E_{0}(z)$ and when we write $\bar{s}$, we are thinking it as an element in $\mathbb{Z}_{2}=\{0,1\}$. Using this convention, (\ref{eq:gl_(1,1)_1dimrep_EFdef}), (\ref{eq:gl_(1,1)_1dimrep_Kdef}) and (\ref{eq:gl_(1,1)chargefunction}) can be rewritten as the following:
\begin{align}
    \begin{split}
    E_{s}(z)[u]_{k,k-\bar{\sigma}}&=\mathcal{E}_{s}\left([u]_{k,k-\bar{\sigma}}\right)\delta\left(\frac{z}{uq_{1}^{k+1}q_{2}^{k+\overline{(\sigma+1)}}}\right) \overline{\delta}_{s+\sigma,1}[u]_{k+\bar{s},k-\overline{(\sigma+1)}},\\
    F_{s}(z)[u]_{k,k-\bar{\sigma}}&=\mathcal{F}_{s}\left([u]_{k,k-\bar{\sigma}}\right)\delta\left(\frac{z}{uq_{1}^{k+1-\bar{s}}q_{2}^{k}}\right)\overline{\delta}_{s+\sigma,0}[u]_{k-\bar{s},k-\overline{(\sigma+1)}},\\
    \Psi^{(1)}_{[u]_{k,k-\bar{\sigma}}}(z)&=\frac{\phi(q_{1}^{-1-k-\overline{(\sigma+1)}}q_{2}^{-k+\overline{(\sigma+1)}};z,u)}{\phi(q_{1}^{-k-\overline{(\sigma+1)}}q_{2}^{-k-\overline{(\sigma+1)}};z,u)},\\
    \Psi^{(2)}_{[u]_{k,k-\bar{\sigma}}}(z)&=\frac{\phi(q_{1}^{-k}q_{2}^{-1-k};z,u)}{\phi(q_{1}^{-1-k}q_{2}^{-k};z,u)},\\
     |[u]_{k,k-\overline{\sigma}}|&=\overline{\sigma}
    \end{split}
\end{align}
We note \begin{align}
    \bar{\delta}_{i,j}\equiv\begin{dcases}
    1,\quad i\equiv j\quad(\text{mod} \;2)\\
    0, \quad i\not\equiv j \quad (\text{mod} \;2).
    \end{dcases}
\end{align}
Let us consider the action of the generators $E_{s}(z),F_{s}(z)$ and $K_{s}(z)$ on the vector \begin{align}
\ket{\lambda,\bar{\sigma}}=\otimes_{i=1}^{N}\left[(q_{1}q_{2}^{-1})^{i-1}u\right]_{\lambda_{i}-1,\lambda_{i}-1-\bar{\sigma_{i}}},
\end{align}
where $\lambda=(\lambda_{1},\lambda_{2},...\lambda_{N})\in\mathbb{Z}^{N}$ and $\bar{\sigma}=(\bar{\sigma}_{1},\bar{\sigma}_{2},....,\bar{\sigma}_{N})\in\mathbb{Z}_{2}^{N}$. $(\lambda,\bar{\sigma})\in\mathbb{Z}^{N}\times\mathbb{Z}_{2}^{N}$ can be naturally embedded in $(\lambda,\bar{\sigma})\in\mathbb{Z}^{N+1}\times\mathbb{Z}_{2}^{N+1}$ by setting $\lambda_{N+1}=0$ and $\bar{\sigma}_{N+1}=0$.

For the action of $E_{s}(z)$, we can naively take the limit $N\rightarrow \infty$ similar to the quantum toroidal $\mathfrak{gl}_{1}$ case. The result is 
\begin{align}
    \begin{split}
        \quad E_{s}(z)\ket{\lambda,\bar{\sigma}}&=\sum_{k=1}^{\ell(\lambda)+1}(-1)^{|s|(\sum_{l=1}^{k-1}\overline{\sigma}_{l}})\prod_{i=1}^{k-1}\left[\Psi^{(s)}_{[u(q_{1}q_{2}^{-1})^{i-1}]_{\lambda_{i}-1,\lambda_{i}-1-\bar{\sigma}_{i}}}(z)\right]_{-}\\
        &\times\mathcal{E}_{s}\left([(q_{1}q_{2}^{-1})^{i-1}u]_{\lambda_{i}-1,\lambda_{i}-1-\bar{\sigma}_{i}}\right)\\
        &\times\overline{\delta}_{s+\sigma_{k},1}\delta\left(\frac{z}{uq_{1}^{k+\lambda_{k}-1}q_{2}^{\lambda_{k}-1+\overline{(\sigma_{k}+1})}}\right)\ket{\lambda+\fbox{$s$}_{k},\overline{\sigma}+\bar{1}_{k}}.
    \end{split}
\end{align}

Next we consider the action of $K_{s}(z)$. In this case, infinite product involves so we have to be careful with the limit $N\rightarrow \infty$. For $K_{1}(z)$, the action is 
\begin{align}
    \begin{split}
        K_{1}(z)\ket{\lambda,\bar{\sigma}}=\prod_{i=1}^{\ell(\lambda)}\Psi^{(1)}_{[u(q_{1}q_{2}^{-1})^{i-1}]_{\lambda_{i}-1,\lambda_{i}-1-\bar{\sigma}_{i}}}(z)\prod_{i=\ell(\lambda)+1}^{\infty}\Psi^{(1)}_{[u(q_{1}q_{2}^{-1})^{i-1}]_{-1,-1}}(z)\ket{\lambda,\overline{\sigma}}.
    \end{split}
\end{align}
The infinite product can be regularized by specifying the order of the product as
\begin{align}
\begin{split}
    \prod_{i=\ell(\lambda)+1}^{\infty}\Psi^{(1)}_{[u(q_{1}q_{2}^{-1})^{i-1}]_{-1,-1}}(z)&=\frac{1}{\phi(q_{1}^{-\ell(\lambda)}q_{2}^{\ell(\lambda)};z,u)}
    \end{split}
\end{align}
and we obtain 
\begin{align}
     \begin{split}
        K_{1}(z)\ket{\lambda,\bar{\sigma}}=\frac{1}{\phi(q_{1}^{-\ell(\lambda)}q_{2}^{\ell(\lambda)};z,u)}\prod_{i=1}^{\ell(\lambda)}\Psi^{(1)}_{[u(q_{1}q_{2}^{-1})^{i-1}]_{\lambda_{i}-1,\lambda_{i}-1-\bar{\sigma}_{i}}}(z)\ket{\lambda,\overline{\sigma}}.
    \end{split}
\end{align}
The same is true for $K_2(z)$ and we obtain
\begin{align}
\begin{split}
    K_{s}(z)\ket{\lambda,\bar{\sigma}}=\frac{\phi(q_{1}^{-\ell(\lambda)+1}q_{2}^{\ell(\lambda)};z,u)^{\delta_{s,2}}}{\phi(q_{1}^{-\ell(\lambda)}q_{2}^{\ell(\lambda)};z,u)^{\delta_{s,1}}}\prod_{i=1}^{\ell(\lambda)}\Psi^{(s)}_{[u(q_{1}q_{2}^{-1})^{i-1}]_{\lambda_{i}-1,\lambda_{i}-1-\bar{\sigma}_{i}}}(z)\ket{\lambda,\overline{\sigma}}.
\end{split}
\end{align}

Next we consider the action of $F_{s}(z)$:
 \begin{align}
     \begin{split}
         &\quad F_{s}(z)\ket{\lambda,\overline{\sigma}}\\
         &=\sum_{k=1}^{\ell(\lambda)}(-1)^{|s|(\sum_{l=1}^{k-1}\overline{\sigma}_{l}})\prod_{i=k+1}^{\ell(\lambda)}\left[\Psi^{(s)}_{[u(q_{1}q_{2}^{-1})^{i-1}]_{\lambda_{i}-1,\lambda_{i}-1-\bar{\sigma}_{i}}}(z)\right]_{+}\prod_{i=\ell(\lambda)+1}^{\infty}\left[\Psi^{(s)}_{[u(q_{1}q_{2}^{-1})^{i-1}]_{-1,-1}}(z)\right]_{+}\\
         &\quad\times\mathcal{F}_{s}\left([u(q_{1}q_{2}^{-1})^{i-1}]_{\lambda_{i}-1,\lambda_{i}-1-\bar{\sigma}_{1}}\right)\delta\left(\frac{z}{uq_{1}^{k+\lambda_{k}-1-\bar{s}}q_{2}^{\lambda_{k}-k}}\right)\overline{\delta}_{s,\sigma} \ket{\lambda-\fbox{$s$}_{k},\bar{\sigma}-\bar{1}_{k} }.
     \end{split}
 \end{align} 
 Also in this case infinite products involve, but we can regularized it similarly and finally obtain 
\begin{align}
    \begin{split}
         F_{s}(z)\ket{\lambda,\overline{\sigma}}&=\frac{\phi(q_{1}^{-\ell(\lambda)+1}q_{2}^{\ell(\lambda)};z,u)^{\delta_{s,2}}}{\phi(q_{1}^{-\ell(\lambda)}q_{2}^{\ell(\lambda)};z,u)^{\delta_{s,1}}}\sum_{k=1}^{\ell(\lambda)}(-1)^{|s|(\sum_{l=1}^{k-1}\overline{\sigma}_{l}})\prod_{i=k+1}^{\ell(\lambda)}\left[\Psi^{(s)}_{[u(q_{1}q_{2}^{-1})^{i-1}]_{\lambda_{i}-1,\lambda_{i}-1-\bar{\sigma}_{i}}}(z)\right]_{+}\\
         &\times\overline{\delta}_{s,\sigma}\mathcal{F}_{s}\left([u(q_{1}q_{2}^{-1})^{i-1}]_{\lambda_{i}-1,\lambda_{i}-1-\bar{\sigma}_{1}}\right)\delta\left(\frac{z}{uq_{1}^{k+\lambda_{k}-1-\bar{s}}q_{2}^{\lambda_{k}-k}}\right) \ket{\lambda-\fbox{$s$}_{k},\bar{\sigma}-\bar{1}_{k} }.\\
     \end{split}
\end{align}

Let us see the action of $K_{s}(z)$ on $\ket{\emptyset}$. The vacuum can be defined as
\begin{align}
    \ket{\emptyset}=\otimes_{i=1}^{\infty}[u(q_{1}q_{2}^{-1})^{i-1}]_{-1,-1}
\end{align}
and the action is 
\begin{align}
    \begin{split}
        K_{s}(z)\ket{\emptyset}=\frac{\phi(q_{1};z,u)^{\delta_{s,2}}}{\phi(1;z,u)^{\delta_{s,1}}}\ket{\emptyset},
    \end{split}
\end{align}
which was expected in (\ref{eq:vacuum_charge_function}). Thus, this will be a representation of the shifted quantum toroidal $\mathfrak{gl}_{1|1}$ with shift parameters
\begin{align}
    r_{1}=-1,\quad r_{2}=1.
\end{align}

\section{Examples}\label{sec:examples}
In this section, we derive one-dimensional and two-dimensional crystal representations which are representations of shifted quantum toroidal algebras defined in the previous section. We use the current realization of this algebra in (\ref{eq:shiftedQuiverAlgebra_for_rep}) and the coproduct in (\ref{eq:tensorproduct_coproduct}). We omit the tilde in $\widetilde{K}^{\pm}(z)$ and write it as $K^{\pm}(z)$. We note that in this section, $K_{i}(z)$ always includes the extra shift parameter. We also omit the subindex of the generalized coproduct $\Delta_{\mathbf{r_{1}},\mathbf{r_{2}}}$ and write it as $\Delta$.

The main strategy is 
\begin{enumerate}
\item Derive the bond factors from the toric diagram, periodic quiver diagram, and quiver diagram. These bond factors define the algebras.
\item List down all perfect matchings of divisors and remove arrows from the original quiver diagram. Removing arrows of the unique perfect matchings of corner divisors gives subquivers of two-dimensional crystals, while removing arrows of the union set of perfect matchings of divisors surrounding the external legs gives subquivers of one-dimensional crystals.
\item Accept the defining relations (\ref{eq:shiftedQuiverAlgebra_for_rep}) and derive the actions of the generators on the one-dimensional crystal. Taking tensor products of these representations gives the two-dimensional crystal representation.  
    \item Shift parameters can be derived from the vacuum charge function.
\end{enumerate}


\subsection{\texorpdfstring{$\mathbb{C}^{3}$}{C3} and quantum toroidal \texorpdfstring{$\mathfrak{gl}_{1}$}{gl1} revisited}

\begin{figure}[H]
    \centering
    \includegraphics[width=6cm]{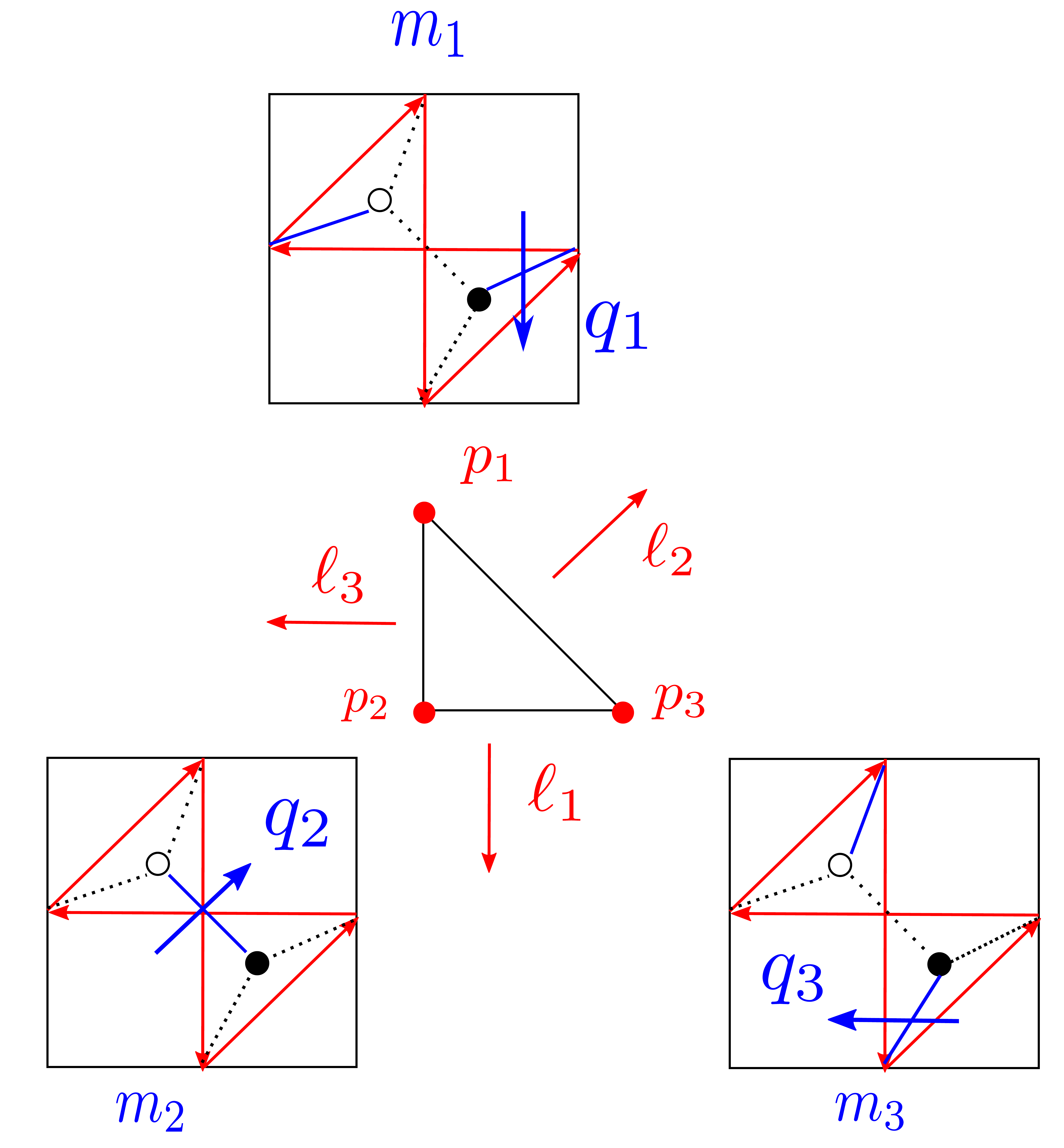}
    \caption{Perfect matchings of $\mathbb{C}^{3}$. There are three corner divisors $p_{i}$ $(i=1,2,3)$, and for each of them there is a unique perfect matching $m_{i}$. We note $m_{i}=\{q_{i}\}$.}
   \label{fig:gl1_perfectmatching}
\end{figure}
Let us derive the subquiver and crystal shape from the perfect matchings. The three divisors $p_{i}$ $(i=1,2,3)$ are all corner divisors and each of them has a unique perfect matching. We denote the perfect matching of divisor $p_{i}$ as $m_{i}$. From Figure \ref{fig:gl1_perfectmatching}, the perfect matching is $m_{i}=\{q_{i}\}$.
\begin{figure}[t]
    \centering
    \includegraphics[width=10cm]{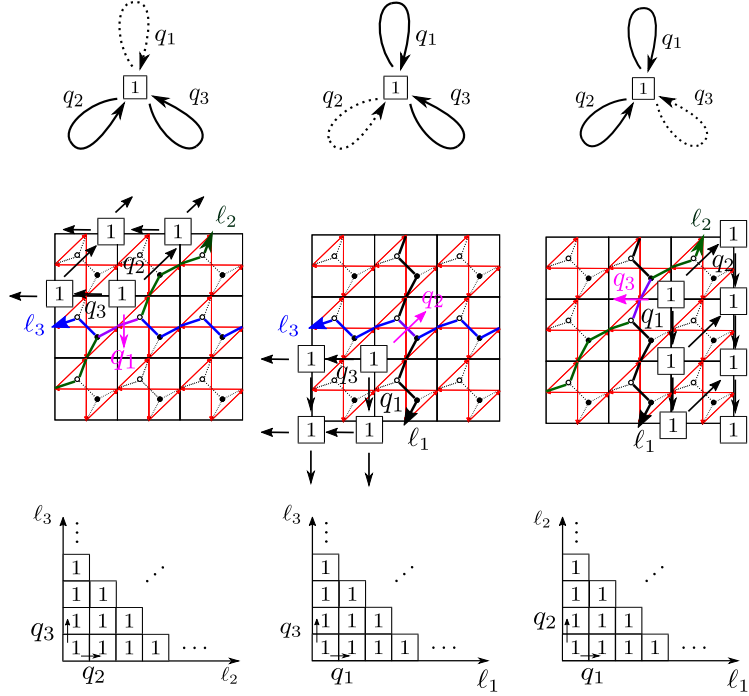}
    \caption{Subquivers and two-dimensional crystals of $\mathbb{C}^{3}$. Left: Crystal of divisor $p_{1}=(0,1)$. arrow removed is $m_{1}=\{q_{1}\}$. The crystal is the region surrounded by external legs $\ell_{2}$ and $\ell_{3}$, which means it is a Young diagram with two coordinates $q_{2}$ and $q_{3}$. The nontrivial central charge is $q_{1}^{1/2}$. Middle: Crystal of divisor $p_{2}=(0,0)$. arrow removed is $m_{2}=\{q_{2}\}$. The crystal is the region surrounded by zig-zag paths of external legs $\ell_{1},\ell_{3}$, which means a Young diagram with coordinates $q_{1},q_{3}$. The central charge is $q_{2}^{1/2}$. Right: Crystal of divisor $p_{3}=(1,0)$. arrow removed is $m_{3}=\{q_{3}\}$. The crystal shape is the region surrounded by $\ell_{1}$ and $\ell_{2}$, which is a Young diagram with two coordinates $q_{1},q_{2}$. The central charge is $q_{3}^{1/2}$.   }
    \label{fig:gl1_2dim_subquiver_crystal}
\end{figure}

Let us consider the subquiver and the shape of the two-dimensional crystal. We already know this should be a Young diagram and that we have three types due to triality. Let us rederive this using the method in section \ref{sec:subcrystal_subquiver}. To make it concrete, we focus on the divisor $p_{3}$ (see Figure \ref{fig:gl1_2dim_subquiver_crystal} for other crystals). The perfect matching associated with this is $m_{3}=\{q_{3}\}$, and thus to obtain the subquiver, we need to remove this arrow. The subquiver we obtain is the right of Figure \ref{fig:gl1_2dim_subquiver_crystal}. Since this divisor is surrounded by two external legs $\ell_{1}$ and $\ell_{2}$, the two-dimensional crystal is the region surrounded by the zig-zag path associated with these two legs. The resulting crystal will be the Young diagram with two coordinates $q_{1}$ and $q_{2}$ as in Figure \ref{fig:gl1_2dim_subquiver_crystal}. The unique edge where the zig-zag path of $\ell_{1}$ and that of $\ell_{2}$ intersect has a parameter $q_{3}$. From (\ref{eq:vacuum_charge_function}), the vacuum charge function is expected to be  
\begin{align}
    K(z)\ket{\emptyset}=\frac{\phi(q_{3}^{-1};z,u)}{\phi(1;z,u)}\ket{\emptyset}.
\end{align}
This is indeed true because of (\ref{eq:gl1_vacuum_charge}). From the definition of the central charge, we can also see $q_{3}^{1/2}$ is the non-trivial central charge of this representation.

\begin{figure}[t]
    \centering
    \includegraphics[width=9cm]{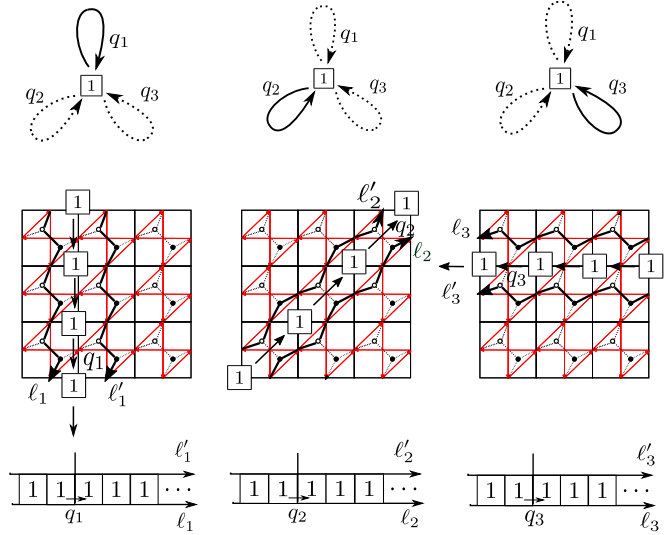}
   \caption{Subquivers and one-dimensional crystals of $\mathbb{C}^{3}$. The subquiver for one-dimensional crystal associated with the external leg $\ell_{i}$ is determined by removing arrows of $m_{i-1}\cup m_{i+1}=\{q_{i-1}, q_{i+1}\}$, where the subindices are understood modulo 3. The resulting quiver has only one loop with parameter $q_{i}$. The crystal shape is determined by the region surrounded with the external legs $\ell_{i}$ and $\ell_{i}'$ in the periodic quiver diagram. $\ell_{i}'$ here is the same external leg $\ell_{i}$, but in a different fundamental region. The crystal can be illustrated as a row of boxes with coordinates $q_{i}$. Left is $i=1$,  middle is $i=2$, and right is $i=3$.}
    \label{fig:gl1_1dim_subquiver_crystal}
\end{figure}

Next, let us consider the one-dimensional subcrystals. Representations associated with these are called vector representations in the literature. We focus on the one-dimensional crystal associated with the external leg $\ell_{1}$ (see Figure \ref{fig:gl1_1dim_subquiver_crystal} for other crystals). This external leg is surrounded by two divisors $p_{2}$ and $p_{3}$. We need to remove arrows $m_{2}\cup m_{3}=\{q_{2},q_{3}\}$ to obtain the subquiver. The subquiver we obtain is the left of Figure \ref{fig:gl1_1dim_subquiver_crystal}. The crystal is the region surrounded by two external legs $\ell_{1}$ and $\ell_{1}'$ in the periodic quiver. $\ell_{1}'$ here is the same external leg as $\ell_{1}$, but in a different fundamental region.\footnote{The periodic quiver diagram drawn in Figure \ref{fig:gl1_2dim_subquiver_crystal} and \ref{fig:gl1_1dim_subquiver_crystal} is the universal covering of the middle tile. We are drawing 9 copies of the middle tile. In deriving the crystal picture of one-dimensional crystals, we are distinguishing $\ell_{1}$ and $\ell_{1}'$ because they are different lines in the periodic quiver diagram.}

\subsection{\texorpdfstring{$(\mathbb{C}^{2}/\mathbb{Z}_{n})\times\mathbb{C}$}{C2ZnC} and quantum toroidal \texorpdfstring{$\mathfrak{gl}_{n}(n\geq 2)$}{gln}}\label{sec:toroidalgln}
In this subsection, we derive subcrystal representations of $(\mathbb{C}^{2}/\mathbb{Z}_{n})\times \mathbb{C}$ $(n\geq 2)$. Some of the representations are already known in the literature \cite{feigin2013representations,negut2021agt}. We will derive one type of one-dimensional crystal representations and two-dimensional crystal representations in this section. Other crystal representations of $(\mathbb{C}^{2}/\mathbb{Z}_{n})\times \mathbb{C}$ $(n\geq 2)$ are obtained in the same way in Appendix \ref{sec:appendixgln}. 
\subsubsection{Definition of the algebra}\label{sec:glndefinition}
\begin{figure}
    \begin{tabular}{c}
      \begin{minipage}{0.95\hsize}
        \centering
      \includegraphics[width=12.5cm]{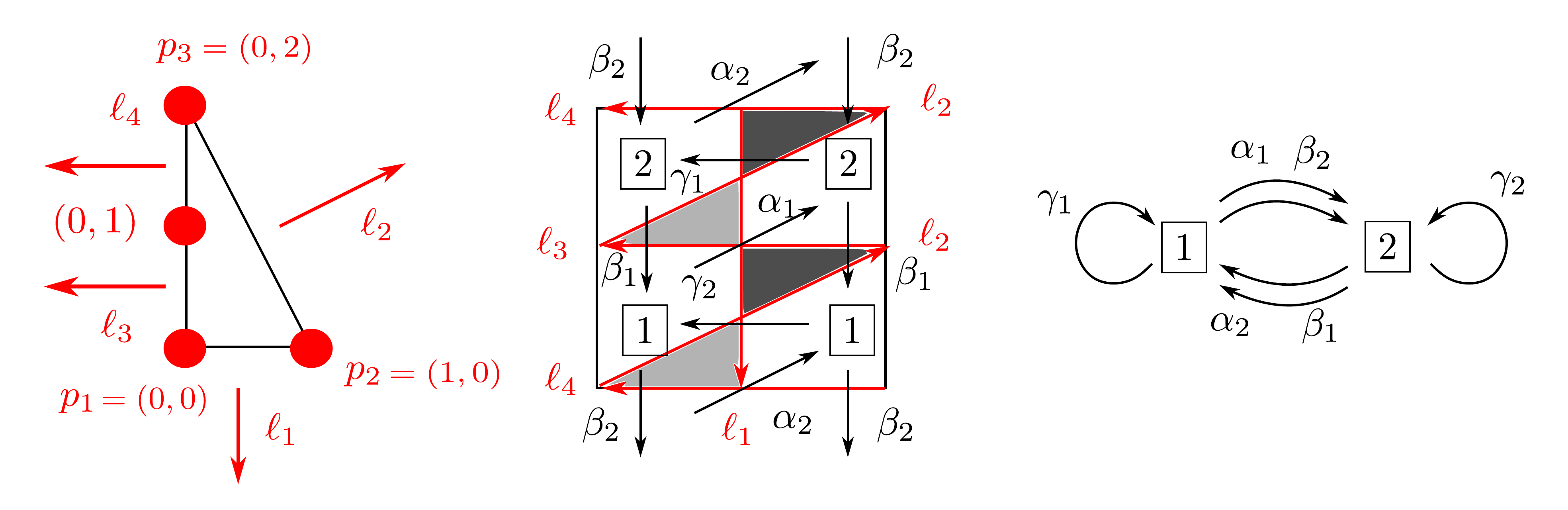} 
        \subcaption{$n=2$}\label{fig:toricgl_2}
      \end{minipage}\\
      \begin{minipage}{0.95\hsize}
        \centering
     \includegraphics[width=12.5cm]{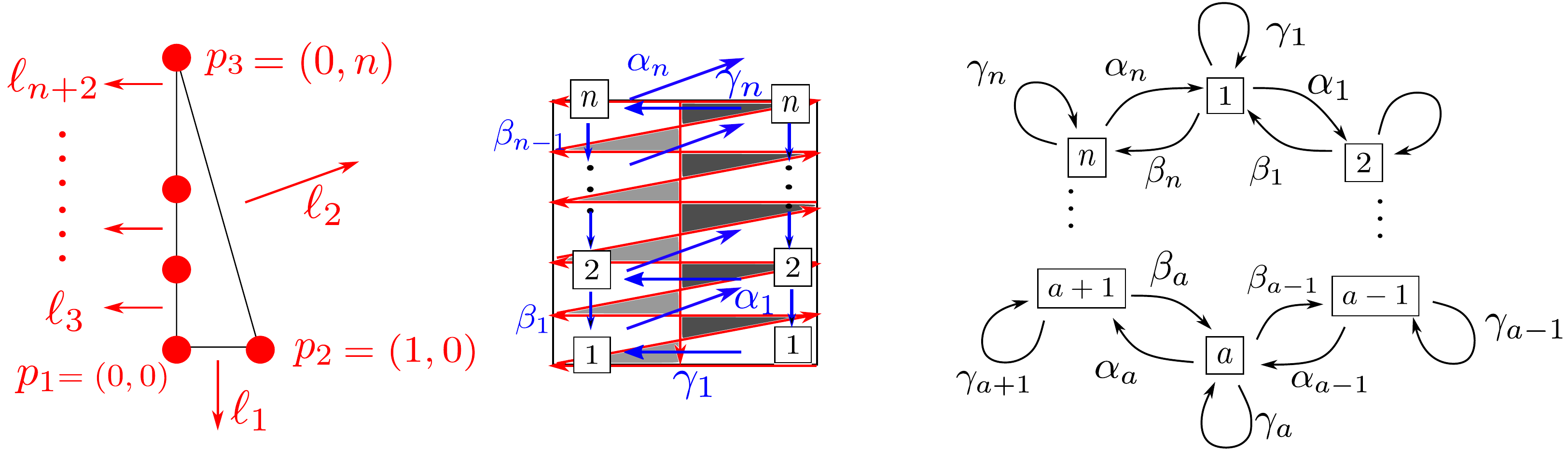} 
       \subcaption{$n\geq 3$}\label{fig:gl_n_toric_quiver}
      \end{minipage}
    \end{tabular}
\caption{Toric diagram, periodic quiver diagram and quiver diagram of $(\mathbb{C}^{2}/\mathbb{Z}_{n})\times\mathbb{C}$ $(n\geq 2)$. The quiver structure depends on the value $n$. (a) When $n=2$, there are four arrows between the two vertices. (b) When $n\geq 3$, there are only two arrows between the adjacent vertices.}\label{fig:gln2toricquiver}
\end{figure}

The toric diagram, periodic quiver diagram, and quiver diagram are as in Figure \ref{fig:gln2toricquiver}. The quiver structure changes whether $n$ is 2 or larger than 2.
Since each vertex has only one loop, all of the vertices are bosonic.\\
We have $3n$ parameters $(\alpha_{a},\beta_{a},\gamma_{a})\;(a=1,2,...,n)$. After imposing the loop conditions and vertex conditions, we obtain

\begin{align}
    \alpha_{a}=q_{3},\quad\beta_{a}=q_{1},\quad\gamma_{a}=q_{2}\quad(a=1,..,n),
\end{align}
where $q_{1}q_{2}q_{3}=1$.
The nontrivial bond factors can be written as, 
\begin{align}
\begin{split}
    &\text{when}\quad n\geq3,\quad \varphi^{a\Rightarrow a+1}(z,w)=\frac{\phi(q_{1};z,w)}{\phi(q_{3}^{-1};z,w)},\quad \varphi^{a+1\Rightarrow a}(z,w)=\frac{\phi(q_{3};z,w)}{\phi(q_{1}^{-1};z,w)},\\
    &\text{when}\quad n=2,\quad \varphi^{1\Rightarrow 2}(z,w)=\varphi^{2\Rightarrow1}(z,w)=\frac{\phi(q_{1};z,w)\phi(q_{3};z,w)}{\phi(q_{1}^{-1};z,w)\phi(q_{3}^{-1};z,w)},\\
    &\text{when}\quad n \geq2,\quad \varphi^{a\Rightarrow a}(z,w)=\frac{\phi(q_{2};z,w)}{\phi(q_{2}^{-1};z,w)}.
    \end{split}
\end{align}
Other bond factors are 
\begin{align}
\varphi^{i\Rightarrow j}(z,w)=1.
\end{align}

\subsubsection{Subquiver and crystal shape}\label{sec:glncrystalshape}
\begin{figure}[h]
    \begin{tabular}{c}
      \begin{minipage}{0.9\hsize}
        \centering
        \includegraphics[width=10cm]{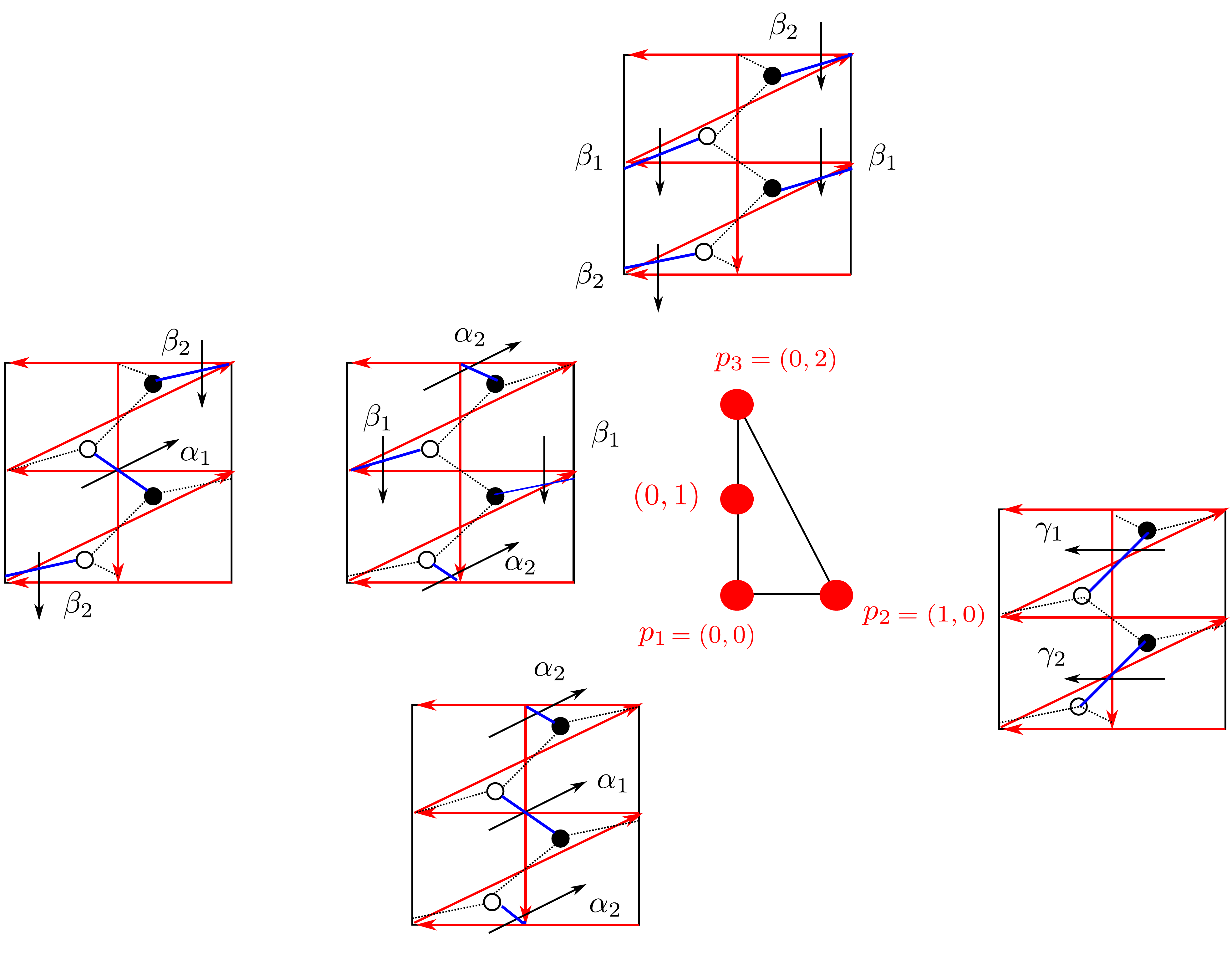}
       \subcaption{Perfect matchings of $(\mathbb{C}^{2}/\mathbb{Z}_{2})\times\mathbb{C}$}\label{fig:gl_2perfectmatching}
      \end{minipage}\\
      \begin{minipage}{0.9\hsize}
        \centering
        \includegraphics[width=10cm]{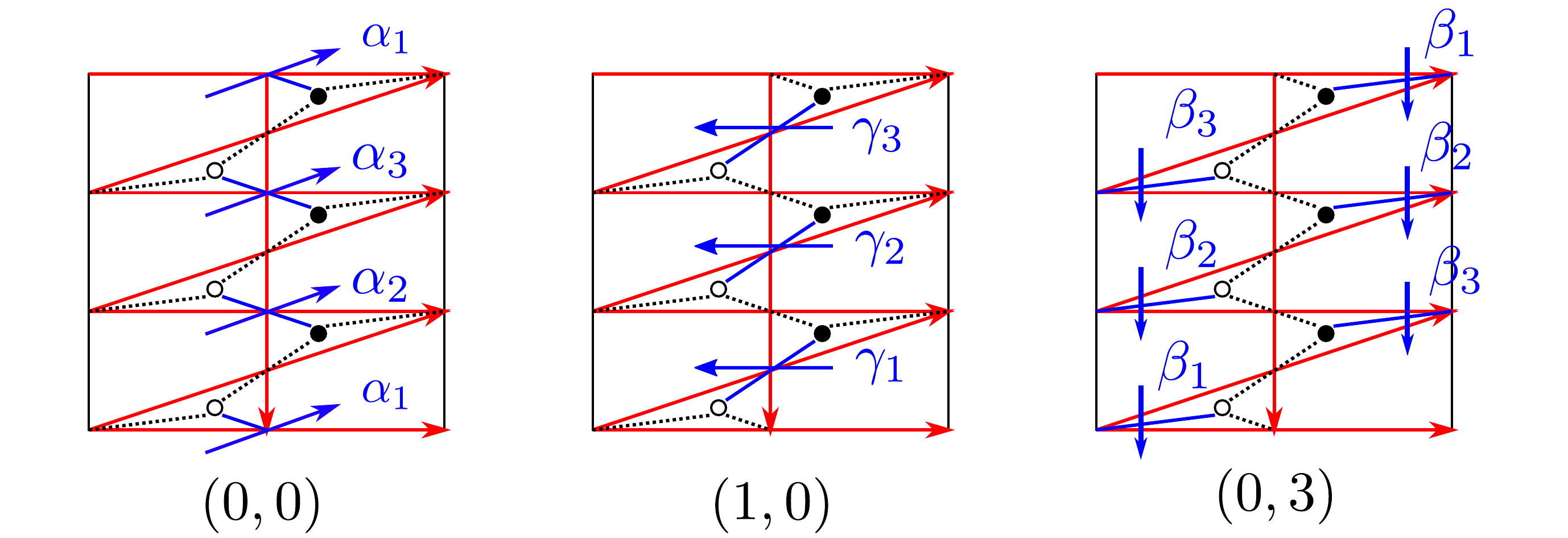}\subcaption{Perfect matchings for corner divisors of $(\mathbb{C}^{2}/\mathbb{Z}_{3})\times\mathbb{C}$.}\label{fig:gl_3_perfectmatching}
      \end{minipage}
    \end{tabular}
\caption{Perfect matchings of divisors of $\mathbb{C}^{2}/\mathbb{Z}_{n}\times\mathbb{C}$ $(n\geq 2)$. (a) For the $n=2$ case, we can explicitly write down all the perfect matchings. Perfect matchings of corner divisors are $m_{1}=\{\alpha_{1},\alpha_{2}\}$, $m_{2}=\{\gamma_{1},\gamma_{2}\}$, and $m_{3}=\{\beta_{1},\beta_{2}\}$. We have two perfect matchings for divisor $(0,1)$: $\{\alpha_{1},\beta_{2}\}$ and $\{\alpha_{2},\beta_{1}\}$. (b) For general $n\geq 3$, the perfect matchings of corner divisors are $m_{1}=\{\alpha_{1},...,\alpha_{n}\}$, $m_{2}=\{\gamma_{1},...,\gamma_{n}\}$, and $m_{3}=\{\beta_{1},...\beta_{n}\}$.}\label{fig:gln2_perfectmatching}
\end{figure}

Let us determine the subquiver and low-dimensional crystal structure of $\mathbb{C}^{2}/\mathbb{Z}_{n}\times\mathbb{C}$ $(n\geq 2)$. We have $n+2$ divisors and three of them are corner divisors. We denote the corner divisors $p_{1}=(0,0)$, $p_{2}=(1,0)$, and $p_{3}=(0,n)$ (see Figure \ref{fig:gl_n_toric_quiver}). The external legs of the diagram are denoted $\ell_{1},..,\ell_{n+2}$. Although the quiver structure slightly differs depending whether $n$ is 2 or larger than 2, we will see that the crystal structures are similar and that the discussions go parallel. \\
The perfect matchings of the divisors are complicated to write down explicitly for general $n$, but the perfect matchings of the corner divisors are easy to write down. The perfect matching for the corner divisor $p_{i}$ is denoted as $m_{i}$. They are
\begin{align}
    m_{1}=\{\alpha_{1},..,\alpha_{n}\},\quad m_{2}=\{\gamma_{1},..,\gamma_{n}\},\quad m_{3}=\{\beta_{1},...,\beta_{n}\}.
\end{align}
See Figure \ref{fig:gl_2perfectmatching} and \ref{fig:gl_3_perfectmatching} for the $n=2,3$ case.

\begin{figure}[h]
\centering
\includegraphics[width=13cm]{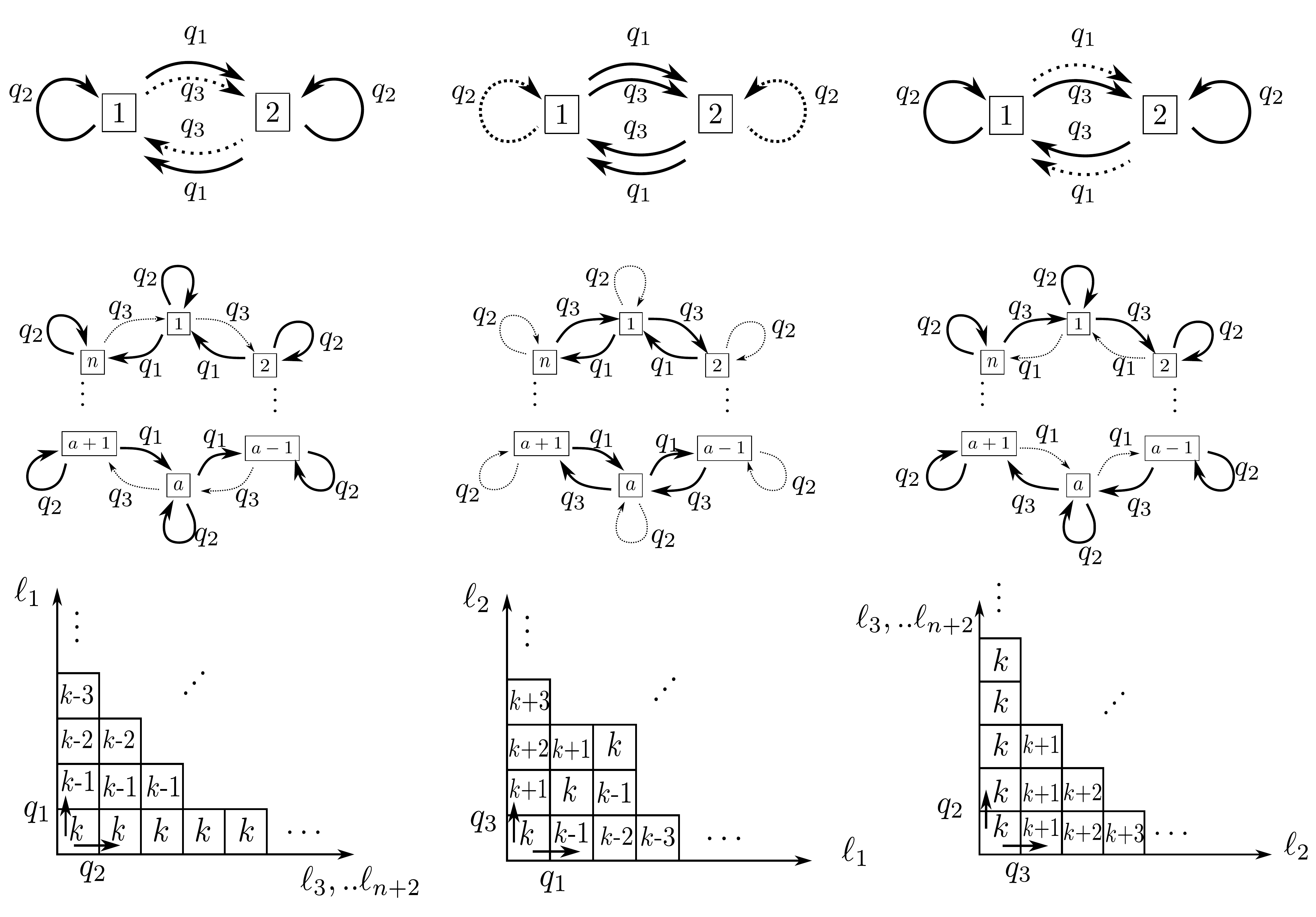}
\caption{We have $n$ colors of atoms and we set the origin to have color $k$. The crystal structure does not change whether $n=2$ or $n>2$, but the quiver structure changes. Left: Subquiver and two-dimensional crystals of corner divisor $p_{1}=(0,0)$. It is obtained by removing arrows of perfect matching $m_{1}=\{\alpha_{1},..,\alpha_{n}\}$, where $\alpha_{a}=q_{3}$. Middle: Subquiver and two-dimensional crystal of corner divisor $p_{2}=(1,0)$. It is obtained by removing $m_{2}=\{\gamma_{1},..,\gamma_{n}\}$, where $\gamma_{a}=q_{2}$.  Right: Subquiver and two-dimensional crystal of corner divisor $p_{3}=(0,n)$. It is obtained by removing $m_{3}=\{\beta_{1},..,\beta_{n}\}$, where $\beta_{a}=q_{1}$. } \label{fig:gln2_subquiver_2dcrystal}
\end{figure}

\begin{figure}[h]
\centering
\includegraphics[width=13cm]{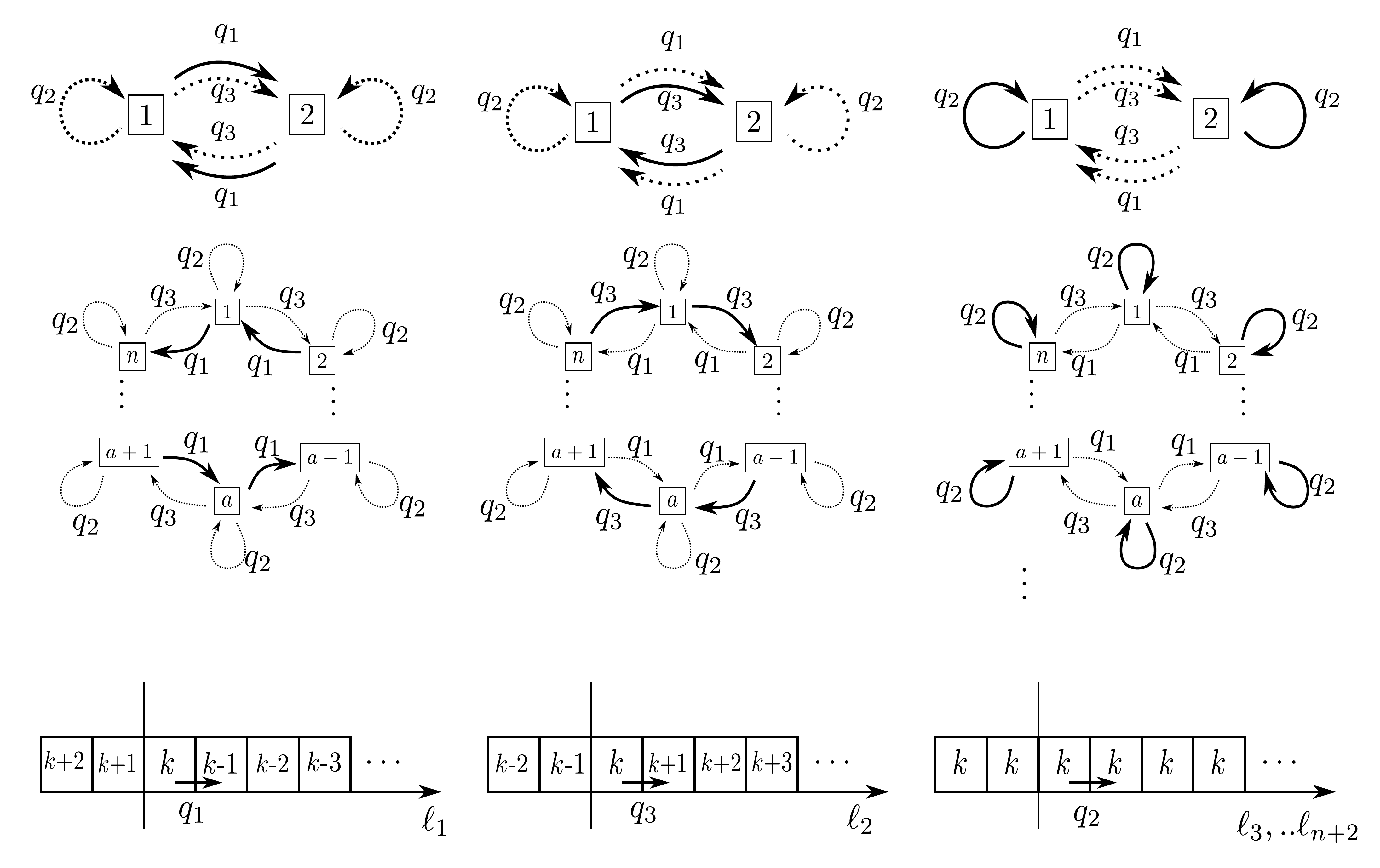}
\caption{Similar to the two-dimensional crystal, the crystal structure does not change whether $n=2$ or $n>2$, but the quiver structure changes. Left: Subquiver and one-dimensional crystal associated with $\ell_{1}$. Middle: Subquiver and one-dimensional crystal associated with $\ell_{2}$. Right: Subquiver and one-dimensional crystal associated with $\ell_{3},..\ell_{n+2}$.}\label{fig:gln2_subquiver_1dcrystal}
\end{figure}

The subquiver and two-dimensional crystals associated with corner divisors can be obtained by removing arrows of the perfect matching of the corner divisors. For $p_{1}=(0,0)$, we need to remove arrows of $m_{1}=\{\alpha_{1},..,\alpha_{n}\}$. For other cases, see Figure \ref{fig:gln2_subquiver_2dcrystal}.

 We can obtain the subquiver and one-dimensional crystal associated with each external legs by removing arrows of the union set of perfect matchings of the divisors surrounding it. To make it concrete, let us consider the case when $n=2$. In this case, we can write down all perfect matchings associated with each of the divisor. Perfect matchings of corner divisors are $m_{1}=\{\alpha_{1},\alpha_{2}\}$, $m_{2}=\{\gamma_{1},\gamma_{2}\}$, and $m_{3}=\{\beta_{1},\beta_{2}\}$, while perfect matchings of divisor $(0,1)$ are $\{\alpha_{1},\beta_{2}\}$ and $\{\alpha_{2},\beta_{1}\}$ (see Figure \ref{fig:gl_2perfectmatching}). Thus, the arrows we need to remove for one-dimensional crystal of $\ell_{1}$ are $m_{1}\cup m_{2}=\{\alpha_{1}, \alpha_{2},\gamma_{1},\gamma_{2}\}$. For $\ell_{2}$, the arrows removed are $m_{2}\cup m_{3}=\{\beta_{1},\beta_{2},\gamma_{1},\gamma_{2}\}$. For $\ell_{3},\ell_{4}$, we need to remove arrows $\{\alpha_{1},\beta_{2}\}\cup\{\alpha_{2},\beta_{1}\}=\{\alpha_{1},\beta_{2},\alpha_{2},\beta_{1}\} $ (see Figure \ref{fig:gln2_subquiver_1dcrystal}).
 
We can do the same discussion for general $n$: for $\ell_{1}$, the arrows we need to remove are $m_{1}\cup m_{2}=\{\alpha_{a}, \gamma_{a}|a=1,..,n\}$. For $\ell_{2}$, the arrows removed are $m_{2}\cup m_{3}=\{\beta_{a}, \gamma_{a}| a=1,..,n\}$. For $\ell_{3},..,\ell_{n+2}$, we need to remove arrows $\{\alpha_{a},\beta_{a}| a=1,..,n\}$ and each of the vertices decouple (see Figure \ref{fig:gln2_subquiver_1dcrystal}). 

Although the quiver structure depends on whether $n=2$ or $n>2$, the crystal picture does not change.

\subsubsection{One-dimensional crystal \texorpdfstring{$\ell_{1}$}{l1}}\label{sec:gln1dl1}
This representation is the one derived in \cite{feigin2013representations}.
The basis of this representation can be illustrated as a semi-infinite row of boxes with coloring due to $\mathbb{Z}_{n}$ $(n\geq2)$ (see the left of Figure \ref{fig:gln2_subquiver_1dcrystal}). We set the origin to have color $k$. $[u]_{j}^{(k)}$ can be illustrated as a semi-infinite row of boxes where there are $j+1$ boxes right to the border and periodically extended left to the border. The boxes are numbered $0,1,,,j$ from the right of the border and are colored $k,k-1,k-2,..$. We denote the vector space of this representation $V^{(\ell_{1})}(u)$.
Index of generators is understood modulo $n$. 
The action of the generators can be written 
\begin{align}
\begin{split}
    E_{s}(z)[u]_{j}^{(k)}&=\mathcal{E}_{s}([u]_{j}^{(k)})\delta\left(\frac{z}{uq_{1}^{j+1}}\right)\bar{\delta}_{k-j-1,s}[u]_{j+1}^{(k)},\\
    F_{s}(z)[u]_{j}^{(k)}&=\mathcal{F}_{s}([u]_{j}^{(k)})\delta\left(\frac{z}{uq_{1}^{j}}\right)\bar{\delta}_{k-j,s}[u]_{j-1}^{(k)},\\
    K_{s}^{\pm}(z)[u]_{j}^{(k)}&=\left[\Psi^{(s)}_{[u]_{j}^{(k)}}(z)\right]_{\pm}[u]_{j}^{(k)},
    \end{split}
\end{align}
where $\bar{\delta}_{i,j}=\begin{cases}
1,\quad i\equiv j\quad (\text{mod}\hspace{2mm}n)\\
0,\quad i\not\equiv j\quad (\text{mod}\hspace{2mm}n)
\end{cases}$. $\mathcal{E}_{s}([u]_{j}^{(k)})$ and $\mathcal{F}_{s}([u]_{j}^{(k)})$ are some coefficients which are determined from the other defining relations of the algebra although we do not write down the explicit expression of them.
By the recursion formula of $\Psi^{(a)}_{[u]_{j}^{(k)}}(z)$ obtained from (\ref{eq:shiftedQuiverAlgebra_for_rep}) and the pole cancellation similar to (\ref{eq:infiniteproductgl1}), we obtain
\begin{align}
    \Psi^{(a)}_{[u]_{j}^{(k)}}(z)=\left(\frac{\phi(q_{2};z,uq_{1}^{j})}{\phi(1;z,uq_{1}^{j})}\right)^{\bar{\delta}_{k,a+j}}\left(\frac{\phi(q_{3};z,uq_{1}^{j})}{\phi(q_{1}^{-1};z,uq_{1}^{j})}\right)^{\bar{\delta}_{k,a+j+1}}.\label{eq:gln_l1_chargefunction}
\end{align}
Since the number of zeros and poles are the same, the shift parameters are 
\begin{align}
    \mathbf{r}=(0,....,0)\in\mathbb{Z}^{n}
\end{align}
and give a representation of the unshifted quantum toroidal $\mathfrak{gl}_{n}$.

\subsubsection{Two-dimensional crystal of  \texorpdfstring{$p_{1}=(0,0)$}{p100}}\label{sec:gln2dp1}
Let us consider the two-dimensional crystal representation of divisor $p_{1}$. The crystal picture is illustrated in the left of Figure \ref{fig:gln2_subquiver_2dcrystal}.
The basis of this representation is defined by the tensor product of one-dimensional representations as
\begin{align}
    \otimes_{i=1}^{N}V^{(\ell_{1})}(uq_{2}^{i-1})\ni\ket{\lambda}=\otimes_{i=1}^{N}[uq_{2}^{i-1}]_{\lambda_{i}-1}^{(k)},\quad \lambda_{1}\geq\lambda_{2}...
\end{align}
and we take the limit $N\rightarrow \infty$.
Here, the $q_2$ shift of parameters $uq_{2}^{i-1}\; (i=1, 2, \cdots, N)$ ensures the melting rule, and this basis forms a submodule.
The actions of generators are written as
\begin{align}
\begin{split}
\begin{split}
E_{s}(z)\ket{\lambda}&=\sum_{i=1}^{\ell(\lambda)+1}\prod_{j=1}^{i-1}\left[\Psi^{(s)}_{[uq_{2}^{j-1}]_{\lambda_{j}-1}^{(k)}}(z)\right]_{-}\mathcal{E}_{s}\left([uq_{2}^{i-1}]^{(k)}_{\lambda_{i}-1}\right)\\
&\times\delta\left(\frac{z}{uq_{2}^{i-1}q_{1}^{\lambda_{i}}}\right)\bar{\delta}_{k-\lambda_{i},s}\ket{\lambda+\fbox{$s$}_{i}},
\end{split}\\
K_{s}(z)\ket{\lambda}&=\frac{\phi(q_{1}q_{2}^{1-\ell(\lambda)};z,u)^{\bar{\delta}_{k,s-1}}}{\phi(q_{2}^{-\ell(\lambda)};z,u)^{\bar{\delta}_{k,s}}}\prod_{i=1}^{\ell(\lambda)}\Psi^{(s)}_{[uq_{2}^{i-1}]^{(k)}_{\lambda_{i}-1}}(z)\ket{\lambda},\\
\begin{split}
F_{s}(z)\ket{\lambda}&=\frac{\phi(q_{1}q_{2}^{1-\ell(\lambda)};z,u)^{\bar{\delta}_{k,s-1}}}{\phi(q_{2}^{-\ell(\lambda)};z,u)^{\bar{\delta}_{k,s}}}\sum_{i=1}^{\ell(\lambda)}\prod_{j=i+1}^{\ell(\lambda)}\left[\Psi^{(s)}_{[uq_{2}^{j-1}]^{(k)}_{\lambda_{j}-1}}(z)\right]_{+}\\
&\times \mathcal{F}_{s}\left([uq_{2}^{i-1}]^{(k)}_{\lambda_{i}-1}\right)\bar{\delta}_{k-\lambda_{i}+1,s}\delta\left(\frac{z}{uq_{2}^{i-1}q_{1}^{\lambda_{i}-1}}\right)\ket{\lambda-\fbox{$s$}_{i}}\end{split},
\end{split}
\end{align}
where we specified the order of infinite products to regularize them as (\ref{eq:gln_l1_chargefunction}) and
\begin{align}
    \prod_{i=\ell(\lambda)+1}^{\infty}\Psi^{(s)}_{[uq_{2}^{i-1}]^{(k)}_{-1}}(z)=\frac{\phi(q_{1}q_{2}^{1-\ell(\lambda)};z,u)^{\bar{\delta}_{k,s-1}}}{\phi(q_{2}^{-\ell(\lambda)};z,u)^{\bar{\delta}_{k,s}}}.
\end{align}
Especially, the action of $K_s(z)$ on the vacuum is
\begin{align}
    K_{s}(z)\ket{\emptyset}=\frac{\phi(q_{3}^{-1};z,u)^{\bar{\delta}_{k,s-1}}}{\phi(1;z,u)^{\bar{\delta}_{k,s}}}\ket{\emptyset}.
\end{align}
By comparing it with (\ref{eq:vacuum_charge_function}), the shift parameters are 
\begin{align}
    r_{k}=-1,\quad r_{k+1}=1,\quad r_{i}=0\hspace{3mm}(i\neq k,k+1).
\end{align}
All of the equations here stand even for the $n=2$ case.

\subsection{Suspended pinch point and quantum toroidal algebra \texorpdfstring{$\mathfrak{gl}_{2|1}$}{gl21}}\label{sec:toroidalgl2|1}
\subsubsection{Definition of the algebra}
The definition of quantum toroidal $\mathfrak{gl}_{m|n}\hspace{1mm}(m\neq n)$ can be obtained similarly as \cite{Li:2020rij} and the result are the ones defined in \cite{bezerra2019quantum,bezerra2021representations}. In this section, we will focus on the simplest case when $m=2, n=1$. Other cases can be derived similarly, although it will be tedious. The crystal picture of the one and two-dimensional crystal representations we derive here seems to be a little different from \cite{bezerra2021representations}. We hope to fill in this gap in the near future. We focus on one type of one-dimensional crystal representation and two-dimensional crystal representation. Other cases are in Appendix \ref{sec:appendixgl2|1}. 

\begin{figure}[H]
    \centering
    \includegraphics[width=14cm]{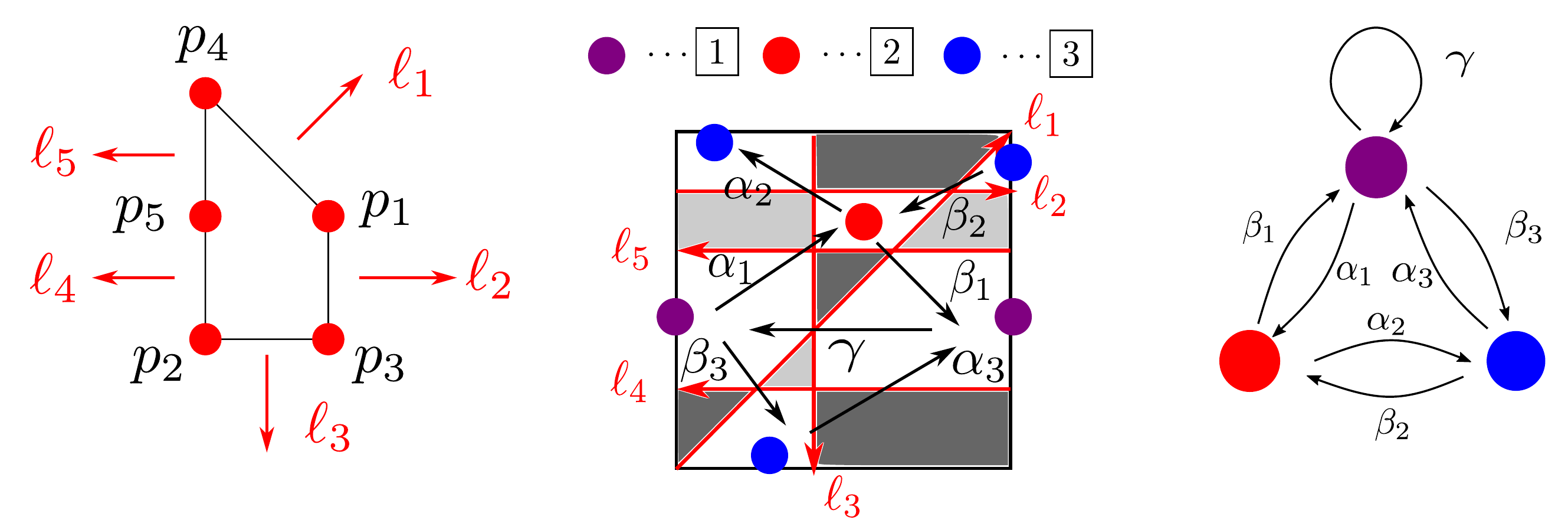}
    \caption{Toric diagram, periodic quiver, and quiver diagram of the suspended pinch point geometry. The lattice points of the toric diagram are $p_{1}=(1,1)$, $p_{2}=(0,0)$, $p_{3}=(1,0)$, $p_{4}=(0,2)$, and $p_{5}=(0,1)$. The external legs are denoted as $\ell_{1},..,\ell_{5}$.}
    \label{fig:gl_(2,1)_toric_periodic_quiver}
\end{figure}
The quantum toroidal algebra corresponding to the suspended pinched point geometry is quantum toroidal $\mathfrak{gl}_{2|1}$. The toric diagram, periodic quiver and quiver diagram of this geometry can be written as Figure \ref{fig:gl_(2,1)_toric_periodic_quiver}. The lattice points of the toric diagram are $p_{1}=(1,1)$, $p_{2}=(0,0)$, $p_{3}=(1,0)$, $p_{4}=(0,2)$, and $p_{5}=(0,1)$. We have 5 external legs, which are denoted as $\ell_{i}$ $(i=1,..5)$.\\
From the loop condition and vertex constraint, we can assign parameters as 
\begin{align}
    \alpha_{1}=\alpha_{3}=q_{1},\quad \beta_{1}=\beta_{3}=q_{3},\quad \alpha_{2}=q_{3}^{-1},\quad \beta_{2}=q_{1}^{-1},\quad \gamma=q_{2},
\end{align}
where $q_{1}q_{2}q_{3}=1$.\\
The bond factors are
\begin{align}
\begin{split}
    &\varphi^{1\Rightarrow 2}(z,w)=\frac{q_{3}^{\frac{1}{2}}z-q_{3}^{-\frac{1}{2}}w}{q_{1}^{-\frac{1}{2}}z-q_{1}^{\frac{1}{2}}w}=\frac{\phi(q_{3};z,w)}{\phi(q_{1}^{-1};z,w)},\; \varphi^{2\Rightarrow 1}(z,w)=\frac{q_{1}^{\frac{1}{2}}z-q_{1}^{-\frac{1}{2}}w}{q_{3}^{-\frac{1}{2}}z^-q_{3}^{\frac{1}{2}}w}=\frac{\phi(q_{1};z,w)}{\phi(q_{3}^{-1};z,w)},\\
    &\varphi^{2\Rightarrow 3}(z,w)=\frac{q_{1}^{-\frac{1}{2}}z-q_{1}^{\frac{1}{2}}w}{q_{3}^{\frac{1}{2}}z-q_{3}^{-\frac{1}{2}}w}=\frac{\phi(q_{1}^{-1};z,w)}{\phi(q_{3};z,w)},\; \varphi^{3\Rightarrow 2}(z,w)=\frac{q_{3}^{-\frac{1}{2}}z-q_{3}^{\frac{1}{2}}w}{q_{1}^{\frac{1}{2}}z-q_{1}^{-\frac{1}{2}}w}=\frac{\phi(q_{3}^{-1};z,w)}{\phi(q_{1};z,w)},\\
    &\varphi^{3\Rightarrow 1}(z,w)=\frac{q_{3}^{\frac{1}{2}}z-q_{3}^{-\frac{1}{2}}w}{q_{1}^{-\frac{1}{2}}z-q_{1}^{\frac{1}{2}}w}=\frac{\phi(q_{3};z,w)}{\phi(q_{1}^{-1};z,w)},\; \varphi^{1\Rightarrow 3}(z,w)=\frac{q_{1}^{\frac{1}{2}}z-q_{1}^{-\frac{1}{2}}w}{q_{3}^{-\frac{1}{2}}z-q_{3}^{\frac{1}{2}}w}=\frac{\phi(q_{1};z,w)}{\phi(q_{3}^{-1};z,w)},\\
    &\varphi^{1\Rightarrow 1}(z,w)=\frac{q_{2}^{\frac{1}{2}}z-q_{2}^{-\frac{1}{2}}w}{q_{2}^{-\frac{1}{2}}z-q_{2}^{\frac{1}{2}}w}=\frac{\phi(q_{2};z,w)}{\phi(q_{2}^{-1};z,w)},\; \varphi^{2\Rightarrow 2}(z,w)=\varphi^{3\Rightarrow 3}(z,w)=1.
\end{split}
\end{align}

The bond factors show that vertex 2 and 3 are fermionic while vertex 1 is bosonic.

\subsubsection{Subquiver and crystal shape}\label{sec:gl2|1subquiver}
\begin{figure}[H]
    \centering
    \includegraphics[width=8cm]{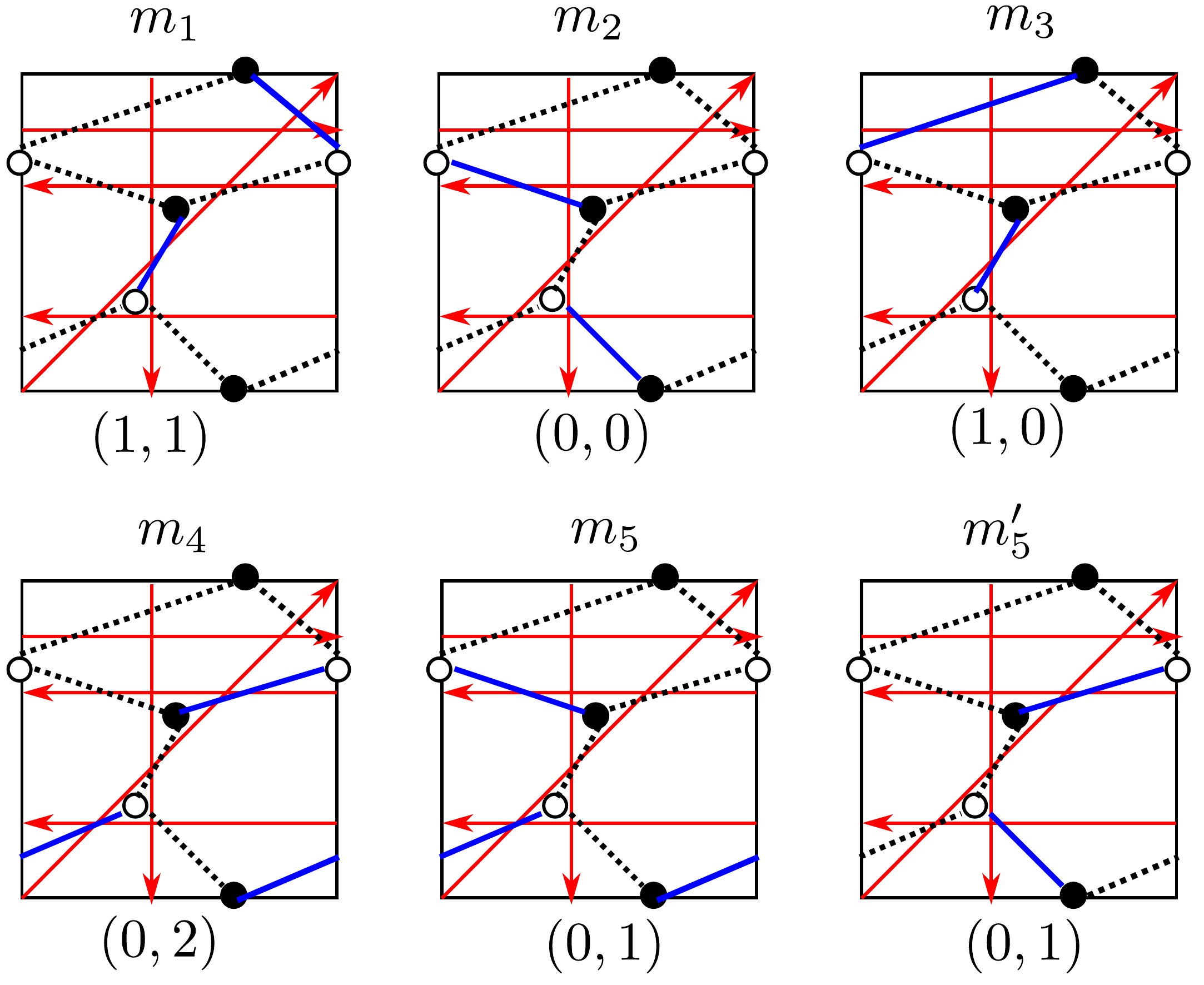}
    \caption{Perfect matchings of the suspended pinch geometry. Perfect matchings for corner divisors $p_{1}, p_{2}, p_{3}, p_{4}$ are $m_{1}, m_{2}, m_{3}, m_{4}$ respectively. Perfect matchings for divisor $p_{5}$ are $m_{5}, m_{5}'$. We note $m_{1}=\{\beta_{2}, \gamma\}, m_{2}=\{\alpha_{1}, \alpha_{3}\},  m_{3}=\{\alpha_{2}, \gamma\}, m_{4}=\{\beta_{1}, \beta_{3}\}, m_{5}=\{\alpha_{1}, \beta_{3}\}, m_{5}'=\{\beta_{1},\alpha_{3}\}$.}
    \label{fig:gl_(2,1)perfectmatching}
\end{figure}
We have four corner divisors $p_{1},p_{2},p_{3},p_{4}$. The perfect matchings of these divisors are unique and denoted $m_{1},m_{2},m_{3},m_{4}$ respectively. For divisor $p_{5}$ we have two perfect matchings and they are $m_{5},m_{5}'$ (see Figure \ref{fig:gl_(2,1)perfectmatching}).

Let us consider the two-dimensional crystal associated with corner divisors $p_{1}$ and $p_{2}$. The subquiver and subcrystal can be obtained by removing the arrows of the corresponding perfect matching. The subquivers and two-dimensional crystals are in Figure \ref{fig:gl_(2,1)2dimcrystal_subquiver}. The crystal of $p_{1}$ is composed of three kinds of atoms included in a box: blue triangle, red triangle, and purple parallelogram (Figure \ref{fig:gl_(2,1)2dcrystal_p1}).
For the crystal $p_{2}$, we also have three kinds of atoms in a box: blue triangle, red triangle, and purple rectangle (Figure \ref{fig:gl_(2,1)2dcrystal_p2}).

\begin{figure}[ht]
    \begin{tabular}{cc}
      \begin{minipage}{0.45\hsize}
        \centering
        \includegraphics[width=6.5cm]{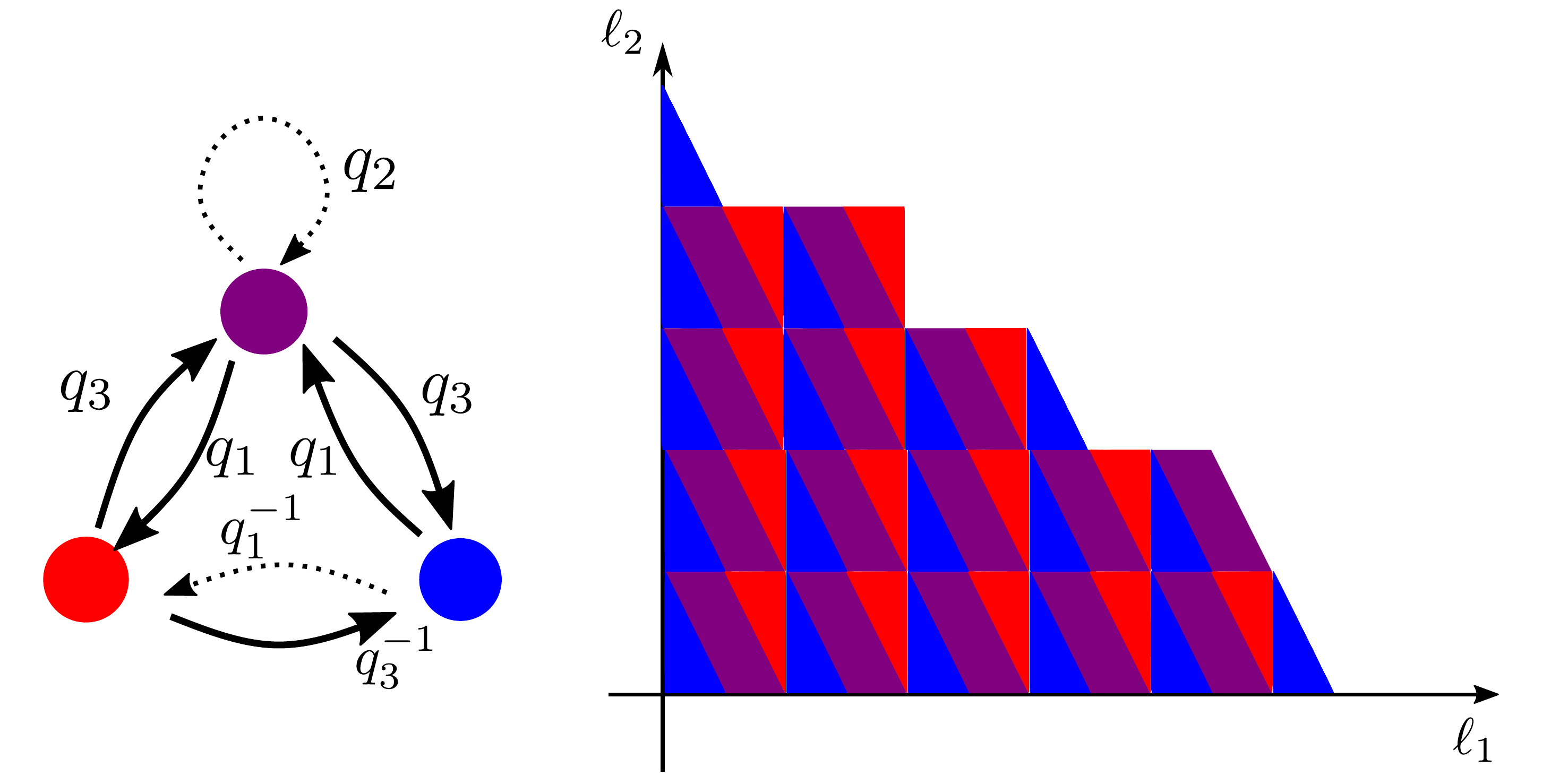}
        \subcaption{Subquiver and shape of two-dimensional crystal of divisor $p_{1}$.}\label{fig:gl_(2,1)2dcrystal_p1}
      \end{minipage}&\hfill
      \begin{minipage}{0.45\hsize}
        \centering
       \includegraphics[width=6.5cm]{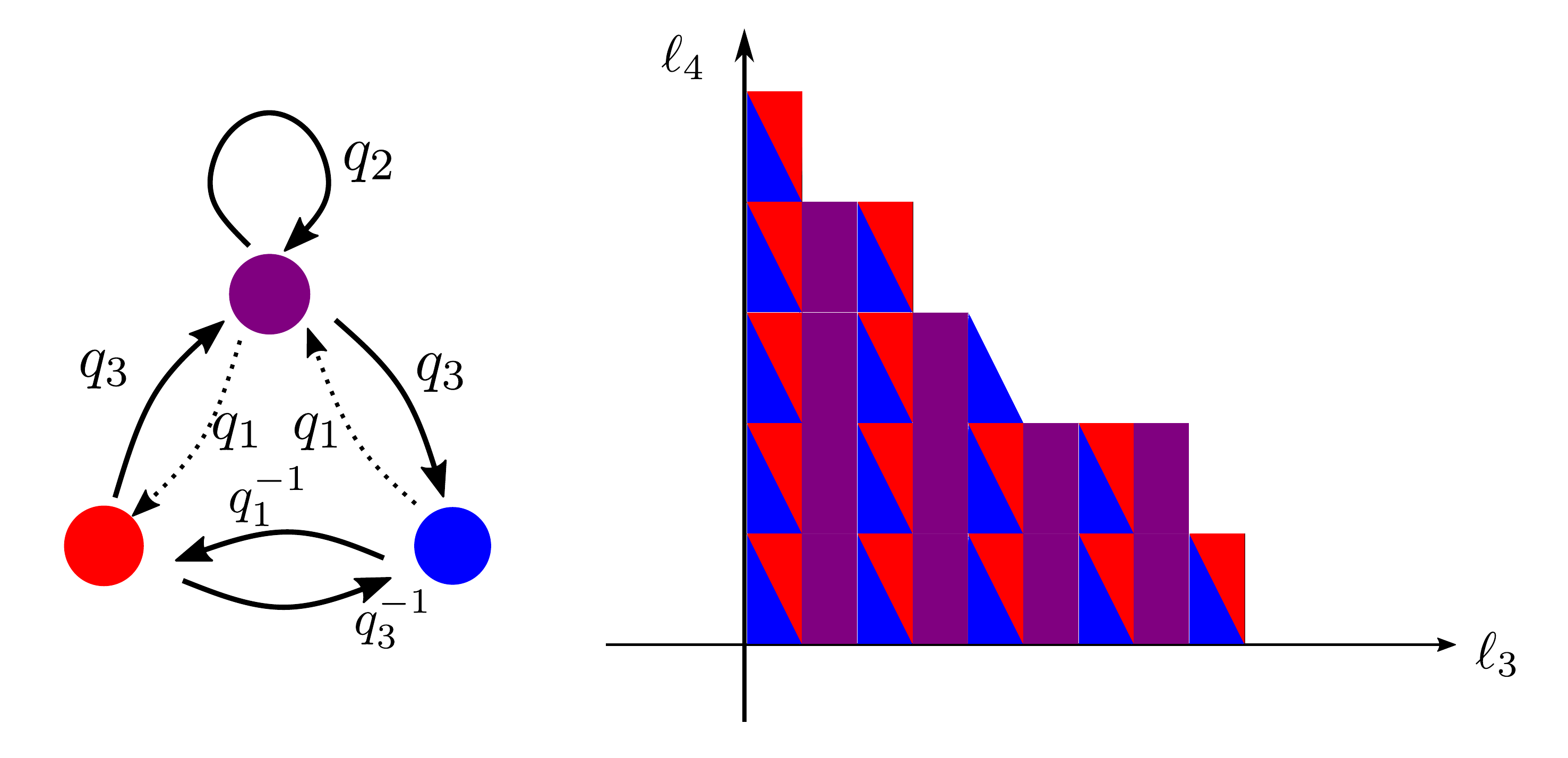}
       \subcaption{Subquiver and shape of two-dimensional crystal of divisor $p_{2}$.}\label{fig:gl_(2,1)2dcrystal_p2}
      \end{minipage}
    \end{tabular}
\caption{Subquivers and two-dimensional crystals of corner divisors $p_{1}$ and $p_{2}$. (a) The arrows removed are $m_{1}=\{\beta_{2},\gamma\}$, where $\beta_{2}=q_{1}^{-1}$ and $\gamma=q_{2}$. (b) The arrows removed are $m_{2}=\{\alpha_{1},\alpha_{3}\}$, where $\alpha_{1}=\alpha_{3}=q_{1}$.}\label{fig:gl_(2,1)2dimcrystal_subquiver}
\end{figure}
\begin{figure}[ht]
    \begin{tabular}{cc}
      \begin{minipage}{0.45\hsize}
        \centering
        \includegraphics[width=6cm]{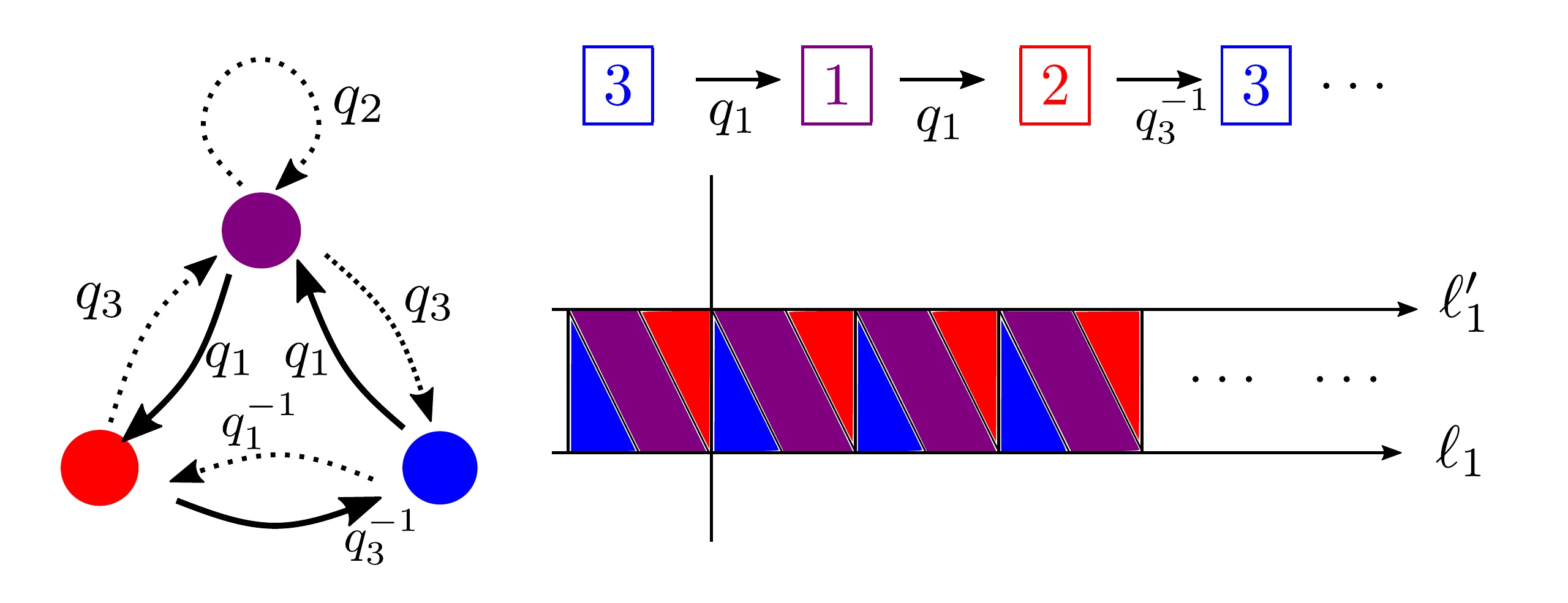}
        \subcaption{Crystal associated with $\ell_{1}$.}\label{fig:gl_(2,1)1dimcrystal_l1}
      \end{minipage}&
      \begin{minipage}{0.45\hsize}
        \centering
      \includegraphics[width=6cm]{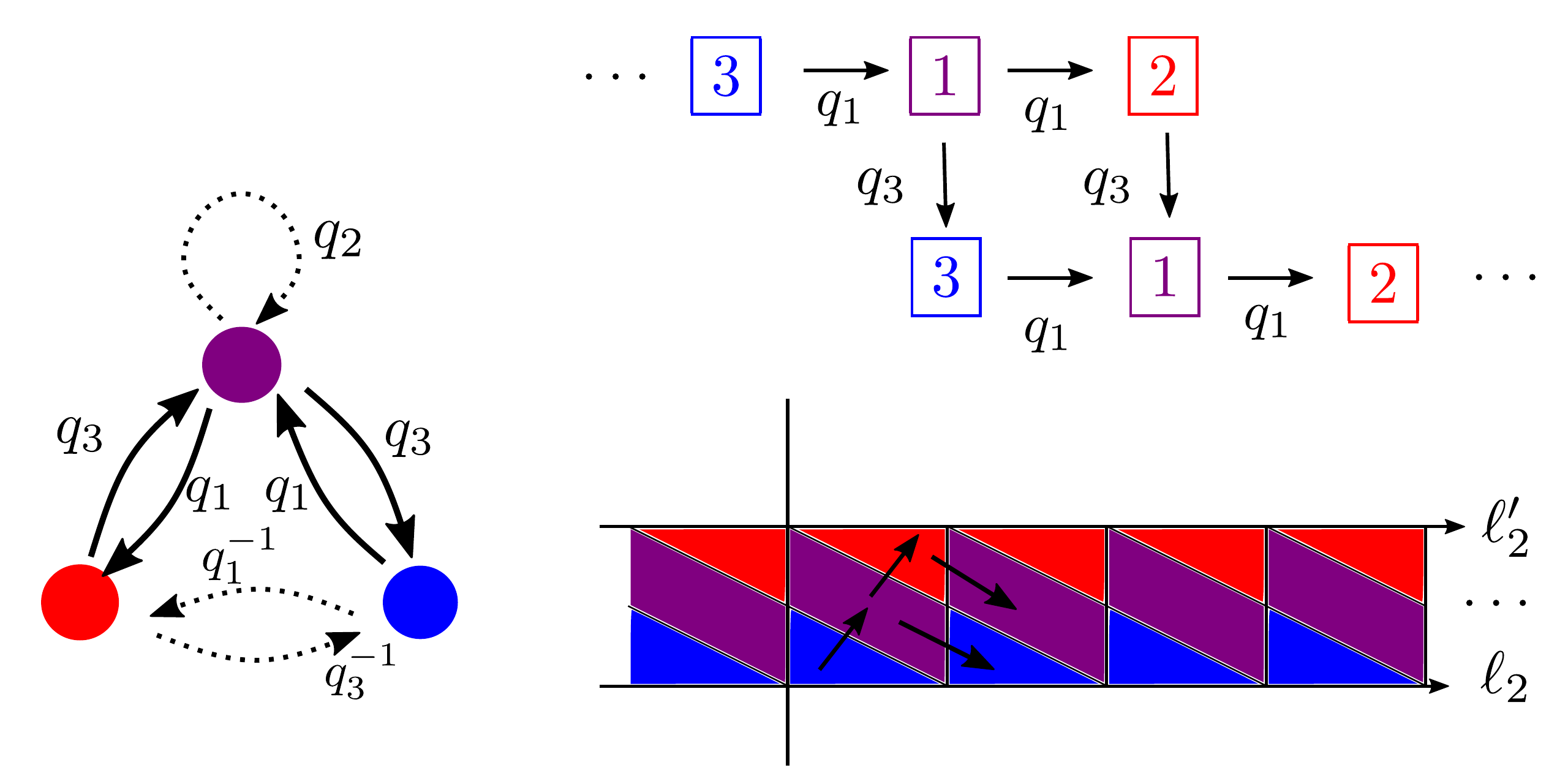}
       \subcaption{Crystal associated with $\ell_{2}$.}\label{fig:gl_(2,1)1dimcrystal_l2}
      \end{minipage}\\
      \begin{minipage}{0.45\hsize}
        \centering
      \includegraphics[width=6cm]{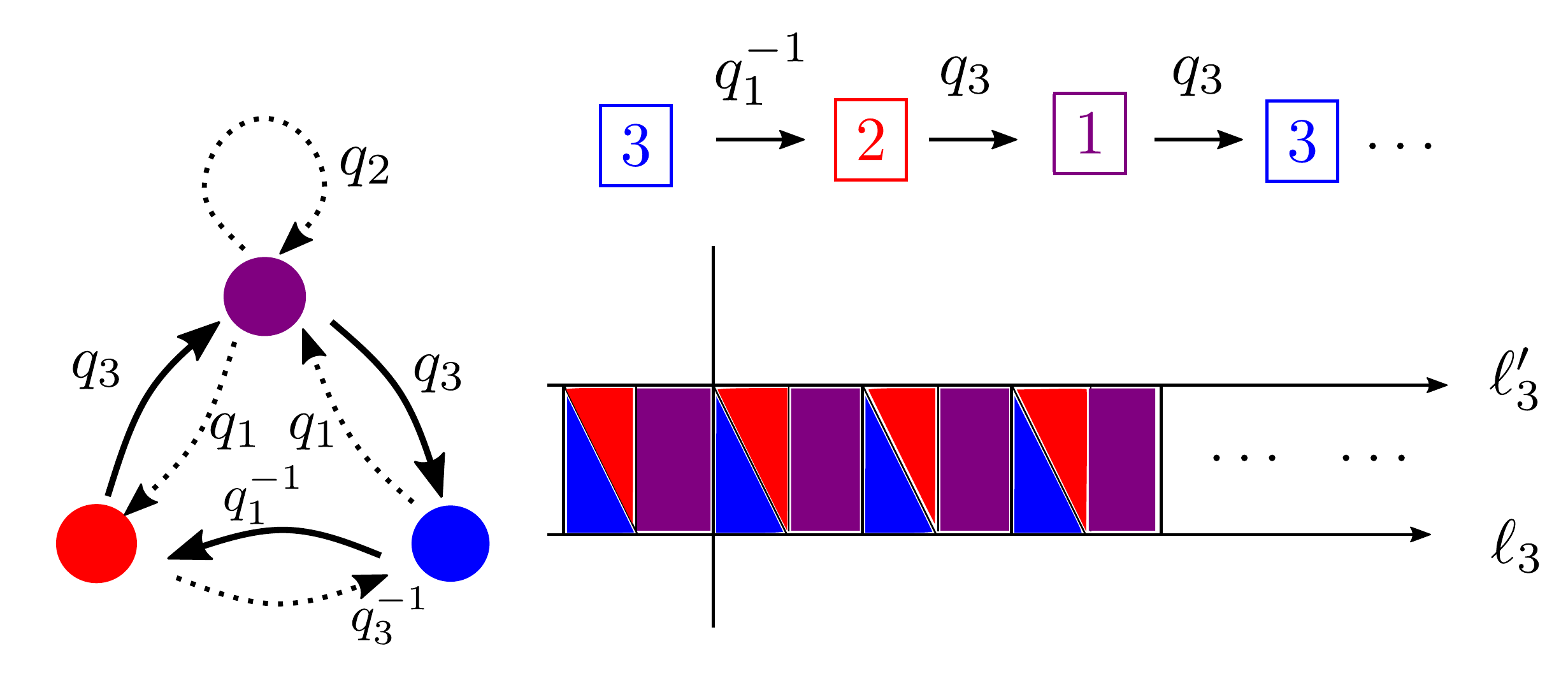}
       \subcaption{Crystal associated with $\ell_{3}$.}\label{fig:gl_(2,1)1dimcrystal_l3}
      \end{minipage}&
      \begin{minipage}{0.45\hsize}
        \centering
      \includegraphics[width=6cm]{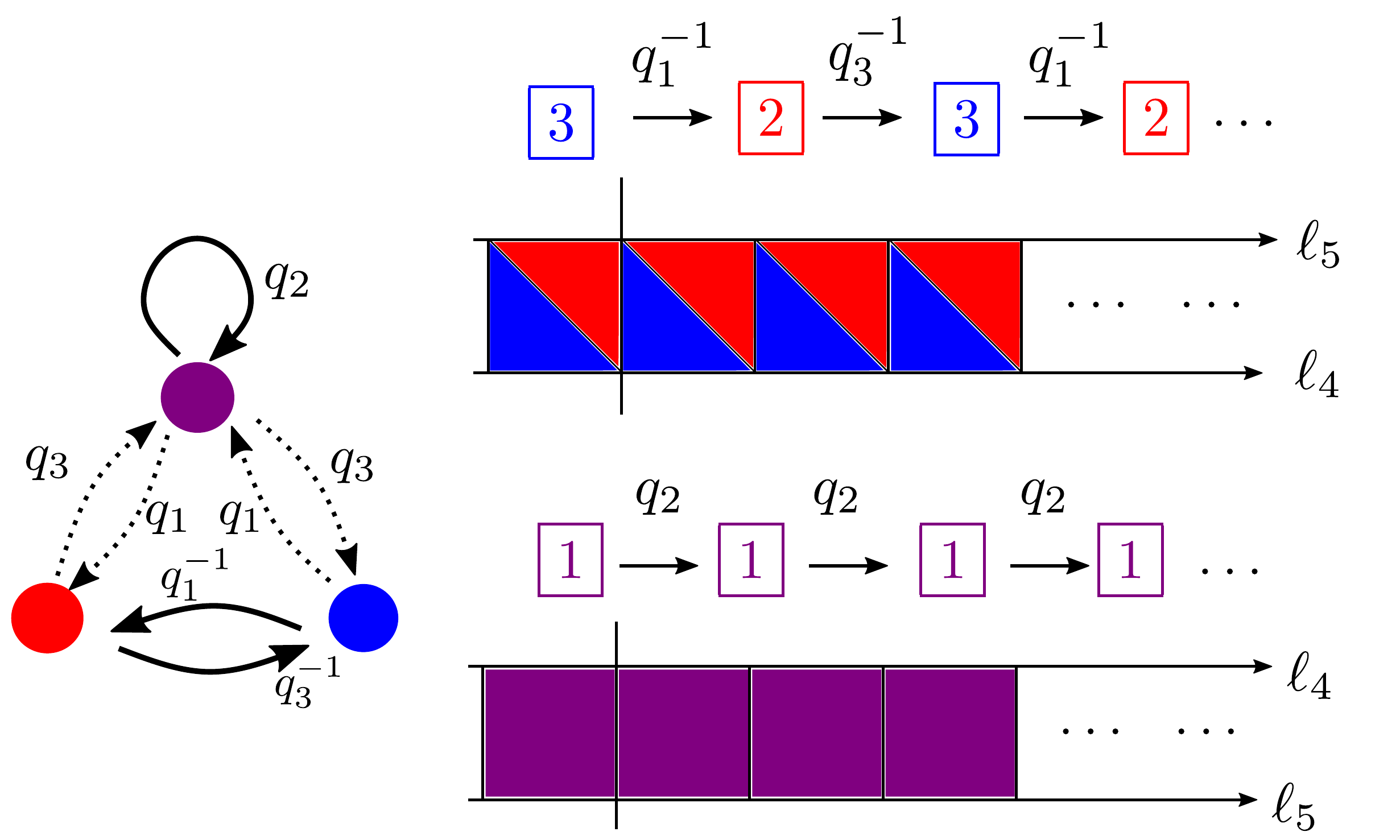}
       \subcaption{Crystals associated with $\ell_{4}$ and $\ell_{5}$.}\label{fig:gl_(2,1)1dimcrystal_l4,5}
      \end{minipage}
    \end{tabular}
\caption{One-dimensional crystals of $\mathfrak{gl}_{2|1}$. (a) The arrows removed are $m_{1}\cup m_{4}=\{\beta_{1},\beta_{2},\beta_{3},\gamma\}$. It is surrounded by the same external leg $\ell_{1}$. $\ell_{1}'$ is in a different fundamental region. (b) The arrows removed are $m_{1}\cup m_{3}=\{\alpha_{2},\beta_{2},\gamma\}$. It is surrounded by the same external leg $\ell_{2}$. (c) The arrows removed are $m_{1}\cup m_{3}=\{\alpha_{2},\beta_{2},\gamma\}$. It is surrounded by the same external leg $\ell_{3}$. (d) After removing arrows of $m_{2}\cup m_{5}\cup m_{5}'=m_{4}\cup m_{5}\cup m_{5}'=\{\alpha_{1},\alpha_{3},\beta_{1},\beta_{3}\}$, two vertices and one vertex decouple. The crystal is surrounded by two different external legs $\ell_{4}$ and $\ell_{5}$. }\label{fig:gl_(2,1)1dimcrystal}
\end{figure}
Let us derive the subquivers and shapes of the one-dimensional crystals. Since we have five external legs, we have five one-dimensional crystals (see Figure \ref{fig:gl_(2,1)1dimcrystal}). We need to remove arrows $m_{1}\cup m_{4}=\{\beta_{1},\beta_{2},\beta_{3},\gamma\}$ for $\ell_{1}$ (Figure \ref{fig:gl_(2,1)1dimcrystal_l1}), $m_{1}\cup m_{3}=\{\alpha_{2},\beta_{2},\gamma\}$ for $\ell_{2}$ (Figure \ref{fig:gl_(2,1)1dimcrystal_l2}), $m_{1}\cup m_{3}=\{\alpha_{2},\beta_{2},\gamma\}$ for $\ell_{3}$ (Figure \ref{fig:gl_(2,1)1dimcrystal_l3}), and $m_{2}\cup m_{5}\cup m_{5}'=m_{4}\cup m_{5}\cup m_{5}'=\{\alpha_{1},\alpha_{3},\beta_{1},\beta_{3}\}$ for $\ell_{4}$, $\ell_{5}$ (Figure \ref{fig:gl_(2,1)1dimcrystal_l4,5}). 


\subsubsection{One-dimensional crystal \texorpdfstring{$\ell_{1}$}{l1}}\label{sec:gl2|11dl1}
\begin{figure}[h]
    \begin{tabular}{cc}
      \begin{minipage}{0.45\hsize}
        \centering
        \includegraphics[width=6.5cm]{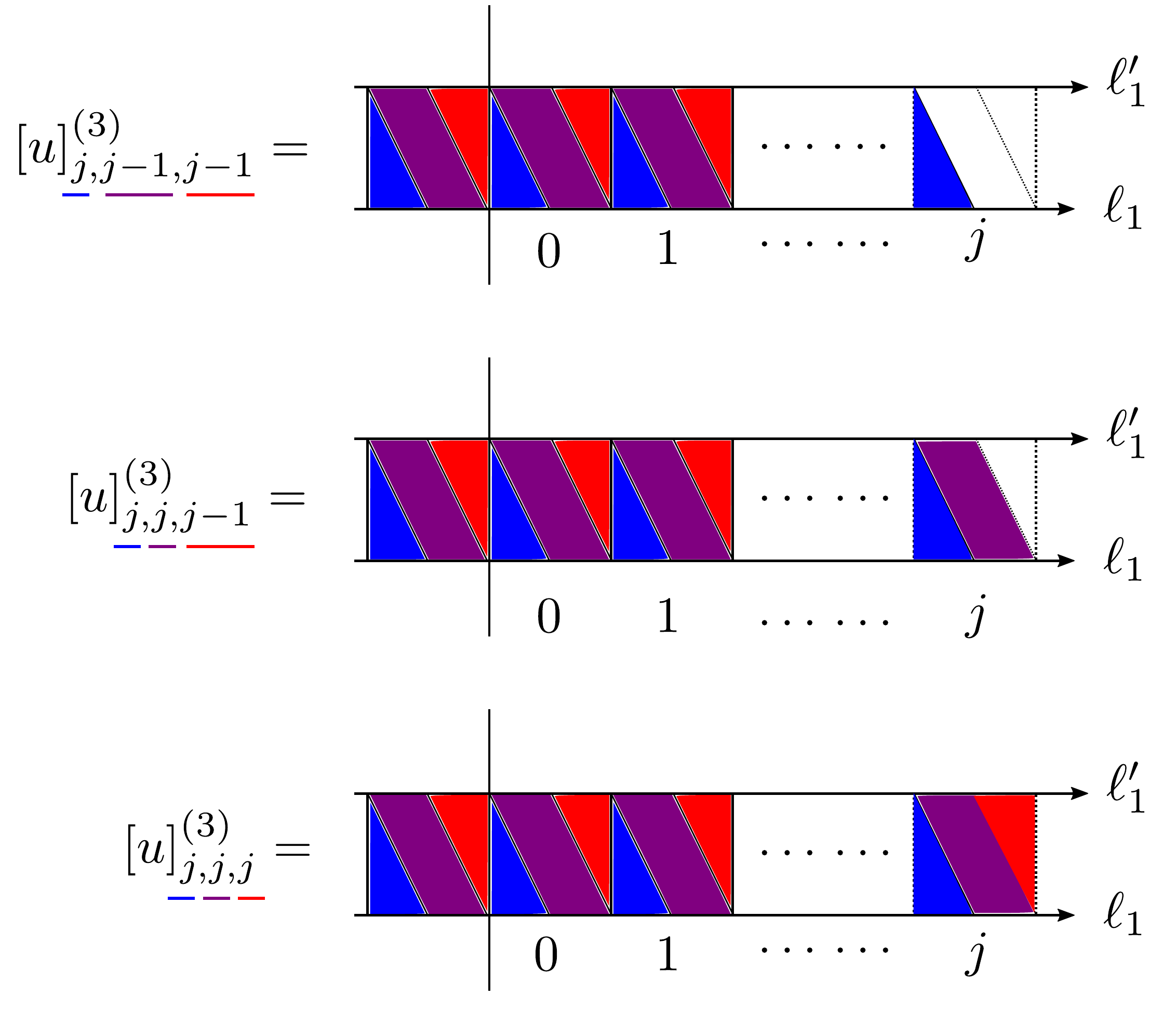}
        \subcaption{Basis of one-dimensional crystal representation $\ell_{1}$.}\label{fig:gl_(2,1)1dimrep_l1_vectorbasis}
      \end{minipage}&
      \begin{minipage}{0.47\hsize}
        \centering
       \includegraphics[width=7cm]{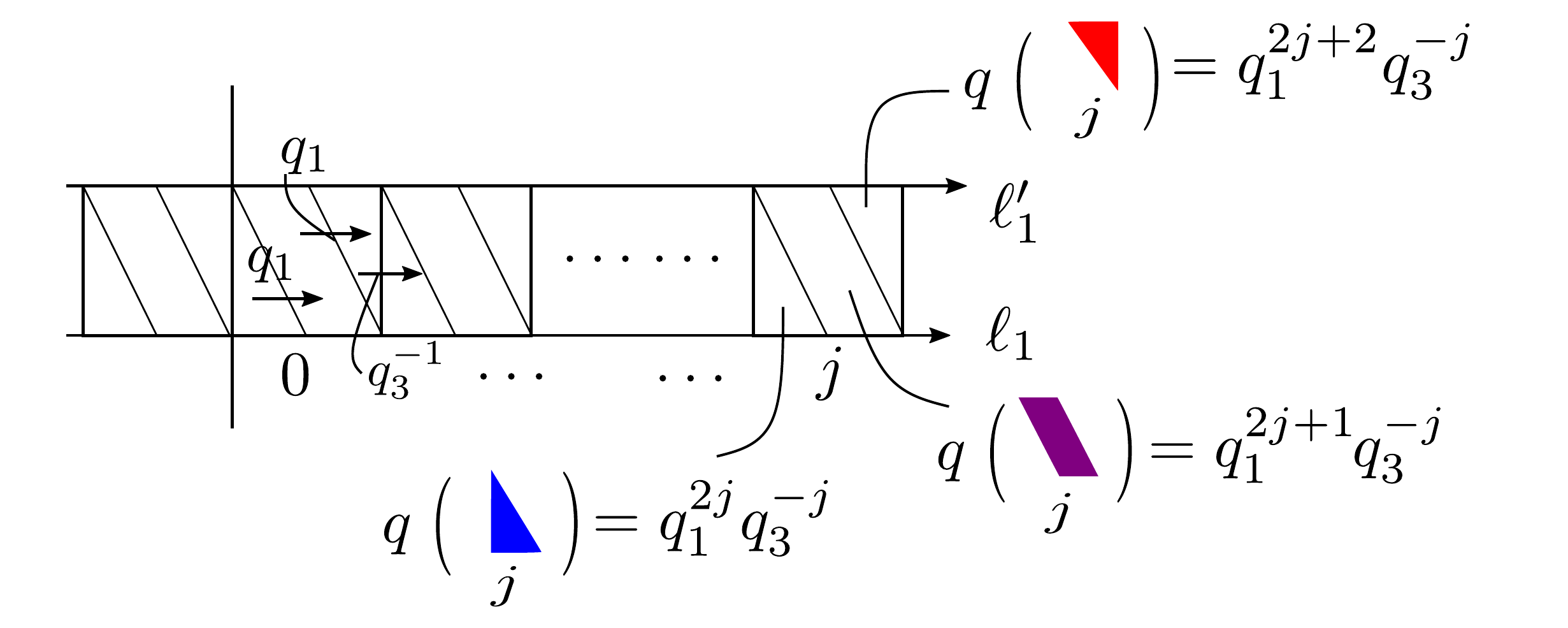}
       \subcaption{Coordinates of atoms.}\label{fig:gl_(2,1)1dimrep_l1_coordinate}
      \end{minipage}
    \end{tabular}
\caption{Basis and coordinates of the one-dimensional crystal representation $V^{(\ell_{1})}(u)$. The one-dimensional crystal is a row of boxes with three-types of atoms in it. Each box is assigned an integer $j\in\mathbb{Z}$, which describes the number of boxes right to the border. We note that it is counted $0,1,..$.}
\end{figure}

Let us consider the representation of crystal in Figure \ref{fig:gl_(2,1)1dimcrystal_l1}. We denote the basis of this representation $[u]^{(3)}_{j,j-1,j-1},[u]^{(3)}_{j,j,j-1}$, and $[u]^{(3)}_{j,j,j}$ $(j\in\mathbb{Z})$. The vector space is denoted $V^{(\ell_{1})}(u)$. Note that we set the origin to have a blue atom. $[u]^{(3)}_{j,j-1,j-1}$ has $j+1$ blue, $j$ purple and $j$ red atoms, $[u]^{(3)}_{j,j,j-1}$ has $j+1$ blue, $j+1$ purple, and $j$ red atoms, while $[u]^{(3)}_{j,j,j}$ has $j+1$ blue, $j+1$ purple, and $j+1$ red atoms. This is illustrated as in Figure \ref{fig:gl_(2,1)1dimrep_l1_vectorbasis}. A blue triangle in a box labeled with $j$ has coordinate $q_{1}^{2j}q_{3}^{-j}$, a purple parallelogram has coordinate $q_{1}^{2j+1}q_{3}^{-j}$, and a red triangle has coordinate $q_{1}^{2j+2}q_{3}^{-j}$. 
The generators $K_{i}(z)$ act diagonally and we can set
\begin{align}
     K_{i}^{\pm}(z)\begin{cases}
    [u]^{(3)}_{j,j-1,j-1}\\
    [u]^{(3)}_{j,j,j-1}\\
    [u]^{(3)}_{j,j,j}
    \end{cases}&=\begin{cases}
    [\Psi_{[u]^{(3)}_{j,j-1,j-1}}^{(i)}(z)]_{\pm}[u]^{(3)}_{j,j-1,j-1}\\
    [\Psi_{[u]^{(3)}_{j,j,j-1}}^{(i)}(z)]_{\pm}[u]^{(3)}_{j,j,j-1}\\
    [\Psi_{[u]^{(3)}_{j,j,j}}^{(i)}(z)]_{\pm}[u]^{(3)}_{j,j,j}
    \end{cases}.
\end{align}
The parity conditions are 
\begin{align}
\begin{split}
    |[u]^{(3)}_{j,j,j}|-|[u]^{(3)}_{j,j,j-1}|=1,\quad |[u]^{(3)}_{j,j,j-1}|-|[u]^{(3)}_{j,j-1,j-1}|=0.
\end{split}
\end{align}
We set the parity condition to be 
\begin{align}
    |[u]^{(3)}_{j,j,j}|=0,\quad |[u]^{(3)}_{j,j,j-1}|=1,\quad |[u]^{(3)}_{j,j-1,j-1}|=1.
\end{align}

The action of $E_{i}(z)$ and $F_{i}(z)$ can be written in a convenient way which we will use in deriving the two-dimensional crystal representations (see Figure \ref{fig:gl_(2,1)1dcrystal_l1_usefulconvention}).
For $\forall\sigma\in\mathbb{Z}$, it can be written as 
\begin{align}
\sigma=3r(\sigma)+s(\sigma),\quad s(\sigma)=0,1,2
\end{align}
where $r(\sigma)$ is the quotient of $\sigma$ by 3 and $s(\sigma)$ is the remainder after being divided by 3.
We unify $[u]^{(3)}_{j,j,j}$, $[u]^{(3)}_{j,j,j-1}$, and $[u]^{(3)}_{j,j-1,j-1}$ by defining new vectors 
\begin{align}
    [u]^{(3)}_{\sigma}=[u]^{(3)}_{(r(\sigma),s(\sigma))}=\begin{dcases}
    [u]^{(3)}_{r(\sigma),r(\sigma)-1,r(\sigma)-1},\quad s(\sigma)=0,\\
    [u]^{(3)}_{r(\sigma),r(\sigma),r(\sigma)-1},\quad s(\sigma)=1,\\
    [u]^{(3)}_{r(\sigma),r(\sigma),r(\sigma)},\quad s(\sigma)=2.\label{eq:gl_(2,1)1dimcrystal_newconvention_l1}
    \end{dcases}
\end{align}
$\sigma$ is the number of atoms counted as $0,1,...$ from the right of the border.

From now on, we write the parity condition as
\begin{align}
 |\sigma|\equiv|[u]^{(3)}_{\sigma}|=\begin{dcases}
 0,\quad s(\sigma)=0,\\
 1,\quad s(\sigma)=1,\\
 1,\quad s(\sigma)=2.
 \end{dcases}   
\end{align}

\begin{figure}[t]
    \centering
    \includegraphics[width=11cm]{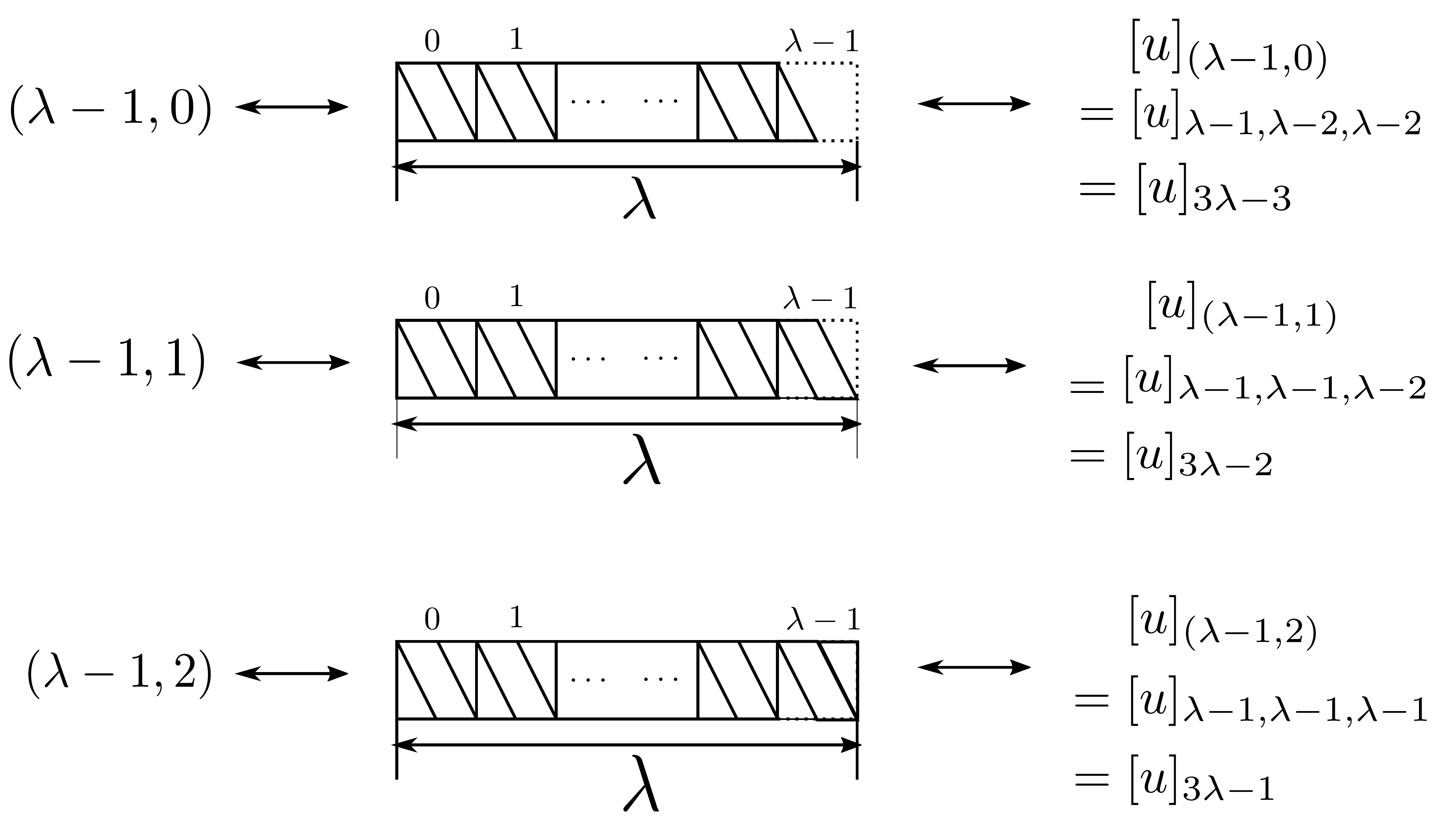}
    \caption{Generalization of Young diagram and correspondence with vectors for one-dimensional crystal $\ell_{1}$. The generalized partition is expressed by two numbers $(\lambda-1,\tau)\in\mathbb{Z}\times\mathbb{Z}_{3}$. We note we set $\tau\in\mathbb{Z}_{3}=\{0,1,2\}$. Using (\ref{eq:gl_(2,1)1dimcrystal_newconvention_l1}), it can be written as $\sigma=3\lambda-3+\tau$, $r(3\lambda-3+\tau)=\lambda-1$, and $s(3\lambda-3+\tau)=\tau$. }
    \label{fig:gl_(2,1)1dcrystal_l1_usefulconvention}
\end{figure}

The action of $E_{s}(z)$ and $F_{s}(z)$ are written as 
\begin{align}
\begin{split}
    E_{s}(z)[u]^{(3)}_{\sigma}&=\mathcal{E}_{s}([u]^{(3)}_{\sigma})\bar{\delta}_{s,\sigma+1}\delta\left(\frac{z}{u(q_{1}^{2}q_{3}^{-1})^{r(\sigma+1)}q^{s(\sigma+1)}_{1}}\right)[u]^{(3)}_{\sigma+1},\\
    F_{s}(z)[u]^{(3)}_{\sigma}&=\mathcal{F}_{s}([u]^{(3)}_{\sigma})\bar{\delta}_{s,\sigma}\delta\left(\frac{z}{u(q_{1}^{2}q_{3}^{-1})^{r(\sigma)}q_{1}^{s(\sigma)}}\right)[u]^{(3)}_{\sigma-1},
    \end{split}
\end{align}
where the explicit expressions of the coefficients $\mathcal{E}_{s}$ and $\mathcal{F}_{s}$ are omitted, and 
\begin{align}
\bar{\delta}_{i,j}=\begin{cases}
1,\quad i\equiv j\mod 3\\
0,\quad i\not\equiv j \mod3
\end{cases}.
\end{align}

The charge functions can be derived from the KE relation as
\begin{align}
\begin{split}
 &\Psi_{[u]^{(3)}_{j,j-1,j-1}}^{(1)}(z)=\frac{\phi(q_{3}^{j+1}q_{1}^{-2j};z,u)}{\phi(q_{1}^{-1-2j}q_{3}^{j};z,u)},\quad \Psi_{[u]^{(3)}_{j,j,j-1}}^{(1)}(z)=\frac{\phi(q_{3}^{j-1}q_{1}^{-2j-2};z,u)}{\phi(q_{1}^{-1-2j}q_{3}^{j};z,u)},\\
     &\Psi_{[u]^{(3)}_{j,j,j-1}}^{(2)}(z)=\frac{\phi(q_{3}^{j+1}q_{1}^{-2j-1};z,u)}{\phi(q_{1}^{-2-2j}q_{3}^{j};z,u)},\quad \Psi_{[u]^{(3)}_{j,j,j}}^{(2)}(z)=\frac{\phi(q_{3}^{j+1}q_{1}^{-2j-1};z,u)}{\phi(q_{1}^{-2-2j}q_{3}^{j};z,u)},\\
     &\Psi_{[u]^{(3)}_{j,j-1,j-1}}^{(3)}(z)=\frac{\phi(q_{3}^{j-1}q_{1}^{-2j-1};z,u)}{\phi(q_{1}^{-2j}q_{3}^{j};z,u)},\quad \Psi_{[u]^{(3)}_{j,j,j}}^{(3)}(z)=\frac{\phi(q_{3}^{j}q_{1}^{-2j-3};z,u)}{\phi(q_{1}^{-2-2j}q_{3}^{j+1};z,u)},\\
      &\Psi_{[u]^{(3)}_{j,j,j}}^{(1)}(z)= \Psi_{[u]^{(3)}_{j,j-1,j-1}}^{(2)}(z)=\Psi_{[u]^{(3)}_{j,j,j-1}}^{(3)}(z)=1.
      \end{split}\label{eq:gl_(2,1)_l1_chargefunction}
\end{align}
Since the charge functions have the same number of poles and zeros, this is a representation of the unshifted quantum toroidal algebra. The shift parameters are 
\begin{align}
    r_{1}=r_{2}=r_{3}=0.
\end{align}

\subsubsection{Two-dimensional crystal of \texorpdfstring{$p_{1}=(1,1)$}{p111}} \label{sec:gl2|12dp1}              
Let us derive the explicit representations of the two-dimensional crystals associated with divisor $p_{1}$ in Figure \ref{fig:gl_(2,1)2dcrystal_p1}. We note we set the origin to be the blue atom. 

We can do two ways to obtain this crystal picture, by using tensor products $V^{(\ell_{1})}(u)\otimes V^{(\ell_{1})}(v)$ or by using tensor products $V^{(\ell_{2})}(u)\otimes V^{(\ell_{2})}(v)$. We discuss here the derivation of the crystal representation by using the former tensor products and charge functions in (\ref{eq:gl_(2,1)_l1_chargefunction}). We set here $v=q_{1}q_{3}u$ and this comes from the crystal picture. The melting rule of this crystal is the following as claimed in \cite{Nishinaka_2014}:
 \begin{itemize}
 \item
 A blue triangle can be removed if and only if its left and lower arrows are not attached to other atoms.
 \item
 A red triangle can be removed if and only if its slope arrow is not attached to other atoms.
 \item 
 A purple parallelogram can be removed if and only if its left and lower arrow are not attached to other atoms.
 \end{itemize}
 \begin{figure}
     \centering
     \includegraphics[width=12.5cm]{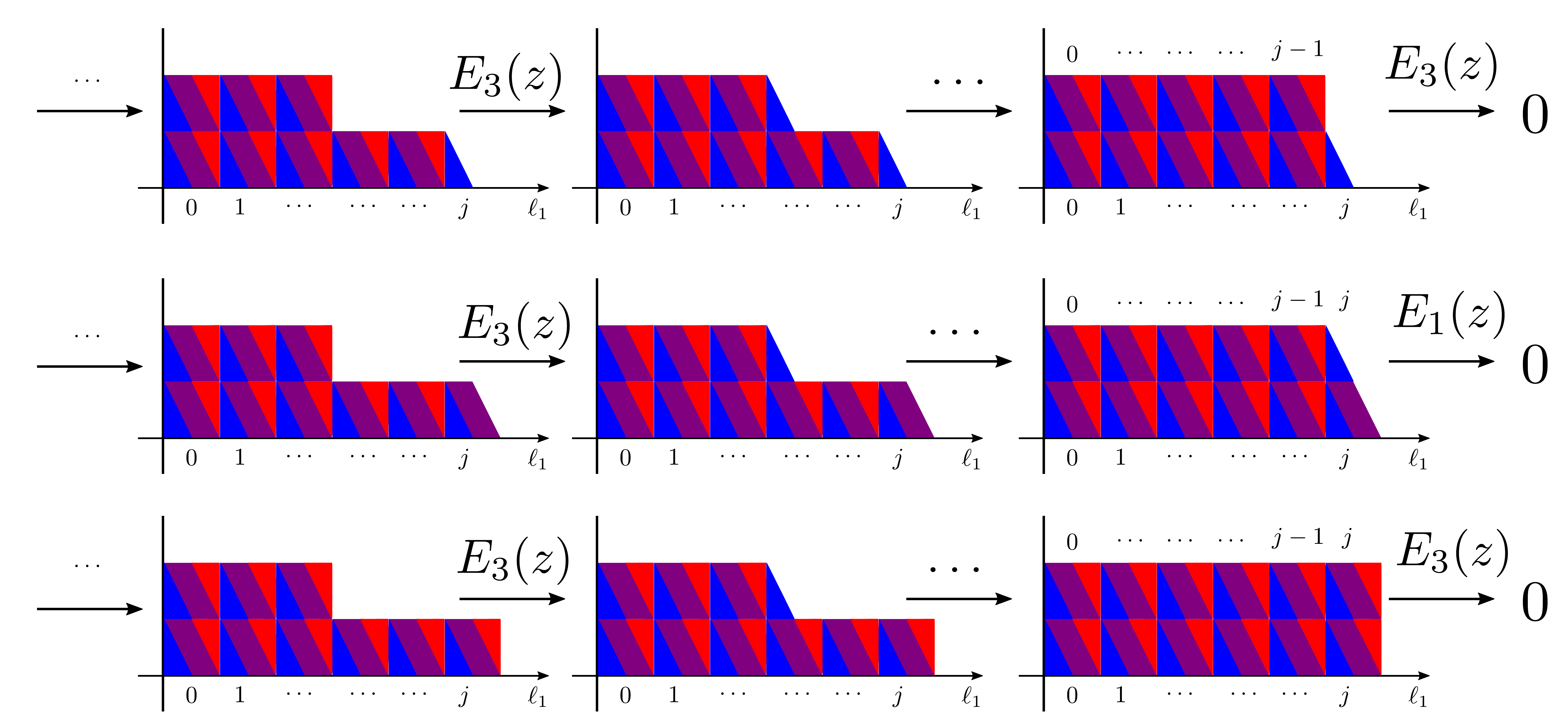}
     \caption{Action of generators on second tensor component of basis of $V^{(\ell_{1})}(u)\otimes V^{(\ell_{1})}(uq_{1}q_{3})$. Top: When the first component is $[u]_{j,j-1,j-1}$, after acting $E_{s}(z)$ many times on the second tensor component and increasing the length of it, it will vanish due to $K_{3}(z)\otimes E_{3}(z) [u]_{j,j-1,j-1}\otimes [uq_{1}q_{3}]_{j-1,j-1,j-1}=0$. Middle: When the first component is $[u]_{j,j,j-1}$, after acting $E_{s}(z)$ many times, it will vanish due to $K_{1}(z)\otimes E_{1}(z) [u]_{j,j,j-1}\otimes [uq_{1}q_{3}]_{j,j-1,j-1}=0$. Bottom: When the first component is $[u]_{j,j,j}$, after acting $E_{s}(z)$ many times, it will vanish due to $K_{3}(z)\otimes E_{3}(z) [u]_{j,j,j}\otimes [uq_{1}q_{3}]_{j,j,j}=0$.}
     \label{fig:gl_(2,1)2dcrystal_p1_meltingrule}
 \end{figure}

 The basis of this representation is defined by the tensor product of one-dimensional representations as\footnote{We omit the superscript $(3)$ in the basis $[u]^{(3)}$.}
 \begin{align}
    \otimes_{i=1}^{N}V^{(\ell_{1})}((q_{1}q_{3})^{i-1}u)\ni \otimes_{i=1}^{N}[(q_{1}q_{3})^{i-1}u]_{\sigma_{i}-1}\equiv\ket{\sigma},\quad \sigma=(\sigma_{1},...\sigma_{N})\in\mathbb{Z}^{N}, 
 \end{align} where we used the convention in (\ref{eq:gl_(2,1)1dimcrystal_newconvention_l1}) and Figure \ref{fig:gl_(2,1)1dcrystal_l1_usefulconvention}.
 The $q_1q_3$ shift of parameters $(q_1q_3)^{i-1}u\; (i=1, 2, \cdots, N)$ ensures the melting rule, and this basis forms a submodule (see Figure \ref{fig:gl_(2,1)2dcrystal_p1_meltingrule}).
 $\sigma_{i}$ here is the number of atoms counted as $1,2,...$ from the right of the border. The shape of the row of the atoms depends on the remainder of $\sigma_{i}-1$ after being divided by 3. The poles and zeros are determined by the quotient $r(\sigma_{i}-1)$ and the remainder $s(\sigma_{i}-1)\in\{0,1,2\}$. The melting rule can be understood in a simple way if we introduce the following conventions:
\begin{align}
\begin{split}
    &(\lambda,\tau)=((\lambda_{1},\tau_{1}),(\lambda_{2},\tau_{2}),.....(\lambda_{N},\tau_{N}))\in\mathbb{Z}^{N}\times\mathbb{Z}_{3}^{N},\quad \mathbb{Z}_{3}=\{0,1,2\}\\
    &\ket{\sigma}=\otimes_{i=1}^{N}[(q_{1}q_{3})^{i-1}u]_{\sigma_{i}-1}=\otimes_{i=1}^{N}[(q_{1}q_{3})^{i-1}u]_{(\lambda_{i}-1,\tau_{i})}=\ket{\lambda,\tau}\\
    &r(\sigma_{i}-1)=\lambda_{i}-1,\quad s(\sigma_{i}-1)=\tau_{i}
    \end{split}
\end{align}
Using this, the melting rule is, for $i<j$ 
\begin{align}
    \begin{split}
    &(\lambda_{i},0)>(\lambda_{j},0),\quad (\lambda_{i},1)>(\lambda_{j},1),\quad (\lambda_{i},2)\geq(\lambda_{j},2),\\
    &(\lambda_{i},0)>(\lambda_{i},1),\quad (\lambda_{i},1)\geq(\lambda_{j},0),\quad (\lambda_{i},0)>(\lambda_{i},2),\\ &(\lambda_{i},2)\geq(\lambda_{j},0),\quad (\lambda_{i},1)>(\lambda_{i},2),\quad (\lambda_{i},2)\geq(\lambda_{j},1),
    \end{split}
\end{align}
where for example $(\lambda_{i},\tau_{i})>(\lambda_{j},\tau_{j})$ means $\lambda_{i}>\lambda_{j}$.

$\sigma\in\mathbb{Z}^{N}$ can naturally be embedded into $\mathbb{Z}^{N+1}$ by setting $\sigma_{N+1}=0$. This is equivalent to embed $(\lambda,\tau)\in\mathbb{Z}^{N}\times\mathbb{Z}_{3}^{N}$ into $\mathbb{Z}^{N+1}\times\mathbb{Z}_{3}^{N+1}$ by setting $\lambda_{N+1}=0$ and $\tau_{N+1}=2.$ From now, let us consider the action of the generators on $\ket{\sigma}$ and take the limit $N\rightarrow \infty$. Using the coproduct in (\ref{eq:tensorproduct_coproduct}) and formally regularizing the products by specifying the order, the actions can be written as 
\begin{align}
    &E_{s}(z)\ket{\sigma}\notag\\
    &=\sum_{i=1}^{\ell(\sigma)+1}(-1)^{|s|(\sum_{l=1}^{i-1}|\sigma_{l}-1|)}\mathcal{E}_{s}([(q_{1}q_{3})^{i-1}]_{\sigma_{i}-1})\bar{\delta}_{s,\sigma_{i}}\prod_{j=1}^{i-1}\left[\Psi^{(s)}_{[(q_{1}q_{3})^{j-1}]_{\sigma_{j}-1}}(z)\right]_{-}\notag\\
    &\times\delta\left(\frac{z}{u(q_{1}q_{3})^{i-1}(q_{1}^{2}q_{3}^{-1})^{r(\sigma_{i})}q_{1}^{s(\sigma_{i})}}\right)\ket{\sigma+\fbox{$s$}_{i}},\\
    &F_{s}(z)\ket{\sigma}\notag\\
    &=\frac{\phi(q_{1}^{1-\ell(\sigma)}q_{3}^{-\ell(\sigma)};z,u)^{\delta_{s,2}}}{\phi(q_{1}^{-\ell(\sigma)}q_{3}^{-\ell(\sigma)};z,u)^{\delta_{s,3}}}\notag\\
    &\times\sum_{i=1}^{\ell(\sigma)}(-1)^{|s|(\sum_{l=1}^{i-1}|\sigma_{l}-1|)}\mathcal{F}_{s}([(q_{1}q_{3})^{i-1}u]_{\sigma_{i}-1})\bar{\delta}_{s,\sigma_{i}-1}
    \prod_{j=i+1}^{\ell(\sigma)}\left[\Psi^{(s)}_{[(q_{1}q_{3})^{j-1}u]_{\sigma_{j}-1}}(z)\right]_{+}\notag\\
    &\times\delta\left(\frac{z}{u(q_{1}q_{3})^{i-1}(q_{1}^{2}q_{3}^{-1})^{r(\sigma_{i}-1)}q_{1}^{s(\sigma_{i}-1)}}\right)\ket{\sigma-\fbox{$s$}_{i}},\\
    &K_{s}(z)\ket{\sigma}=\frac{\phi(q_{1}^{1-\ell(\sigma)}q_{3}^{-\ell(\sigma)};z,u)^{\delta_{s,2}}}{\phi(q_{1}^{-\ell(\sigma)}q_{3}^{-\ell(\sigma)};z,u)^{\delta_{s,3}}}\prod_{i=1}^{\ell(\sigma)}\Psi^{(s)}_{[(q_{1}q_{3})^{i-1}u]_{\sigma_{i}-1}}(z)\ket{\sigma}.
\end{align}
Especially, the action of $K_s(z)$ on the vacuum is
\begin{align}
    K_{s}(z)\ket{\emptyset}=\frac{\phi(q_{1};z,u)^{\delta_{s,2}}}{\phi(1;z,u)^{\delta_{s,3}}}\ket{\emptyset}
\end{align}
as expected in (\ref{eq:vacuum_charge_function}).

\subsection{ \texorpdfstring{$\mathbb{C}^{3}/(\mathbb{Z}_{2}\times\mathbb{Z}_{2})$}{C3Z2Z2} and quantum toroidal \texorpdfstring{$D(2,1;\alpha)$}{D21}}
\subsubsection{Definition of the algebra}
\begin{figure}[H]
    \begin{tabular}{cc}
      \begin{minipage}{0.4\hsize}
        \centering
       \includegraphics[width=4cm]{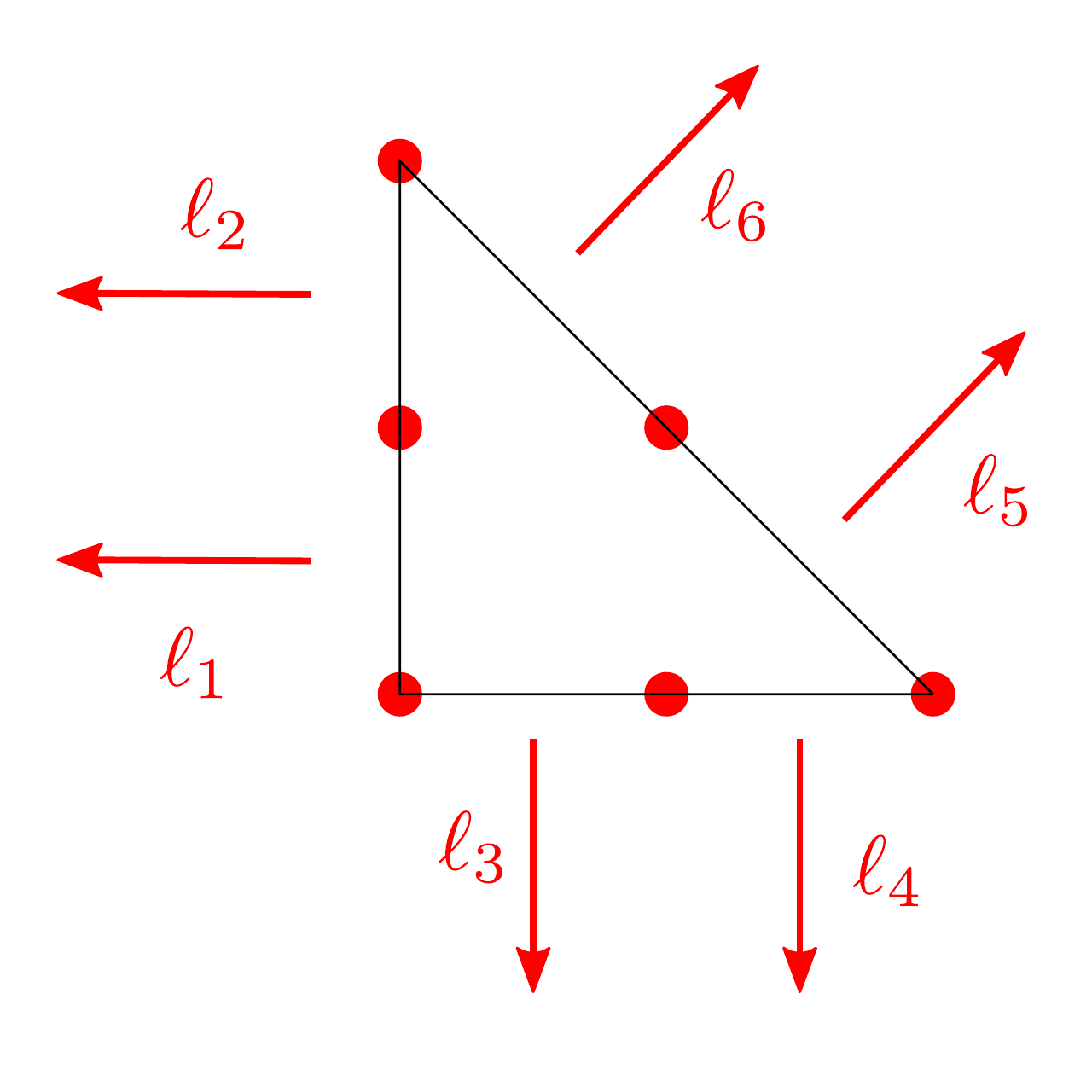}
        \subcaption{Toric diagram }\label{fig:D(2,1)toric_web}
      \end{minipage}&\hfill
      \begin{minipage}{0.5\hsize}
        \centering
     \includegraphics[width=7cm]{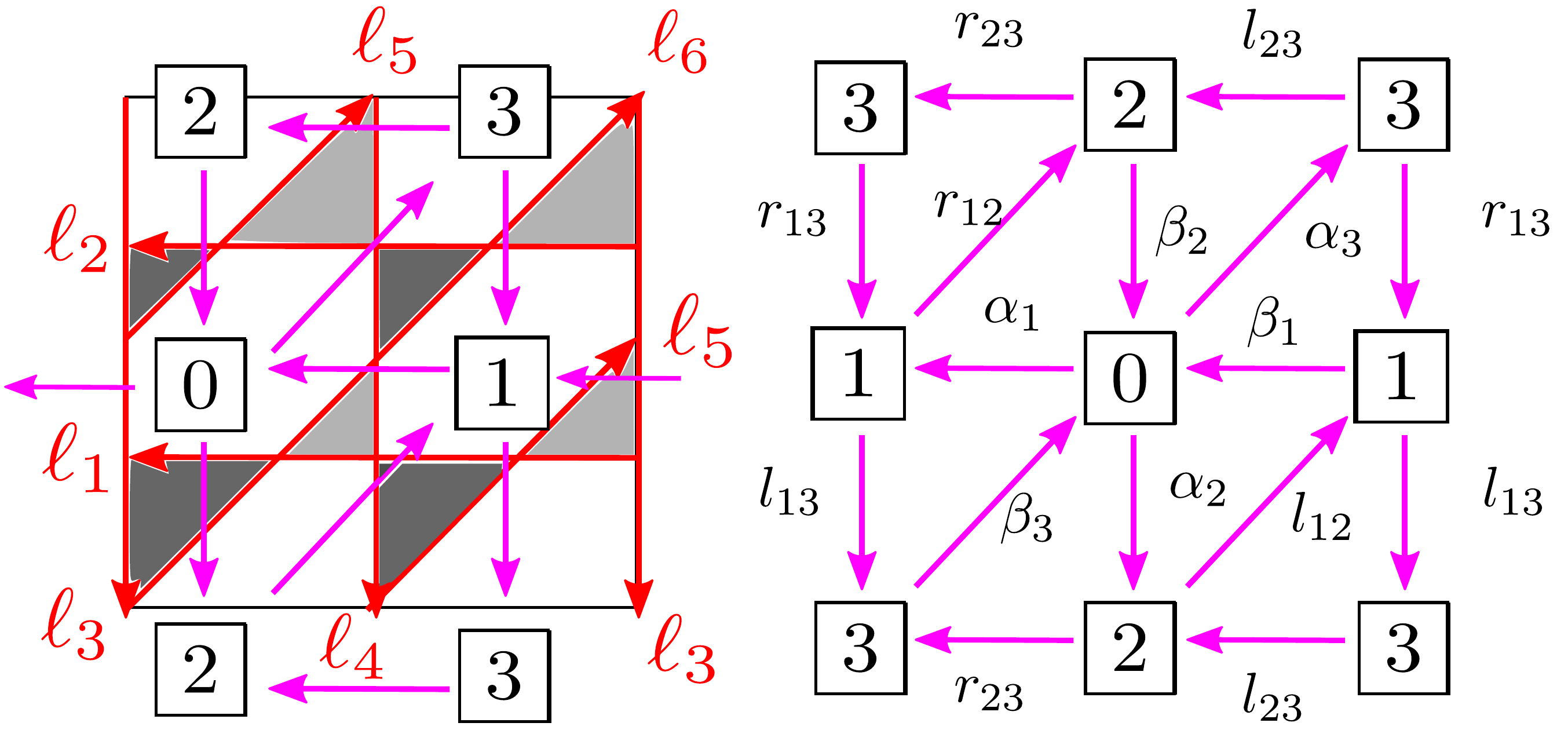}
       \subcaption{Periodic quiver}\label{fig:D(2,1)periodic_quiver}
      \end{minipage}
    \end{tabular}
\caption{(a) Toric diagram and (b) periodic quiver of $\mathbb{C}^{3}/(\mathbb{Z}_{2}\times\mathbb{Z}_{2})$. The left bottom lattice point of the toric diagram is $p_{1}=(0,0)$. Other lattice points are $ p_{2}=(2,0),\; p_{3}=(0,2),\;p_{4}=(1,0),\;p_{5}=(0,1)$, and $p_{6}=(1,1)$. External legs are denoted $\ell_{i}\;(i=1,..,6).$}\label{fig:D(2,1)toric_periodicquiver}
\end{figure}

\begin{figure}[ht]
    \begin{tabular}{cc}
      \begin{minipage}{0.55\hsize}
        \centering
      \includegraphics[width=8cm]{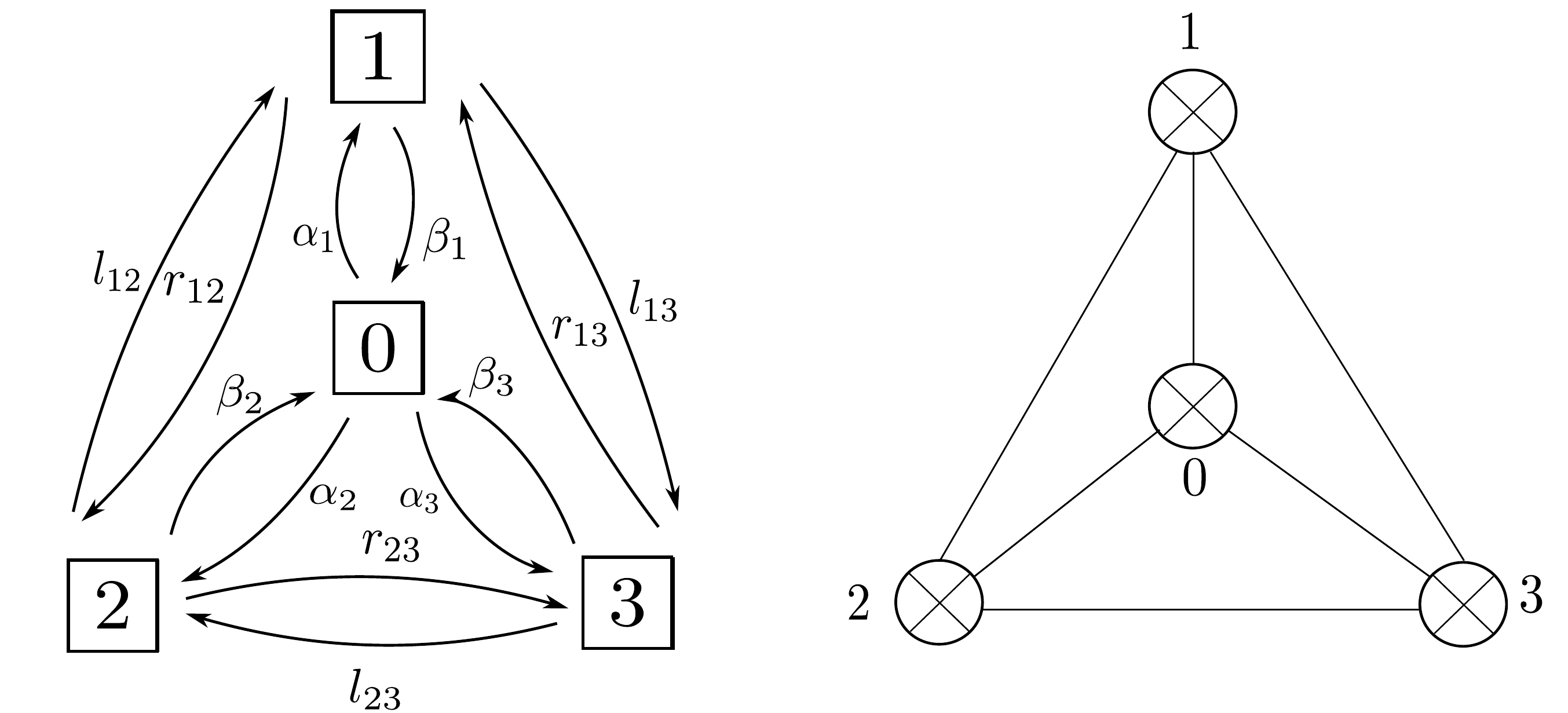}
      \subcaption{Quiver diagram and the associated Dynkin diagram.}\label{fig:D(2,1)quiver}
      \end{minipage}&\hfill
      \begin{minipage}{0.35\hsize}
        \centering
     \includegraphics[width=5cm]{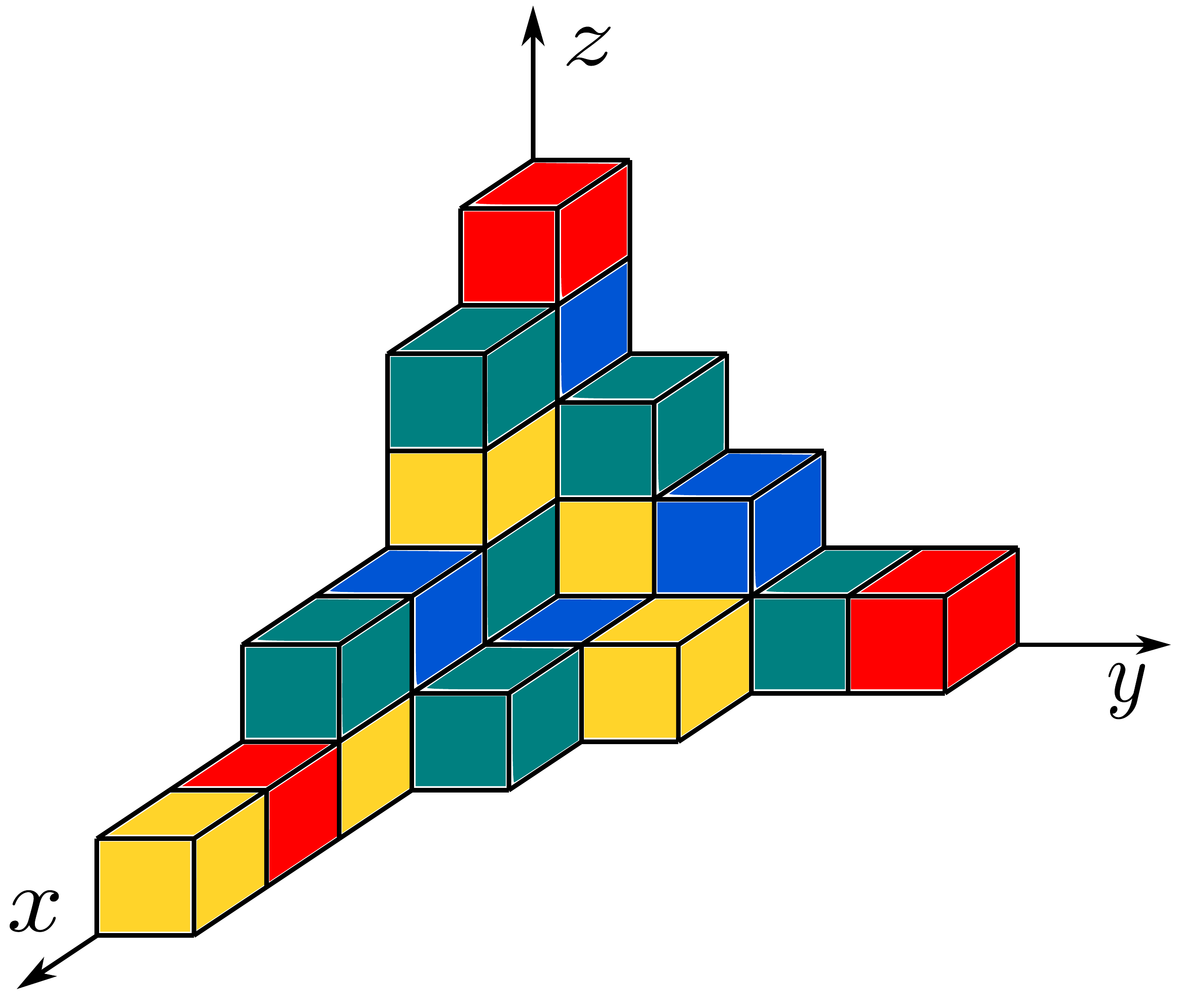}
       \subcaption{Three-dimensional crystal.}\label{fig:D(2,1)three-dimensionalcrystal}
      \end{minipage}
    \end{tabular}
\caption{Quiver diagram, Dynkin diagram, and three-dimensional crystal of $\mathbb{C}^{3}/(\mathbb{Z}_{2}\times\mathbb{Z}_{2})$ \cite{Noshita:2021ldl}. The quiver diagram is the same as the Dynkin diagram of the affine superalgebra $\hat{D}(2,1;\alpha)$ (see Figure \ref{fig:D(2,1)quiver}). The three-dimensional crystal is a plane partition, which is the same as the quantum toroidal $\mathfrak{gl}_{1}$, but the coloring is different. There are four colors: red, blue, yellow, and green. Each of them corresponds to the four vertices of the quiver diagram. The origin box is red.} \label{fig:D(2,1)quiver_three-dimensionalcrystal}
\end{figure}

The quantum toroidal algebra of $\mathbb{C}^{3}/(\mathbb{Z}_{2}\times\mathbb{Z}_{2})$ was defined in \cite{Noshita:2021ldl} (see also \cite{Feigin_2019,feigin2021combinatorics,heckenberger2008drinfeld}). The toric diagram and periodic quiver is in Figure \ref{fig:D(2,1)toric_periodicquiver}. 
We denote the six lattice points of the toric diagram as $p_{1}=(0,0)$, $p_{2}=(2,0)$, $p_{3}=(0,2)$, $p_{4}=(1,0)$, $p_{5}=(0,1)$, and $p_{6}=(1,1)$. The external legs of the toric diagram are denoted $\ell_{1},..,\ell_{6}$. 

The quiver diagram and three-dimensional crystal obtained from them are in Figure \ref{fig:D(2,1)quiver_three-dimensionalcrystal}. Its shape is the same as the plane partition representation of quantum toroidal $\mathfrak{gl}_{1}$, but the colors of the boxes are different. The boxes are colored with four colors so that no two adjacent boxes have the same color.

Using the periodic quiver (Figure \ref{fig:D(2,1)periodic_quiver}), quiver diagram (Figure \ref{fig:D(2,1)quiver}), loop constraint (\ref{eq:loop_constraint}), and vertex constraint (\ref{eq:vertex_constraint}) we obtain the following:

\begin{align}
\begin{split}
\alpha_{1}=\beta_{1}=l_{23}=r_{23}=q_{1},\\
\alpha_{2}=\beta_{2}=l_{13}=r_{13}=q_{2},\\
\alpha_{3}=\beta_{3}=l_{12}=r_{12}=q_{3},
\end{split}
\end{align}
with the condition $q_{1}q_{2}q_{3}=1$.
The bond factors are read of
\begin{align}
    \varphi^{i\Rightarrow j}(z,w)=\frac{\phi(q_{ij};z,w)}{\phi(q_{ij}^{-1};z,w)},
\end{align}
where we set
\begin{align}
    q_{ij}=q_{ji}=\begin{cases} 
    q_{1}\quad(i,j)=(0,1),(2,3),\\
    q_{2}\quad(i,j)=(0,2),(1,3),\\
    q_{3}\quad(i,j)=(0,3),(1,2).
    \end{cases}
\end{align}

\subsubsection{Subquiver and crystal shape}
\begin{figure}
    \centering
    \includegraphics[width=10cm]{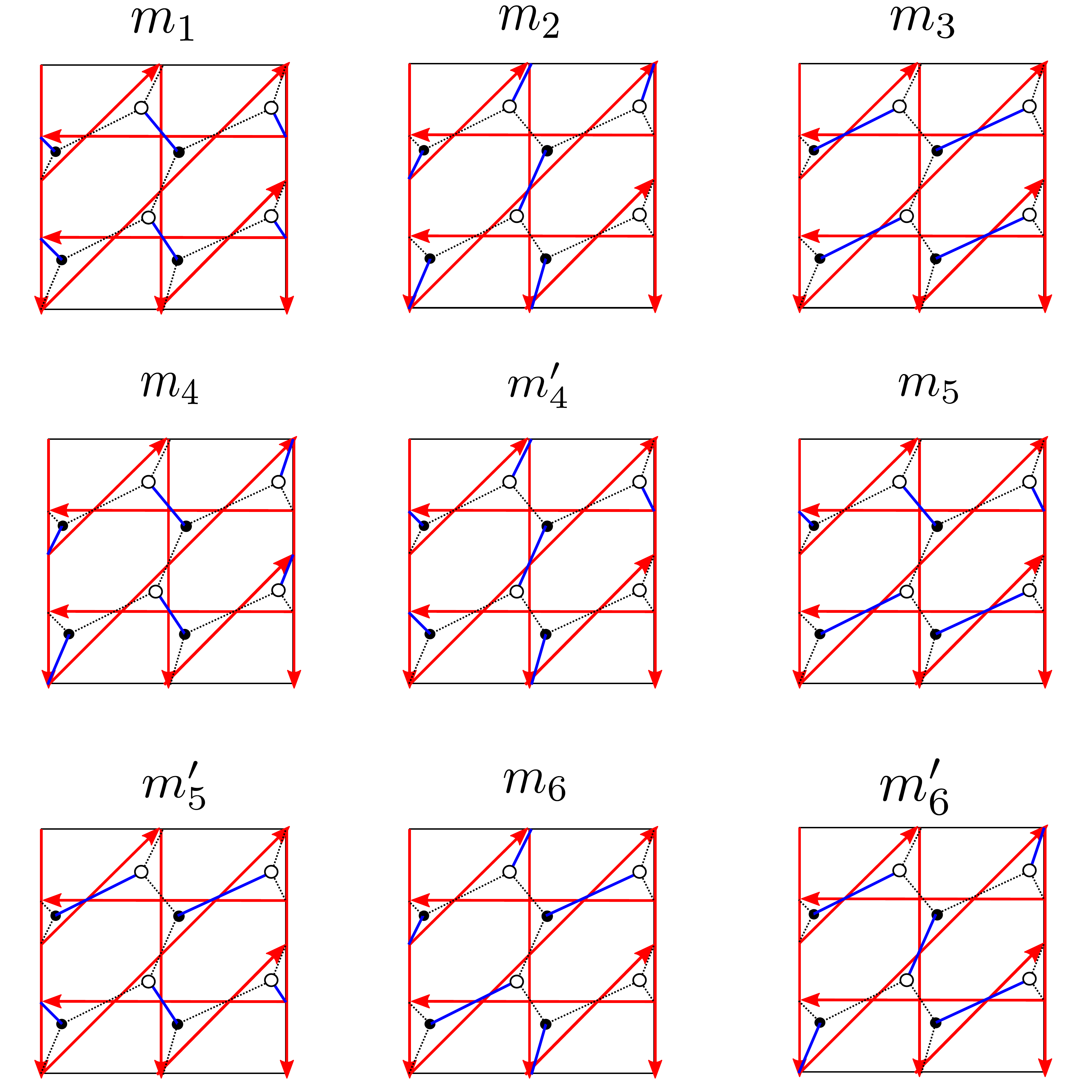}
    \caption{Perfect matchings of $\mathbb{C}^{3}/(\mathbb{Z}_{2}\times\mathbb{Z}_{2})$. Perfect matchings $m_{1},m_{2}$, and $m_{3}$ correspond to corner divisors $p_{1},p_{2}$, and $p_{3}$ respectively. $(m_{4},m_{4}')$ are perfect matchings of $p_{4}$, $(m_{5},m_{5}')$ are perfect matchings of $p_{5}$, and $(m_{6},m_{6}')$ are perfect matchings of $p_{6}$. 
    We note $m_{1}=\{\alpha_{3},\beta_{3},l_{12},r_{12}\}$, $m_{2}=\{\alpha_{1},\beta_{1},l_{23},r_{23}\}$, $m_{3}=\{\alpha_{2},\beta_{2},l_{13},r_{13}\}$, $m_{4}=\{\alpha_{1},\alpha_{3},l_{12},r_{23}\}$, $m_{4}'=\{\beta_{1},\beta_{3},l_{23},r_{12}\}$, $m_{5}=\{\alpha_{2},\alpha_{3},l_{13},r_{12}\}$, $m_{5}'=\{\beta_{2},\beta_{3},l_{12},r_{13}\}$, $m_{6}=\{\beta_{1},\beta_{3},l_{23},r_{12}\}$, and $m_{6}'=\{\beta_{1},\beta_{2},l_{13},r_{23}\}$.}
    \label{fig:D(2,1)perfectmatching}
\end{figure}
Let us consider the subquiver and two-dimensional crystal of the orbifold $\mathbb{C}^{3}/(\mathbb{Z}_{2}\times\mathbb{Z}_{2})$. Perfect matchings of each lattice points are in Figure \ref{fig:D(2,1)perfectmatching}. Since all of the two-dimensional crystals of the corner divisors can be obtained in a similar way, let us focus on corner divisor $p_{1}=(0,0)$. The external legs surrounding the toric divisor are $\ell_{1}$ and $\ell_{3}$. Since the perfect matching of $p_{1}$ is $m_{1}=\{\alpha_{3},\beta_{3},l_{12},r_{12}\}$, which is unique, the subquiver can be obtained by removing arrows $m_{1}$ from the original quiver. The subquiver and crystal shape is as in Figure \ref{fig:D(2,1)2dimcrystal_subquiver}. Other subquivers associated with corner divisors $p_{2}$ and $p_{3}$ can be obtained similarly. They indeed correspond to the $(x,y)$ plane, $(y,z)$ plane, and $(z,x)$ plane of the three dimensional crystal shown in Figure \ref{fig:D(2,1)three-dimensionalcrystal}.

\begin{figure}
    \begin{tabular}{c}
      \begin{minipage}{0.95\hsize}
        \centering
      \includegraphics[width=10cm]{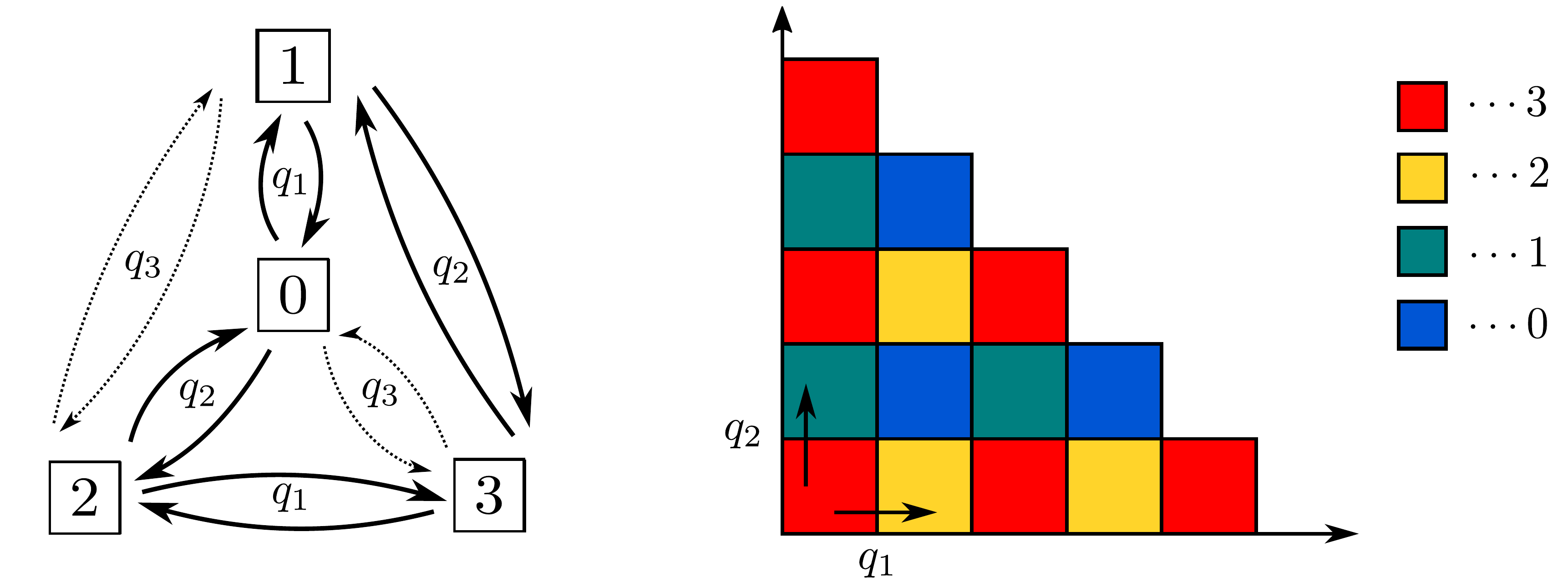}
        \subcaption{Subquiver and two-dimensional crystal of divisor $p_{1}=(0,0)$. They are obtained by removing arrows $m_{1}=\{\alpha_{3},\beta_{3},l_{12},r_{12}\}$. We note $\alpha_{3}=\beta_{3}=l_{12}=r_{12}=q_{3}$.} \label{fig:D(2,1)2dimcrystal_subquiver}
      \end{minipage}\\
      \begin{minipage}{0.95\hsize}
        \centering
     \includegraphics[width=10cm]{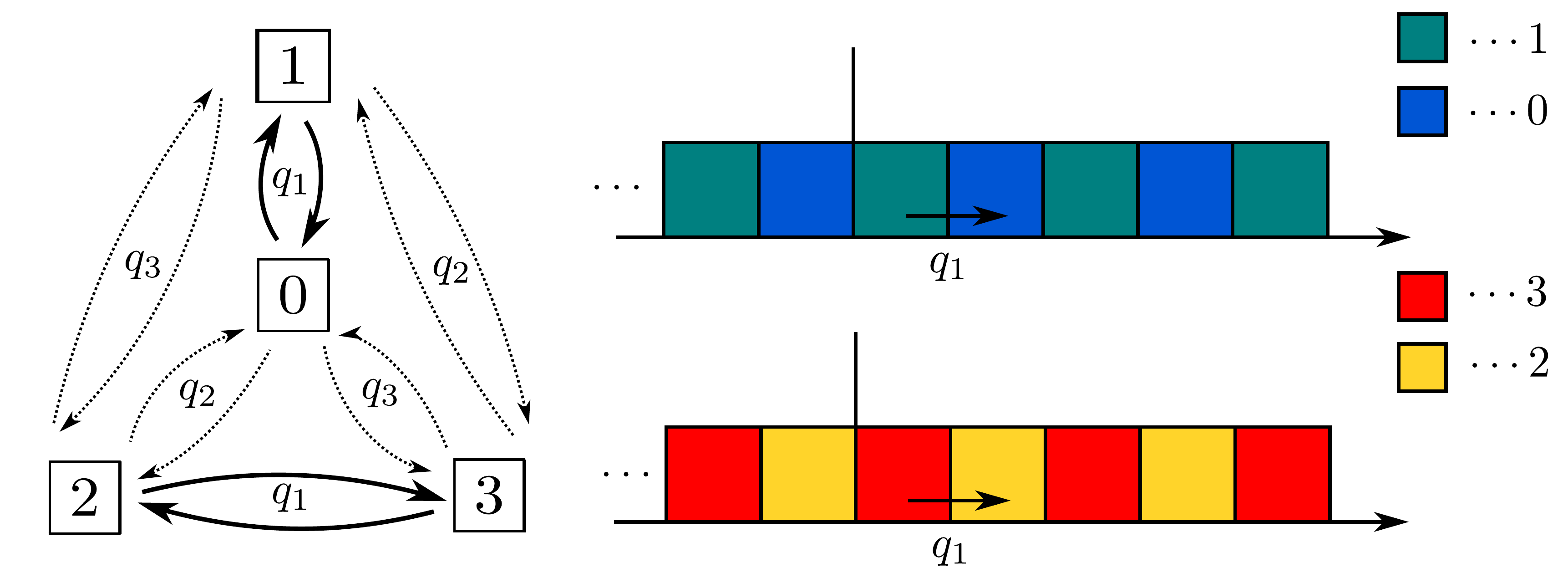}
    \subcaption{Subquiver and one-dimensional crystal associated with external legs $\ell_{1}$ and $\ell_{2}$. They are obtained by removing arrows $\{\alpha_{2},\alpha_{3},\beta_{2},\beta_{3},l_{12},l_{13},r_{12},r_{13}\}$. We note $\alpha_{2}=\beta_{2}=l_{13}=r_{13}=q_{2}$ and $\alpha_{3}=\beta_{3}=l_{12}=r_{12}=q_{3}$. Two pairs of vertices decouple.}\label{fig:D(2,1)1dimcrystal_subquiver}
      \end{minipage}
    \end{tabular}
\caption{Subquiver and crystal shape of one-dimensional and two-dimensional crystal. }\label{fig:D(2,1)1d_2d}
\end{figure}

Next, let us consider the one-dimensional crystal representation associated with $\ell_{1}$ and $\ell_{2}$. $\ell_{1}$ is surrounded by two divisors $p_{1}$ and $p_{5}$. For $\ell_{1}$, the union set of the perfect matchings is $m_{1}\cup m_{5}\cup m_{5}'=\{\alpha_{2},\alpha_{3},\beta_{2},\beta_{3},l_{12},l_{13},r_{12},r_{13}\}$. For $\ell_{2}$, we get the same $m_{3}\cup m_{5}\cup m_{5}'=\{\alpha_{2},\alpha_{3},\beta_{2},\beta_{3},l_{12},l_{13},r_{12},r_{13}\}$. After removing arrows $\{\alpha_{2},\alpha_{3},\beta_{2},\beta_{3},l_{12},l_{13},r_{12},r_{13}\}$ from the original quiver diagram, we obtain two decoupled subquiver diagrams. They are the one-dimensional crystal representations associated with the external legs $\ell_{1}$ and $\ell_{2}$ (see Figure \ref{fig:D(2,1)1dimcrystal_subquiver}).

This analysis shows that one-dimensional crystal representations can be obtained by choosing two colors from the four colors and lining boxes with their colors alternately, which means we have $6$ types.

\subsubsection{One-dimensional crystal representations}
Let us construct the one-dimensional crystal representations.
These representations can be determined by choosing two vertices of the four vertices as mentioned in the previous subsection. We choose two different numbers $a$ and $b$ ($a,b=0,1,2,3, a\neq b$). We denote $V^{(a;b)}(u)$ the complex vector space with bases $[u]_{j}^{(a;b)}$, $j\in\mathbb{Z}$. 
We picture $[u]_{l}^{(a;b)}$ as a semi-infinite row of boxes in colors $a,b,a,b,a,b...$ from the divider to the right and continuing the pattern to the left (see Figure \ref{fig:D(2,1)1dvectordef}). The coordinates of each boxes will be labeled $q(\Abox)=q_{ab}^{l}(l\in{\mathbb{Z}})$ where $l$ is counted $0,1,2...$ from the right of the border.\\
We use $\bar{\delta}_{i,j}=\begin{dcases}
1\quad i\equiv j\mod 2,\\0 \quad i\not\equiv j\mod 2
\end{dcases}$.
\begin{figure}[H]
    \centering
    \includegraphics[width=10cm]{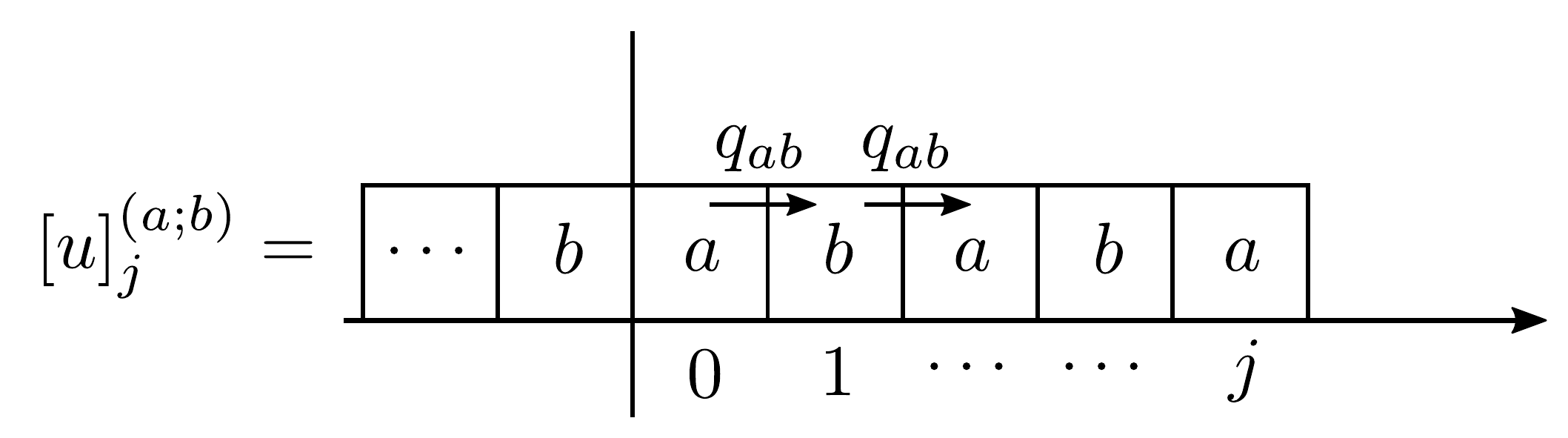}
    \caption{Basis of one-dimensional crystal representation. }
    \label{fig:D(2,1)1dvectordef}
\end{figure}
The box in $q(\Abox)=q_{ab}^{l}$ is color $a$ when $l\equiv 0\mod2$ and $b$ when $l\equiv1 \mod 2$.
The action of the algebra can be written as 
\begin{align}
    \begin{split}
        &K_{i}^{\pm}(z)[u]_{l}^{(a;b)}=\left[\Psi_{[u]_{l}^{(a;b)}}^{(i)}(z)\right]_{\pm}[u]_{l}^{(a;b)},
        \end{split}\\
            \begin{split}
        &E_{a}(z)
        [u]_{l}^{(a;b)}
       =\begin{dcases}
        0\quad l\equiv 0 ,\\
        \mathcal{E}_{a}([u]_{l}^{(a;b)})\delta\left(\frac{z}{uq_{ab}^{l+1}}\right)[u]_{l+1}^{(a;b)}\quad l\equiv1 ,
        \end{dcases}\\
        &E_{b}(z)
        [u]_{l}^{(a;b)}
       =\begin{dcases}
        \mathcal{E}_{b}([u]_{l}^{(a;b)})\delta\left(\frac{z}{uq_{ab}^{l+1}}\right)[u]_{l+1}^{(a;b)}\quad l\equiv 0,\\
        0\quad l\equiv 1,
        \end{dcases}
        \end{split}
        \end{align}
        \begin{align}
            \begin{split}
        &F_{a}(z)[u]_{l}^{(a;b)}=\begin{dcases}
        \mathcal{F}_{a}([u]_{l}^{(a;b)})\delta\left(\frac{z}{uq_{ab}^{l}}\right)[u]_{l-1}^{(a;b)}\quad l\equiv 0,\\
        0\quad l\equiv 1,\\
        \end{dcases}\\
        &F_{b}(z)[u]_{l}^{(a;b)}=\begin{dcases}
        0\quad l\equiv 0,\\
        \mathcal{F}_{b}([u]_{l}^{(a;b)})\delta\left(\frac{z}{uq_{ab}^{l}}\right)[u]_{l-1}^{(a;b)}\quad l\equiv 1,
        \end{dcases}\\
        \begin{split}
            &E_{i}(z), F_{i}(z)[u]_{l}^{(a;b)}=0\quad i\neq a,b.
        \end{split}\label{eq:D(2,1)1dcrystalrep}
    \end{split}
\end{align}
Explicit expressions of the coefficients $\mathcal{E}_{s}([u]_{l}^{(a;b)}),\;\mathcal{F}_{s}([u]_{l}^{(a;b)})$ are omitted.
The charge functions can be derived from the KE relation as
\begin{align}
    \Psi_{[u]_{2p+1}^{(a;b)}}^{(i)}(z)=\begin{dcases}
    \frac{1}{\phi(q_{ab}^{-2p-2};z,u)}\quad i=a,\\
    \frac{1}{\phi(q_{ab}^{-2p-1};z,u)}\quad i=b,\\
    \phi(q_{bi}q_{ab}^{-2p-1};z,u)\quad i\neq a,b,
    \end{dcases},
    \Psi_{[u]_{2p}^{(a;b)}}^{(i)}(z)=\begin{dcases}
    \frac{1}{\phi(q_{ab}^{-2p};z,u)}\quad i=a,\\
    \frac{1}{\phi(q_{ab}^{-2p-1};z,u)}\quad i=b,\\
    \phi(q_{ai}q_{ab}^{-2p};z,u)\quad i\neq a,b,
    \end{dcases}
\end{align}
where we used the loop condition and vertex condition $q_{ai}q_{ab}q_{bi}=1$ when $i\neq a,b$, during the calculation. We can see that the charge functions do not have the same numbers of poles and zeros, which mean they are representations of shifted quantum toroidal algebra. The shift parameters are determined as 
\begin{align}
\begin{split}
    r_{a}=-1,\quad r_{b}=-1,\\
    r_{i}=1,\quad i\neq a,b.
    \end{split}
\end{align}

From the above relation, we obtain the following parity condition
\begin{align}
 |[u]_{l+1}^{(a;b)}|-|[u]_{l}^{(a;b)}|=1,   
\end{align}
where we used $|a|=|b|=1$.

\subsubsection{Two-dimensional crystal representations}\label{sec:D(2,1)2dcrystal}
Let us construct the two-dimensional representations of the corner divisor $p_{1}=(0,0)$. Other representations of other divisors can be obtained similarly by choosing the one-dimensional crystal representations properly. We can see from Figure \ref{fig:D(2,1)1d_2d} that the two-dimensional crystal can be obtained by taking tensor products of $V^{(3;2)}(u)$ and $V^{(1;0)}(v)$. 
We assign the parities of the vectors of $V^{(3;2)}(u)$ and $V^{(1;0)}(v)$ as the following
\begin{align}
    |[u]_{j}^{(3;2)}|=\begin{dcases}
    0\quad j\equiv 1\mod2,\\
    1\quad j\equiv 0\mod2,
    \end{dcases},\\
    |[v]_{j}^{(1;0)}|=\begin{dcases}
    1\quad j\equiv 1\mod2,\\
    0\quad j\equiv 0\mod2
    \end{dcases}.
\end{align}
We use the notation 
\begin{align}
    |j|=\begin{dcases}
    0,\quad j\equiv0\mod2,\\
    1,\quad j\equiv1\mod2,\\
    \end{dcases}
\end{align}
and then obtain 
\begin{align}
    |[u]_{j}^{(3;2)}|=|j+1|,\quad |[v]_{j}^{(1;0)}|=|j|.
\end{align}

Let us consider the action of $E_{s}(z)$ on $V^{(3;2)}(u)\otimes V^{(1;0)}(v)$. The bases are $[u]_{j}^{(3;2)}\otimes [v]_{k}^{(1;0)}$, where $j,k\in\mathbb{Z}$.
We set $v=uq_{2}$ to form a submodule and satisfy the melting rule.  
In this case, the melting rule is the same as the Young diagram condition with colors.

To construct two-dimensional crystal representations we need to take infinite tensor products $\otimes_{i=1}^{\infty}(V^{(3;2)}(uq_{2}^{2i-2})\otimes V^{(1;0)}(uq_{2}^{2i-1}))$.
The basis of this representation is defined by the tensor product of one-dimensional representations as
\begin{align}
\begin{split}
\otimes_{i=1}^{N}(V^{(3;2)}(uq_{2}^{2i-2})\otimes V^{(1;0)}(uq_{2}^{2i-1}))&\ni \ket{\lambda}=\otimes_{i=1}^{N}\left([uq_{2}^{2i-2}]_{\lambda_{2i-1}-1}^{(3;2)}\otimes [uq_{2}^{2i-1}]_{\lambda_{2i}-1}^{(1;0)}\right),\\
\quad\lambda=(\lambda_{1},\lambda_{2},...,\lambda_{2N-2},\lambda_{2N-1})\in \mathbb{Z}^{2N},&\quad \lambda_{1}\geq \lambda_{2}\geq...\geq\lambda_{2N-2}\geq\lambda_{2N-1}.
\end{split}
\end{align}

The actions of generators after taking the limit $N\rightarrow \infty$ properly are written as follows (see Appendix \ref{sec:appendix_D(2,1)}):
\begin{align}
    \begin{split}
        &E_{s}(z)\ket{\lambda}\\
        =&\sum_{i=1}^{\lfloor\frac{\ell(\lambda)+1}{2}\rfloor}\prod_{l=1}^{i-1}\left[\Psi^{(s)}_{[uq_{2}^{2l-2}]^{(3;2)}_{\lambda_{2l-1}-1}}(z)\Psi^{(s)}_{[uq_{2}^{2l-1}]^{(1;0)}_{\lambda_{2l}-1}}(z)\right]_{-}(-1)^{|s|(\sum_{l=1}^{i-1}(|\lambda_{2l-1}|+|\lambda_{2l}-1|))}\\
        &\times\left\{\mathcal{E}_{s}\left([uq_{2}^{2i-2}]_{\lambda_{2i-1}-1}^{(3;2)}\right)\left(\delta_{s,3}\bar{\delta}_{\lambda_{2i-1},0}+\delta_{s,2}\bar{\delta}_{\lambda_{2i-1},1}\right)\delta\left(\frac{z}{uq_{2}^{2i-2}q_{1}^{\lambda_{2i-1}}}\right)\ket{\lambda+\fbox{$s$}_{2i-1}}\right.\\
        &\quad+(-1)^{|s||\lambda_{2i-1}|}\left[\Psi^{(s)}_{[uq_{2}^{2i-2}]_{\lambda_{2i-1}-1}^{(3;2)}}(z)\right]_{-}\mathcal{E}_{s}\left([uq_{2}^{2i-1}]_{\lambda_{2i}-1}^{(1;0)}\right)\\
        &\hspace{1cm}\left.\times\left(\delta_{s,1}\bar{\delta}_{\lambda_{2i},0}+\delta_{s,0}\bar{\delta}_{\lambda_{2i},1}\right)\delta\left(\frac{z}{uq_{2}^{2i-1}q_{1}^{\lambda_{2i}}}\right)\ket{\lambda+\fbox{$s$}_{2i}}\right\}\\
        &+\bar{\delta}_{\ell(\lambda),0}
        \prod_{l=1}^{\frac{\ell(\lambda)}{2}}\left[\Psi^{(s)}_{[uq_{2}^{2l-2}]^{(3;2)}_{\lambda_{2l-1}-1}}(z)\Psi^{(s)}_{[uq_{2}^{2l-1}]^{(1;0)}_{\lambda_{2l}-1}}(z)\right]_{-}(-1)^{|s|(\sum_{l=1}^{\frac{\ell(\lambda)}{2}}(|\lambda_{2l-1}|+|\lambda_{2l}-1|))}\\
        &\quad\times \delta_{s,3}\mathcal{E}_{s}\left([uq_{2}^{\ell(\lambda)}]_{-1}^{(3;2)}\right)\delta\left(\frac{z}{uq_{2}^{\ell(\lambda)}}\right)\ket{\lambda+\fbox{s}_{\ell(\lambda)+1}},
    \end{split}
\end{align}
\begin{align}
    \begin{split}
    &F_{s}(z)\ket{\lambda}\\
    =&\sum_{i=1}^{\lfloor\frac{\ell(\lambda)+1}{2} \rfloor}(-1)^{|s|(\sum_{l=1}^{i-1}(|\lambda_{2l-1}|+|\lambda_{2l}-1|))}\beta_{s}^{(\lfloor\frac{\ell(\lambda)+1}{2}\rfloor)}(z)\prod_{j=i+1}^{\lfloor\frac{\ell(\lambda)+1}{2}\rfloor}\Psi^{(s)}_{[uq_{2}^{2j-2}]^{(3;2)}_{\lambda_{2j-1}-1}}(z)\Psi^{(s)}_{[uq_{2}^{2j-1}]^{(1;0)}_{\lambda_{2j}-1}}(z)\\
    &\times\left\{\mathcal{F}_{s}([uq_{2}^{2i-2}]^{(3;2)}_{\lambda_{2i-1}-1})\Psi^{(s)}_{[uq_{2}^{2i-1}]^{(1;0)}_{\lambda_{2i}-1}}(z)\delta\left(\frac{z}{uq_{2}^{2i-2}q_{1}^{\lambda_{2i-1}-1}}\right)\right.\\
    &\hspace{1cm}\times\left(\delta_{s,3}\bar{\delta}_{\lambda_{2i-1},1}+\delta_{s,2}\bar{\delta}_{\lambda_{2i-1},0}\right)\ket{\lambda-\fbox{$s$}_{2i-1}}\\
    &\quad\left.+(-1)^{|s||\lambda_{2i-1}|} \mathcal{F}_{s}([uq_{2}^{2i-1}]^{(1;0)}_{\lambda_{2i}-1})\delta\left(\frac{z}{uq_{2}^{2i-1}q_{1}^{\lambda_{2i}-1}}\right)\left(\delta_{s,1}\bar{\delta}_{\lambda_{2i},1}+\delta_{s,0}\bar{\delta}_{\lambda_{2i},0}\right)\ket{\lambda-\fbox{$s$}_{2i}}        \right\},
\end{split}
\end{align}
\begin{align}
&K_{s}(z)\ket{\lambda}=\beta_{s}^{\left(\lfloor\frac{\ell(\lambda)+1}{2} \rfloor\right)}(z)\prod_{i=1}^{\lfloor\frac{\ell(\lambda)+1}{2} \rfloor}\left(\Psi^{(s)}_{[uq_{2}^{2i-2}]^{(3;2)}_{\lambda_{2i-1}-1}}(z)\Psi^{(s)}_{[uq_{2}^{2i-1}]^{(1;0)}_{\lambda_{2i}-1}}(z) \right)\ket{\lambda},
\end{align}
where 
\begin{align}
\begin{split}
\beta_{0}^{(N)}(z)&=q_{2}^{\frac{N}{2}}\phi(q_{1}q_{2}^{-2N+1};z,u),\quad \beta_{1}^{(N)}(z)=q_{2}^{\frac{N}{2}},\\
\beta_{2}^{(N)}(z)&=q_{2}^{-\frac{N}{2}},\quad \beta_{3}^{(N)}(z)=q_{2}^{-\frac{N}{2}}\frac{1}{\phi(q_{2}^{-2N};z,u)}.
\end{split}
\end{align}

Especially, the action on the vacuum is 
\begin{align}
    K_{s}(z)\ket{\emptyset}=\frac{\phi(q_{3}^{-1};z,u)^{\delta_{s,0}}}{\phi(1;z,u)^{\delta_{s,3}}}\ket{\emptyset}
\end{align}
as expected in (\ref{eq:vacuum_charge_function}).

\section{Conclusion and discussion}\label{sec:summary}
We introduced shifted quiver quantum toroidal algebra (shifted QQTA), a generalized version of the QQTA. These algebras are expected to act on subcrystals of the original three-dimensional BPS crystal. Motivated by \cite{Nishinaka_2011,Nishinaka_2012, Nishinaka_2014}, we defined one and two-dimensional subcrystals and showed that they are derived from subquivers of the original quiver diagram. We also showed the relation between the subquiver and perfect matchings.

Shifted QQTA has a generalized Hopf algebra structure: coproduct, counit, and antipode. The generalized coproduct and antipode are maps between algebras with \textit{different} shift parameters. In particular, the coproduct was essential, and we used it to derive one and two-dimensional sub-crystal representations in various examples.  

Let us list down possible directions for future work.
\begin{itemize}
    \item Although we focused on one-dimensional and two-dimensional crystals, it is possible to study general subcrystals and derive their representations from one and two-dimensional crystals we constructed. A general formula was already announced in \cite{galakhov2021shifted}. Deriving this formula from lower-dimensional crystals might help us understand various truncations of the mother algebra. How to take tensor products to derive general three-dimensional crystals seems to be the difficult part in the $\mathfrak{gl_{m|n}}$ case. Studying the relation with \cite{bezerra2019quantum,bezerra2021representations} might help.
    \item The three-dimensional and two-dimensional crystals have physical interpretations. The three-dimensional crystal is a crystal melting model for D6-D2-D0 branes on a toric Calabi-Yau singularity. The two-dimensional crystal is that for D4-D2-D0 branes, and it is constructed by defining a corner divisor on a toric diagram. The one-dimensional crystal is constructed by removing the union set of two perfect matchings, but its physical interpretation is not clear yet. It should be studied in the near future.
    \item Similar to our previous paper \cite{Noshita:2021ldl}, we mainly focused on the quiver quantum toroidal algebra associated with toric Calabi-Yau manifolds not including compact 4-cycles and when one of the central elements is trivial $C=1$. As already mentioned there, by modifying the bond factors with $(-1)$ factors and changing the KK relations slightly, we can obtain algebras for CYs including compact 4-cycles. Although we did not derive the representations explicitly in this paper, we can do the same analysis and derive one and two-dimensional crystal representations for general toric CYs.
    \item The two-dimensional crystal representations are expected to be related to supersymmetric gauge theories including surface operators \cite{Kanno_2011}. The middle subquiver of Figure \ref{fig:gln2_subquiver_2dcrystal} is the quiver of ALE space \cite{kronheimer_yang-mills_1990}, while the left and right quivers of Figure \ref{fig:gln2_subquiver_2dcrystal} are chain-saw quivers \cite{finkelberg2014quantization,feigin2011yangians}. Other subcrystal representations are generalizations of these quivers and should have similar 2d/4d correspondences.
    \item Studying horizontal representations ($C\neq1$) \cite{bershtein2018plane, Shiraishi:1995rp,Feigin:1995sf,Awata:1995zk,Awata:1996dx, FHSSY:2010, Miki2007, Kojima2019, Kojima2021, Harada:2021xnm, Awata2019} of shifted QQTA is also one of the studies that must be done. Studying generalized intertwiners \cite{Iqbal_2009,Aganagic:2003db,Bourgine:2017jsi,Bourgine_2020,Bourgine_2019,Bourgine_2016,Bourgine_2017,Bourgine:2021nyw,Bourgine:2015szm,Bourgine:2021yba,Bourgine:2021gnb,Awata_2016,Awata_2017RTT,Awata_2017,Awata_2018,Awata:2011ce,zenkevich2019mathfrakgln,zenkevich2020mixed,zenkevich2021higgsed,Cheewaphutthisakun:2021cud,Mironov:2016yue,Ghoneim:2020sqi} with shift parameters as $\Phi_{\mathbf{r},\mathbf{r'}}:(\text{vertical})_{\mathbf{r}}\otimes(\text{horizontal})_{\mathbf{r'}}\rightarrow(\text{horizontal})_{\mathbf{r+r'}}$  is interesting. Actually the original motivation of this work was to study two-dimensional crystal representations that might enter in the vertical representation part of the intertwiner. The R matrix intertwining representations of different shifted algebras is also an interesting problem \cite{Awata:2016mxc,Awata2019,Prochazka:2019dvu,Fukuda:2017qki,Garbali:2020sll,Negut:2020npc,Awata_2017RTT}. All of these studies might help us understand the complete picture of the \textit{algebraic engineering} of supersymmetric gauge theories.   
\end{itemize}

\acknowledgments 
The authors thank Koichi Harada and Yutaka Matsuo for useful discussions. GN is supported in part by FoPM, the University of Tokyo. AW is supported in part by JSPS fellowship, MEXT, and JSR Fellowship, the University of Tokyo.

\appendix

\section{Convention}\label{sec:appendix_notation}
In this section, we list down the conventions and few residue formulas we used in this paper. The convention we use is the same with the former paper \cite{Noshita:2021ldl}.
\begin{align}
    \begin{split}
        &\phi(a;z,w)\equiv a^{1/2}z-a^{-1/2}w,\\
        &\frac{\phi(a;z,w)}{\phi(b;z,w)}=\frac{a^{1/2}z-a^{-1/2}w}{b^{1/2}z-b^{-1/2}w},\\
        &\frac{\phi(a;z,pw)}{\phi(b;z,pw)}=\frac{\phi(ap^{-1};z,u)}{\phi(bp^{-1};z,u)},\\
        &\frac{\phi(a;pz,w)}{\phi(b;pz,w)}=\frac{\phi(ap;z,w)}{\phi(bp;z,w)}.
    \end{split}
\end{align}
The formal expansion of the delta function is 
\begin{align}
    \delta(z)=\sum_{n\in\mathbb{Z}}z^{n}.
\end{align}

Two formal expansions $\left[\quad\right]_{\pm}$ are defined as
\begin{align}
\left[\frac{1}{\phi(p;z,w)}\right]_{+}\equiv\frac{1}{p^{1/2}z(1-p^{-1}w/z)}= \frac{1}{p^{1/2}z}\sum_{n\geq0}\left(\frac{w}{pz}\right)^{n}, \label{eq:plusexpansion}\\
\left[\frac{1}{\phi(p;z,w)}\right]_{-}\equiv\frac{1}{-p^{-1/2}w(1-\frac{pz}{w})}=-\frac{1}{p^{-1/2}w}\sum_{n\geq 0}\left(\frac{pz}{w}\right)^{n}\label{eq:minusexpansion}.
\end{align}
Let $f(z)$ be a general rational function. Then $[f(z)]_{+}$ is a formal expansion of $z^{-1}$ (expanded for $|z|\gg1$) and $[f(z)]_{-}$ is an expansion of $z$ (expanded for $|z|\ll1$). We obtain the following identity:
\begin{align}
    \left[f(z)\right]_{+}-\left[f(z)\right]_{-}=\sum_{i=1}^{d}\delta\left(\frac{z}{\alpha_{i}}\right)\underset{z=\alpha_{i}}{\Res}f(z)\footnotemark,
\end{align}
where $\alpha_{i}\;(i=1,\ldots,d)$ are poles of $f(z)$ different from $0,\;\infty$.
\footnotetext{Note that our convention of the residue slightly differs from the original residue. If we use the original residue, the right hand side should be written as $\sum_{i=1}^{d}\delta\left(\frac{z}{\alpha_{i}}\right)\underset{z=\alpha_{i}}{\Res}\frac{f(z)}{z}$. They differ by a constant coming from $\alpha_{i}$, which is not so important.}
As an example, we can obtain the following formula:
\begin{align}
\begin{split}
    \left[\frac{\phi(p;z,w)}{\phi(q;z,w)}\right]_{+}-\left[\frac{\phi(p;z,w)}{\phi(q;z,w)}\right]_{-}&=\phi(pq^{-1};1,1)\delta\left(\frac{z}{wq^{-1}}\right)\\
    &\equiv\Res_{z=wq^{-1}}\frac{\phi(p;z,w)}{\phi(q;z,w)}\delta\left(\frac{z}{q^{-1}w}\right).
    \end{split}\label{eq:residue}
\end{align}
We note 
\begin{align}
    \left[\phi(p;z,w)\right]_{+}-\left[\phi(p;z,w)\right]_{-}=0.
\end{align}
Other useful formulas are 
\begin{align}
\begin{split}
    &\frac{1}{z^{2}}\frac{1}{\phi(a;1,\frac{u}{z})\phi(b;1,\frac{u}{z})}-\frac{1}{\phi(a;z,u)\phi(b;z,u)}\\
    &=\frac{1}{u}\left\{\frac{a}{u}\frac{1}{\phi(ba^{-1};1,1)}\delta\left(\frac{u}{az}\right)+\frac{b}{u}\frac{1}{\phi(ab^{-1};1,1)}\delta\left(\frac{u}{bz}\right)\right\},
    \end{split}\\
    &\frac{1}{z}\frac{1}{\phi(a;1,u/z)}-\frac{1}{\phi(a;z,u)}=\frac{a^{\frac{1}{2}}}{u}\delta\left(\frac{u}{az}\right).
\end{align}
\section{Derivation of shift parameters}\label{sec:shiftderiv}
Let us derive the shift parameters of the following situation by using the formulas in Appendix \ref{sec:appendix_notation}.
The Drinfeld currents are defined as $K_{i}^{\pm}(z)=\sum_{r\geq 0}K^{\pm}_{i,\pm r}z^{\mp r}$ and the action on the vacuum configuration is 
\begin{align}
\begin{split}
    z^{r_{s}}K^{+}_{s}(z)\ket{\emptyset}=\left[\frac{\phi(q_{\mathfrak{m}}^{-1};z,u)^{\delta_{s,b}}}{\phi(1;z,u)^{\delta_{s,a}}}\right]_{+}\ket{\emptyset},\\
    K^{-}_{s}(z)\ket{\emptyset}=\left[\frac{\phi(q_{\mathfrak{m}}^{-1};z,u)^{\delta_{s,b}}}{\phi(1;z,u)^{\delta_{s,a}}}\right]_{-}\ket{\emptyset}.
\end{split} 
\end{align}
We consider the case when $a\neq b$. 

When $s=a$, using (\ref{eq:plusexpansion}) and (\ref{eq:minusexpansion}) 
\begin{align}
\left[\frac{1}{\phi(1;z,u)}\right]_{+}&= \frac{1}{z}\sum_{n\geq0}\left(\frac{u}{z}\right)^{n},\label{eq:plusexpansionshifta} \\
\left[\frac{1}{\phi(1;z,u)}\right]_{-}&=-\frac{1}{u}\sum_{n\geq 0}\left(\frac{z}{u}\right)^{n}\label{eq:minusexpansionshifta}.
\end{align}
Comparing the degrees of $z$, we need $r_{a}=-1$, because (\ref{eq:plusexpansionshifta}) is an expansion in $z^{-(n+1)}$ $(n\geq0)$. On the other hand, since (\ref{eq:minusexpansionshifta}) is still an expansion in $z^{n}\;(n\geq 0)$, we do not need any shift parameter.

When $s=b$, $[\phi(q_{\mathfrak{m}}^{-1};z,u)]_{\pm}$ is a formal expansion of $z^{n}\;(n\geq0)$ (a polynomial with $z^{0}$ and $z^{1}$). Since $z^{r_{b}}K^{+}_{b}(z)$ has modes $z^{-r}\;(r\geq -r_{b})$, to match the degrees of both side, we need $r_{b}=1$. On the other hand, $K_{b}^{-}(z)$ has modes $z^{r}\;(r\geq 0)$ and we do not need any shift parameter. 

Therefore, the shift parameters are determined as 
\begin{align}
    r_{a}=-1,\quad r_{b}=1,\quad r_{c}=0\;(c\neq a,b).
\end{align}

\section{Brane tiling and quiver gauge theory}\label{sec:appendix-branetiling}
In Section \ref{sec:QQTA_review}, the quiver quantum toroidal algebra was defined by a quiver diagram $Q=(Q_0, Q_1, Q_2)$.
The quiver diagram is a combination of a set of vertices $Q_0$, a set of arrows between vertices $Q_1$, and a set of closed loops $Q_2$.
All of $Q_0, Q_1, Q_2$ can be constructed from a toric diagram, and all examples in Section \ref{sec:examples} also start from a toric diagram.
See \cite{Ooguri_2009,Li:2020rij, Noshita:2021ldl} for more details.
In this section, we summarize how to derive the quiver diagram and brane tiling from the toric diagram.
\begin{figure}[h]
    \begin{tabular}{cc}
      \begin{minipage}{0.45\hsize}
        \centering
        \includegraphics[width=3cm]{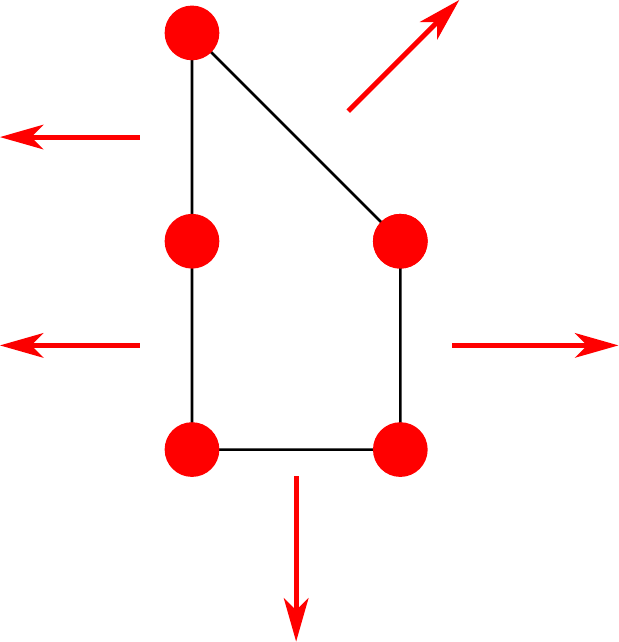}
        \subcaption{Toric diagram}
      \end{minipage} &
      \begin{minipage}{0.45\hsize}
        \centering
        \includegraphics[width=3.5cm]{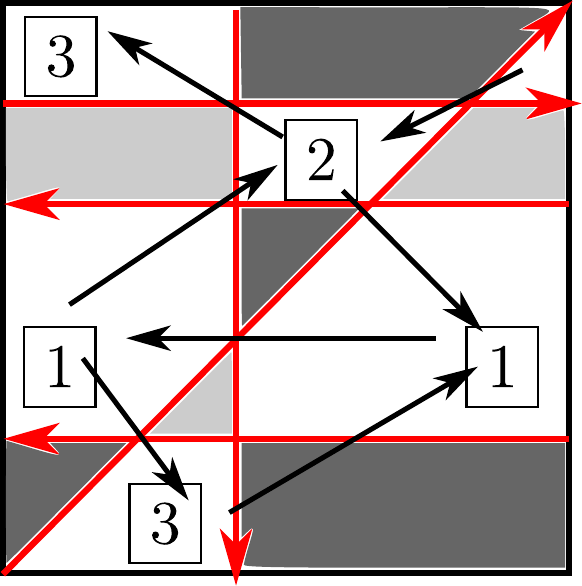}
        \subcaption{Brane configuration}
      \end{minipage}\\
      \begin{minipage}{0.45\hsize}
        \centering
        \includegraphics[width=4cm]{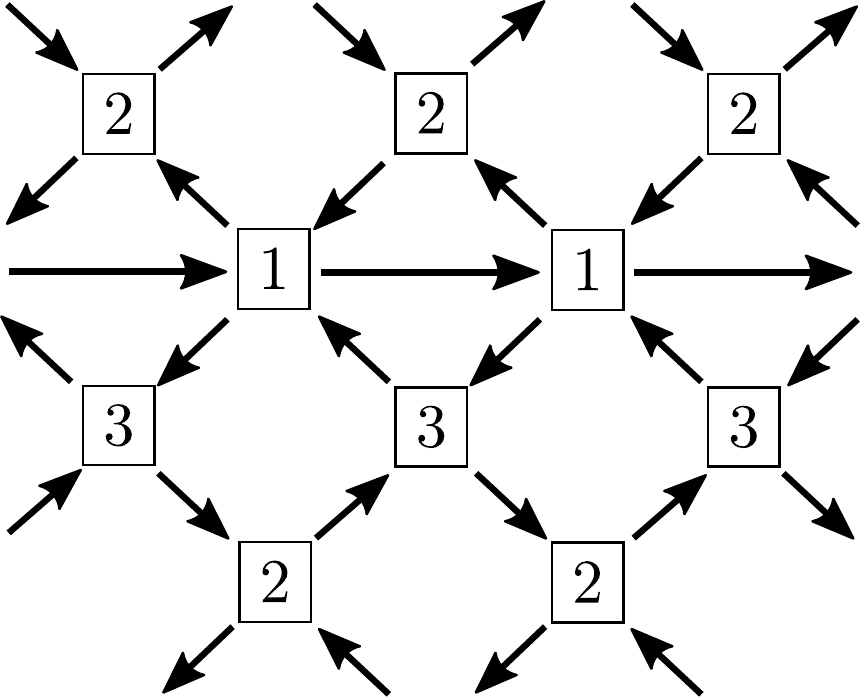}
        \subcaption{Periodic quiver diagram}
      \end{minipage} &
      \begin{minipage}{0.45\hsize}
        \centering
        \includegraphics[width=3.5cm]{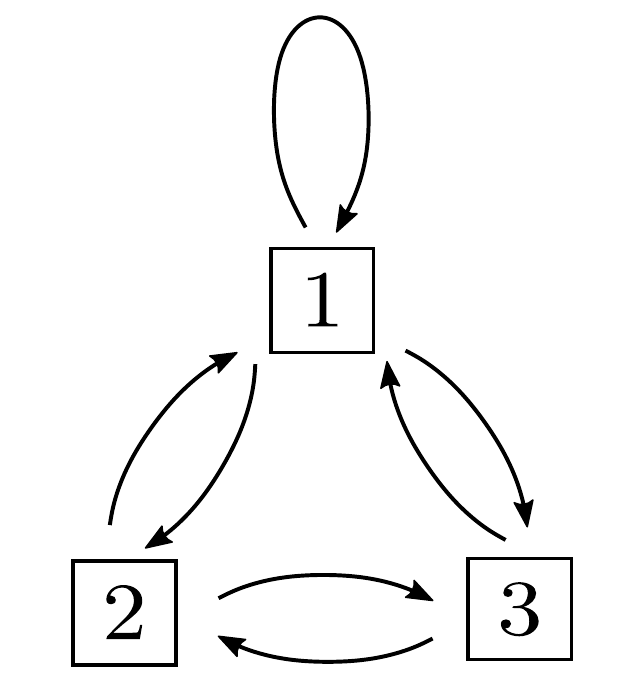}
        \subcaption{Quiver diagram}
      \end{minipage}
    \end{tabular}
\caption{Toric diagram, brane configuration, periodic quiver, and quiver diagram for the Suspended Pinch Point (SPP) singularity. (a) Toric diagram for the SPP. (b) The rectangle is a fundamental domain of the torus $\mathbb{T}^{2}$, and the boundary is identified periodically. The red lines divides the fundamental domain into several domains, and the white regions are assigned vertices of (c) and (d). (c) Periodic quiver diagram obtained by extracting vertices and arrows from the brane configuration.
(d) Quiver diagram obtained from the brane configuration.}
\label{fig:gl21_3d}
\end{figure}

We start with a toric diagram and draw an outward red line perpendicular to each arrow of the toric diagram, which is called the external leg.
Physically, each external leg corresponds to an NS5-brane, and its direction means the direction of the NS5-brane.
\paragraph{Brane configuration and periodic quiver diagram}
Now that we know how many NS5-branes we have and which direction they face, we consider their configuration on a torus $\mathbb{T}^2$. This brane configuration drawn on the torus is called brane tiling and was originally invented in \cite{Franco:2005rj}. The fundamental region of a torus is expressed by a square region, where the top and bottom, left and right, are identical.
We place the red lines corresponding to the external leg on the square region. 
Because of the constraint on NS5-charge, arbitrary configurations of NS5-branes are not allowed.
To describe the constraint, we paint each region white, dark gray, or light gray to indicate the NS5-branes orientation.
We paint dark gray if the red boundary lines are all counter-clockwise, light gray if all clockwise, and white otherwise.
The constraints on the placement of the red lines are as follows:
\begin{itemize}
    \item Two lines can intersect, but three or more lines must not intersect at a single point.
    \item White regions can connect by points, but not by lines.
\end{itemize}
The configuration of red lines satisfying the above constraints always exists as in Figure \ref{fig:gl21_3d}(b).
\paragraph{Periodic quiver diagram and quiver diagram}
The white regions in the brane configuration correspond to the vertices in the quiver diagram we obtain later, and we assign numbers $1, 2, \cdots, |Q_0|$ to each of them.
We denote the set of these vertices by $Q_0$.
In the example of Figure \ref{fig:gl21_3d},
\begin{equation}
    Q_0=\{1,2,3\}.
\end{equation}
Next, to obtain the arrows in the quiver diagram, we connect all neighboring white regions with arrows.
The orientation of the arrows is chosen so that the dark (resp. light) gray region is always on the right (resp. left) of the arrowhead. We denote the set of these arrows by $Q_1$.
In the example of Figure \ref{fig:gl21_3d},
\begin{equation}
    Q_1=\{1\to1, 1\to2, 1\to3, 2\to1, 2\to3, 3\to1, 3\to2\}.
\end{equation}
There may be some arrows whose two ends are identical, and in such cases, we distinguish them by adding an extra index over the arrow as $i\xrightarrow{a} j$.
By extracting such vertices and arrows from the brane configuration and considering the periodic boundary condition, we obtain the periodic quiver diagram as in Figure \ref{fig:gl21_3d}(c).
One may call the brane configuration the periodic quiver diagram because we can obtain the latter easily from the former.
Sets $Q_0$ and $Q_1$ are not enough to explain the shape of the periodic quiver diagram. The loops contained in the periodic quiver diagram characterize its shape, so we denote the set of loops as $Q_2$.
In the example of Figure \ref{fig:gl21_3d}, the loops $1\to1\to3\to1$ is expressed as $1\to1\to3$, and $Q_2$ are as following:
\begin{equation}
    Q_2=\{1\to1\to2, 1\to1\to3, 1\to2\to3\to2, 1\to3\to2\to3\}.
\end{equation}
We can also say that the faces (dark and light gray regions) of the brane tiling are identified with these loops.

By these procedures, we obtain the combination $(Q_0, Q_1, Q_2)$.
The information of $Q_0$ and $Q_1$ can be expressed more simply, and it is (also) called quiver diagram.
In the quiver diagram, we write the vertices of $Q_0$ and then connect them by arrows of $Q_1$ as in Figure \ref{fig:gl21_3d}(d).

In the gauge theory language, elements in $Q_{0}, Q_{1}$ correspond to gauge groups and bifundamental chiral fields, respectively. Elements of $Q_{2}$ determine the superpotential of the quiver gauge theory, which is an essential element in determining the theory. The general formula for the superpotential is 
\begin{align}
    W=\sum_{F:\text{light gray face}}\text{Tr}(\prod_{e\in F}X_{e})-\sum_{\Tilde{F}:\text{dark gray face}} \text{Tr} (\prod_{e\in \Tilde{F}}X_{e}),\label{eq:superpotentialnotation}
\end{align}
 where $F,\Tilde{F}$ are elements in $Q_{2}$. The sign in front of each term is just a convention we will use in this paper.

\section{Consistency check of the generalized coproduct for shifted QQTA}\label{sec:appendix_welldef}
Since the generalized coproduct (\ref{eq:shifted_coproduct_tilde}) is crucial in deriving representations, let us check that this map is consistent. The nontrivial part is the EF relation. We use the original expression (without the tilde) because it is easy to keep track of the shift parameters. In this notation, the coproduct of (\ref{eq:shifted_coproduct_tilde}) is 
\begin{align}
\begin{split}
     &\Delta_{\mathbf{r},\mathbf{r'}} E_{i}(z)=E_{i}(z)\otimes 1+K_{i}^{-}(z)\otimes E_{i}(z),\\
    &\Delta_{\mathbf{r},\mathbf{r'}} F_{i}(z)=F_{i}(z)\otimes z^{r'_{i}}K_{i}^{+}(z)+1\otimes F_{i}(z),\\
    &\Delta_{\mathbf{r},\mathbf{r'}} K_{i}^{+}(z)=K_{i}^{+}(z)\otimes K_{i}^{+}(z),\\
    &\Delta_{\mathbf{r},\mathbf{r'}} K_{i}^{-}(z)=K_{i}^{-}(z)\otimes K_{i}^{-}(z).\label{eq:shifted_coproduct_notilde}
    \end{split}
\end{align}
What we want to prove is 
\begin{align}
    \Delta_{\mathbf{r},\mathbf{r'}}([E_{i}(z),F_{j}(w)])=\delta_{i,j}\delta\left(\frac{w}{z}\right)\left(z^{r_{i}+r'_{i}}\Delta_{\mathbf{r},\mathbf{r'}}(K_{i}^{+}(z))-\Delta_{\mathbf{r},\mathbf{r'}}(K_{i}^{-}(z))\right).
\end{align}
Let us calculate the left hand side. From
\begin{align}
\begin{split}
    &\Delta_{\mathbf{r},\mathbf{r'}}(E_{i}(z))\Delta_{\mathbf{r},\mathbf{r'}}(F_{j}(w))\\
    &=E_{i}(z)F_{j}(w)\otimes w^{r'_{j}}K_{j}^{+}(w)+E_{i}(z)\otimes F_{j}(w)\\
    &\quad+(-1)^{|i||j|}K_{i}^{-}(z)F_{j}(w)\otimes w^{r'_{j}}E_{i}(z)K_{j}^{+}(w)+K_{i}^{-}(z)\otimes E_{i}(z)F_{j}(w),\\[5pt]
       &-(-1)^{|i||j|}\Delta_{\mathbf{r},\mathbf{r'}}(F_{j}(w))\Delta(E_{i}(z))\\
       &= -(-1)^{|i||j|}\left(F_{j}(w)E_{i}(z)\otimes w^{r'_{j}}K_{j}^{+}(w)+(-1)^{|i||j|}E_{i}(z)\otimes F_{j}(w)\right.\\
        &\quad\left.+F_{j}(w)K_{i}^{-}(z)\otimes w^{r'_{i}}K_{i}^{+}(w)E_{i}(z)+K_{i}^{-}(z)\otimes F_{j}(w)E_{i}(z)\right),
    \end{split}
\end{align}
and (\ref{eq:defofshiftedQuiverAlgebra}), we obtain
\begin{align}
    \begin{split}
        \Delta_{\mathbf{r},\mathbf{r'}}([E_{i}(z),F_{j}(w)])&=[E_{i}(z),F_{j}(w)]\otimes w^{r'_{j}}K^{+}_{j}(w)+K_{i}^{-}(z)\otimes [E_{i}(z),F_{j}(w)]\\
        &=\delta_{i,j}\delta\left(\frac{w}{z}\right)\left(z^{r_{i}}K^{+}_{i}(z)\otimes w^{r'_{j}}K_{j}^{+}(w)-K^{-}_{i}(z)\otimes w^{r'_{j}}K_{j}^{+}(w)\right.\\
        &\left.+K_{i}^{-}(z)\otimes z^{r'_{i}}K_{i}^{+}(z)-K_{i}^{-}(z)\otimes K_{i}^{-}(z)\right)\\
        &=\delta_{i,j}\delta\left(\frac{w}{z}\right)\left(z^{r_{i}+r'_{i}}K_{i}^{+}(z)\otimes K_{i}^{+}(z)-K_{i}^{-}(z)\otimes K_{i}^{-}(z)\right)\\
        &=\delta_{i,j}\delta\left(\frac{w}{z}\right)\left(z^{r_{i}+r'_{i}}\Delta_{\mathbf{r},\mathbf{r'}}(K_{i}^{+}(z))-\Delta_{\mathbf{r},\mathbf{r'}}(K_{i}^{-}(z))\right).
    \end{split}
\end{align}

\section{Quantum toroidal \texorpdfstring{$\mathfrak{gl}_{n}\;(n\geq 2)$}{glnn2}}\label{sec:appendixgln}
We show other examples not presented in the main text of section \ref{sec:toroidalgln}. See section \ref{sec:toroidalgln} for figures and formulas. Note that in this section, the Kronecker delta $\bar{\delta}_{i,j}$ is defined as
 \begin{align}
 \bar{\delta}_{i,j}=\begin{cases}
1,\quad i\equiv j\quad (\text{mod}\hspace{2mm}n)\\
0,\quad i\not\equiv j\quad (\text{mod}\hspace{2mm}n)
\end{cases}.
\end{align}
\subsection{One-dimensional crystals}\label{sec:appendixgln1d}
\subsubsection{One-dimensional crystal \texorpdfstring{$\ell_{2}$}{l2}}\label{sec:appendixgln1dl2}
This is also the representation derived in \cite{feigin2013representations}. The crystal shape is in the middle of Figure \ref{fig:gln2_subquiver_1dcrystal}. The origin atom is colored with $k$. Other boxes are colored $k,k+1,...$ from the right of the border and periodically extended left to the border. We also consider the sub-index modulo $n$. The basis $[u]^{(k)}_{j}$ also can be illustrated as a semi-infinite row of boxes where there are $j+1$ boxes right to the border. The boxes are numbered $0,1...j$ from the right of the border as in the crystal associated with $\ell_{1}$ in section \ref{sec:gln1dl1} but the coloring is different. The vector space is denoted $V^{(\ell_{2})}(u)$.

The actions of generators can be written as 
\begin{align}
    \begin{split}
        E_{s}(z)[u]_{j}^{(k)}&=\mathcal{E}_{s}([u]_{j}^{(k)})\delta\left(\frac{z}{uq_{3}^{j+1}}\right)\bar{\delta}_{k+j+1,s}[u]_{j+1}^{(k)},\\
        F_{s}(z)[u]_{j}^{(k)}&=\mathcal{F}_{s}([u]_{j}^{(k)})\delta\left(\frac{z}{uq_{3}^{j}}\right)\bar{\delta}_{k+j,s}[u]_{j-1}^{(k)},\\
        K_{s}^{\pm}(z)[u]_{j}^{(k)}&=\left[\Psi^{(s)}_{[u]_{j}^{(k)}}(z)\right]_{\pm}[u]_{j}^{(k)}.
    \end{split}
\end{align}
By the recursion formula of $\Psi^{(a)}_{[u]_{j}^{(k)}}(z)$ obtained from (\ref{eq:shiftedQuiverAlgebra_for_rep}) and the pole cancellation similar to (\ref{eq:infiniteproductgl1}), we obtain
\begin{align}
   \Psi_{[u]_{j}^{(k)}}^{(a)}(z)=\left(\frac{\phi(q_{1}q_{3}^{-j};z,u)}{\phi(q_{3}^{-1-j};z,u)}\right)^{\bar{\delta}_{k+j,a-1}}\left(\frac{\phi(q_{1}^{-1}q_{3}^{-j-1};z,u)}{\phi(q_{3}^{-j};z,u)}\right)^{\bar{\delta}_{k+j,a}}.\label{eq:gln_l2_chargefunction}
\end{align}
This is also a representation of the unshifted quantum toroidal $\mathfrak{gl}_{n}$ and the shift parameter is 
\begin{align}
    \mathbf{r}=(0,....,0)\in\mathbb{Z}^{n}.
\end{align}

\subsubsection{One-dimensional crystal \texorpdfstring{$\ell_{k+3}$}{l3}}\label{sec:appendixgln1dlk+3}
Let us consider the representation whose crystal picture is the right of Figure \ref{fig:gln2_subquiver_1dcrystal}. $[u]_{j}^{(k)}$ here denotes a semi-infinite row of boxes with $j+1$ boxes right to the border as section \ref{sec:gln1dl1} and Appendix \ref{sec:appendixgln1dl2}, but the coloring is different. In this case, all of the boxes have the same color $k$ as the origin. This representation seems to be not studied in the previous literature.  

The action of the generators can be written as
\begin{align}
\begin{split}
    E_{s}(z)[u]_{j}^{(k)}&=\mathcal{E}_{s}([u]_{j}^{(k)})\delta\left(\frac{z}{uq_{2}^{j+1}}\right)\bar{\delta}_{s,k}[u]_{j+1}^{(k)},\\
    F_{s}(z)[u]_{j}^{(k)}&=\mathcal{F}_{s}([u]_{j}^{(k)})\delta\left(\frac{z}{uq_{2}^{j}}\right)\bar{\delta}_{s,k}[u]_{j-1}^{(k)},\\
    K_{s}^{\pm}(z)[u]_{j}^{(k)}&=\left[\Psi^{(s)}_{[u]_{j}^{(k)}}(z)\right]_{\pm}[u]_{j}^{(k)}.
    \end{split}
\end{align}
By using an analogue of (\ref{eq:infiniteproductgl11d}), we obtain 
\begin{align}
\Psi^{(a)}_{[u]_{j}^{(k)}}(z)=\frac{\phi(q_{2}^{-j}q_{3};z,u)^{\bar{\delta}_{a,k-1}}\phi(q_{1}q_{2}^{-j};z,u)^{\bar{\delta}_{a,k+1}}}{\phi(q_{2}^{-1-j};z,u)^{\bar{\delta}_{a,k}}\phi(q_{2}^{-j};z,u)^{\bar{\delta}_{a,k}}}.
\end{align}
Compared to the cases in section \ref{sec:gln1dl1} and Appendix \ref{sec:appendixgln1dl2}, the charge function have different number of zeros and poles, and thus this is a representation of the shifted quantum toroidal algebra with shift parameters
\begin{align}
\begin{split}
r_{k}=-2,\quad r_{k\pm 1}=1,\\
r_{i}=0\quad (i\neq k,k\pm1).
\end{split}
\end{align}
Note that when $n=2$, because of $k-1\equiv k+1$ (\hspace{-2mm}$\mod2$), this should be understood as
\begin{align}
    r_{k}=-2,\quad r_{k+1}=2,
\end{align}
where the subindex is understood modulo 2.

\subsection{Two-dimensional crystal representations}\label{sec:appendigln2d}
\subsubsection{Two-dimensional crystal of \texorpdfstring{$p_{2}=(1,0)$}{p210}}\label{sec:appendixgln2dp2}
The crystal shape and subquiver of this representation is in the middle of Figure \ref{fig:gln2_subquiver_2dcrystal}.
The basis of this representation is defined by the tensor product of one-dimensional representations as
\begin{align}
    \otimes_{i=1}^{N}V^{(\ell_{1})}(uq_{2}^{-i+1})\ni\ket{\lambda}=\otimes_{i=1}^{N}[uq_{2}^{-i+1}]_{\lambda_{i}-i}^{(k)},\quad \lambda_{1}\geq\lambda_{2}\geq.....
\end{align}
and we take the limit $N\rightarrow \infty$.
This basis forms a submodule.
Almost all is the same as section \ref{sec:gln2dp1}, but we note that the shift of parameters is $q_2^{-1}$, not $q_2$.
The actions of generators are written as
\begin{align}
\begin{split}
\begin{split}
E_{s}(z)\ket{\lambda}&=\sum_{i=1}^{\ell(\lambda)+1}\prod_{j=1}^{i-1}\left[\Psi^{(s)}_{[uq_{2}^{-j+1}]_{\lambda_{j}-j}^{(k)}}(z)\right]_{-}\mathcal{E}_{s}\left([uq_{2}^{-i+1}]^{(k)}_{\lambda_{i}-i}\right)\\
&\times\delta\left(\frac{z}{uq_{2}^{-i+1}q_{1}^{\lambda_{i}-i+1}}\right)\bar{\delta}_{k-\lambda_{i}+i-1,s}\ket{\lambda+\fbox{$s$}_{i}},
\end{split}\\
K_{s}(z)\ket{\lambda}&=\frac{\phi(q_{2}^{-1}q_{3}^{-\ell(\lambda)};z,u)^{\bar{\delta}_{k,s-\ell(\lambda)}}}{\phi(q_{3}^{-\ell(\lambda)};z,u)^{\bar{\delta}_{k,s-\ell(\lambda)}}}\prod_{i=1}^{\ell(\lambda)}\Psi^{(s)}_{[uq_{2}^{-i+1}]^{(k)}_{\lambda_{i}-i}}(z)\ket{\lambda},\\
\begin{split}
F_{s}(z)\ket{\lambda}&=\frac{\phi(q_{2}^{-1}q_{3}^{-\ell(\lambda)};z,u)^{\bar{\delta}_{k,s-\ell(\lambda)}}}{\phi(q_{3}^{-\ell(\lambda)};z,u)^{\bar{\delta}_{k,s-\ell(\lambda)}}}\sum_{i=1}^{\ell(\lambda)}\prod_{j=i+1}^{\ell(\lambda)}\left[\Psi^{(s)}_{[uq_{2}^{-j+1}]^{(k)}_{\lambda_{j}-j}}(z)\right]_{+}\\
&\times \mathcal{F}_{s}\left([uq_{2}^{-i+1}]^{(k)}_{\lambda_{i}-i}\right)\bar{\delta}_{k-\lambda_{i}+i,s}\delta\left(\frac{z}{uq_{2}^{-i+1}q_{1}^{\lambda_{i}-i}}\right)\ket{\lambda-\fbox{$s$}_{i}},\end{split}
\end{split}
\end{align}
where we used 
\begin{align}
   \prod_{i=\ell(\lambda)+1}^{\infty}\Psi^{(s)}_{[uq_{2}^{-i+1}]^{(k)}_{\lambda_{i}-i}}(z)= \frac{\phi(q_{2}^{-1}q_{3}^{-\ell(\lambda)};z,u)^{\bar{\delta}_{k,s-\ell(\lambda)}}}{\phi(q_{3}^{-\ell(\lambda)};z,u)^{\bar{\delta}_{k,s-\ell(\lambda)}}},
\end{align}
which comes from an analogue of (\ref{eq:infiniteproductgl1}).
Especially, the action of $K_s(z)$ on the vacuum is
\begin{align}
    K_{s}(z)\ket{\emptyset}=\left(\frac{\phi(q_{2}^{-1};z,u)}{\phi(1;z,u)}\right)^{\bar{\delta}_{k,s}}
\end{align} as expected in (\ref{eq:vacuum_charge_function}).
This is a representation of the unshifted quantum toroidal algebra $\mathfrak{gl}_{n}$ and the shift parameter is
\begin{align}
    \mathbf{r}=(0,...,0)\in\mathbb{Z}^{n}.
\end{align}
We note these equations are true even for the $n=2$ case.


\subsubsection{Two-dimensional crystal of \texorpdfstring{$p_{3}=(0,n)$}{p30n}}\label{sec:appendixgln2dp3}
The crystal shape and subquiver of this representation is in the right of Figure \ref{fig:gln2_subquiver_2dcrystal}. We consider the action of the algebra on $V^{(\ell_{2})}(u)\otimes V^{(\ell_{2})}(v)$. 
The basis of this representation is defined as
\begin{align}
    \otimes_{i=1}^{N}V^{(\ell_{2})}(uq_{2}^{i-1})\ni\ket{\lambda}=\otimes_{i=1}^{N}[uq_{2}^{i-1}]_{\lambda_{i}-1}^{(k)}.\quad \lambda_{1}\geq\lambda_{2}\geq.....
\end{align}
It forms a submodule, and  we take the limit $N\rightarrow \infty$.
The actions of generators are written as
\begin{align}
\begin{split}
\begin{split}
E_{s}(z)\ket{\lambda}&=\sum_{i=1}^{\ell(\lambda)+1}\prod_{j=1}^{i-1}\left[\Psi^{(s)}_{[uq_{2}^{j-1}]_{\lambda_{j}-1}^{(k)}}(z)\right]_{-}\mathcal{E}_{s}\left([uq_{2}^{i-1}]^{(k)}_{\lambda_{i}-1}\right)\\
&\times\delta\left(\frac{z}{uq_{2}^{i-1}q_{3}^{\lambda_{i}}}\right)\bar{\delta}_{k+\lambda_{i},s}\ket{\lambda+\fbox{$s$}_{i}},
\end{split}\\
K_{s}(z)\ket{\lambda}&=\frac{\phi(q_{1}^{-1}q_{2}^{-\ell(\lambda)};z,u)^{\bar{\delta}_{k,s+1}}}{\phi(q_{2}^{-\ell(\lambda)};z,u)^{\bar{\delta}_{k,s}}}\prod_{i=1}^{\ell(\lambda)}\Psi^{(s)}_{[uq_{2}^{i-1}]^{(k)}_{\lambda_{i}-1}}(z)\ket{\lambda},\\
\begin{split}
F_{s}(z)\ket{\lambda}&=\frac{\phi(q_{1}^{-1}q_{2}^{-\ell(\lambda)};z,u)^{\bar{\delta}_{k,s+1}}}{\phi(q_{2}^{-\ell(\lambda)};z,u)^{\bar{\delta}_{k,s}}}\sum_{i=1}^{\ell(\lambda)}\prod_{j=i+1}^{\ell(\lambda)}\left[\Psi^{(s)}_{[uq_{2}^{j-1}]^{(k)}_{\lambda_{j}-1}}(z)\right]_{+}\\
&\times \mathcal{F}_{s}\left([uq_{2}^{i-1}]^{(k)}_{\lambda_{i}-1}\right)\bar{\delta}_{k+\lambda_{i}-1,s}\delta\left(\frac{z}{uq_{2}^{i-1}q_{3}^{\lambda_{i}-1}}\right)\ket{\lambda-\fbox{$s$}_{i}}\end{split},
\end{split}
\end{align}
where we used an analogue of (\ref{eq:infiniteproductgl1}).
Especially, the action of $K_s(z)$ on the vacuum is
\begin{align}
    K_{s}(z)\ket{\emptyset}=\frac{\phi(q_{1}^{-1};z,u)^{\bar{\delta}_{k,s+1}}}{\phi(1;z,u)^{\bar{\delta}_{k,s}}}\ket{\emptyset}
\end{align}
as expected in (\ref{eq:vacuum_charge_function}).
The shift is 
\begin{align}
    r_{k-1}=1,\quad r_{k}=-1,\quad r_{i}=0\hspace{2mm}(i\neq k,k-1).
\end{align}
We note this representation is also true for the $n=2$ case.

\section{Quantum toroidal \texorpdfstring{$\mathfrak{gl}_{2|1}$}{gl21}}\label{sec:appendixgl2|1}
In this section, we construct other examples of the subcrystal representations defined in section \ref{sec:toroidalgl2|1}. The subquiver and crsytal picture are illustrated in Figure \ref{fig:gl_(2,1)2dimcrystal_subquiver} and Figure \ref{fig:gl_(2,1)1dimcrystal}.
\subsection{One-dimensional crystal representations}
\subsubsection{One-dimensional crystal \texorpdfstring{$\ell_{2}$}{l2}}
\begin{figure}[h]
    \begin{tabular}{cc}
      \begin{minipage}{0.45\hsize}
        \centering
       \includegraphics[width=5.5cm]{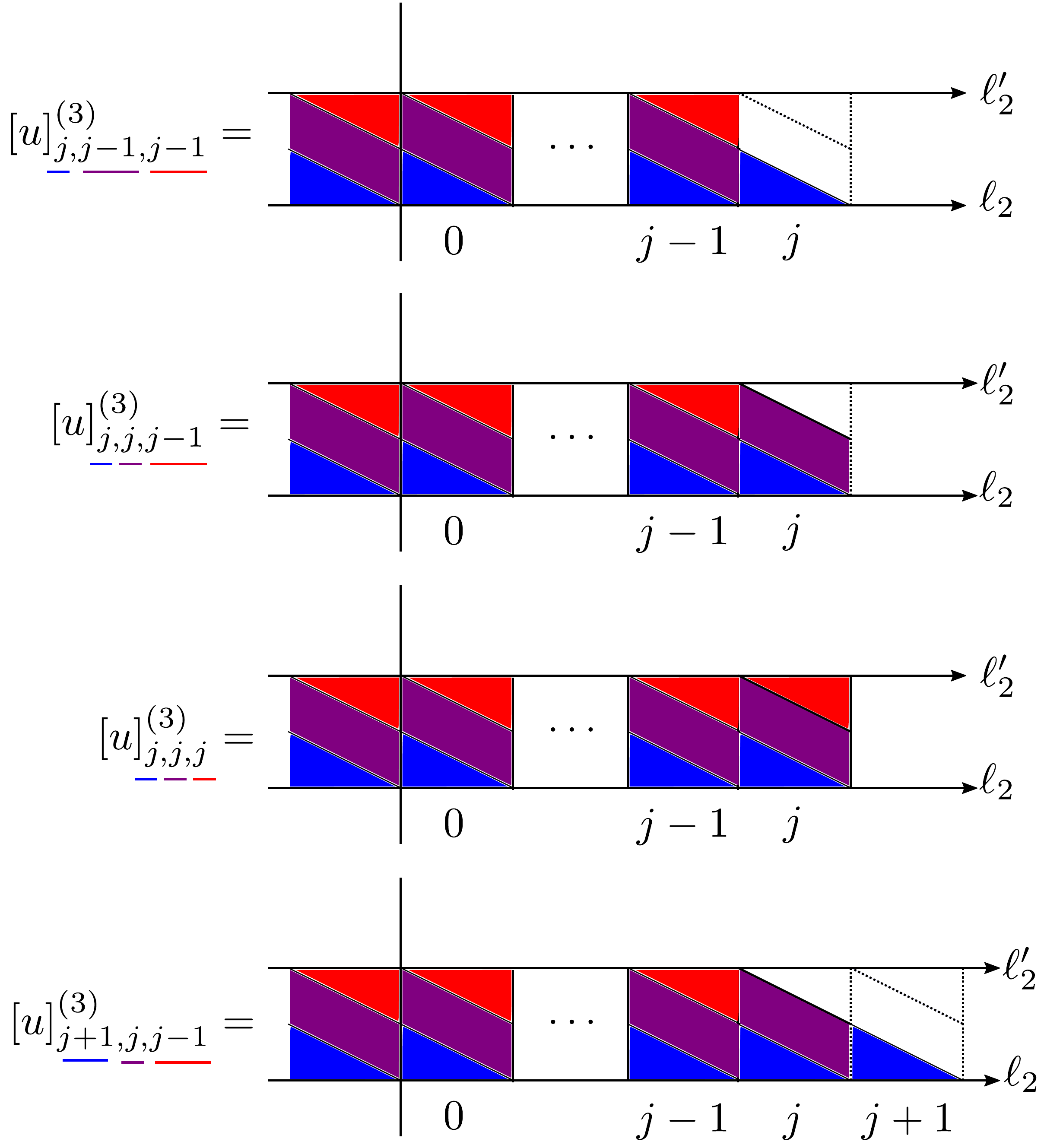}
        \subcaption{Basis of one-dimensional crystal representation $\ell_{2}$.}\label{fig:gl_(2,1)1dimrep_l2_vectorbasis}
      \end{minipage}&
      \begin{minipage}{0.45\hsize}
        \centering
       \includegraphics[width=6.5cm]{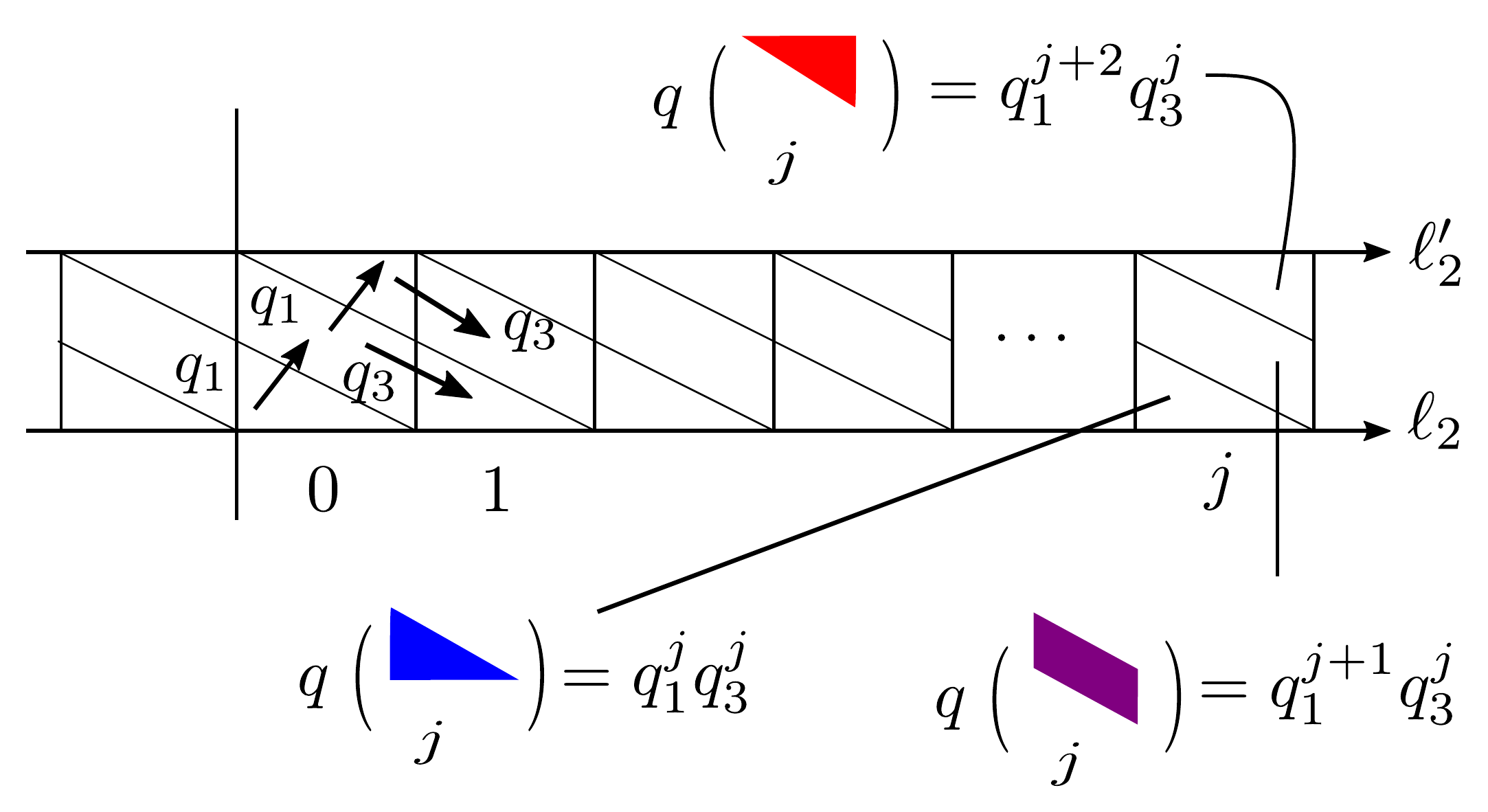}
       \subcaption{Coordinates of atoms.}\label{fig:gl_(2,1)1dimrep_l2_coordinate}
      \end{minipage}
    \end{tabular}
\caption{Basis and coordinates of the one-dimensional crystal representation $\ell_{2}$.}
\end{figure}

Let us consider the crystal associated with the external leg $\ell_{2}$. The subquiver and shape of the crystal are illustrated in Figure \ref{fig:gl_(2,1)1dimcrystal_l2}.
We denote $V^{(l_{2})}(u)$ to be a complex vector space with four types of bases $[u]_{j,j-1,j-1}^{(3)}$, $[u]_{j,j,j-1}^{(3)}$, $[u]_{j,j,j}^{(3)}$ and $[u]^{(3)}_{j+1,j,j-1}$, where $j\in \mathbb{Z}$. The basis can be pictured as Figure \ref{fig:gl_(2,1)1dimrep_l2_vectorbasis}. We have a semi-infinite row of boxes with coordinates counted 0,1,2,... from the border and extended to the left of the border. We have three types of atoms in each box: blue triangle, purple parallelogram, and red triangle. The subscripts of the vectors are the coordinates of the rightest box in which the three types of atoms are. The first subscript is the index for the blue triangle, the second is for the purple parallelogram, and the third is for the red triangle. We can assign coordinates to the atoms as Figure \ref{fig:gl_(2,1)1dimrep_l2_coordinate}.

Since $K_{i}(z)$ acts diagonally we obtain
 \begin{align}
            \begin{split}
        K_{i}^{\pm}(z)\begin{cases}
        [u]^{(3)}_{j,j-1,j-1}\\
        [u]^{(3)}_{j,j,j-1}\\
        [u]^{(3)}_{j,j,j}\\
        [u]^{(3)}_{j+1,j,j-1}
        \end{cases}&=\begin{cases}
        [\Psi^{(i)}_{[u]^{(3)}_{j,j-1,j-1}}(z)]_{\pm}[u]^{(3)}_{j,j-1,j-1}\\
        [\Psi^{(i)}_{[u]^{(3)}_{j,j,j-1}}(z)]_{\pm}[u]^{(3)}_{j,j,j-1}\\
        [\Psi^{(i)}_{[u]^{(3)}_{j,j,j}}(z)]_{\pm}[u]^{(3)}_{j,j,j}\\
        [\Psi^{(i)}_{[u]^{(3)}_{j+1,j,j-1}}(z)]_{\pm}[u]^{(3)}_{j+1,j,j-1}.
        \end{cases}
   \end{split}
\end{align}
For the actions of $E_{i}(z)$ and $F_{i}(z)$, the non-vanishing contributions are 
\begin{align}
    \begin{split}
        E_{1}(z)[u]^{(3)}_{j,j-1,j-1}
        &=\mathcal{E}_{1}([u]^{(3)}_{j,j-1,j-1})\delta\left(\frac{z}{uq_{1}^{j+1}q_{3}^{j}}\right)[u]_{j,j,j-1}^{(3)},
        \\
        E_{2}(z)\begin{cases}
        [u]^{(3)}_{j,j,j-1}\\
        [u]^{(3)}_{j+1,j,j-1}
        \end{cases}
        &=\begin{cases}
        \mathcal{E}_{2}([u]^{(3)}_{j,j,j-1})\delta\left(\frac{z}{uq_{1}^{j+2}q_{3}^{j}}\right)[u]_{j,j,j}^{(3)},\\
        \mathcal{E}_{2}([u]^{(3)}_{j+1,j,j-1})\delta\left(\frac{z}{uq_{1}^{j+2}q_{3}^{j}}\right)[u]_{j+1,j,j}^{(3)},
        \end{cases}\\
        E_{3}(z)\begin{cases}
        [u]^{(3)}_{j,j,j-1}\\
        [u]^{(3)}_{j,j,j}\\
        \end{cases}&=\begin{cases}
        \mathcal{E}_{3}([u]^{(3)}_{j,j,j-1})\delta\left(\frac{z}{uq_{1}^{j+1}q_{3}^{j+1}}\right)[u]^{(3)}_{j+1,j,j-1},\\
        \mathcal{E}_{3}( [u]^{(3)}_{j,j,j})\delta\left(\frac{z}{uq_{1}^{j+1}q_{3}^{j+1}}\right)[u]_{j+1,j,j}^{(3)},\\
        \end{cases}
        \end{split}
        \end{align}
\begin{align}
    \begin{split}
        F_{1}(z)[u]^{(3)}_{j,j,j-1}&=\mathcal{F}_{1}([u]^{(3)}_{j,j,j-1})\delta\left(\frac{z}{uq_{1}^{j+1}q_{3}^{j}}\right)[u]^{(3)}_{j,j-1,j-1},\\
        F_{2}(z)\begin{cases}
        [u]^{(3)}_{j,j-1,j-1}\\
        [u]^{(3)}_{j,j,j}\\
        \end{cases}&=\begin{cases}
        \mathcal{F}_{2}([u]^{(3)}_{j,j-1,j-1})\delta\left(\frac{z}{uq_{1}^{j+1}q_{3}^{j-1}}\right)[u]_{j,j-1,j-2}^{(3)},\\
        \mathcal{F}_{2}([u]^{(3)}_{j,j,j})\delta\left(\frac{z}{uq_{1}^{j+2}q_{3}^{j}}\right)[u]^{(3)}_{j,j,j-1},\end{cases}\\
        F_{3}(z)\begin{cases}
        [u]^{(3)}_{j,j-1,j-1}\\
        [u]^{(3)}_{j+1,j,j-1}
        \end{cases}&=\begin{cases}
        \mathcal{F}_{3}([u]^{(3)}_{j,j-1,j-1})\delta\left(\frac{z}{uq_{1}^{j}q_{3}^{j}}\right)[u]^{(3)}_{j-1,j-1.j-1},\\
        \mathcal{F}_{3}([u]^{(3)}_{j+1,j,j-1})\delta\left(\frac{z}{uq_{1}^{j+1}q_{3}^{j+1}}\right)[u]_{j,j,j-1}^{(3)}.
        \end{cases}
    \end{split}
\end{align}
The parity conditions are 
\begin{align}
    \begin{split}
        &|[u]^{(3)}_{j,j,j-1}|=|[u]^{(3)}_{j,j-1,j-1}|,\quad |[u]^{(3)}_{j,j,j}|=|[u]^{(3)}_{j,j,j-1}|+1,\\
        &|[u]^{(3)}_{j+1,j,j}|=|[u]^{(3)}_{j+1,j,j-1}|+1,\quad |[u]^{(3)}_{j+1,j,j-1}|=|[u]^{(3)}_{j,j,j-1}|+1,\\
        &|[u]^{(3)}_{j+1,j,j}|=|[u]^{(3)}_{j,j,j}|+1
    \end{split}
\end{align}

Other actions vanish. The coefficients written in $\mathcal{E}_{s},\;\mathcal{F}_{s}$ are nonzero coefficients that can be derived from the defining relations of the algebra. They are not necessary, so we do not write down the explicit formulas.
These actions of the generators of the algebra can be summarized as Figure \ref{fig:gl_(2,1)_l2_action}. The charge function can be derived by using the KE relations and we obtain:

\begin{figure}[th]
    \centering
    \includegraphics[width=15cm]{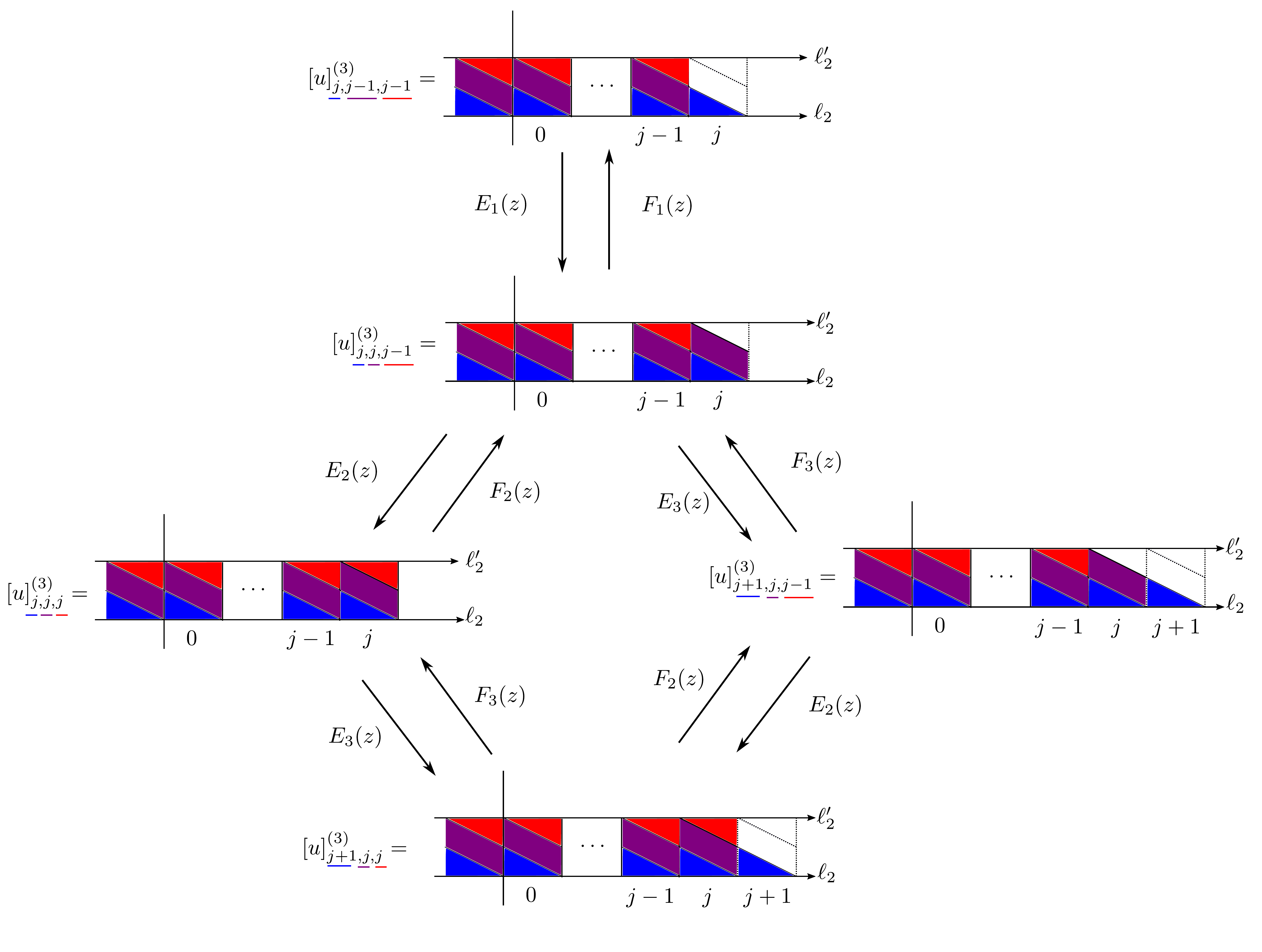}
    \caption{Action of the generators $E_{s}(z),F_{s}(z)$ in one-dimensional crystal representation of $\ell_{2}$.}
    \label{fig:gl_(2,1)_l2_action}
\end{figure}
\begin{align}
\begin{split}
    \Psi^{(1)}_{[u]^{(3)}_{j+1,j,j}}(z)&=\frac{\phi(q_{1}^{-j-1}q_{3}^{-j};z,u)}{\phi(q_{1}^{-j-2}q_{3}^{-1-j};z,u)},\quad
  \Psi^{(1)}_{[u]^{(3)}_{j,j,j-1}}(z)=\frac{\phi(q_{1}^{-j-2}q_{3}^{-j-1};z,u)}{\phi(q_{3}^{-j}q_{1}^{-j-1};z,u)},\\
  \Psi^{(1)}_{[u]^{(3)}_{j+1,j,j-1}}(z)&=\Psi^{(1)}_{[u]^{(3)}_{j,j,j}}(z)=1,\\
        \Psi^{(2)}_{[u]^{(3)}_{j+1,j,j}}(z)&=\frac{\phi(q_{3}^{-j-2}q_{1}^{-j-1};z,u)}{\phi(q_{1}^{-j-2}q_{3}^{-j};z,u)},\quad
        \Psi^{(2)}_{[u]^{(3)}_{j,j,j-1}}(z)=\frac{\phi(q_{1}^{-j}q_{3}^{-j-1};z,u)}{\phi(q_{1}^{-j-2}q_{3}^{-j};z,u)},\\
        \Psi^{(2)}_{[u]^{(3)}_{j+1,j,j-1}}(z)&=\frac{\phi(q_{1}^{-j-1}q_{3}^{-j-2};z,u)}{\phi(q_{1}^{-j-2}q_{3}^{-j};z,u)},\quad
        \Psi^{(2)}_{[u]^{(3)}_{j,j,j}}(z)=\frac{\phi(q_{1}^{-j}q_{3}^{-j-1};z,u)}{\phi(q_{1}^{-j-2}q_{3}^{-j};z,u)},\\
        \Psi^{(3)}_{[u]^{(3)}_{j+1,j,j}}(z)&=\frac{\phi(q_{1}^{-3-j}q_{3}^{-j};z,u)}{\phi(q_{1}^{-j-1}q_{3}^{-j-1};z,u)},\quad
        \Psi^{(3)}_{[u]^{(3)}_{j,j,j-1}}(z)=\frac{\phi(q_{1}^{-j-2}q_{3}^{-j+1};z,u)}{\phi(q_{1}^{-j-1}q_{3}^{-j-1};z,u)},\\
        \Psi^{(3)}_{[u]^{(3)}_{j+1,j,j-1}}(z)&=\frac{\phi(q_{1}^{-j-2}q_{3}^{-j+1};z,u)}{\phi(q_{3}^{-1-j}q_{1}^{-j-1};z,u)},\quad
        \Psi^{(3)}_{[u]^{(3)}_{j,j,j}}(z)=\frac{\phi(q_{1}^{-3-j}q_{3}^{-j};z,u)}{\phi(q_{1}^{-j-1}q_{3}^{-j-1};z,u)}.
    \end{split}\label{eq:gl_(2,1)_l2_chargefunction}
\end{align}
As one can see, the charge functions have the same number of zeros and poles, which means this is a representation of the unshifted quantum toroidal algebra. The shift parameters are determined as 
\begin{align}
    r_{1}=r_{2}=r_{3}=0.
\end{align}

\subsubsection{One-dimensional crystal \texorpdfstring{$\ell_{3}$}{l3}}
\begin{figure}[H]
    \begin{tabular}{cc}
      \begin{minipage}{0.45\hsize}
        \centering
       \includegraphics[width=6.5cm]{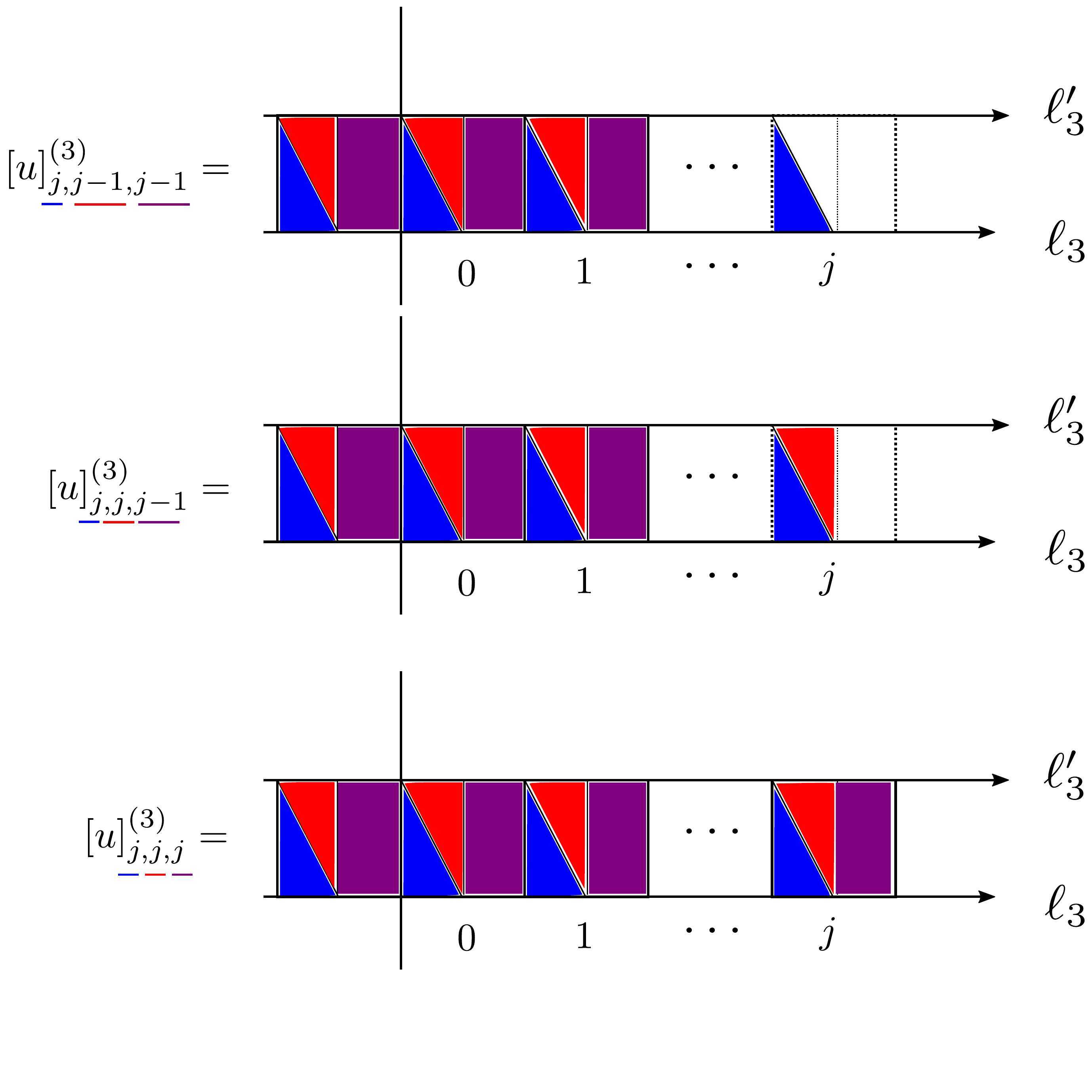}
        \subcaption{Basis of one-dimensional crystal representation $\ell_{3}$.}\label{fig:gl_(2,1)1dimrep_l3_vectorbasis}
      \end{minipage}&
      \begin{minipage}{0.45\hsize}
        \centering
      \includegraphics[width=6.5cm]{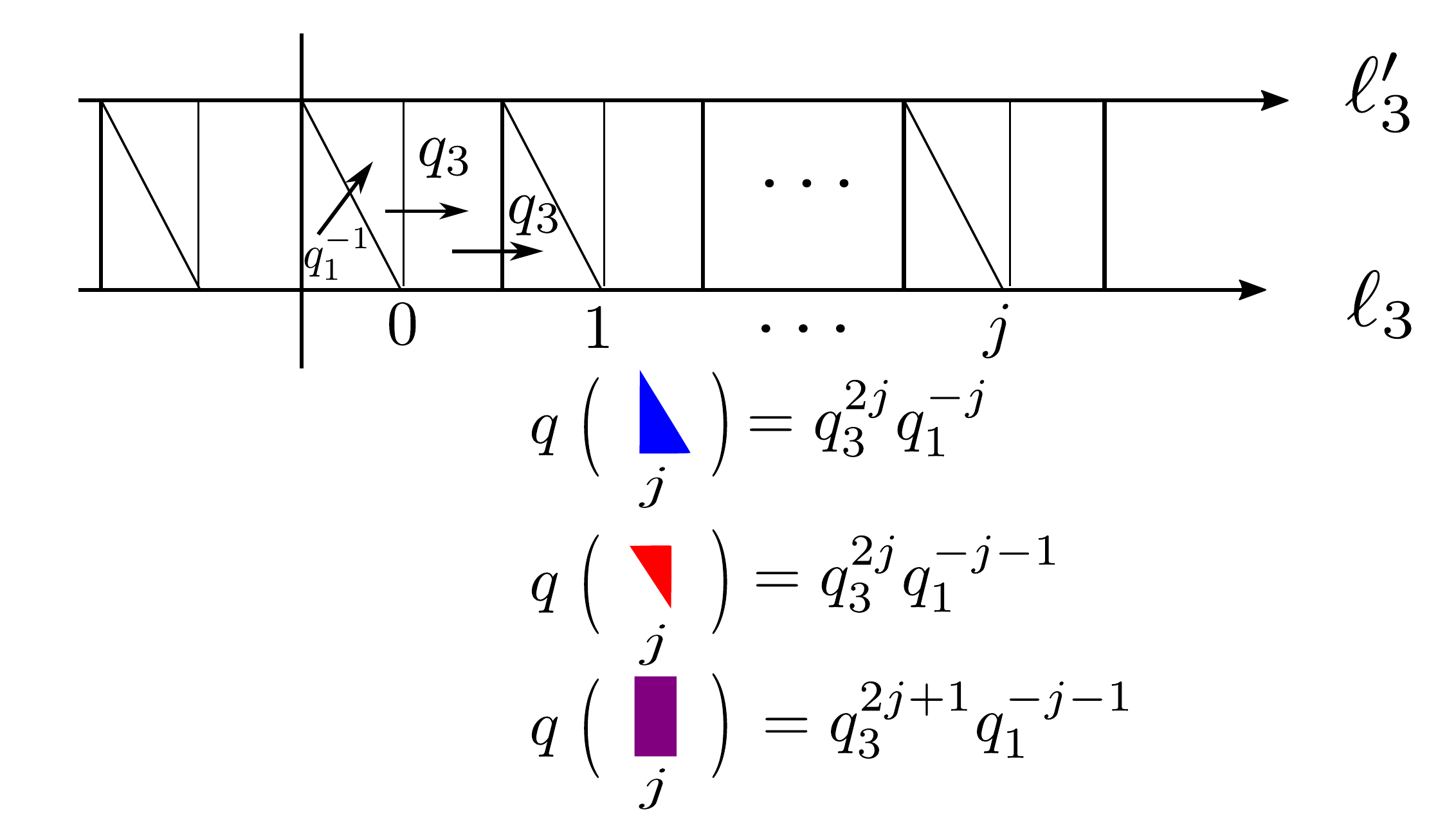}
       \subcaption{Coordinates of atoms.}\label{fig:gl_(2,1)1dimrep_l3_coordinate}
      \end{minipage}
    \end{tabular}
\caption{Basis and coordinates of the one-dimensional crystal representation $\ell_{3}$.}
\end{figure}

Let us construct the representation associated with the external leg $\ell_{3}$. The subquiver and crystal shape are in Figure \ref{fig:gl_(2,1)1dimcrystal_l3}. We denote the vector space of this representation as $V^{(\ell_{3})}(u)$. It has three types of vectors, which we denote $[u]^{(3)}_{j,j-1,j-1},[u]^{(3)}_{j,j,j-1}$, and $[u]^{(3)}_{j,j,j}$. The vectors and coordinates are in Figure \ref{fig:gl_(2,1)1dimrep_l3_vectorbasis} and \ref{fig:gl_(2,1)1dimrep_l3_coordinate}.
Similar to the one-dimensional crystal representation associated with $\ell_{1}$, the action of $K_{i}(z)$ can be written as 
\begin{align}
    K_{i}^{\pm}(z)\begin{cases}
    [u]^{(3)}_{j,j-1,j-1}\\
    [u]^{(3)}_{j,j,j-1}\\
    [u]^{(3)}_{j,j,j}
    \end{cases}&=\begin{cases}
    [\Psi_{[u]^{(3)}_{j,j-1,j-1}}^{(i)}(z)]_{\pm}[u]^{(3)}_{j,j-1,j-1}\\
    [\Psi_{[u]^{(3)}_{j,j,j-1}}^{(i)}(z)]_{\pm}[u]^{(3)}_{j,j,j-1}\\
    [\Psi_{[u]^{(3)}_{j,j,j}}^{(i)}(z)]_{\pm}[u]^{(3)}_{j,j,j}.
    \end{cases}
    \end{align}
    
The actions of generators $E_{s}(z), F_{s}(z)$ can be written in a simple way similar to the vector representation associated with $\ell_{1}$. We use the following notations:
\begin{align}
   & [u]^{(3)}_{\sigma}=[u]^{(3)}_{(r(\sigma),s(\sigma))}=\begin{cases}
    [u]^{(3)}_{r(\sigma),r(\sigma)-1,r(\sigma)-1},\quad s(\sigma)=0,\\
    [u]^{(3)}_{r(\sigma),r(\sigma),r(\sigma)-1},\qquad s(\sigma)=1,\\
    [u]^{(3)}_{r(\sigma),r(\sigma),r(\sigma)},\qquad s(\sigma)=2,
    \end{cases}\\
    &\sigma\in\mathbb{Z},\quad \sigma=3r(\sigma)+s(\sigma),\quad \label{eq:gl_(2,1)1dimcrystal_newconvention_l3}
\end{align}
where $r(\sigma)$ is the quotient of $\sigma$ by 3 and $s(\sigma)\in\mathbb{Z}_{3}=\{0,1,2\}$ is the remainder after $\sigma$ is divided by 3.

We assign the parity condition as 
\begin{align}
 |\sigma|\equiv|[u]_{\sigma}^{(3)}|=\begin{dcases}
 1\quad s(\sigma)=0\\
 0\quad s(\sigma)=1\\
 0\quad s(\sigma)=2
 \end{dcases}   
\end{align}

\begin{figure}[H]
    \centering
    \includegraphics[width=11cm]{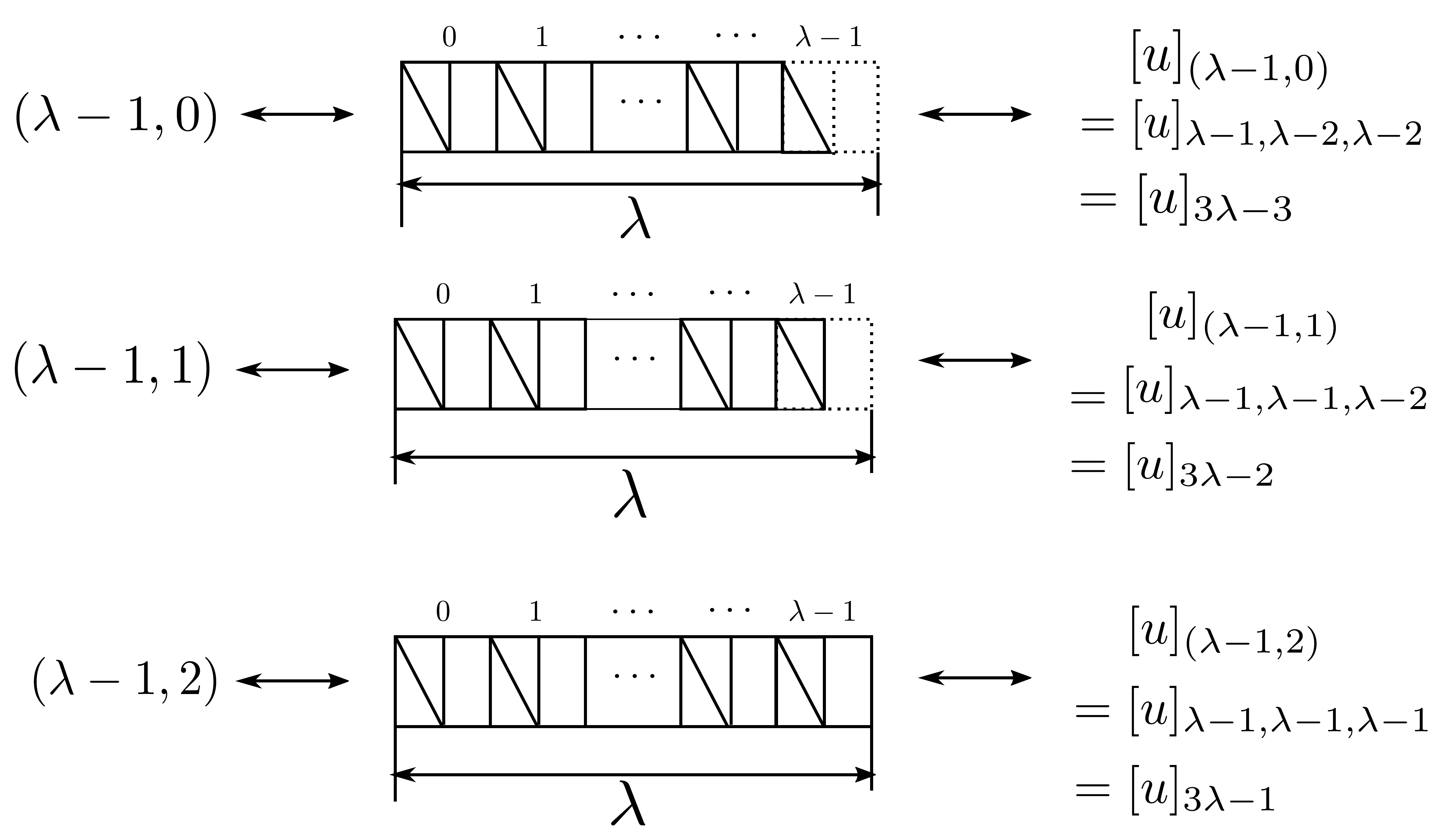}
    \caption{Generalization of Young diagram and correspondence with vectors for one-dimensional crystal of $\ell_{3}$. The generalized partition is expressed by two numbers $(\lambda-1,\tau)\in\mathbb{Z}\times\mathbb{Z}_{3}$. We note we set $\tau\in\mathbb{Z}_{3}=\{0,1,2\}$. Using (\ref{eq:gl_(2,1)1dimcrystal_newconvention_l3}), it can be written as $\sigma=3\lambda-3+\tau$, $r(3\lambda-3+\tau)=\lambda-1$, and $s(3\lambda-3+\tau)=\tau$. }
    \label{fig:gl_(2,1)1dcrystal_l3_usefulconvention}
\end{figure}
The action of $E_{s}(z)$ is 
\begin{align}
    E_{s}(z)[u]^{(3)}_{\sigma}=\mathcal{E}_{s}([u]^{(3)}_{\sigma})\delta\left(\frac{z}{uq_{1}^{-1}(q_{3}^{2}q_{1}^{-1})^{r(\sigma)}q_{3}^{s(\sigma)}}\right)\bar{\delta}_{s+s(\sigma),2}[u]^{(3)}_{\sigma+1},
\end{align}
and the action of $F_{s}(z)$ is 
\begin{align}
    F_{s}(z)[u]^{(3)}_{\sigma}=\mathcal{F}_{s}([u]^{(3)}_{\sigma})\delta\left(\frac{z}{uq_{1}^{-1}(q_{3}^{2}q_{1}^{-1})^{r(\sigma-1)}q_{3}^{s(\sigma-1)}}\right)\bar{\delta}_{s+s(\sigma-1),0}[u]^{(3)}_{\sigma-1},
\end{align}
where $\bar{\delta}_{i,j}=\begin{cases}
1,\quad i\equiv j\;(\text{mod}\; 3)\\
0,\quad i\not\equiv j\;(\text{mod}\; 3)
\end{cases}$. $\mathcal{E}_{s}([u]^{(3)}_{\sigma}),\;\mathcal{F}_{s}([u]^{(3)}_{\sigma})$ are coefficients which can be determined by the defining relations.\\
The charge functions can be determined by the KE relations:
\begin{align}
\begin{split}
    &\Psi^{(1)}_{[u]^{(3)}_{j,j-1,j-1}}(z)=\Psi^{(2)}_{[u]^{(3)}_{j,j,j}}(z)=\Psi^{(3)}_{[u]^{(3)}_{j,j,j-1}}(z)=1,\\
    &\Psi^{(1)}_{[u]^{(3)}_{j,j,j-1}}(z)=\frac{\phi(q_{1}^{2-j}q_{3}^{-2j};z,u)}{\phi(q_{3}^{-1-2j}q_{1}^{j+1};z,u)},\quad\Psi^{(1)}_{[u]^{(3)}_{j,j,j}}(z)=\frac{\phi(q_{1}^{j}q_{3}^{-2j-2};z,u)}{\phi(q_{3}^{-1-2j}q_{1}^{j+1};z,u)},\\
    &\Psi^{(2)}_{[u]^{(3)}_{j,j-1,j-1}}(z)=\frac{\phi(q_{1}^{j}q_{3}^{-1-2j};z,u)}{\phi(q_{1}^{j+1}q_{3}^{-2j};z,u)},\quad\Psi^{(2)}_{[u]^{(3)}_{j,j,j-1}}(z)=\frac{\phi(q_{1}^{j}q_{3}^{-1-2j};z,u)}{\phi(q_{1}^{j+1}q_{3}^{-2j};z,u)},\\
    &\Psi^{(3)}_{[u]^{(3)}_{j,j-1,j-1}}(z)=\frac{\phi(q_{1}^{j+1}q_{3}^{-2j+1};z,u)}{\phi(q_{1}^{j}q_{3}^{-2j};z,u)},\quad\Psi^{(3)}_{[u]^{(3)}_{j,j,j}}(z)=\frac{\phi(q_{1}^{j+2}q_{3}^{-2j-1};z,u)}{\phi(q_{1}^{j+1}q_{3}^{-2-2j};z,u)}.
    \end{split}\label{eq:gl_(2,1)_l3_chargefunction}
\end{align}
As one can see, the charge functions have the same numbers of zeros and poles, which means they are representations of unshifted quantum toroidal algebra. The shift parameters are
\begin{align}
    r_{1}=r_{2}=r_{3}=0.
\end{align}
\subsubsection{One-dimensional crystals of \texorpdfstring{$\ell_{4}$}{l4} and \texorpdfstring{$\ell_{5}$}{l5}}
\begin{figure}[H]
    \begin{tabular}{cc}
      \begin{minipage}{0.45\hsize}
        \centering
       \includegraphics[width=6.5cm]{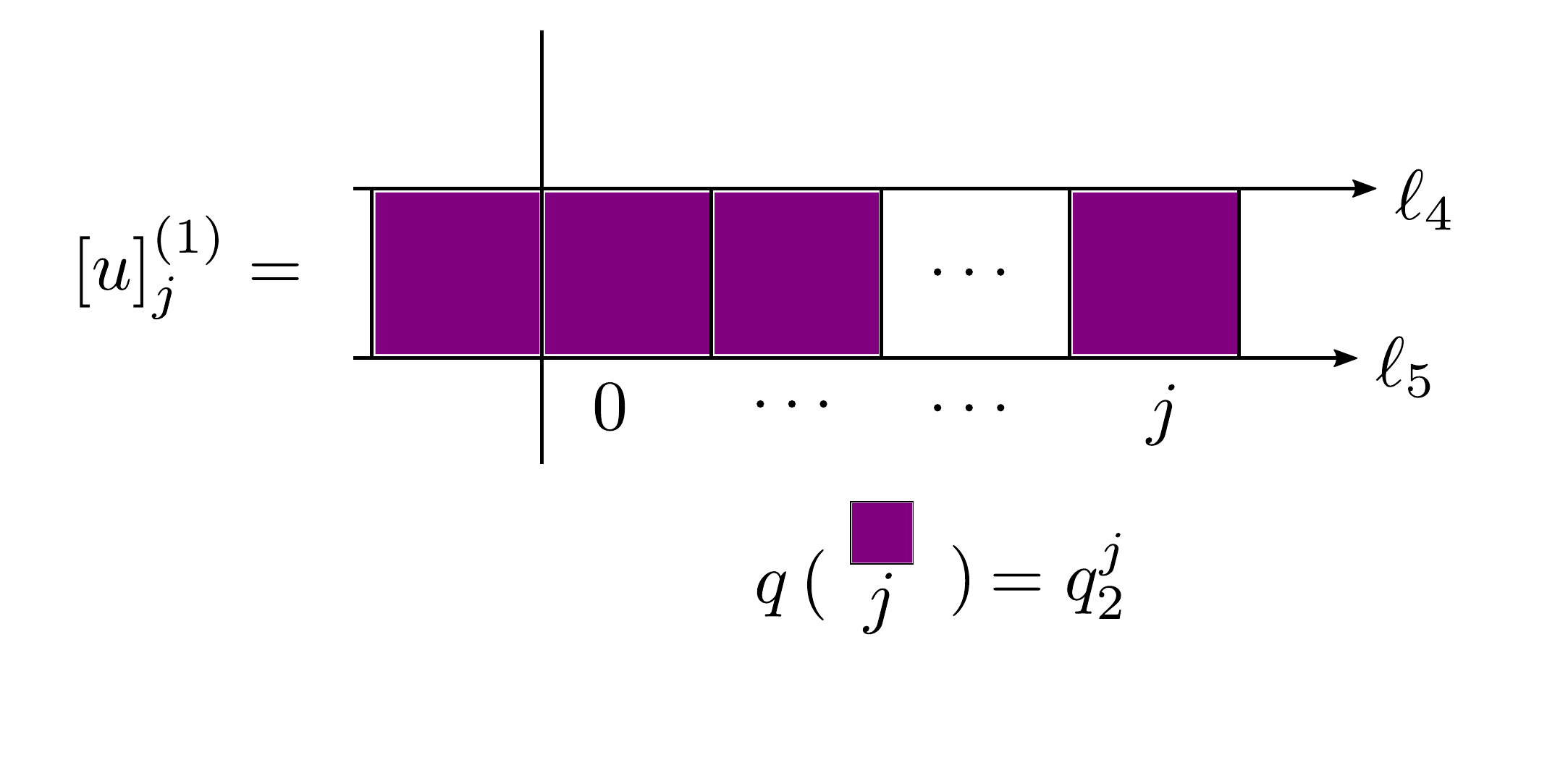}
        \subcaption{Basis and coordinates of one-dimensional crystal representation of $\ell_{4},\ell_{5}$. The atom in the origin is in purple.}\label{fig:gl_(2,1)1dimrep_l4,l5_part1}
      \end{minipage}&\hfill
      \begin{minipage}{0.45\hsize}
        \centering
      \includegraphics[width=6.5cm]{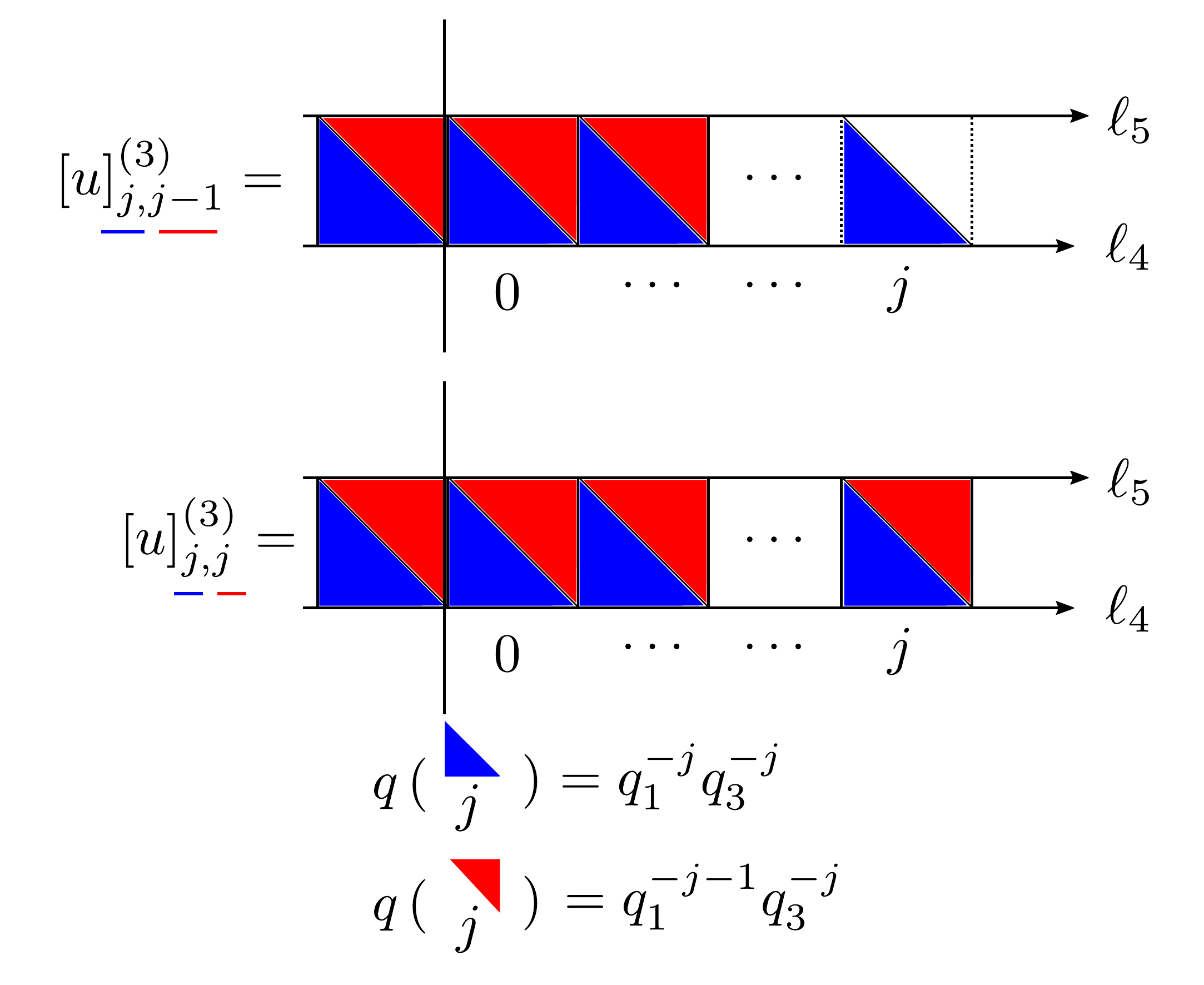}
       \subcaption{Basis and coordinates of one-dimensional crystal representation of $\ell_{4},\ell_{5}$. The atom in the origin is chosen to be blue.}\label{fig:gl_(2,1)1dimrep_l4,l5_part2}
      \end{minipage}
    \end{tabular}
\caption{Basis and coordinates of one-dimensional crystal representation of $\ell_{4},\ell_{5}$.}
\end{figure}

We have two one-dimensional crystal representations associated with the external legs $\ell_{4}$ and $\ell_{5}$. The crystal picture is in Figure \ref{fig:gl_(2,1)1dimcrystal_l4,5}. One of them has the purple atom in the origin, while the other has two choices (blue or red) to be in the atom. We denote the vector space of the former one as $V^{(\ell_{5};\ell_{4})}(u)$ and the latter one  as $V^{(\ell_{4};\ell_{5})}(u)$. The basis and coordinates of the former vector space $V^{(\ell_{5};\ell_{4})}(u)$ are in Figure \ref{fig:gl_(2,1)1dimrep_l4,l5_part1}. For the latter one see Figure \ref{fig:gl_(2,1)1dimrep_l4,l5_part2}.

The action of the generators on $V^{(\ell_{5};\ell_{4})}(u)$ can be written as,
\begin{align}
\begin{split}
    &K_{i}^{\pm}(z)[u]^{(1)}_{j}=[\Psi_{[u]_{j}^{(1)}}^{(i)}]_{\pm}[u]_{j}^{(1)},\\
    &E_{1}(z)[u]^{(1)}_{j}=\mathcal{E}_{1}([u]^{(1)}_{j})\delta\left(\frac{z}{uq_{2}^{j+1}}\right)[u]^{(1)}_{j+1},\\
    &F_{1}(z)[u]^{(1)}_{j+1}=\mathcal{F}_{1}([u]^{(1)}_{j+1})\delta\left(\frac{z}{uq_{2}^{j+1}}\right)[u]^{(i)}_{j},\\
    &E_{i}(z), F_{i}(z) [u]^{(1)}_{j}=0,\quad (i=2,3),
    \end{split}
\end{align}
where
\begin{align}
\begin{split}
    &\Psi^{(1)}_{[u]_{j}^{(1)}}(z)=\frac{1}{\phi(q_{2}^{-1-j};z,u)\phi(q_{2}^{-j};z,u)},\\
    &\Psi^{(2)}_{[u]_{j}^{(1)}}(z)=\phi(q_{3}q_{2}^{-j};z,u),\quad\Psi^{(3)}_{[u]_{j}^{(1)}}(z)=\phi(q_{1}q_{2}^{-j};z,u).
    \end{split}\label{eq:gl_(2,1)_l4,5_chargefunction1}
\end{align}
The parity conditions are
\begin{align}
    |[u]^{(1)}_{j+1}|=|[u]^{(1)}_{j}|.
\end{align}

Since the charge function has different number of poles and zeros, this is a representation of the shifted quantum toroidal algebra $\mathfrak{gl}_{2|1}$ with shift parameters
\begin{align}
    r_{1}=-2,\quad r_{2}=1,\quad r_{3}=1.
\end{align}
For the action of the $K_{s}^{\pm}(z)$ on $V^{(\ell_{4};\ell_{5})}(u)$, we obtain 
\begin{align}
    K_{i}^{\pm}(z)\begin{cases}
    [u]_{j,j-1}^{(3)}\\
    [u]_{j,j}^{(3)}
    \end{cases}&=\begin{cases}
    \left[\Psi^{(i)}_{[u]_{j,j-1}^{(3)}}(z)\right]_{\pm}[u]_{j,j-1}^{(3)},\\
    \left[\Psi^{(i)}_{[u]_{j,j}^{(3)}}(z)[u]_{j,j}^{(3)}\right]_{\pm}[u]_{j,j}^{(3)}.\end{cases}
    \end{align}
    The nonvanishing action of $E_{i}(z)$ and $F_{i}(z)$ can be written as 
\begin{align}
    \begin{split}
    E_{2}(z)[u]_{j,j-1}^{(3)}&=\mathcal{E}_{2}([u]_{j,j-1}^{(3)})\delta\left(\frac{z}{uq_{1}^{-j-1}q_{3}^{-j}}\right)[u]_{j,j}^{(3)}
    ,\\
    F_{2}(z)[u]_{j,j}^{(3)}&=\mathcal{F}_{2}([u]_{j,j}^{(3)})\delta\left(\frac{z}{uq_{1}^{-j-1}q_{3}^{-j}}\right)[u]_{j,j-1}^{(3)},\\
    E_{3}(z)[u]_{j,j}^{(3)}&=\mathcal{E}_{3}([u]_{j,j}^{(3)})\delta\left(\frac{z}{uq_{2}^{j+1}}\right)[u]_{j+1,j}^{(3)},\\
    F_{3}(z)[u]_{j,j-1}^{(3)}&=\mathcal{F}_{3}([u]_{j,j-1}^{(3)})\delta\left(\frac{z}{uq_{2}^{j}}\right)[u]_{j-1,j-1}^{(3)}.
    \end{split}
\end{align}
\begin{align}
 |[u]^{(3)}_{j,j}|=|[u]^{(3)}_{j,j-1}|+1,\quad |[u]^{(3)}_{j,j-1}|=|[u]^{(3)}_{j-1,j-1}|+1   
\end{align}

The charge functions are
\begin{align}
\begin{split}
     &\Psi^{(1)}_{[u]_{j,j-1}^{(3)}}(z)=\phi(q_{1}^{j+1}q_{3}^{j-1};z,u)\phi(q_{1}^{j}q_{3}^{j+1};z,u),\quad\Psi^{(1)}_{[u]_{j,j}^{(3)}}(z)=\phi(q_{1}^{j+2}q_{3}^{j};z,u)\phi(q_{1}^{j}q_{3}^{j+1};z,u),\\
      & \Psi^{(2)}_{[u]_{j,j-1}^{(3)}}(z)=\frac{1}{\phi(q_{1}^{j+1}q_{3}^{j};z,u)},\quad \Psi^{(2)}_{[u]_{j,j}^{(3)}}(z)=\frac{1}{\phi(q_{1}^{j+1}q_{3}^{j};z,u)},\\
      & \Psi^{(3)}_{[u]_{j,j-1}^{(3)}}(z)=\frac{1}{\phi(q_{2}^{-j};z,u)},\quad\Psi^{(3)}_{[u]_{j,j}^{(3)}}(z)=\frac{1}{\phi(q_{2}^{-j};z,u)}.
      \end{split}\label{gl_(2,1)_l4,5_chargefunction2}
\end{align}
and this is a representation of the shifted quantum toroidal algebra with shift parameters
\begin{align}
    r_{1}=2,\quad r_{2}=r_{3}=-1.
\end{align}

\subsection{Two-dimensional crystal of \texorpdfstring{$p_{2}=(0,0)$}{p200}}
The subquiver and crystal shape is in Figure \ref{fig:gl_(2,1)2dcrystal_p2}. 
The basis of this representation is defined by the tensor product of one-dimensional representations as
\begin{align}
     \otimes_{i=1}^{N}V^{(\ell_{3})}((q_{1}^{-1}q_{3}^{-1})^{i-1}u)\ni \otimes_{i=1}^{N}[u(q_{1}^{-1}q_{3}^{-1})^{i-1}]_{\sigma_{i}-1}\equiv\ket{\sigma},\quad \sigma=(\sigma_{1},...\sigma_{N})\in\mathbb{Z}^{N},
\end{align}
where we used the conventions in (\ref{eq:gl_(2,1)1dimcrystal_newconvention_l3}) and Figure \ref{fig:gl_(2,1)1dcrystal_l3_usefulconvention}.
The $q_1^{-1}q_3^{-1}$ shift of parameters $u(q_1^{-1}q_3^{-1})^{i-1}$ ensures the melting rule proposed in \cite{Nishinaka_2011,Nishinaka_2012,Nishinaka_2014}, and this basis forms a submodule (see Figure \ref{fig:gl_(2,1)2dcrystal_p2_meltingrule}).
 \begin{figure}
     \centering
     \includegraphics[width=12.5cm]{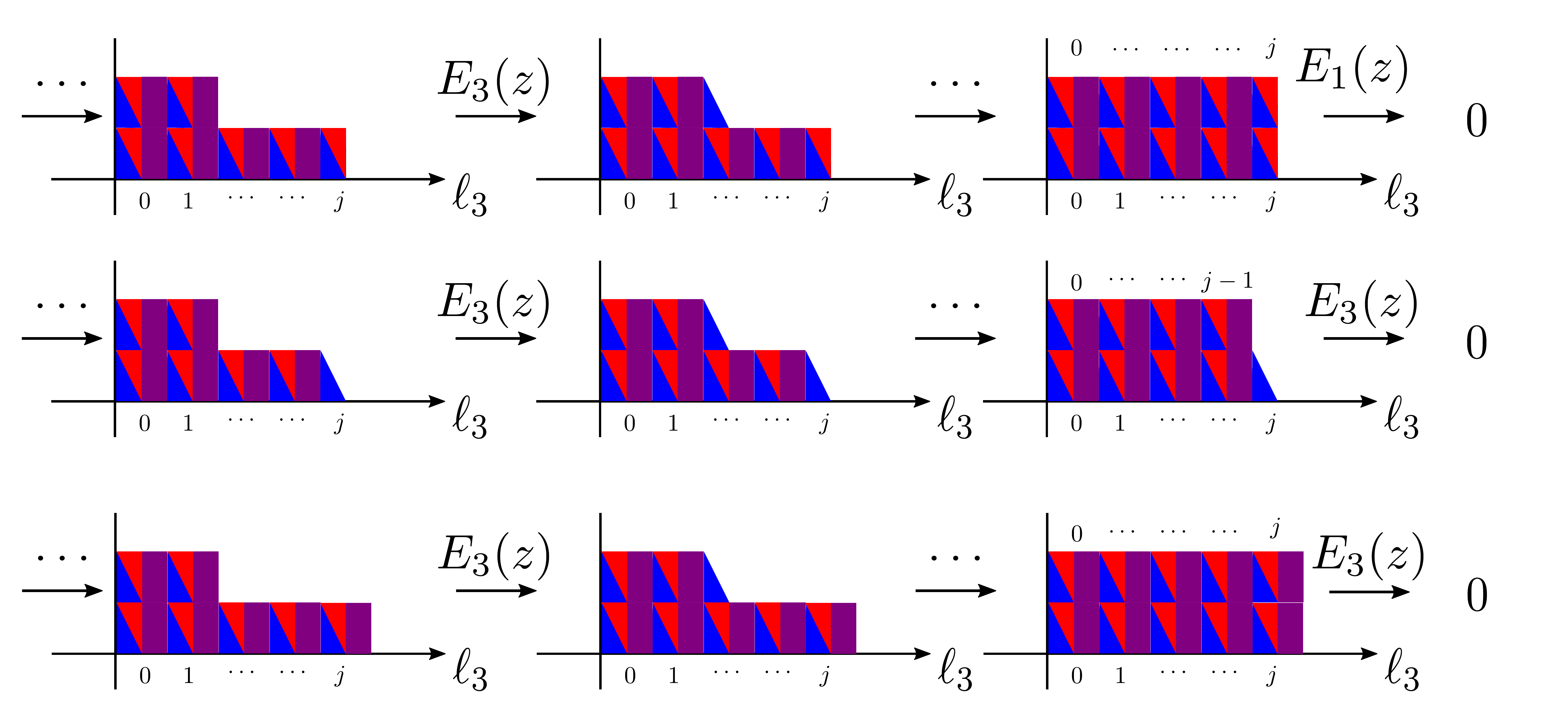}
     \caption{Action of generators on second tensor component of $V^{(\ell_{3})}(u)\otimes V^{(\ell_{3})}(q_{1}^{-1}q_{3}^{-1}u)$. Top: When the first tensor component is $[u]_{j,j,j-1}$, after acting $E_{s}(z)$ several times, the action on the second tensor component will be zero due to $K_{1}(z)\otimes E_{1}(z) [u]_{j,j,j-1}\otimes [q_{1}^{-1}q_{3}^{-1}u]_{j,j,j-1}=0$. Middle: When the first tensor component is $[u]_{j,j-1,j-1}$, after acting $E_{s}(z)$ several times, the action on the second tensor component will be zero due to $K_{3}(z)\otimes E_{3}(z) [u]_{j,j-1,j-1}\otimes [q_{1}^{-1}q_{3}^{-1}u]_{j-1,j-1,j-1}=0$. Bottom: When the first tensor component is $[u]_{j,j,j}$, after acting $E_{s}(z)$ several times, the action on the second tensor component will be zero due to $K_{3}(z)\otimes E_{3}(z) [u]_{j,j,j}\otimes [q_{1}^{-1}q_{3}^{-1}u]_{j,j,j}=0$.  }
     \label{fig:gl_(2,1)2dcrystal_p2_meltingrule}
 \end{figure}
The melting rule can be understood in a simple way if we introduce the following conventions:
\begin{align}
\begin{split}
    &(\lambda,\tau)=((\lambda_{1},\tau_{1}),(\lambda_{2},\tau_{2}),.....(\lambda_{N},\tau_{N}))\in\mathbb{Z}^{N}\times\mathbb{Z}_{3}^{N},\quad \mathbb{Z}_{3}=\{0,1,2\}\\
    &\ket{\sigma}=\otimes_{i=1}^{N}[(q_{1}^{-1}q_{3}^{-1})^{i-1}u]_{\sigma_{i}-1}=\otimes_{i=1}^{N}[(q_{1}^{-1}q_{3}^{-1})^{i-1}u]_{(\lambda_{i}-1,\tau_{i})}=\ket{\lambda,\tau}\\
    &r(\sigma_{i}-1)=\lambda_{i}-1,\quad s(\sigma_{i}-1)=\tau_{i}
    \end{split}
\end{align}
The melting rule is for $i<j$
\begin{align}
\begin{split}
    (\lambda_{i},0)>(\lambda_{j}>0),\quad (\lambda_{i},0)>(\lambda_{j},1),\quad (\lambda_{i},0)>(\lambda_{j},2)\\
    (\lambda_{i},1)\geq(\lambda_{j},0),\quad (\lambda_{i},1)\geq(\lambda_{j},1),\quad (\lambda_{i},1)>(\lambda_{j},2),\\
    (\lambda_{i},2)\geq (\lambda_{j},0),\quad (\lambda_{i},2)\geq (\lambda_{j},1),\quad (\lambda_{i},2)\geq(\lambda_{j},2).
    \end{split}
\end{align}
We can take the limit $N\rightarrow \infty$ by embedding $\sigma\in\mathbb{Z}^{N}$ to $\mathbb{Z}^{N+1}$ by setting $\sigma_{N+1}=0$ or equivalently by setting $(\lambda_{N+1},\tau_{N+1})=(0,2)$. The result is 
\begin{align}
\begin{split}
\begin{split}
    E_{s}(z)\ket{\sigma}&=\sum_{i=1}^{\ell(\sigma)+1}(-1)^{|s|(\sum_{l=1}^{i-1}|\sigma_{i}-1|)}\mathcal{E}_{s}([u(q_{1}^{-1}q_{3}^{-1})^{i-1}]_{\sigma_{i}-1})\prod_{j=1}^{i-1}\left[\Psi^{(s)}_{[u(q_{1}^{-1}q_{3}^{-1})^{i-1}]}(z)\right]_{-}\\
    &\times \delta\left(\frac{z}{u(q_{1}^{-1}q_{3}^{-1})^{i-1}(q_{3}^{2}q_{1}^{-1})^{r(\sigma)}q_{1}^{-1}q_{3}^{s(\sigma)}}\right)\bar{\delta}_{s+s(\sigma),2}\ket{\sigma+\fbox{$s$}_{i}}
    \end{split},\\
    K_{s}(z)\ket{\sigma}&=\frac{\phi(q_{1}^{\ell(\sigma)-1}q_{3}^{\ell(\sigma)};z,u)^{\delta_{s,1}}}{\phi(q_{1}^{\ell(\sigma)}q_{3}^{\ell(\sigma)};z,u)^{\delta_{s,3}}}\prod_{i=1}^{\ell(\sigma)}\Psi^{(s)}_{[(q_{1}^{-1}q_{3}^{-1})^{i-1}u]_{\sigma_{i}-1}}(z)\ket{\sigma},\\
    \begin{split}
        F_{s}(z)\ket{\sigma}&=\frac{\phi(q_{1}^{\ell(\sigma)-1}q_{3}^{\ell(\sigma)};z,u)^{\delta_{s,1}}}{\phi(q_{1}^{\ell(\sigma)}q_{3}^{\ell(\sigma)};z,u)^{\delta_{s,3}}}\sum_{i=1}^{\ell(\sigma)}\mathcal{F}_{s}\left([u(q_{1}^{-1}q_{3}^{-1})^{i-1}]_{\sigma_{i}-1}\right)\\
        &\times(-1)^{|s|(\sum_{l=1}^{i-1}|\sigma_{i}-1|)}\prod_{j=i+1}^{\ell(\sigma)}\left[\Psi^{(s)}_{[u(q_{1}^{-1}q_{3}^{-1})^{j-1}]_{\sigma_{j}-1}}(z)\right]_{+}\\
        &\times\bar{\delta}_{s+s(\sigma-1),0}\delta\left(\frac{z}{u(q_{1}^{-1}q_{3}^{-1})^{i-1}(q_{3}^{2}q_{1}^{-1})^{r(\sigma-1)}q_{1}^{-1}q_{3}^{s(\sigma-1)}}\right)\ket{\sigma-\fbox{s}_{i}} ,
    \end{split}
    \end{split}
\end{align}
where we used 
\begin{align}
    \prod_{i=\ell(\sigma)+1}^{\infty}\Psi^{(s)}_{[u(q_{1}^{-1}q_{3}^{-1})^{i-1}]_{-1}}(z)=\frac{(q_{1}^{\ell(\sigma)-1}q_{3}^{\ell(\sigma)};z,u)^{\delta_{s,1}}}{(q_{1}^{\ell(\sigma)}q_{3}^{\ell(\sigma)};z,u)^{\delta_{s,3}}}
\end{align}
and 
$\bar{\delta}_{i,j}=\begin{cases}
1,\quad i\equiv j\;(\text{mod}\; 3)\\
0,\quad i\not\equiv j\;(\text{mod}\; 3)
\end{cases}$.
Especially, the action of $K_{s}(z)$ on the vacuum is
\begin{align}
    K_{s}(z)\ket{\emptyset}=\frac{\phi(q_{1}^{-1};z,u)^{\delta_{s,1}}}{\phi(1;z,u)^{\delta_{s,3}}}\ket{\emptyset}.
\end{align}
as expected in (\ref{eq:vacuum_charge_function}).
Since the vacuum charge function has different number of zeros and poles, this is a representation of the shifted quantum toroidal algebra with shift parameters
\begin{align}
    r_{1}=1,\quad r_{2}=0,\quad r_{3}=-1.
\end{align}

\section{Quantum toroidal \texorpdfstring{$D(2,1;\alpha)$}{D21a}}\label{sec:appendix_D(2,1)}
Let us consider the action of $K_{s}(z)$ and $F_{s}(z)$ on the two-dimensional crystals of $\mathbb{C}^{3}/(\mathbb{Z}_{2}\times\mathbb{Z}_{2})$. See section \ref{sec:D(2,1)2dcrystal} for the notation. The review part in \cite{Awata2019} is a good reference for the regularization procedure we perform in this section.

 $\lambda\in\mathbb{Z}^{2N}$ can be naturally embedded into $\lambda\in\mathbb{Z}^{2N+2}$ by setting $\lambda_{2N+1}=\lambda_{2N+2}=0$. We want the action of $K_{s}(z)$ to be the same for $\forall 2N\geq \ell(\lambda)$. To do this, we have to modify the action of $K_{s}(z)$ by multiplying a factor $\beta_{s}^{(N)}(z)$:
\begin{align}
    \beta_{s}^{(N)}(z)\Delta^{(2N-1)}(K_{s}(z))\ket{\lambda}=\beta_{s}^{(N+1)}(z)\Delta^{(2N+1)}(K_{s}(z))\ket{\lambda}.
\end{align}
Note that the $\ket{\lambda}$ in the right hand side is understood as an embedding into $\mathbb{Z}^{2N+2}$. This gives the recursion formula 
\begin{align}
    \frac{\beta_{s}^{(N+1)}(z)}{\beta_{s}^{(N)}(z)}=\frac{1}{\Psi^{(s)}_{[uq_{2}^{2N}]_{-1}^{(3;2)}}(z)\Psi^{(s)}_{[uq_{2}^{2N+1}]^{(1;0)}_{-1}}(z)  }
\end{align}
and we obtain
\begin{align}
\beta_{0}^{(N)}(z)&=q_{2}^{\frac{N}{2}}\frac{\phi(q_{1}q_{2}^{-2N+1};z,u)}{\phi(q_{1}q_{2};z,u)}\beta_{0}^{(0)}(z),\quad \beta_{1}^{(N)}(z)=q_{2}^{\frac{N}{2}}\beta_{1}^{(0)}(z),\\
\beta_{2}^{(N)}(z)&=q_{2}^{-\frac{N}{2}}\beta_{2}^{(0)}(z),\quad \beta_{3}^{(N)}(z)=q_{2}^{-\frac{N}{2}}\frac{\phi(1;z,u)}{\phi(q_{2}^{-2N};z,u)}\beta_{3}^{(0)}(z).
\end{align}
The initial conditions $\beta_{s}^{(0)}(z)$ are related to the vacumm structure. We define the vacuum as 
\begin{align}
\ket{\emptyset}\equiv \otimes_{i=1}^{N}\left([uq_{2}^{2i-2}]^{(3;2)}_{-1}\otimes [uq_{2}^{2i-1}]^{(1;0)}_{-1}\right).
\end{align}
The action of $K_{s}(z)$ is 
\begin{align}
    \Delta^{(2N-1)}(K_{s}(z))\ket{\emptyset}=\begin{dcases}
    q_{2}^{-\frac{N}{2}}\frac{\phi(q_{1}q_{2};z,u)}{\phi(q_{1}q_{2}^{1-2N};z,u)}\ket{\emptyset},\quad s=0\\
    q_{2}^{-\frac{N}{2}}\ket{\emptyset},\quad s=1\\
    q_{2}^{\frac{N}{2}}\ket{\emptyset},\quad s=2\\
    q_{2}^{\frac{N}{2}}\frac{\phi(q_{2}^{-2N};z,u)}{\phi(1;z,u)}\ket{\emptyset}. \quad s=3
    \end{dcases}
\end{align}
We can choose the initial conditions to be 
\begin{align}
\begin{split}
    \beta_{0}^{(0)}(z)=\phi(q_{1}q_{2};z,u),\quad \beta_{1}^{(0)}(z)=1,\\
    \beta_{2}^{(0)}(z)=1,\quad \beta_{3}^{(0)}(z)=\frac{1}{\phi(1;z,u)}
    \end{split}
\end{align}
and then obtain 
\begin{align}
\begin{split}
\beta_{0}^{(N)}(z)&=q_{2}^{\frac{N}{2}}\phi(q_{1}q_{2}^{-2N+1};z,u),\quad \beta_{1}^{(N)}(z)=q_{2}^{\frac{N}{2}},\\
\beta_{2}^{(N)}(z)&=q_{2}^{-\frac{N}{2}},\quad \beta_{3}^{(N)}(z)=q_{2}^{-\frac{N}{2}}\frac{1}{\phi(q_{2}^{-2N};z,u)}.
\end{split}
\end{align}
The limit $N\rightarrow \infty$ can be defined as $K_{s}(z)\ket{\lambda}=\lim_{N\rightarrow \infty}\Delta^{(2N-1)}(K_{s}(z))\beta_{s}^{(N)}(z)\ket{\lambda}$ and we obtain
\begin{align}
    K_{s}(z)\ket{\lambda}=\beta_{s}^{\left(\lfloor\frac{\ell(\lambda)+1}{2} \rfloor\right)}(z)\prod_{i=1}^{\lfloor\frac{\ell(\lambda)+1}{2} \rfloor}\left(\Psi^{(s)}_{[uq_{2}^{2i-2}]^{(3;2)}_{\lambda_{2i-1}-1}}(z)\Psi^{(s)}_{[uq_{2}^{2i-1}]^{(1;0)}_{\lambda_{2i}-1}}(z) \right)\ket{\lambda}
\end{align}

The action of $F_{s}(z)$ can be obtained by defining it as
\begin{equation}
    F_{s}(z)\ket{\lambda}=\lim_{N\rightarrow \infty}\Delta^{(2N-1)}(F_{s}(z))\beta_{s}^{(N)}(z)\ket{\lambda}.
\end{equation}
The result is 
\begin{align}
\begin{split}
    &F_{s}(z)\ket{\lambda}\\
    =&\sum_{i=1}^{\lfloor\frac{\ell(\lambda)+1}{2} \rfloor}(-1)^{|s|(\sum_{l=1}^{i-1}(|\lambda_{2l-1}|+|\lambda_{2l}-1|))}\beta_{s}^{(\lfloor\frac{\ell(\lambda)+1}{2}\rfloor)}(z)\prod_{j=i+1}^{\lfloor\frac{\ell(\lambda)+1}{2}\rfloor}\Psi^{(s)}_{[uq_{2}^{2j-2}]^{(3;2)}_{\lambda_{2j-1}-1}}(z)\Psi^{(s)}_{[uq_{2}^{2j-1}]^{(1;0)}_{\lambda_{2j}-1}}(z)\\
    &\times\left\{\mathcal{F}_{s}([uq_{2}^{2i-2}]^{(3;2)}_{\lambda_{2i-1}-1})\Psi^{(s)}_{[uq_{2}^{2i-1}]^{(1;0)}_{\lambda_{2i}-1}}(z)\delta\left(\frac{z}{uq_{2}^{2i-2}q_{1}^{\lambda_{2i-1}-1}}\right)\right.\\
    &\hspace{1cm}\times\left(\delta_{s,3}\bar{\delta}_{\lambda_{2i-1},1}+\delta_{s,2}\bar{\delta}_{\lambda_{2i-1},0}\right)\ket{\lambda-\fbox{$s$}_{2i-1}}\\
    &\left.\quad+(-1)^{|s||\lambda_{2i-1}|} \mathcal{F}_{s}([uq_{2}^{2i-1}]^{(1;0)}_{\lambda_{2i}-1})\delta\left(\frac{z}{uq_{2}^{2i-1}q_{1}^{\lambda_{2i}-1}}\right)\left(\delta_{s,1}\bar{\delta}_{\lambda_{2i},1}+\delta_{s,0}\bar{\delta}_{\lambda_{2i},0}\right)\ket{\lambda-\fbox{$s$}_{2i}}        \right\}.
\end{split}
\end{align}

\bibliographystyle{ytphys}
\bibliography{QQTA}
\end{document}